\documentclass[12pt]{article}

\usepackage{amsthm}
\usepackage{mathtools}
\usepackage{amsmath}
\usepackage{amssymb}
\DeclareMathAlphabet{\mathbbold}{U}{bbold}{m}{n}
\newcommand*{\boldone}{\mathbbold{1}}
\allowdisplaybreaks

\usepackage[T1]{fontenc}
\usepackage[utf8]{inputenc}
\usepackage{newtxtext, newtxmath}
\usepackage[protrusion=true, expansion=true]{microtype}
\usepackage{setspace}

\usepackage[english]{babel}
\usepackage{csquotes}

\usepackage[a4paper, margin=1in]{geometry}

\usepackage{booktabs}
\usepackage[hang, small, labelfont=bf, textfont=it, up]{caption}
\usepackage{enumerate}
\usepackage{graphicx}
\usepackage{threeparttable}

\newtheorem{assumption}{Assumption}
\newtheorem{corollary}{Corollary}
\newtheorem{example}{Example}
\newtheorem{remark}{Remark}
\newtheorem{proposition}{Proposition}
\newtheorem{theorem}{Theorem}

\usepackage{abstract}

\usepackage[dvipsnames]{xcolor}

\usepackage[colorlinks=true, linkcolor=blue, urlcolor=blue, citecolor=blue, hypertexnames=false]{hyperref}

\usepackage[authordate, backend=biber]{biblatex-chicago}
\usepackage{titling}

\usepackage{tikz}
\usetikzlibrary{calc}

\title{
{\Large\bfseries
(Debiased) Inference for Fixed Effects Estimators with Three-Dimensional Panel and Network Data
}\thanks{This paper supersedes  the paper entitled ``Debiased Fixed Effects Estimation of Binary Logit Models with Three-Dimensional Panel Data'' by \textcite{s2023}. We are grateful to Christoph Breunig, Joachim Freyberger, Ayden Higgins, Jannis Kück, Claudia Noack, Cavit Pakel, Martin Schumann, Joschka Wanner, and Martin Weidner, as well as seminar and conference participants for valuable suggestions, feedback, and discussions throughout the development of this paper. The authors used generative AI tools to assist with language editing. All output was reviewed carefully, and the authors bear full responsibility for the content and any remaining errors. This work was supported by the Deutsche Forschungsgemeinschaft (DFG) under Grant 462020252.}
}
\author{
\normalsize
Daniel Czarnowske
\thanks{
Heinrich-Heine-Universität Düsseldorf, Universitätsstr. 1, 40225 Düsseldorf, Germany; e-mail: \texttt{\href{mailto:daniel.czarnowske@hhu.de}{daniel.czarnowske@hhu.de}}
}
\and
\normalsize
Amrei Stammann
\thanks{
Universität Bayreuth, Universitätsstr. 30, 95447  Bayreuth, Germany; e-mail: \texttt{\href{mailto:amrei.stammann@uni-bayreuth.de}{amrei.stammann@uni-bayreuth.de}}
}
}
\date{\small\today}

\DeclareMathOperator{\argmin}{\arg\,\min\;}
\DeclareMathOperator{\argmax}{\arg\,\max\;}
\DeclareMathOperator{\diag}{\text{diag}}
\DeclareMathOperator{\bdiag}{\text{bdiag}}
\DeclareMathOperator{\iid}{\text{iid.}\;}
\DeclareMathOperator{\ind}{\boldone}
\DeclareMathOperator{\N}{\mathcal{N}}

\DeclareMathOperator{\Real}{\mathbb{R}}

\newcommand{\abs}[1]{\lvert #1 \rvert}
\newcommand{\bigabs}[1]{\big\lvert #1 \big\rvert}
\newcommand{\biggabs}[1]{\bigg\lvert #1 \bigg\rvert}
\newcommand{\Bigabs}[1]{\Big\lvert #1 \Big\rvert}

\newcommand{\norm}[1]{\lVert #1 \rVert}
\newcommand{\bignorm}[1]{\big\lVert #1 \big\rVert}

\newcommand{\EX}[1]{\mathbb{E}\left[ #1 \right]}

\newcommand{\LEX}[1]{\overline{\mathbb{E}}\left[ #1 \right]}
\newcommand{\Prob}[1]{\mathbb{P}\left(#1\right)}
\newcommand{\dop}[1]{\mathrm{d}^{#1}}
\theoremstyle{definition}

\newtheorem{lemma}{Lemma}

\begin{document}

\maketitle

\thispagestyle{empty}
\renewcommand{\abstractname}{\vspace{-5em}}
\begin{abstract}
    \noindent Inference for fixed effects estimators is often unreliable due to Nickell- and incidental parameter biases. While these issues are well understood for classical two-dimensional panels, little is known about three-dimensional panel structures (e.g., sender $\times$ receiver $\times$ time). We develop inferential theory for a broad class of linear and nonlinear fixed effects M-estimators in this setting, covering bipartite, directed, and undirected network panel data, multiple specifications of additively separable unobserved effects, and both strictly exogenous and predetermined regressors. Our analysis reveals fundamentally different asymptotic properties compared to two-dimensional panels. In particular, we find a sharp dichotomy across specifications: (i) when unobserved effects vary along a single panel dimension, estimators are asymptotically unbiased, (ii) when they vary along two panel dimensions, estimators may suffer from a severe inference problem characterized by a degenerate asymptotic distribution. We resolve the latter by deriving explicit bias formulas and proposing analytically debiased estimators with nondegenerate, correctly centered asymptotic distributions. An empirical application studies dynamic network formation in a directed panel of bilateral trade relationships.\\[1em]
\end{abstract}

\clearpage
\onehalfspacing

\setcounter{page}{1}

\section{Introduction}

The inconsistency of fixed effects estimators for linear and nonlinear panel models, first noted by \textcite{ns1948} and \textcite{n1981}, remains a highly studied topic in panel data econometrics. Although empirical researchers increasingly use more granular, \textit{multi-dimensional} panel data with layers of variation beyond the classical cross-sectional and time dimensions, available solutions to these inference problems typically focus on \textit{two-dimensional} panels.

This paper contributes to closing this gap by developing an inferential theory for fixed effects M-estimators in a generic class of (non)linear models with \textit{three-dimensional} panels, i.e., panels with two cross-sectional dimensions and one time dimension, covering \textit{bipartite}, \textit{directed}, and \textit{undirected network panels}. An example is data on bilateral network activities observed over time, widely used in international trade research to study flows between each $i = 1, \dots, N_{1}$ exporting country and each $j = 1, \dots, N_{1}$ importing country for each $t = 1, \dots, T$ year. This multidimensionality allows researchers to control for multiple, richer sources of unobserved heterogeneity varying along one or two panel dimensions. We call heterogeneity arising from the interaction of two panel indices \textit{interacted} and the one-dimensional counterpart \textit{non-interacted}. For example, empirical analyses in international trade typically control for interacted unobserved heterogeneity of the form exporter-by-year, importer-by-year, and importer-by-exporter, using the three-way fixed effects specification $\alpha_{it} + \gamma_{jt} + \rho_{ij}$.\footnote{Controlling for this type of unobserved heterogeneity is, among others, recommended in \textcite{hm2014}.}

We make the following main contributions. First, we develop the first generic asymptotic framework for inference on fixed effects M-estimators in three-dimensional panel models in which all dimensions grow jointly. This framework covers different data structures (bipartite, directed, and undirected network panel data), several types of fixed effects, and accommodates both strictly exogenous and predetermined regressors. In our main analysis, we derive the asymptotic properties of fixed effects estimators for models with three sets of unobserved heterogeneity in interacted and non-interacted form.\footnote{In a bipartite panel, the fixed effects enter the linear index as $\alpha_{it} + \gamma_{jt} + \rho_{ij}$ in the interacted case and as $\alpha_{i} + \gamma_{j} + \rho_{t}$ in the non-interacted case.} These two specifications are the most general cases and nest a wide class of one- and two-way fixed effects models, allowing us to characterize the asymptotic properties of the corresponding estimators across all such settings. For cases in which inference would otherwise be invalid, we develop appropriate debiased estimators.\footnote{The estimators are computationally tractable and are implemented in our R-package \href{https://cran.r-project.org/web/packages/alpaca/index.html}{alpaca}.} Second, we develop new analytical tools that address the specific technical challenges of the multi-dimensional panel setting but which also apply to two-dimensional settings. Treating unobserved effects as incidental parameters creates a non-standard inference problem because the dimension of the parameter vector grows with all three panel dimensions. Moreover, in multi-way interacted specifications, shared indices of two unobserved effects induce a high-dimensional incidental parameter Hessian with a nontrivial sparsity pattern. These features render standard approaches from two-dimensional panels inadequate and motivate new analytical tools, which may also prove useful for future work on multi-dimensional panels. In particular, the uniform approximation argument for the inverse of the expected incidental parameter Hessian developed by \textcite{fw2016} for models with individual and time effects does not apply under this sparsity pattern. We hence develop a generic strategy for approximating the inverse of the high-dimensional expected incidental parameter Hessian in sparse settings.\footnote{Our strategy nests that of \textcite{fw2016} as a special case.} The approximation argument is crucial to show that the asymptotic bias \textit{decouples} into one component per incidental parameter. Moreover, \textcite{fw2016} used a third-order expansion. However, the faster rate of convergence in three-dimensional panels requires expanding the profile objective function to fourth order rather than third. In addition, controlling the terms in the expansion requires a new bound on the $p$-th moment of the spectral norm of a random matrix whose rows are independent, mean-zero, weakly dependent processes. These challenges may explain the substantial gap between the literature on fixed effects estimators for two-dimensional versus multi-dimensional panels. Third, our findings clarify that standard results for two-dimensional panel data models do not simply carry over to three-dimensional settings.

Our first main finding concerns models with three-way interacted unobserved heterogeneity. Except for some special cases, we identify an unusually severe inference problem caused by a stark imbalance between the order of the leading bias and the order of the standard deviation. Since the former exceeds the latter, the asymptotic distribution \textit{degenerates} and the inference problem grows worse asymptotically.\footnote{For bipartite panel data, the order of the leading bias is $N_{2}^{-1} + N_{1}^{-1} + T^{-1}$ and the order of the standard deviation is $(N_{1} N_{2} T)^{-1/2}$, where $N_{1}$ and $N_{2}$ denote the lengths of the two cross-sections and $T$ the number of time periods.}  OLS and (pseudo-)Poisson maximum likelihood estimators with strictly exogenous regressors are the exception. For those, we show that the asymptotic distribution remains nondegenerate and correctly centered, so that no bias correction is needed. For nonlinear models such as probit and logit, however, the degeneracy result appears even under strict exogeneity, and it reemerges for OLS and (pseudo-)Poisson maximum likelihood once predetermined regressors enter the model. This general degeneracy result contrasts with the typical result in the two-dimensional literature, where the orders of bias and standard deviation can be balanced by choosing appropriate rates, yielding a \textit{nondegenerate} asymptotic distribution with a distorted first moment. Although the inference problem in our degenerate case is more severe, the bias can be estimated at a sufficient rate, and we develop an analytically debiased estimator with a nondegenerate asymptotic distribution centered at zero.

Our second set of results concerns models with three-way non-interacted unobserved heterogeneity. This is the opposite extreme: the fixed effects estimator is asymptotically free of any incidental parameter bias, even though the dimension of the fixed effects grows with the sample size. This is because the leading bias of the fixed effects estimator always shrinks faster than the standard deviation.\footnote{For bipartite panel data, the order of the bias is $(N_{1} T)^{-1} + (N_{2} T)^{-1} + (N_{1} N_{2})^{-1}$ and the order of the standard deviation is $(N_{1} N_{2} T)^{-1/2}$.}

Our simulation experiments confirm the severity of the inference problem for the interacted specification: confidence intervals around the uncorrected estimator almost never cover the true parameter. Our proposed bias correction substantially improves inferential accuracy. For the non-interacted specification, we confirm empirically that the fixed effects estimator is asymptotically unbiased.

In our empirical application, we apply our debiased estimator to study the dynamic formation of export and import relationships between countries, using a probit model with three-way interacted unobserved heterogeneity. We adapt the dynamic network formation model of \textcite{g2016} to a directed network setting. We find strong evidence for state dependence: any past trade relationship, regardless of direction, increases the probability of an export link today. We also find evidence for two triadic transitivity channels, indirect trade linkages through intermediary countries and shared upstream suppliers, each of which raises the probability of a direct bilateral trade link forming.

\paragraph{Related literature.} Our work contributes to the large-$T$ literature on bias corrections for the inconsistency of fixed effects estimators identified by \textcite{n1981} and \textcite{ns1948}.\footnote{Another important strand focuses on fixed-$T$ asymptotics, where estimators typically eliminate unobserved effects by differencing or conditioning on sufficient statistics. Examples for binary logit models include \textcite{r1960}, \textcite{a1970}, \textcite{c1980}, \textcite{hk2000}, and \textcite{hw2025}.} The largest part of this literature has focused on classical panel data models with individual effects. For linear models, see among others \textcite{hk2002} and \textcite{dj2015}. For nonlinear models, see among others \textcite{l2002}, \textcite{w2002}, \textcite{s2003}, \textcite{hn2004}, \textcite{c2007}, \textcite{ab2009}, \textcite{bh2009}, \textcite{f2009}, \textcite{hk2011}, \textcite{dj2015}, \textcite{ks2016}, \textcite{p2019}, \textcite{sst2021}, \textcite{schumann2023}, and \textcite{hj2024}. These solutions differ in their assumptions, techniques, and types of corrections. We refer the reader to \textcite{ah2007} and \textcite{fw2018} for comprehensive reviews. For linear models, \textcite{hm2006} show that additional time effects do not introduce a further inference problem, so standard Nickell-bias corrections developed for individual effects carry over. In nonlinear models, time effects introduce an additional incidental parameter bias. \textcite{fw2016} and \textcite{yjfl2018} advance this literature by developing asymptotic theory and bias corrections for nonlinear models with two types of fixed effects. Whereas \textcite{yjfl2018} focus on logit models with sender and receiver fixed effects for directed network data, \textcite{fw2016} cover nonlinear models with individual and time effects, accommodating time dependence and predetermined regressors.\footnote{The analysis of \textcite{fw2016} also applies to directed networks where the second panel dimension is not time but another cross-section, such as countries or industries. Their theory therefore nests the structural parameter results of \textcite{yjfl2018}.} Using a likelihood correction approach, \textcite{jo2019} propose an alternative to the ex-post bias correction of \textcite{fw2016}. Building on \textcite{fw2016}, \textcite{d2019} extends the framework to directed network models with sender and receiver fixed effects, developing specific inference procedures and specification tests, and \textcite{h2026} introduces an improved jackknife bias correction for network models. For undirected network data, \textcite{g2017} develops a bias-corrected estimator for network formation with degree heterogeneity.\footnote{Another strand develops conditional logit estimators in the spirit of \textcite{r1960} and \textcite{c1980}. See for example \textcite{g2017} for undirected network data, or \textcite{c2017} and \textcite{j2018} for directed network data.}

Despite the growing use of multi-dimensional panels in empirical work, \textcite{wz2021} is the only other paper in the bias correction literature that studies fixed effects estimators for three-dimensional panel data. They focus on a special case: the fixed effects  (pseudo-)Poisson maximum
likelihood estimator  for the gravity model with three-way interacted heterogeneity and strictly exogenous regressors. Due to a property specific to this setting, their problem reduces essentially to a two-way fixed effects analysis, allowing them to rely substantially on \textcite{fw2016}.
\textcite{wz2021} consider asymptotics in which the two cross-sectional dimensions grow while $T$ remains fixed. They show that their fixed effects estimator has a nondegenerate asymptotic distribution that is not correctly centered around zero, and recenter it with a bias correction.
Under asymptotics in which all three panel dimensions grow jointly, we show that for the general class of M-estimators with three-way interacted heterogeneity, the asymptotic distribution degenerates. The Poisson case with strictly exogenous regressors is the exception: there, the estimator is correctly centered around zero and requires no bias correction. However, the degeneracy problem re-emerges in the Poisson case once regressors are predetermined. Our framework is thus considerably broader than that of \textcite{wz2021}, covering linear, Poisson, and binary choice models with both strictly exogenous and predetermined regressors and several fixed effects structures, at the cost of requiring all three panel dimensions to grow jointly.
\textcite{h2026_jackknife} develops a jackknife $t_{q}$-statistic for a broad class of fixed effects models, including multi-dimensional panels. Applying it in our setting requires knowledge of the asymptotic bias structure, which we provide in this paper.

A growing interest in methods for three-dimensional panel models has also emerged in other strands of the econometrics literature. For example, \textcite{g2016} develops a conditional fixed effects logit estimator for dynamic network formation in a directed network panel model with pair-specific unobserved heterogeneity. \textcite{mp2025} recently introduce a conditional fixed effects logit estimator for a triadic network model with three-way interacted unobserved heterogeneity. \textcite{yh2023} provide an estimator for a gravity model with three multiplicative unobserved effects, extending the GMM strategy of \textcite{j2017}. \textcite{f2022} and \textcite{jls2025} propose interactive fixed effects estimators for linear panel models. \textcite{kss2021} extend \textcite{p2006}'s correlated
effects approach to linear panel models with heterogeneous structural parameters and a three-way error component featuring a hierarchical multi-factor structure.

\paragraph{Outline.} Section \ref{sec:data_structures_models_estimators} introduces the data structures, model, and fixed effects estimators. Section \ref{sec:asymptotic_theory} presents the asymptotic theory. Section \ref{sec:simulation_experiments} reports simulation results. Section \ref{sec:empirical_application} presents the empirical application. Section \ref{sec:concluding_remarks} concludes.

\paragraph{Notation.} Throughout, $\Prob{\cdot}$ and $\EX{\cdot}$ denote probability and expectation, and a superscript ``0'' on a parameter denotes its true population value. We abbreviate almost surely and with probability approaching one by a.\,s.\ and wpa1, respectively. We write $o_{P}(1)$ for a sequence of random variables that converges in probability to zero, and $\mathcal{O}_{P}(1)$ for a sequence that is bounded in probability. For two positive sequences $a_{m}$ and $b_{m}$, $a_{m} \asymp b_{m}$ means that the two sequences are of the same order of magnitude. We write $\xrightarrow{d}$ and $\xrightarrow{p}$ for convergence in distribution and in probability. Unless stated otherwise, all stochastic statements are almost-sure statements conditional on $\Phi$, the sigma-algebra generated by the unobserved effects and the initial conditions.

\section{Data Structures, Models, and Estimators}
\label{sec:data_structures_models_estimators}

\subsection{Data Structures}
\label{sec:data_structures}
 
We observe three-dimensional panel data $\{(y_{ijt}, x_{ijt}) \colon (i, j) \in \mathcal{D}_{s}, \, t \in \{1, \ldots, T\}\}$, where $y_{ijt}$ is an outcome variable, $x_{ijt}$ is a $K$-dimensional vector of explanatory variables,  $\mathcal{D}_{s}$ is the set of observed pairs $(i, j)$ corresponding to one of the three panel structures indexed by $s \in \{1, 2, 3\}$, and $T$ is the number of time periods.\footnote{The time dimension can also be replaced by a third cross-sectional
dimension.} Let $\mathcal{I} \coloneqq \{1, \ldots, N_{1}\}$ and $\mathcal{J} \coloneqq \{1, \ldots, N_{2}\}$ denote the index sets for the two cross-sectional dimensions.
 
In \textit{bipartite} panel data ($s = 1$), the indices $i$ and $j$ refer to two distinct samples of cross-sectional units, and the set of observed pairs is $\mathcal{D}_{1} = \{(i, j) \colon i \in \mathcal{I}, j \in \mathcal{J}\}$. In network panel data ($s \in \{2, 3\}$), both indices refer to the same set of $N_{1}$ agents, and self-ties are excluded ($i \neq j$). Network panels can be directed or undirected. In \textit{directed} network panels ($s = 2$), the relationship between agents is asymmetric: $(y_{ijt}, x_{ijt})$ need not equal $(y_{jit}, x_{jit})$. The set of observed pairs is $\mathcal{D}_{2} = \{(i, j) \colon i \in \mathcal{I}, j \in \mathcal{I}, i \neq j\}$. In \textit{undirected} network panels ($s = 3$), the relationship is symmetric: $(y_{ijt}, x_{ijt}) = (y_{jit}, x_{jit})$. To avoid duplicates, the observed set is restricted to $\mathcal{D}_{3} = \{(i, j) \colon i \in \mathcal{I}, j \in \mathcal{I}, i < j\}$. The total number of observations is $n_{s} \coloneqq \lvert \mathcal{D}_{s} \rvert \, T$, giving sample sizes $n_{1} = N_{1} N_{2} T$, $n_{2} = N_{1} (N_{1} - 1) T$, and $n_{3} = N_{1} (N_{1} - 1) T / 2$.
 
\begin{example}[Bipartite Panel Data]
    Firm-level panel data often contain information beyond the standard firm and time identifiers that can be used to form a bipartite panel. Examples include data on products or industries, export destinations, and the locations of subsidiaries or investments.
\end{example}
 
\begin{example}[Network Panel Data]
    Data on countries observed over time can often be used to form a network panel. Country pairs may represent origin and destination in bilateral trade or migration data, initiator and signatory in data on bilateral agreements, or aggressor and target in data on (military) conflicts.
\end{example}

\begin{remark}[Unbalancedness]
    The exclusion of self-ties is one source of unbalancedness in network panels. All three data structures can be further unbalanced along the cross-sectional and time dimensions. Given that the source of attrition is conditionally random, this does not cause any theoretical problems (see \textcite{fw2018}). For notational simplicity, we treat self-tie exclusion as the only source of unbalancedness and otherwise work with balanced data.
\end{remark}

\subsection{Model}
\label{sec:model}

We consider the following generic semiparametric unobserved effects model for $(i, j) \in \mathcal{D}_{s}$, $s \in \{1, 2, 3\}$, and $t \in \{1, \dots, T\}$,
\begin{equation}
    \label{eq:model}
    \EX{y_{ijt}  \mid \mathcal{X}_{ij}^{t}} = g(\pi_{ijt}(\beta^{0}, \mu_{ijt}(\phi^{0}))) \, ,
\end{equation}
where $\mathcal{X}_{ij}^{t} \coloneqq \sigma(\{x_{ijt^{\prime}} \colon t^{\prime} \in \{1, \ldots, t\}\})$, $g(\pi)$ is a known link function, and $\pi_{ijt}(\beta, \mu) \coloneqq x_{ijt}^{\prime} \beta + \mu$ is the linear index. Here, $\beta$ is a $K$-dimensional vector of model parameters and $\mu$ is a scalar. Let $\phi$ be an $L$-dimensional (block) vector stacking $\mathcal{M} \in \{1, 2, 3\}$ vectors of unobserved effects. The function $\mu_{ijt}(\phi)$ sums a specific subset of up to $\mathcal{M}$ component vectors of $\phi$, where the subset may vary across observations. We impose no restrictions on the relationship between the unobserved effects and the explanatory variables, and we do not assume a parametric distribution for the unobserved effects.

Table \ref{tab:linear_combinations_phi} presents 17 identifiable linear combinations through which unobserved heterogeneity can enter the linear index.
\begin{table}[!htbp]
\centering
\caption{Explicit Expressions for $\mu_{ijt}(\phi)$}
\begin{threeparttable}
\begin{tabular}{llll}
\toprule
$\mathcal{M}$ & \multicolumn{3}{c}{Unobserved Heterogeneity} \\
\cmidrule(lr){2-4}
& Non-interacted   & Interacted  & Mixed \\
\midrule
1 & $\alpha_{i}^{\star}$ \quad (1.1.a) & $\alpha_{it}$ \quad (1.1.b) &  \\
& $\gamma_{j}^{\star}$ \quad (1.2.a) & $\gamma_{jt}$ \quad (1.2.b) &  \\
& $\rho_{t}^{\star}$ \quad (1.3.a) & $\rho_{ij}$ \quad (1.3.b)& \\
\midrule
2 & $\alpha_{i}^{\star} + \gamma_{j} ^{\star}$  \quad (2.1.a) & $\alpha_{it} + \gamma_{jt}$ \quad (2.1.b)  & $\alpha_{i}^{\star} + \gamma_{jt}$ \quad (2.1.c)  \\
& $\alpha_{i}^{\star} + \rho_{t}^{\star}$ \quad (2.2.a)  & $\alpha_{it} + \rho_{ij}$  \quad (2.2.b)  &  $\gamma_{j}^{\star} + \alpha_{it}$ \quad (2.2.c) \\
& $\gamma_{j}^{\star} + \rho_{t}^{\star}$ \quad (2.3.a)  & $\gamma_{jt} + \rho_{ij}$ \quad (2.3.b)  &  $\rho_{t}^{\star} + \rho_{ij}$ \quad (2.3.c)  \\
\midrule
3 & $\alpha_{i}^{\star} + \gamma_{j}^{\star} + \rho_{t}^{\star}$ \quad (3.a) & $\alpha_{it} + \gamma_{jt} + \rho_{ij}$ \quad (3.b)  &  \\
\bottomrule
\end{tabular}
\begin{tablenotes}
\footnotesize
\item\emph{Note:} For undirected network panels ($s = 3$), $\gamma_{j}^{\star} = \alpha_{j}^{\star}$ and $\gamma_{jt} = \alpha_{jt}$. In addition, $\mathcal{M}$ is reduced by one for some combinations.
\end{tablenotes}
\end{threeparttable}
\label{tab:linear_combinations_phi}
\end{table}
The specifications in Table \ref{tab:linear_combinations_phi} differ along two dimensions. First, they differ in $\mathcal{M}$, the number of unobserved-effect vectors stacked in $\phi$. Second, each component vector may capture heterogeneity varying along one panel dimension (\textit{non-interacted}) or two (\textit{interacted}). For undirected network panels, the symmetry condition $(y_{ijt}, x_{ijt}) = (y_{jit}, x_{jit})$ imposes additional restrictions on the unobserved effects. For instance, in a directed network panel one might specify $\mu_{ijt}(\phi) = \alpha_{it} + \gamma_{jt}$, with $\alpha_{it}$ and $\gamma_{jt}$ as distinct effects, and $L = 2 N_1 T$. In an undirected panel, symmetry requires $\gamma_{jt} = \alpha_{jt}$, so that $\mu_{ijt}(\phi) = \alpha_{it} + \alpha_{jt}$ and $L =  N_1 T$.

The following three examples illustrate model \eqref{eq:model} across different specifications and data structures.

\begin{example}[Linear Model for Bipartite Panel Data]
\label{example:ols_bipartite}
Let $y$ be a dependent variable with a conditional mean:
\begin{equation*}
    \EX{y_{ijt}  \mid \mathcal{X}_{ij}^{t}} = x_{ijt}^{\prime} \beta^{0} + \mu_{ijt}(\phi^{0}) \, , \quad \big( (i,j) \in \mathcal{D}_1, t \in \{1, \dots, T\}\big) \, .
\end{equation*}
For example, \textcite{ss2015technology} use a country-industry-time panel to analyze how technological characteristics interact with business cycle contractions to affect industry growth. They employ the interacted three-way specification (3.b) in Table \ref{tab:linear_combinations_phi} with an OLS estimator applied to the properly within-transformed model equation. Our theoretical results show that, under strict exogeneity, the OLS estimator they employ does not require a bias correction. At the same time, our framework would also allow one to relax this assumption and treat the contraction indicator as a predetermined regressor, in which case OLS would suffer from a Nickell-type bias that our debiased estimator corrects for.
\end{example}

\begin{example}[Exponential Model for Directed Network Panel Data]
\label{example:exp_model_directed}
Let $y$ be a positive outcome variable with a conditional mean:
\begin{equation*}
    \EX{y_{ijt}  \mid \mathcal{X}_{ij}^{t}} = \exp\big(x_{ijt}^{\prime} \beta^{0} + \mu_{ijt}(\phi^{0})\big) \, , \quad \big( (i,j) \in \mathcal{D}_2, t \in \{1, \dots, T\}\big) \, .
\end{equation*}
In international trade research, this specification is commonly used to estimate the impact of trade policy variables, such as joint WTO membership, on bilateral trade flows. Researchers using directed network panel data typically model the relationship between exporter $i$ and importer $j$ at time $t$ with an exponential conditional mean, as recommended by \textcite{st2006} and \textcite{lsy2025}, among others. Standard practice is to account for exporter-by-time, importer-by-time, and exporter-by-importer unobserved heterogeneity, as in specification (3.b) of Table \ref{tab:linear_combinations_phi}, and,  following \textcite{wz2021}, to correct the resulting (pseudo-)Poisson maximum likelihood (PPML) estimator for its incidental parameter bias of order $N_1^{-1}$. Their result relies on strict exogeneity and fixed-$T$ asymptotics. Our results show that, under the same strict exogeneity assumption but under asymptotics in which $N_1, T \rightarrow \infty$ jointly, the PPML estimator is in fact free of this bias, hence complementing \textcite{wz2021}'s fixed-$T$ results. More importantly, as recently advocated by \textcite{swz2026}, many trade policy variables should be treated as predetermined rather than strictly exogenous, since they respond to past trade outcomes. \textcite{wz2021}'s bias correction does not extend to this setting, whereas our framework accommodates predetermined regressors and provides a suitable bias correction that  addresses the resulting Nickell-type bias.
\end{example}

\begin{example}[Binary Response Model for Undirected Network Panel Data]
\label{example:binary_logit_undirected}
Let $y$ be a binary dependent variable with a conditional mean:
\begin{equation*}
    \EX{y_{ijt}  \mid \mathcal{X}_{ij}^{t}} = F_{\epsilon}(x_{ijt}^{\prime} \beta^{0} + \mu_{ijt}(\phi^{0})) \, , \quad \big( (i,j) \in \mathcal{D}_3, t \in \{1, \dots, T\}\big) \, ,
\end{equation*}
where $F_{\epsilon}(\pi)$ is a suitable cumulative distribution function, such as the standard normal CDF in the probit case. This specification can be used to study the formation of regional trade agreements between country pairs over time. \textcite{bb2007} study this question using network data on country pairs $(i,j)$, without a time dimension. Since both the outcome and typical regressors, such as country-pair similarity measures, are symmetric in $i$ and $j$, the data naturally form an undirected network. Our framework extends the cross-sectional setting of \textcite{bb2007} to a network panel. The richest unobserved heterogeneity structure available for this model is the symmetric counterpart of specification (3.b), namely $\alpha_{it}^{0} + \alpha_{jt}^{0} + \rho_{ij}^{0}$.  To the best of our knowledge, this has not been employed in empirical research yet, likely due to the absence of a suitable estimator and inferential theory.
\end{example}

\subsection{Fixed Effects Estimators}
\label{sec:estimators}

We follow a fixed effects approach and treat the unobserved effects as nuisance parameters to be estimated jointly with the model parameters.

For each $s \in \{1, 2, 3\}$, we assume the true parameter values are identified by the population problem:
\begin{equation}
    \label{eq:true_parameters}
    (\beta_{s}^{0}, \phi_{s}^{0}) = \underset{\{\beta \in \Real^{K}, \; \phi \in \Real^{L}\}}{\argmin}  \EX{\mathcal{L}_{n, s}(\beta, \phi)} \, ,
\end{equation}
where the objective function,
\begin{equation}
    \label{eq:objective_function_general}
    \mathcal{L}_{n, s}(\beta, \phi) \coloneqq  \frac{1}{w_{n, s}} \bigg\{ \sum_{(i, j) \in \mathcal{D}_{s}} \sum_{t = 1}^{T} \psi_{ijt}(\pi_{ijt}(\beta, \mu_{ijt}(\phi))) + \frac{\phi^{\prime} V V^{\prime} \phi}{2} \bigg\} \, ,
\end{equation}
consists of a criterion function $\psi_{ijt}(\pi) \coloneqq \psi(y_{ijt}, \pi)$ and a penalty term that ensures unique identification of $\phi_{s}^{0}$ by imposing suitable linear constraints. 

We tie the criterion function to the model by requiring that the (pseudo-)score of $\psi_{ijt}(\pi)$ be affine in $y_{ijt}$.
\begin{equation}
    \label{eq:affine_score}
    \frac{\partial \psi_{ijt}(\pi)}{\partial \pi} = a(\pi)\,(g(\pi) - y_{ijt}) \, ,
\end{equation}
where $a(\pi)$ is a known weight. The same link $g(\pi)$ therefore enters both the conditional mean in \eqref{eq:model} and the criterion function in \eqref{eq:objective_function_general}. Integrating gives $\psi_{ijt}(\pi) = \int^{\pi} a(z)\,(g(z) - y_{ijt}) \, \mathrm{d} z$, so $g(\pi)$ and $a(\pi)$ determine the criterion function up to a $\pi$-independent term. For example, let $g(\pi) = \exp(\pi)$. Then $a(\pi) = 1$ and $a(\pi) = \exp(\pi)$ yield the Poisson pseudo-maximum likelihood and nonlinear least-squares criterion functions, respectively. The affine form in \eqref{eq:affine_score} is a natural choice in our setting. The objective function is normalized by a model-specific scaling factor $w_{n, s}$ that grows at a suitable rate with the sample size $n_{s}$. We restrict attention to objective functions that are strictly convex in a neighborhood of the true parameter values and sufficiently smooth in all parameters. This class encompasses a wide range of estimators, including many popular (pseudo-)maximum likelihood and (non)linear least-squares estimators, but it excludes nonsmooth criterion functions, such as the check function.

We estimate $\beta_{s}^{0}$ and $\phi_{s}^{0}$ by minimizing the sample analogue of \eqref{eq:objective_function_general}. For each $s \in \{1, 2, 3\}$, the M-estimator is:
\begin{equation}
    \label{eq:uncorrected_estimators}
    (\hat{\beta}_{s}, \hat{\phi}_{s}) = \underset{\{\beta \in \Real^{K}, \; \phi \in \Real^{L}\}}{\argmin}  \mathcal{L}_{n, s}(\beta, \phi) \, .
\end{equation}

\begin{remark}[Penalty Term]
    To understand the identification problem and the role of the penalty term in \eqref{eq:objective_function_general}, consider a bipartite panel with three-way interacted unobserved heterogeneity (specification (3.b) in Table \ref{tab:linear_combinations_phi}), so that $\pi_{ijt}(\beta, \mu_{ijt}(\phi)) = x_{ijt}^{\prime} \beta + \alpha_{it} + \gamma_{jt} + \rho_{ij}$. Because the incidental parameters enter the linear index additively, the objective function is invariant to certain parameter transformations. For example, adding a constant $c_{t}$ to all $\alpha_{it}$ while subtracting it from all $\gamma_{jt}$ leaves the linear index unchanged. Likewise, subtracting a constant $c_{i}$ from all $\alpha_{it}$ while adding it to all $\rho_{ij}$, or adding a constant $c_{j}$ to all $\gamma_{jt}$ while subtracting it from all $\rho_{ij}$, leaves the linear index unchanged. To resolve this identification problem, we impose a system of linear constraints on the incidental parameters. In particular, we require: $\sum_{i = 1}^{N_{1}} \alpha_{it} = \sum_{j = 1}^{N_{2}} \gamma_{jt}$ for all $t \in \{1, \ldots, T\}$, $\sum_{t = 1}^{T} \alpha_{it} = \sum_{j = 1}^{N_{2}} \rho_{ij}$ for all $i \in \{1, \ldots, N_{1}\}$, and $\sum_{t = 1}^{T} \gamma_{jt} = \sum_{i = 1}^{N_{1}} \rho_{ij}$ for all $j \in \{1, \ldots, N_{2}\}$. Together, these $T + N_{1} + N_{2}$ restrictions are encoded as $V^{\prime} \phi = \mathbf{0}_{T + N_{1} + N_{2}}$, where $V$ is an $L \times (T + N_{1} + N_{2})$ matrix defined as
    \begin{equation*}
	   V = \begin{pmatrix}
        \iota_{N_{1}} \otimes I_{T}   & I_{N_{1}} \otimes \iota_{T}     & 0	\\
        - \iota_{N_{2}} \otimes I_{T} & 0         & I_{N_{2}} \otimes \iota_{T} \\
        0      & - I_{N_{1}} \otimes \iota_{N_{2}}   & - \iota_{N_{1}} \otimes I_{N_{2}}
	   \end{pmatrix} \, .
    \end{equation*}
\end{remark}

\begin{remark}[Undirected Network Panels]
    An undirected network panel model, for example with linear index $\pi_{ijt}(\beta^{0}, \mu_{ijt}(\phi^{0})) = x_{ijt}^{\prime} \beta^{0} + \alpha_{it}^{0} + \alpha_{jt}^{0}$, is typically estimated on a sample of unique pairs $(i, j)$. Alternatively, one can use a doubled sample that includes both $(i, j)$ and $(j, i)$ and estimate the more general specification $\pi_{ijt}(\beta^{0}, \mu_{ijt}(\phi^{0})) = x_{ijt}^{\prime} \beta^{0} + \alpha_{it}^{0} + \gamma_{jt}^{0}$. Both approaches yield identical estimates of $\beta^{0}$.\footnote{A similar equivalence was recently stated by \textcite{h2026} and \textcite{lsw2025} in the context of network data (without time dimension).} We exploit this equivalence to derive our asymptotic results for this data structure.
\end{remark}

\begin{remark}[Computation]
    In practice, joint estimation of $\beta^{0}$ and $\phi^{0}$ may involve a very large number of parameters, making \eqref{eq:uncorrected_estimators} a high-dimensional optimization problem. Standard software routines that rely on dummy variables are therefore often computationally infeasible, even for moderately sized panels. We recommend algorithms such as those of \textcite{gp2010}, \textcite{g2013}, \textcite{b2018}, and \textcite{s2018}, which are designed for high-dimensional problems of this type and can handle unbalanced data.\footnote{When \eqref{eq:uncorrected_estimators} reduces to OLS, closed-form within-group transformations exist in two specific cases: (i) one-way fixed effects structures, and (ii) two- or three-way fixed effects structures for balanced panels.} We provide computationally efficient implementations of the (debiased) estimators studied in this paper in the R package \href{https://cran.r-project.org/web/packages/alpaca/index.html}{alpaca}.
\end{remark}

\section{Asymptotic Theory}
\label{sec:asymptotic_theory}

\subsection{Assumptions}
\label{sec:assumptions}

Before stating our assumptions, we introduce additional notation. For each $s \in \{1, 2, 3\}$, let $X$ denote the $n_{s} \times K$ matrix of explanatory variables, where $x_{ijt}^{\prime}$ is its $ijt$-th row. Let $(\dop{\mathrm{r}} \psi(\beta, \phi))_{ijt}$ denote the $\mathrm{r}$-th derivative of the criterion function with respect to the linear index, $\partial^{\mathrm{r}} \psi_{ijt}(\pi_{ijt}(\beta, \mu_{ijt}(\phi))) / \left(\partial \pi_{ijt}\right)^{\mathrm{r}}$. When evaluated at the true parameter values, we suppress the arguments and write $(\dop{\mathrm{r}} \psi)_{ijt}$. Let $(\dop{r}_{\mathcal{X}} \psi(\beta, \phi))_{ijt} \coloneqq \mathbb{E}\big[(\dop{\mathrm{r}} \psi(\beta, \phi))_{ijt} \mid \mathcal{X}_{ij}^{t}\big]$ denote the corresponding conditional expectation.

We further define population weighted least-squares projections. For each $s \in \{1, 2, 3\}$ and $k \in \{1, \ldots, K\}$, let
\begin{align}
    &\xi_{k}^{0} \coloneqq \label{eq:population_wls_program} \\
    &\quad\underset{\xi \in \Real^{L}}{\argmin} \frac{1}{w_{n, s}} \Bigg\{ \sum_{(i, j) \in \mathcal{D}_{s}} \sum_{t = 1}^{T} \EX{(\dop{2}_{\mathcal{X}} \psi)_{ijt}}\left(\frac{\EX{x_{ijt, k} (\dop{2}_{\mathcal{X}} \psi)_{ijt}}}{\EX{(\dop{2}_{\mathcal{X}} \psi)_{ijt}}} - \mu_{ijt}(\xi)\right)^{2} + \frac{\xi^{\prime} V V^{\prime} \xi}{2} \Bigg\} \, , \nonumber
\end{align}
where $x_{ijt, k}$ denotes the $k$-th element of $x_{ijt}$. The resulting $n_{s} \times K$ matrix of fitted values is denoted by $\mathfrak{X}$, with its $ijt$-th row defined as
\begin{equation}
    \label{eq:population_wls_fitted}
    \mathfrak{x}_{ijt} \coloneqq \big(\mu_{ijt}(\xi_{1}^{0}), \ldots, \mu_{ijt}(\xi_{K}^{0})\big) \, .
\end{equation}
Thus, the fitted values in \eqref{eq:population_wls_fitted} are the weighted least-squares projection of \linebreak $\EX{x_{ijt} (\dop{2}_{\mathcal{X}} \psi)_{ijt}} / \EX{(\dop{2}_{\mathcal{X}} \psi)_{ijt}}$ onto the subspace spanned by the incidental parameters. For OLS, with $g(\pi) = \pi$, $a(\pi) = 1$, $\psi_{ijt}(\pi) = (y_{ijt} - \pi)^2 / 2$, and $(\dop{2}_{\mathcal{X}} \psi)_{ijt} = 1$, \eqref{eq:population_wls_fitted} reduces to the familiar population between-transformation. For example, for a bipartite panel ($s = 1$) with three-way interacted unobserved heterogeneity (specification (3.b) in Table \ref{tab:linear_combinations_phi}), the corresponding between-transformation follows from \textcite{bmw2018}:
\begin{align*}
    \mathfrak{x}_{ijt} &= \frac{1}{T} \sum_{t^{\prime}=1}^{T} \EX{x_{ijt^{\prime}}} + \frac{1}{N_{1}} \sum_{i^{\prime}=1}^{N_{1}} \EX{x_{i^{\prime}jt}} + \frac{1}{N_{2}} \sum_{j^{\prime}=1}^{N_{2}} \EX{x_{ij^{\prime}t}}\\
    & - \frac{1}{N_{1} N_{2}} \sum_{i^{\prime}=1}^{N_{1}} \sum_{j^{\prime}=1}^{N_{2}} \EX{x_{i^{\prime}j^{\prime}t}} - \frac{1}{N_{1} T} \sum_{i^{\prime}=1}^{N_{1}} \sum_{t^{\prime}=1}^{T} \EX{x_{i^{\prime}jt^{\prime}}} -  \frac{1}{N_{2} T} \sum_{j^{\prime}=1}^{N_{2}} \sum_{t^{\prime}=1}^{T} \EX{x_{ij^{\prime}t^{\prime}}}\\
    & +  \frac{1}{N_{1} N_{2} T} \sum_{i^{\prime}=1}^{N_{1}} \sum_{j^{\prime}=1}^{N_{2}} \sum_{t^{\prime}=1}^{T} \EX{x_{i^{\prime}j^{\prime}t^{\prime}}} \, .
\end{align*}
We define the population residuals as $\ddot{X} \coloneqq X - \mathfrak{X}$.

We make the following assumptions.
\begin{assumption}[Sampling and Regularity Conditions]
    \label{assumption:general}
    Let $z_{ijt} = (y_{ijt}, x_{ijt})$, $\nu > \check{\nu} > 0$, and $\varphi > 10 (20 + \nu) / \nu > 10$. Furthermore, for each $s \in \{1, 2, 3\}$, let $\varepsilon > 0$ and let $\Theta_{s}^{0}(\varepsilon)$ be a subset of $\Real^{K + 1}$ that contains an $\varepsilon$-neighborhood of $(\beta_{s}^{0}, \mu_{ijt}(\phi_{s}^{0}))$ for all $i, j, t, \lvert \mathcal{D}_{s} \rvert, T$.
    \begin{enumerate}[(i)]
        \item \textit{Asymptotics}: We consider joint limits in which all panel dimensions diverge proportionally. For $s = 1$: $N_{1}, N_{2}, T \rightarrow \infty$ with $N_{1} / T \rightarrow \tau_{1} \in (0, \infty)$ and $N_{2} / T \rightarrow \tau_{2} \in (0, \infty)$. For $s \in \{2, 3\}$: $N_{1}, T \rightarrow \infty$ with $N_{1} / T \rightarrow \tau_{1} \in (0, \infty)$.
        \item \textit{Sampling:} For each $s \in \{1, 2, 3\}$, conditional on $\Phi$, $\{ \{z_{ijt}\}_{t = 1}^{T} \colon (i, j) \in \mathcal{D}_{s}\}$ is independent across $(i, j)$, and, for each $(i, j)$, $\{z_{ijt}\}_{t = 1}^{T}$ is $\alpha$-mixing with mixing coefficients satisfying $\sup_{ij} a_{ij}(m) = \mathcal{O}(m^{- \varphi})$ a.\,s. as $m \rightarrow \infty$, where $\mathcal{A}_{ij}^{t}$ is the sigma-algebra generated by $(z_{ijt}, z_{ij(t - 1)}, \ldots)$, $\mathcal{B}_{ij}^{t}$ is the sigma-algebra generated by $(z_{ijt}, z_{ij(t + 1)}, \ldots)$, and
        \begin{equation*}
            a_{ij}(m) \coloneqq \sup_{t} \sup_{A \in \mathcal{A}_{ij}^{t}, B \in \mathcal{B}_{ij}^{t + q}} \left\lvert \Prob{A \cap B} - \Prob{A} \Prob{B} \right\rvert \quad \text{a.\,s.}
        \end{equation*}
        \item \textit{Model}: For each $s \in \{1, 2, 3\}$, we assume that for all $i, j, t, \lvert \mathcal{D}_{s} \rvert, T$, $\mathbb{E}\big[y_{ijt}  \mid \mathcal{X}_{ij}^{t}\big] = g(\pi_{ijt}(\beta_{s}^{0}, \mu_{ijt}(\phi_{s}^{0})))$. The unobserved effects $\phi_{s}^{0}$ are normalized to $V V^{\prime} \phi_{s}^{0} = \mathbf{0}_{L}$.
        \item \textit{Smoothness and Moments}: For each $s \in \{1, 2, 3\}$, the map $(\beta, \mu) \mapsto \psi_{ijt}(\pi_{ijt}(\beta, \mu))$ is five times continuously differentiable over $\Theta_{s}^{0}(\varepsilon)$ a.\,s. Each element of $x_{ijt}$, and the partial derivatives of $\psi_{ijt}(\pi_{ijt}(\beta, \mu))$ with respect to the elements of $(\beta, \mu)$ up to fifth order, are bounded in absolute value uniformly over $(\beta, \mu) \in \Theta_{s}^{0}(\varepsilon)$ by a function $\Psi(z_{ijt}) > 0$ a.\,s. In addition, $\sup_{ijt} \EX{\Psi(z_{ijt})^{20 + \nu}}$ is a.\,s.\ uniformly bounded over $i, j, t, \lvert \mathcal{D}_{s} \rvert, T$.
        \item \textit{Convexity}: For each $s \in \{1, 2, 3\}$, there exists a constant $c_{H, s}$ such that $\EX{(\dop{2}_{\mathcal{X}} \psi)_{ijt}} \geq c_{H, s} > 0$ a.\,s.\ uniformly over $i, j, t, \lvert \mathcal{D}_{s} \rvert, T$. Furthermore, there exists a constant $c_{W, s} > 0$ such that for all $s \in \{1, 2, 3\}$,
        \begin{equation*}
            \underset{\{v \in \mathbb{R}^{K} \colon \norm{v}_{2} = 1\}}{\min} \; \frac{1}{n_{s}} \sum_{(i, j) \in \mathcal{D}_{s}} \sum_{t = 1}^{T} \EX{(\dop{2}_{\mathcal{X}} \psi)_{ijt} \big\{(\ddot{X} v)_{ijt} \big\}^{2}} \geq c_{W, s} \quad \text{a.\,s.}
        \end{equation*}
    \end{enumerate}
\end{assumption}

\begin{remark}[Assumption \ref{assumption:general}]
\label{remark:assumption_general}
    We comment on each part in turn, noting differences from Assumption 4.1 of \textcite{fw2016} where relevant.
    \begin{enumerate}[(i)]
        \item The asymptotics condition specifies a joint limit in which all panel dimensions diverge at proportional rates. This extends the approach of \textcite{fw2016} to the multi-dimensional panel setting. For bipartite panels ($s = 1$), two proportionality conditions are required: $N_{1} / T \rightarrow \tau_{1}$ and $N_{2} / T \rightarrow \tau_{2}$. For network panels ($s \in \{2, 3\}$), both cross-sectional dimensions index the same set of $N_{1}$ agents, so the single condition $N_{1} / T \rightarrow \tau_{1}$ suffices.
        \item The sampling condition restricts dependence in two ways: it imposes independence across dyads conditional on $\Phi$, and $\alpha$-mixing within each dyad with coefficients decaying at a polynomial rate. We use $\alpha$-mixing because it is the weakest among the standard mixing conditions in econometrics and is preserved under measurable transformations. The $\alpha$-mixing condition could be replaced by any other form of weak dependence that ensures the implied regularity conditions underlying our results remain satisfied. Compared with \textcite{fw2016}, who impose independence across individuals conditional on $\Phi$, we impose independence across dyadic pairs $(i, j)$. This formulation accommodates both bipartite and network panel data. In the directed network case ($s = 2$), the framework can be extended to allow for reciprocity or pairwise clustering (see Remark \ref{remark:reciprocity_directed_network} below).
        \item The model condition imposes a conditional mean restriction. Conditioning on $\mathcal{X}_{ij}^{t}$ allows the regressors to be predetermined, accommodating dynamic models. Beyond the first moment, the conditional distribution of $y_{ijt}$ is left unrestricted. The normalization $V V^{\prime} \phi_{s}^{0} = \mathbf{0}_{L}$ is required for identification, since constants can be shifted between the components of $\phi$ without affecting $\pi_{ijt}$. If one additionally specifies a fully parametric model, for example,
        \begin{equation*}
            y_{ijt} \mid \mathcal{X}_{ij}^{t} \sim f_{Y}\big(y_{ijt} \mid \pi_{ijt}(\beta_{s}^{0}, \mu_{ijt}(\phi_{s}^{0})) \big)
        \end{equation*}
        for all $i, j, t, \lvert \mathcal{D}_{s} \rvert, T$, with $f_{Y}(\cdot)$ a known conditional density or probability mass function, then one can also exploit Bartlett identities (see \cite{b1953}). These identities can simplify bias and variance expressions and improve the finite-sample performance of debiased estimators, as noted by \textcite{f2009}. We return to this point in Remark \ref{remark:theorem_asymptotic_distributions_interacted} (ii). In contrast, \textcite{fw2016} assume a fully parametric model for their main results (see Assumption 4.1 (iii)) and present results for conditional mean models only in Remark 3. We therefore adopt the more general formulation as our baseline.
        \item The smoothness and moment condition supports the higher-order asymptotic expansion underlying our results. Our expansion is of higher order than that of \textcite{fw2016}, so we require $\psi_{ijt}(\pi_{ijt}(\beta, \mu))$ to be five times continuously differentiable, rather than four. The dominating function $\Psi(z_{ijt})$ uniformly bounds both the regressors and all derivatives of the criterion function up to fifth order. Requiring $\sup_{ijt} \EX{\Psi(z_{ijt})^{20 + \nu}}$ to be a.\,s.\ uniformly bounded ensures these envelopes have sufficiently many finite moments: $20 + \nu$ here, compared with $8 + \nu$ in \textcite{fw2016}.
        \item The convexity condition has two components. The first, $\EX{(\dop{2} \psi)_{ijt}} \geq c_{H, s} > 0$, requires the expected second derivative of the criterion function to be uniformly bounded away from zero. This ensures local strict convexity in $\phi$, which is needed to apply our asymptotic expansions. The second component is a generalized non-collinearity condition: after projecting out the incidental parameters, the within-transformed regressors $\ddot{X}$ must display sufficient variation in all directions $v$. Together, the two components imply strict convexity of the expected objective function over the relevant part of the parameter space, guaranteeing that \eqref{eq:true_parameters} has a unique solution. Rather than directly assuming global strict convexity as in \textcite{fw2016}, we impose this more primitive non-collinearity condition.
    \end{enumerate}
\end{remark}

\subsection{A Heuristic Discussion of the Dichotomy}
\label{sec:heuristic_discussion}

Before presenting our theoretical results for the three-way fixed effects specifications in Section \ref{sec:asymptotic_theory_3way} and for the two- and one-way fixed effects specifications in Section \ref{sec:asymptotic_theory_other}, we provide some intuition for the sharp dichotomy in our results: interacted specifications may yield a degenerate asymptotic distribution, whereas non-interacted specifications yield a nondegenerate, correctly centered one. For illustration, we focus on the bipartite case $s=1$. The argument for $s\in\{2,3\}$ is analogous.

The central difference between interacted and non-interacted specifications lies in the variation of the fixed effects. Consider, for example, the fixed effect $\alpha_{it}^{0}$ versus the fixed effect $\alpha_i^{\star,0}$. Each $\alpha_{it}^{0}$ is constant across $j$ and hence uses the corresponding $N_2$ observations for estimation, while $\alpha_i^{\star,0}$ is constant across $j$ and $t$ and hence uses the corresponding $N_2 T$ observations. To assess the order of the incidental parameter bias, a useful heuristic relates the number of incidental parameters to the sample size (see \textcite{fw2018}):
\begin{equation*}
    \text{order of bias} \;\approx\; \frac{\text{number of incidental parameters}}{\text{sample size}} .
\end{equation*}
Since we have $N_1 T$ parameters $\alpha_{it}^{0}$, relative to the sample size $N_1 N_2 T$, this suggests a bias component of order $N_1T/(N_1N_2T) = N_2^{-1}$, the reciprocal of the $N_2$ observations available to estimate each $\alpha_{it}^{0}$. Since we have only $N_1$ parameters $\alpha_i^{\star,0}$, this suggests a bias component of order $(N_2T)^{-1}$, the reciprocal of the $N_2 T$ observations available to estimate each $\alpha_i^{\star,0}$.\footnote{By the same logic, the remaining fixed effects, $\gamma_{jt}^{0}$ and $\rho_{ij}^{0}$ (interacted), as well as $\gamma_j^{\star,0}$ and $\rho_t^{\star,0}$ (non-interacted), contribute analogous bias components of order $N_1^{-1}$ and $T^{-1}$, and of order $(N_1T)^{-1}$ and $(N_1N_2)^{-1}$, respectively.} The same logic applies in the classical two-dimensional panel model literature, where, for example, an individual effect that is constant over $T$ periods contributes a bias component of order $T^{-1}$.

In our setting, $\beta^{0}$ is estimated from $N_1 N_2 T$ observations, so its standard deviation is of order $(N_1 N_2 T)^{-1/2}$ in both the interacted and the non-interacted case. The two specifications therefore differ only in the order of the bias, which we now compare with this common order of the standard deviation. In the three-way interacted case, the order of the bias, $N_{1}^{-1} + N_{2}^{-1} + T^{-1}$, resembles that of a classical two-dimensional panel estimator (see \textcite{hk2011}, \textcite{fw2016}), yet it is now contrasted with a standard deviation of strictly smaller order, $(N_1 N_2 T)^{-1/2}$. This imbalance between the order of the standard deviation and the order of the bias generates the degenerating asymptotic distribution result of Theorem \ref{theorem:asymptotic_distributions_interacted} below. In the three-way non-interacted case, by contrast, the order of the bias, $(N_{1}N_{2})^{-1} + (N_{1}T)^{-1} + (N_{2}T)^{-1}$, is strictly smaller than that of the standard deviation. The bias is hence asymptotically negligible, which yields the correctly centered asymptotic distribution result of Theorem \ref{theorem:asymptotic_distributions_noninteracted} below.

For comparison, consider a classical two-dimensional panel model with $N$ individual and $T$ time effects, as studied in \textcite{fw2016}. There, the bias of the fixed effects estimator is of order $T^{-1} + N^{-1}$ and, under proportional growth of the cross-section with $T$, the standard deviation is of the same order, $(NT)^{- 1 / 2}$. Bias and standard deviation are hence balanced, yielding a nondegenerate but incorrectly centered asymptotic distribution.

The heuristic discussion above does not replace our theoretical analysis, for several reasons. First, the heuristic of \textcite{fw2018} assesses only the order of the leading bias. It is silent on whether other terms in the asymptotic expansion remain non-negligible, and it cannot capture special cases in which the leading bias vanishes. For instance, we show that fixed effects estimators are asymptotically unbiased in the interacted case for linear or (pseudo-)Poisson models with strictly exogenous regressors (see Remark \ref{remark:strictly_exogenous} below). Identifying such special cases requires a formal characterization of the bias components. Second, our debiased inference procedures are built on explicit analytical expressions for the bias and variance components (see Theorem \ref{theorem:asymptotic_distributions_interacted} below), which the heuristic arguments cannot provide.

\subsection{Three-way Fixed Effects Specifications}
\label{sec:asymptotic_theory_3way}

Before presenting our results for the three-way fixed effects specifications (3.a) and (3.b) in Table \ref{tab:linear_combinations_phi}, we introduce $\delta_{b}$ to denote a degenerate distribution concentrated at $b$. We also write $\LEX{\cdot}$ for the probability limit of the enclosed expression, where the limit is taken as $N_{1}, N_{2}, T \rightarrow \infty$ for $s = 1$ and as $N_{1}, T \rightarrow \infty$ for $s \in \{2, 3\}$. By construction, $\EX{\ddot{x}_{ijt} (\dop{2}_{\mathcal{X}} \psi)_{ijt}}$ denotes the weighted residuals of the weighted least-squares problem \eqref{eq:population_wls_program}. As a consequence, for specification (3.b) and all $(i, j) \in \mathcal{D}_{s}$, for example, $\sum_{t = 1}^{T} \EX{\ddot{x}_{ijt} (\dop{2}_{\mathcal{X}} \psi)_{ijt}} = \mathbf{0}_{K}$ for each $s \in \{1, 2, 3\}$.

\paragraph{Interacted Specification.} Theorem \ref{theorem:asymptotic_distributions_interacted} establishes that, for specification (3.b), estimators $\hat{\beta}_{s}$ have \textit{degenerate} asymptotic distributions for any $s \in \{1, 2, 3\}$.\footnote{More precisely, $\hat{\beta}_{s}$ does not have a nondegenerate asymptotic distribution for any $s \in \{1, 2, 3\}$. We use \textit{degenerate} asymptotic distributions to avoid double negation.}
\begin{theorem}[Asymptotic Distributions of $\hat{\beta}_{s}$ - Interacted Specification]
    \label{theorem:asymptotic_distributions_interacted}
    Let Assumptions \ref{assumption:general} hold. Then, for any $s \in \{1, 2, 3\}$,
    \begin{equation*}
        \frac{\sqrt{n_{s}}}{\sqrt{T}} \, (\hat{\beta}_{s} - \beta_{s}^{0}) \xrightarrow{d} \delta_{\overline{b}_{s, \infty}} \, ,
    \end{equation*}
    where 
    \begin{align*}
          \overline{b}_{1, \infty} \coloneqq& \, - \overline{W}_{1, \infty}^{- 1} \big( \tau_{1}^{\frac{1}{2}} \, \tau_{2}^{- \frac{1}{2}} \, \overline{B}_{1, \alpha, \infty} + \tau_{1}^{- \frac{1}{2}} \, \tau_{2}^{\frac{1}{2}} \, \overline{B}_{1, \gamma, \infty} + \tau_{1}^{\frac{1}{2}} \, \tau_{2}^{\frac{1}{2}} \, \overline{B}_{1, \rho, \infty} \big) \, , \\
          \overline{b}_{2, \infty} \coloneqq& \, - \overline{W}_{2, \infty}^{- 1} \big( \overline{B}_{2, \alpha, \infty} + \overline{B}_{2, \gamma, \infty} + \tau_{1} \, \overline{B}_{2, \rho, \infty} \big) \, , \\
          \overline{b}_{3, \infty} \coloneqq& \, - \overline{W}_{3, \infty}^{- 1} \big( \overline{B}_{3, \alpha, \infty} + \tau_{1} \, \overline{B}_{3, \rho, \infty} \big) \, ,
    \end{align*}
    with
    \begin{align*}
        & \overline{B}_{1, \alpha, \infty} \coloneqq \overline{\mathbb{E}}\left[ - \frac{1}{N_{1} T} \sum_{i = 1}^{N_{1}} \sum_{t = 1}^{T} \frac{\sum_{j = 1}^{N_{2}} \EX{\ddot{x}_{ijt} (\dop{2} \psi)_{ijt} (\dop{1} \psi)_{ijt}}}{\sum_{j = 1}^{N_{2}} \EX{(\dop{2}_{\mathcal{X}} \psi)_{ijt}}} \, + \right. \\
        & \quad \left. \frac{1}{2 \, N_{1} T} \sum_{i = 1}^{N_{1}} \sum_{t = 1}^{T} \frac{\big\{\sum_{j = 1}^{N_{2}} \EX{\ddot{x}_{ijt} (\dop{3}_{\mathcal{X}} \psi)_{ijt}}\big\} \sum_{j = 1}^{N_{2}} \EX{\big\{(\dop{1} \psi)_{ijt}\big\}^{2}}}{\big\{\sum_{j = 1}^{N_{2}} \EX{(\dop{2}_{\mathcal{X}} \psi)_{ijt}}\big\}^{2}} \right]  \, , \\
        & \overline{B}_{1, \gamma, \infty} \coloneqq \overline{\mathbb{E}}\left[ - \frac{1}{N_{2} T} \sum_{j = 1}^{N_{2}} \sum_{t = 1}^{T} \frac{\sum_{i = 1}^{N_{1}} \EX{\ddot{x}_{ijt} (\dop{2} \psi)_{ijt} (\dop{1} \psi)_{ijt}}}{\sum_{i = 1}^{N_{1}} \EX{(\dop{2}_{\mathcal{X}} \psi)_{ijt}}} \right. \, + \\
        & \quad \left. \frac{1}{2 \, N_{2} T} \sum_{j = 1}^{N_{2}} \sum_{t = 1}^{T} \frac{\big\{\sum_{i = 1}^{N_{1}} \EX{\ddot{x}_{ijt} (\dop{3}_{\mathcal{X}} \psi)_{ijt}}\big\} \sum_{i = 1}^{N_{1}} \EX{\big\{(\dop{1} \psi)_{ijt}\big\}^{2}}}{\big\{\sum_{i = 1}^{N_{1}} \EX{(\dop{2}_{\mathcal{X}} \psi)_{ijt}}\big\}^{2}} \right] \, , \\
        & \overline{B}_{1, \rho, \infty} \coloneqq \overline{\mathbb{E}}\left[ - \frac{1}{N_{1} N_{2}} \sum_{i = 1}^{N_{1}} \sum_{j = 1}^{N_{2}} \frac{\sum_{t = 1}^{T} \sum_{t^{\prime} = t}^{T} \EX{\ddot{x}_{ijt^{\prime}} (\dop{2} \psi)_{ijt^{\prime}} (\dop{1} \psi)_{ijt}}}{\sum_{t = 1}^{T} \EX{(\dop{2}_{\mathcal{X}} \psi)_{ijt}}} \right. \, + \\
        & \quad \left. \frac{1}{2 \, N_{1} N_{2}} \sum_{i = 1}^{N_{1}} \sum_{j = 1}^{N_{2}} \frac{\big\{\sum_{t = 1}^{T} \EX{\ddot{x}_{ijt} (\dop{3}_{\mathcal{X}} \psi)_{ijt}}\big\} \Big\{\sum_{t = 1}^{T} \EX{\big\{(\dop{1} \psi)_{ijt}\big\}^{2}}\Big\}}{\big\{\sum_{t = 1}^{T} \EX{(\dop{2}_{\mathcal{X}} \psi)_{ijt}}\big\}^{2}} \right] \, ,  \\
        & \overline{W}_{1, \infty} \coloneqq \LEX{\frac{1}{N_{1} N_{2} T} \sum_{i = 1}^{N_{1}} \sum_{j = 1}^{N_{2}} \sum_{t = 1}^{T} \EX{(\dop{2}_{\mathcal{X}} \psi)_{ijt} \, \ddot{x}_{ijt} \, \ddot{x}_{ijt}^{\prime}}} \, , \\
        & \overline{\Sigma}_{1, \infty} \coloneqq \LEX{\frac{1}{N_{1} N_{2} T} \sum_{i = 1}^{N_{1}} \sum_{j = 1}^{N_{2}} \sum_{t = 1}^{T} \EX{\big\{(\dop{1} \psi)_{ijt}\big\}^{2} \ddot{x}_{ijt} \, \ddot{x}_{ijt}^{\prime}}} \, .
    \end{align*}
    The expressions for $s = 2$ and $s = 3$ are stated in Appendix \ref{appendix:main_results_distributions_interacted}.
\end{theorem}

\begin{remark}[Theorem \ref{theorem:asymptotic_distributions_interacted}]
\label{remark:theorem_asymptotic_distributions_interacted}
To keep the following discussion concise, we focus on bipartite panel data ($s = 1$).
\begin{enumerate}[(i)]
    \item Each component of the leading incidental parameter bias terms, $\overline{B}_{\alpha, \infty}, \overline{B}_{\gamma, \infty}, \overline{B}_{\rho, \infty}$, originates from estimating the incidental parameters $\alpha^{0}$, $\gamma^{0}$, and $\rho^{0}$, respectively. The corresponding estimators $\hat{\alpha}, \hat{\gamma}$, and $\hat{\rho}$ converge at rates much slower than $\sqrt{N_{1} N_{2} T}$, which causes the inference problem. The biases $\overline{B}_{\alpha, \infty}$ and $\overline{B}_{\gamma, \infty}$ occur even under strict exogeneity. By contrast, $\overline{B}_{\rho, \infty}$ contain an additional component that arises when regressors are predetermined, besides those that also arise under strict exogeneity. Given the similarity to \textcite{n1981}, we refer to this additional bias component as Nickell-type bias. We show
    \begin{equation*}
        \sqrt{N_{1} N_{2} T}\big(\hat{\beta} - \beta^{0}\big) \approx \overline{W}_{\infty}^{- 1} \partial_{\beta} \mathcal{L}_{n}\big(\beta^{0}, \phi^{0}\big) - \overline{W}_{\infty}^{- 1} \overline{B}_{\infty} \, , 
    \end{equation*}
    where
    \begin{equation*}
        \overline{B}_{\infty} \coloneqq \overline{B}_{\alpha, \infty} \sqrt{(N_{1} T) / N_{2}}+\overline{B}_{\gamma, \infty} \sqrt{(N_{2} T) / N_{1}} + \overline{B}_{\rho, \infty} \sqrt{(N_{1} N_{2}) / T} \, .
    \end{equation*}
    The first term obeys a central limit theorem, i.e., $\partial_{\beta} \mathcal{L}_{n}(\beta^{0}, \phi^{0}) \xrightarrow{d} \mathcal{N}(0, \overline{\Sigma}_{\infty})$. Under Assumption \ref{assumption:general} (i), $N_{1} / T \rightarrow \tau_{1} \in (0, \infty)$ and $N_{2} / T \rightarrow \tau_{2} \in (0, \infty)$ as $N_{1}, N_{2}, T \rightarrow \infty$, the second term is of order $\sqrt{N_{1}} + \sqrt{N_{2}} + \sqrt{T}$. Hence, one cannot balance the orders of the bias and variance to obtain a nondegenerate asymptotic distribution. This is a key difference from the existing large-$T$ bias-correction literature, in which uncorrected estimators have nondegenerate asymptotic distributions. Intuitively, the incidental parameters in a three-dimensional panel are estimated from the same number of effective observations as in a standard two-dimensional panel with individual and time fixed effects, even though the overall sample is much larger. Hence, the increased sample size does not improve the convergence rates of the estimators of the incidental parameters.
    \item Using the fully parametric model assumption (e.g., as outlined in Remark \ref{remark:assumption_general}) would allow us to simplify the bias and variance components using Bartlett identities. In particular, by the second Bartlett identity, $\EX{\{(\dop{1} \psi)_{ijt}\}^{2}} = \EX{(\dop{2}_{\mathcal{X}} \psi)_{ijt}}$ and $\mathbb{E}\big[\{(\dop{1} \psi)_{ijt}\}^{2} \ddot{x}_{ijt} \, \ddot{x}_{ijt}^{\prime}\big] = \mathbb{E}\big[(\dop{2}_{\mathcal{X}} \psi)_{ijt} \ddot{x}_{ijt} \, \ddot{x}_{ijt}^{\prime}\big]$ for all $i, j, t, N_{1}, N_{2}, T$. Hence, $\overline{\Sigma}_{\infty} = \overline{W}_{\infty}$, $\sum_{j = 1}^{N_{2}} \EX{\{(\dop{1} \psi)_{ijt}\}^{2}} = \sum_{j = 1}^{N_{2}} \EX{(\dop{2}_{\mathcal{X}} \psi)_{ijt}}$, $\sum_{i = 1}^{N_{1}} \EX{\{(\dop{1} \psi)_{ijt}\}^{2}} = \sum_{i = 1}^{N_{1}} \EX{(\dop{2}_{\mathcal{X}} \psi)_{ijt}}$, and $\sum_{t = 1}^{T} \EX{\{(\dop{1} \psi)_{ijt}\}^{2}} = \sum_{t = 1}^{T} \EX{(\dop{2}_{\mathcal{X}} \psi)_{ijt}}$ for all $i, j, t, N_{1}, N_{2}, T$.
\end{enumerate}
\end{remark}

\begin{remark}[(Pseudo-) ML Estimators for Binary Response Models]
    \label{remark:pml_binary_response}
    If $y_{ijt} \in \{0, 1\}$ for all $i, j, t, \lvert \mathcal{D}_{s} \rvert, T$, the simplifications from the second Bartlett identity in Remark \ref{remark:theorem_asymptotic_distributions_interacted} (ii) follow automatically from the conditional mean assumption. Let $F_{\epsilon}(\pi)$ be a cumulative distribution function, and define $g_{ijt}(\beta, \mu) \coloneqq F_{\epsilon}(\pi_{ijt}(\beta, \mu))$, $g_{ijt}^{\prime}(\beta, \mu) \coloneqq \partial_{\mu} g_{ijt}(\beta, \mu)$, $a_{ijt}(\beta, \mu) \coloneqq g_{ijt}^{\prime}(\beta, \mu) / (g_{ijt}(\beta, \mu) (1 - g_{ijt}(\beta, \mu)))$, and $a_{ijt}^{\prime}(\beta, \mu) \coloneqq \partial_{\mu} a_{ijt}(\beta, \mu)$. Then,
    \begin{equation*}
        \psi_{ijt}(\pi_{ijt}(\beta, \mu_{ijt}(\phi))) = - \big(y_{ijt} \log(g_{ijt}(\beta, \mu_{ijt}(\phi))) + (1 - y_{ijt}) \log(1 - g_{ijt}(\beta, \mu_{ijt}(\phi)))\big) \, ,
    \end{equation*}
    so that $(\dop{1} \psi)_{ijt} = - a_{ijt} (y_{ijt} - g_{ijt})$ and $(\dop{2} \psi)_{ijt} = a_{ijt} g_{ijt}^{\prime} - a_{ijt}^{\prime} (y_{ijt} - g_{ijt})$. By the tower property of conditional expectations, the conditional mean assumption $\mathbb{E}\big[y_{ijt} \mid \mathcal{X}_{ij}^{t}\big] = g_{ijt}$, and the fact that $y_{ijt}^{2} = y_{ijt}$, it follows that
    \begin{align*}
        &\EX{\{(\dop{1} \psi)_{ijt}\}^{2}} = \EX{a_{ijt}^{2} (y_{ijt}^{2} - g_{ijt}^{2})} = \EX{a_{ijt}^{2} (y_{ijt} - g_{ijt}^{2})} = \EX{a_{ijt}^{2} g_{ijt} (1 - g_{ijt})} = \\
        & \quad \EX{a_{ijt} g_{ijt}^{\prime}} = \EX{(\dop{2}_{\mathcal{X}} \psi)_{ijt}} \, .
    \end{align*}
\end{remark}

\begin{remark}[No Predetermined Regressors]
    \label{remark:strictly_exogenous}
    When none of the regressors is predetermined (i.e., all regressors are strictly exogenous), the conditional mean assumption in Assumption \ref{assumption:general} (iii) can be strengthened to
    \begin{equation*}
        \EX{y_{ijt} \mid \mathcal{X}_{ij}^{T}} = g(\pi_{ijt}(\beta_{s}^{0}, \mu_{ijt}(\phi_{s}^{0})))
    \end{equation*}
    for all $i, j, t, \lvert \mathcal{D}_{s} \rvert, T$. Accordingly, the bias term $\overline{B}_{1, \rho, \infty}$ simplifies to
    \begin{align*}
        & \overline{B}_{1, \rho, \infty} = \overline{\mathbb{E}}\left[ - \frac{1}{N_{1} N_{2}} \sum_{i = 1}^{N_{1}} \sum_{j = 1}^{N_{2}} \frac{\sum_{t = 1}^{T} \EX{\ddot{x}_{ijt} (\dop{2} \psi)_{ijt} (\dop{1} \psi)_{ijt}}}{\sum_{t = 1}^{T} \EX{(\dop{2}_{\mathcal{X}} \psi)_{ijt}}} \right. \, + \\
        & \quad \left. \frac{1}{2 \, N_{1} N_{2}} \sum_{i = 1}^{N_{1}} \sum_{j = 1}^{N_{2}} \frac{\big\{\sum_{t = 1}^{T} \EX{\ddot{x}_{ijt} (\dop{3}_{\mathcal{X}} \psi)_{ijt}}\big\} \Big\{\sum_{t = 1}^{T} \EX{\big\{(\dop{1} \psi)_{ijt}\big\}^{2}}\Big\}}{\big\{\sum_{t = 1}^{T} \EX{(\dop{2}_{\mathcal{X}} \psi)_{ijt}}\big\}^{2}} \right] \, .
    \end{align*}
    Strict exogeneity is a natural assumption when the time dimension is replaced by a third cross-sectional dimension, as in a tripartite data structure.
\end{remark}

\begin{remark}[OLS and (Pseudo-)Poisson ML Estimators]
\label{remark:corollary_theorem_interacted}
    For OLS and (pseudo-)Poisson maximum likelihood (PPML) estimators, the bias expressions in Theorem \ref{theorem:asymptotic_distributions_interacted} simplify. 
    \begin{enumerate}[(i)]
        \item Several terms vanish because $\sum_{j = 1}^{N_{2}} \EX{\ddot{x}_{ijt} (\dop{3}_{\mathcal{X}} \psi)_{ijt}} = \mathbf{0}_{K}$, $\sum_{i = 1}^{N_{1}} \EX{\ddot{x}_{ijt} (\dop{3}_{\mathcal{X}} \psi)_{ijt}} = \mathbf{0}_{K}$, and $\sum_{t = 1}^{T} \EX{\ddot{x}_{ijt} (\dop{3}_{\mathcal{X}} \psi)_{ijt}} = \mathbf{0}_{K}$ for all $i, j, t, \lvert \mathcal{D}_{s} \rvert, T$. For OLS, this follows because $(\dop{2} \psi)_{ijt} = 1$, and hence $(\dop{3} \psi)_{ijt} = 0$ for all $i, j, t, \lvert \mathcal{D}_{s} \rvert, T$. For PPML, it follows from the weighted least-squares problem \eqref{eq:population_wls_program} together with $(\dop{3}_{\mathcal{X}} \psi)_{ijt} = (\dop{2}_{\mathcal{X}} \psi)_{ijt}$ for all $i, j, t, \lvert \mathcal{D}_{s} \rvert, T$. These results also hold for $s = 2$ and $s = 3$.
        \item If none of the regressors is predetermined (see Remark \ref{remark:strictly_exogenous}), we obtain correctly centered, nondegenerate asymptotic distributions. The first terms of each of the three bias expressions, $\overline{B}_{1, \alpha, \infty}$, $\overline{B}_{1, \gamma, \infty}$, and $\overline{B}_{1, \rho, \infty}$, vanish by the law of iterated expectations, since $\mathbb{E}\big[\ddot{x}_{ijt} (\dop{2}_{\mathcal{X}} \psi)_{ijt} \mid \mathcal{X}_{ij}^{T}\big] = \ddot{x}_{ijt} (\dop{2}_{\mathcal{X}} \psi)_{ijt}$ and $\mathbb{E}\big[(\dop{1} \psi)_{ijt} \mid \mathcal{X}_{ij}^{T}\big] = 0$. These results also hold for $s = 2$ and $s = 3$. Hence, for OLS and PPML estimators under strict exogeneity and any $s \in \{1, 2, 3\}$,
        \begin{equation*}
            \sqrt{n_{s}} (\hat{\beta}_{s} - \beta_{s}^{0}) \xrightarrow{d} \N\big(0, \overline{W}_{s, \infty}^{- 1} \overline{\Sigma}_{s, \infty} \overline{W}_{s, \infty}^{- 1}\big) \, .
            \end{equation*}
        The expressions for $s = 2$ and $s = 3$ are stated in Appendix \ref{appendix:main_results_distributions_interacted}. This also verifies Remark 2 in \textcite{wz2021}.
    \end{enumerate}
\end{remark}

\begin{remark}[Reciprocity in Directed Networks]
    \label{remark:reciprocity_directed_network}
    In the directed network case ($s = 2$), the framework can be extended to allow for reciprocity, that is, pairwise dependence between the dyads $(i, j)$ and $(j, i)$. Let $z_{ijt}^{\ast} \coloneqq (z_{ijt}, z_{jit})$, $x_{ijt}^{\ast} \coloneqq (x_{ijt}, x_{jit})$, and $\mathcal{X}_{ij}^{t, \ast} \coloneqq \sigma(\{x_{ijt^{\prime}}^{\ast} \colon t^{\prime} \in \{1, \ldots, t\}\})$. We impose Assumption \ref{assumption:general} (ii) and (iii) on the reciprocal pair rather than the dyad: conditional on $\Phi$, $\{\{z_{ijt}^{\ast}\}_{t = 1}^{T} \colon (i, j) \in \mathcal{D}_{2}, \, i < j\}$ is independent across pairs and $\alpha$-mixing in $t$ at the stated rate, and the conditional mean restriction holds with respect to $\mathcal{X}_{ij}^{t, \ast}$. Reciprocity does not affect the asymptotic bias, but it does affect $\overline{\Sigma}_{2,\infty}$, which is replaced everywhere by
    \begin{equation*}
        \overline{\Sigma}_{2, \infty}^{\ast} \coloneqq \LEX{\frac{1}{N_{1} (N_{1} - 1) T} \sum_{i = 1}^{N_{1}} \sum_{j \neq i} \sum_{t = 1}^{T} \EX{\big\{(\dop{1} \psi)_{ijt} \ddot{x}_{ijt} + (\dop{1} \psi)_{jit} \ddot{x}_{jit}\big\} \, (\dop{1} \psi)_{ijt} \ddot{x}_{ijt}^{\prime}}} \, .
    \end{equation*}
\end{remark}

Theorem \ref{theorem:asymptotic_distributions_interacted} shows that uncorrected estimators under the interacted specification have degenerate asymptotic distributions under Assumption \ref{assumption:general}, except for the special cases in Remark \ref{remark:corollary_theorem_interacted}. Hence, any inference based on these uncorrected estimators is generally invalid. To address this, we propose debiased estimators that enable asymptotically valid hypothesis tests and confidence intervals.

For each $s \in \{1, 2, 3\}$, we adopt an analytical bias correction: the uncorrected estimator $\hat{\beta}_{s}$ is adjusted by an estimate of the bias terms. This estimate is constructed using sample analogues of the expressions in Theorem \ref{theorem:asymptotic_distributions_interacted}, with the true parameter values replaced by the corresponding fixed effects estimates. In particular, for $\mathrm{r} \in \{1, 2\}$, let $(\widehat{\dop{\mathrm{r}} \psi})_{ijt}$ denote the sample analogue of $(\dop{\mathrm{r}} \psi)_{ijt}$. For $\mathrm{r} \in \{2, 3\}$, let $(\widehat{\dop{r}_{\mathcal{X}} \psi})_{ijt}$ denote the sample analogue of $(\dop{r}_{\mathcal{X}} \psi)_{ijt}$. Moreover, for each $k \in \{1, \ldots, K\}$, let
\begin{equation}
    \label{eq:sample_wls_program}
    \hat{\xi}_{k} \coloneqq \underset{\xi \in \Real^{L}}{\argmin} \frac{1}{w_{n, s}} \bigg\{ \sum_{(i, j) \in \mathcal{D}_{s}} \sum_{t = 1}^{T} (\widehat{\dop{2}_{\mathcal{X}} \psi})_{ijt} \left(x_{ijt, k} - \mu_{ijt}(\xi)\right)^{2} + \frac{\xi^{\prime} V V^{\prime} \xi}{2} \bigg\} 
\end{equation}
be the sample analogue of \eqref{eq:population_wls_program}. The resulting $n_{s} \times K$ matrix of fitted values is denoted by $\widehat{\mathfrak{X}}$, with its $ijt$-th row defined as $\hat{\mathfrak{x}}_{ijt} \coloneqq (\mu_{ijt}(\hat{\xi}_{1}), \ldots, \mu_{ijt}(\hat{\xi}_{K}))$. We set $\hat{\ddot{X}} \coloneqq X - \widehat{\mathfrak{X}}$, so that $\hat{\ddot{x}}_{ijt} = x_{ijt} - \hat{\mathfrak{x}}_{ijt}$. We denote the debiased estimator by $\tilde{\beta}_{s}$.

Theorem \ref{theorem:asymptotic_distributions_interacted_debiased} establishes that the debiased estimators have nondegenerate asymptotic distributions and are asymptotically unbiased. It also establishes the consistency of the components of the asymptotic variance. To estimate the spectral expectations in $\overline{B}_{s, \rho, \infty}$ (from Theorem \ref{theorem:asymptotic_distributions_interacted}), we adapt the truncated spectral density estimator of \textcites{hk2007}{hk2011} and include the finite-sample adjustment of \textcite{fw2016}. The corresponding bandwidth parameter is denoted by $h$.
\begin{theorem}[Asymptotic Distributions of $\tilde{\beta}_{s}$ - Interacted Specification]
    \label{theorem:asymptotic_distributions_interacted_debiased}
    Let Assumptions \ref{assumption:general} hold. Then,
    \begin{equation*}
        \widehat{\Sigma}_{s} \xrightarrow{p} \overline{\Sigma}_{s, \infty} \quad \text{and} \quad \widehat{W}_{s}^{- 1} \xrightarrow{p} \overline{W}_{s, \infty}^{- 1} \, .
    \end{equation*}
    If, in addition, $h \asymp T^{1 / 10}$, then, for any $s \in \{1, 2, 3\}$,
    \begin{equation*}
        \sqrt{n_{s}} \, (\tilde{\beta}_{s} - \beta_{s}^{0}) \xrightarrow{d} \N\big(0, \overline{W}_{s, \infty}^{- 1} \overline{\Sigma}_{s, \infty} \overline{W}_{s, \infty}^{- 1}\big) \, ,
    \end{equation*}
    where 
    \begin{align*}
        \tilde{\beta}_{1} =& \, \hat{\beta}_{1} + \widehat{W}_{1}^{- 1} \big( N_{2}^{- 1} \, \widehat{B}_{1, \alpha} + N_{1}^{- 1} \, \widehat{B}_{1, \gamma} + T^{- 1} \, \widehat{B}_{1, \rho} \big) \, , \\
        \tilde{\beta}_{2} =& \, \hat{\beta}_{2} + \widehat{W}_{2}^{- 1} \big( N_{1}^{- 1} \, \widehat{B}_{2, \alpha} + N_{1}^{- 1} \, \widehat{B}_{2, \gamma} + T^{- 1} \, \widehat{B}_{2, \rho} \big) \, , \\
        \tilde{\beta}_{3} =& \, \hat{\beta}_{3} + \widehat{W}_{3}^{- 1} \big( N_{1}^{- 1} \, \widehat{B}_{3, \alpha} + T^{- 1} \, \widehat{B}_{3, \rho} \big) \, , 
    \end{align*}
    with
    \begin{align*}
        & \widehat{B}_{1, \alpha} \coloneqq  - \frac{1}{N_{1} T} \sum_{i = 1}^{N_{1}} \sum_{t = 1}^{T} \frac{\sum_{j = 1}^{N_{2}} \hat{\ddot{x}}_{ijt} (\widehat{\dop{2} \psi})_{ijt} (\widehat{\dop{1} \psi})_{ijt}}{\sum_{j = 1}^{N_{2}} (\widehat{\dop{2}_{\mathcal{X}} \psi})_{ijt}} \, + \\
        & \quad \frac{1}{2 \, N_{1} T} \sum_{i = 1}^{N_{1}} \sum_{t = 1}^{T} \frac{\big\{\sum_{j = 1}^{N_{2}} \hat{\ddot{x}}_{ijt} (\widehat{\dop{3}_{\mathcal{X}} \psi})_{ijt} \big\} \sum_{j = 1}^{N_{2}} \big\{(\widehat{\dop{1} \psi})_{ijt}\big\}^{2}}{\big\{\sum_{j = 1}^{N_{2}} (\widehat{\dop{2}_{\mathcal{X}} \psi})_{ijt}\big\}^{2}}  \, , \\
        & \widehat{B}_{1, \gamma} \coloneqq  - \frac{1}{N_{2} T} \sum_{j = 1}^{N_{2}} \sum_{t = 1}^{T} \frac{\sum_{i = 1}^{N_{1}} \hat{\ddot{x}}_{ijt} (\widehat{\dop{2} \psi})_{ijt} (\widehat{\dop{1} \psi})_{ijt}}{\sum_{i = 1}^{N_{1}} (\widehat{\dop{2}_{\mathcal{X}} \psi})_{ijt}}  \, + \\
        & \frac{1}{2 \, N_{2} T} \sum_{j = 1}^{N_{2}} \sum_{t = 1}^{T} \frac{\big\{\sum_{i = 1}^{N_{1}} \hat{\ddot{x}}_{ijt} (\widehat{\dop{3}_{\mathcal{X}} \psi})_{ijt} \big\} \sum_{i = 1}^{N_{1}} \big\{(\widehat{\dop{1} \psi})_{ijt}\big\}^{2}}{\big\{\sum_{i = 1}^{N_{1}} (\widehat{\dop{2}_{\mathcal{X}} \psi})_{ijt}\big\}^{2}} \, , \\
        & \widehat{B}_{1, \rho} \coloneqq  - \frac{1}{N_{1} N_{2}} \sum_{i = 1}^{N_{1}} \sum_{j = 1}^{N_{2}} \frac{\sum_{m = 0}^{h} T / (T - m) \sum_{t = m + 1}^{T} \hat{\ddot{x}}_{ijt} (\widehat{\dop{2} \psi})_{ijt} (\widehat{\dop{1} \psi})_{ij(t - q)}}{\sum_{t = 1}^{T} (\widehat{\dop{2}_{\mathcal{X}} \psi})_{ijt}} \, + \\
        & \quad \frac{1}{2 \, N_{1} N_{2}} \sum_{i = 1}^{N_{1}} \sum_{j = 1}^{N_{2}} \frac{\big\{\sum_{t = 1}^{T} \hat{\ddot{x}}_{ijt} (\widehat{\dop{3}_{\mathcal{X}} \psi})_{ijt} \big\} \Big\{\sum_{t = 1}^{T} \big\{(\widehat{\dop{1} \psi})_{ijt}\big\}^{2}\Big\}}{\big\{\sum_{t = 1}^{T} (\widehat{\dop{2}_{\mathcal{X}} \psi})_{ijt}\big\}^{2}} \, , \\
        & \widehat{W}_{1} \coloneqq \frac{1}{N_{1} N_{2} T} \sum_{i = 1}^{N_{1}} \sum_{j = 1}^{N_{2}} \sum_{t = 1}^{T} (\widehat{\dop{2}_{\mathcal{X}} \psi})_{ijt} \, \hat{\ddot{x}}_{ijt} \, \hat{\ddot{x}}_{ijt}^{\prime} \, , \\
        & \widehat{\Sigma}_{1} \coloneqq \frac{1}{N_{1} N_{2} T} \sum_{i = 1}^{N_{1}} \sum_{j = 1}^{N_{2}} \sum_{t = 1}^{T} \big\{(\widehat{\dop{1} \psi})_{ijt}\big\}^{2} \, \hat{\ddot{x}}_{ijt} \, \hat{\ddot{x}}_{ijt}^{\prime} \, .
    \end{align*}
    The expressions for $s = 2$ and $s = 3$ are stated in Appendix \ref{appendix:main_results_distributions_interacted_debiased}.
\end{theorem}

\begin{remark}[Theorem \ref{theorem:asymptotic_distributions_interacted_debiased}]
    \label{remark:theorem_asymptotic_distributions_interacted_debiased}
    The condition $h \asymp T^{1 / 10}$ can be relaxed at the cost of stronger moment restrictions. Following \textcite{fw2016}, we recommend reporting results for several bandwidths. Importantly, constructing the debiased estimator does not require knowing whether regressors are predetermined, i.e., the proposed estimators for the bias and variance components are consistent in either case. Finally, the bias formulas in Theorem \ref{theorem:asymptotic_distributions_interacted_debiased} formally justify the conjecture of \textcite{hsw2020}, who proposed a bias correction for three-way fixed effects binary choice models based on the heuristic of \textcite{fw2018} without formal derivation.
\end{remark}

\begin{remark}[Jackknife Inference]
    As an alternative to the analytical bias correction, one may conduct inference using the jackknife $t_{q}$-statistic of \textcite{h2026_jackknife}, at the cost of an additional unconditional homogeneity assumption. Under this assumption, the asymptotic expansion underlying Theorems \ref{theorem:asymptotic_distributions_interacted} and \ref{theorem:asymptotic_distributions_interacted_debiased} satisfies both Assumption AD$^{\dagger}$ and Assumption JK$^{\dagger}$ of \textcite{h2026_jackknife}, and the result $J_{q} \xrightarrow{d} t_{q}$ for $q \in \{1, 2, 3\}$ follows from his Example 3, where $J_{q}$ denotes the jackknife $t_{q}$-statistic. We refer to Appendix \ref{supplement:jackknife_inference} for further details. The jackknife approach is tuning-parameter-free, i.e., no bandwidth selection is necessary, and does not require the explicit bias formulas derived in this paper. However, it still requires knowledge of the asymptotic bias structure, derived in this paper, to form suitable subsamples.
\end{remark}
\vspace{0.5em}

\paragraph{Non-interacted Specification.} Theorem \ref{theorem:asymptotic_distributions_noninteracted} establishes that, for specification (3.a), estimators $\hat{\beta}_{s}$ have nondegenerate asymptotic distributions and are asymptotically unbiased for any $s \in \{1, 2, 3\}$.
\begin{theorem}[Asymptotic Distributions of $\hat{\beta}_{s}$ - Non-interacted Specification]
    \label{theorem:asymptotic_distributions_noninteracted}
    Let Assumptions \ref{assumption:general} hold. Then, for any $s \in \{1, 2, 3\}$,
    \begin{equation*}
        \sqrt{n_{s}} \, (\hat{\beta}_{s} - \beta_{s}^{0}) \xrightarrow{d} \N\big(0, \overline{W}_{s, \infty}^{- 1} \overline{\Sigma}_{s, \infty} \overline{W}_{s, \infty}^{- 1}\big)
    \end{equation*}
    and
    \begin{equation*}
        \widehat{\Sigma}_{s} \xrightarrow{p} \overline{\Sigma}_{s, \infty} \quad \text{and} \quad \widehat{W}_{s}^{- 1} \xrightarrow{p} \overline{W}_{s, \infty}^{- 1} \, .
    \end{equation*}
\end{theorem}
Theorem \ref{theorem:asymptotic_distributions_noninteracted} shows that debiasing is unnecessary for non-interacted specifications. Notably, this result holds even for nonlinear models such as binary choice models, where incidental parameter bias is typically an issue in two-dimensional panels, and for models with predetermined regressors that would otherwise require a Nickell-type bias correction. To explain this finding heuristically, we consider the case of a bipartite panel ($s = 1$). Although $\hat{\alpha}^{\star}$, $\hat{\gamma}^{\star}$, and $\hat{\rho}^{\star}$ converge at rates slower than $\sqrt{N_{1} N_{2} T}$, the bias of $\hat{\beta}$ is of order $1 / (N_{2} T) + 1 / (N_{1} T) + 1 / (N_{1} N_{2})$, which shrinks faster than $1 / \sqrt{N_{1} N_{2} T}$. The bias is therefore asymptotically negligible. This is consistent with the conjecture of \textcite{fw2018} (Example 11).

\subsection{Other Fixed Effects Specifications}
\label{sec:asymptotic_theory_other}

The three-way fixed effects specifications are the most general cases. The asymptotic results for the remaining 15 one- and two-way structures in Table \ref{tab:linear_combinations_phi} follow as special cases. We discuss each in turn.

\paragraph{Interacted Specifications (one- and two-way).} For specifications with only interacted unobserved heterogeneity (Table \ref{tab:linear_combinations_phi} (1.1.b)--(1.3.b), (2.1.b)--(2.3.b)), the asymptotic distribution remains degenerate, and the bias components from Theorem \ref{theorem:asymptotic_distributions_interacted} induced by the respective incidental parameters carry over. For illustration, consider specification (2.1.b), $\alpha_{it}^{0} + \gamma_{jt}^{0}$. Let Assumptions \ref{assumption:general} hold. For any $s \in \{1, 2, 3\}$,
\begin{equation*}
    \frac{\sqrt{n_{s}}}{\sqrt{T}} \, (\hat{\beta}_{s} - \beta_{s}^{0}) \xrightarrow{d} \delta_{\overline{b}_{s, \infty}} \, ,
\end{equation*}
where
\begin{align*}
    &\overline{b}_{1, \infty} \coloneqq - \, \overline{W}_{1, \infty}^{- 1} \big( \tau_{1}^{\frac{1}{2}} \, \tau_{2}^{- \frac{1}{2}} \, \overline{B}_{1, \alpha, \infty} + \tau_{1}^{- \frac{1}{2}} \, \tau_{2}^{\frac{1}{2}} \, \overline{B}_{1, \gamma, \infty} \big) \, , \\
    &\overline{b}_{2, \infty} \coloneqq - \, \overline{W}_{2, \infty}^{- 1} \big( \overline{B}_{2, \alpha, \infty} + \overline{B}_{2, \gamma, \infty}  \big) \, , \quad \overline{b}_{3, \infty} \coloneqq - \, \overline{W}_{3, \infty}^{- 1} \big( \overline{B}_{3, \alpha, \infty}  \big) \, .
\end{align*}
These expressions are the same as in Theorem \ref{theorem:asymptotic_distributions_interacted}, except that the $\rho^{0}$-component is absent and quantities such as $\xi_{k}^{0}$ are adjusted for the new specification. Bias correction and inference proceed as in Theorem \ref{theorem:asymptotic_distributions_interacted_debiased}, with all components adjusted accordingly.

\paragraph{Non-interacted Specifications (one- and two-way).} For specifications with only non-interacted unobserved heterogeneity (Table \ref{tab:linear_combinations_phi} (1.1.a)--(1.3.a), (2.1.a)--(2.3.a)), the asymptotic distributions are nondegenerate and centered at zero, as in Theorem \ref{theorem:asymptotic_distributions_noninteracted}. For any $s \in \{1, 2, 3\}$,
\begin{equation*}
    \sqrt{n_{s}} \, (\hat{\beta}_{s} - \beta_{s}^{0}) \xrightarrow{d} \N\big(0, \overline{W}_{s, \infty}^{- 1} \overline{\Sigma}_{s, \infty} \overline{W}_{s, \infty}^{- 1}\big) \, .
\end{equation*}

\paragraph{Mixed Specifications.} For specifications with both non-interacted and interacted unobserved heterogeneity (Table \ref{tab:linear_combinations_phi} (2.1.c)--(2.3.c)), the slowly decaying bias from the interacted component dominates the faster-decaying bias from the non-interacted component. Hence, the estimators of $\beta_{s}^{0}$ have a degenerate asymptotic distribution, as in Theorem \ref{theorem:asymptotic_distributions_interacted}. As an example, consider specification (2.3.c), $\rho_{t}^{\star,0} + \rho_{ij}^{0}$, for which the bias component for $\rho^{0}$ from Theorem \ref{theorem:asymptotic_distributions_interacted} remains. Let Assumptions \ref{assumption:general} hold. For any $s \in \{1, 2, 3\}$,
\begin{equation*}
    \frac{\sqrt{n_{s}}}{\sqrt{T}} \, (\hat{\beta}_{s} - \beta_{s}^{0}) \xrightarrow{d} \delta_{\overline{b}_{s, \infty}} \, ,
\end{equation*}
where
\begin{equation*}
    \overline{b}_{1, \infty} \coloneqq - \, \overline{W}_{1, \infty}^{- 1} \big( \tau_{1}^{\frac{1}{2}} \, \tau_{2}^{\frac{1}{2}} \, \overline{B}_{1, \rho, \infty} \big) \, , \quad \overline{b}_{2, \infty} \coloneqq - \, \overline{W}_{2, \infty}^{- 1} \big(  \tau_{1} \, \overline{B}_{2, \rho, \infty} \big) \, , \quad \overline{b}_{3, \infty} \coloneqq - \, \overline{W}_{3, \infty}^{- 1} \big( \tau_{1} \, \overline{B}_{3, \rho, \infty} \big) \, .
\end{equation*}
We again follow Theorem \ref{theorem:asymptotic_distributions_interacted_debiased} to form the debiased estimator, with bias and variance components adjusted for the respective specification.

\section{Simulation Experiments}
\label{sec:simulation_experiments}

We study the finite-sample behavior of the uncorrected (FE) and the debiased three-way fixed effects maximum likelihood estimators in a series of simulation experiments. For debiasing, we consider the analytical bias corrections (ABC) and the jackknife (JK) approaches proposed by \textcite{h2026_jackknife}. Among the latter, we consider the minimum-variance unbiased (MVU) jackknife and the jackknife $t_{3}$-statistic. We report relative bias (in percent of the true value), bias relative to standard deviation, and coverage rates of confidence intervals with 95\% nominal level.\footnote{We calculate the bias as $R^{-1} \sum_{r=1}^{R} (b_{r} - \beta^{0})$, where $R$ is the number of Monte Carlo replications and $b_{r}$ is the estimate in the $r$-th replication.}

We adapt the dynamic data generating process of \textcite{fw2016} to a probit model for bipartite panels with three sets of unobserved effects. For $i \in \{1, \ldots, N_{1}\}$, $j \in \{1, \ldots, N_{2}\}$, and $t \in \{0, \ldots, T\}$,
\begin{equation*}
    y_{ijt} = \begin{cases}
        \ind \{ \beta_y^{0} \, y_{ij(t-1)} + \beta_x^{0} \, x_{ijt} + \mu_{ijt}(\phi^{0}) \geq u_{ijt}\} & \text{if } t \geq 1 \\
        \ind \{ \beta_x^{0} \, x_{ij0} +  \mu_{ij0}(\phi^{0}) \geq u_{ij0}\} & \text{if } t = 0
    \end{cases} \, ,
\end{equation*}
where $u_{ijt}, v_{ijt} \sim \iid \N(0, 1)$ and $x_{ijt} = \mu_{ijt}(\phi^{0}) + v_{ijt}$. DGP I has interacted unobserved effects, $\mu_{ijt}(\phi^{0}) = \alpha_{it}^{0} + \gamma_{jt}^{0} + \rho_{ij}^{0}$, with $\alpha_{it}^{0}, \gamma_{jt}^{0}, \rho_{ij}^{0} \sim \iid \N(0, 1 / 24)$. DGP II has non-interacted unobserved effects, $\mu_{ijt}(\phi^{0}) = \alpha_{i}^{\star,0} + \gamma_{j}^{\star,0} + \rho_{t}^{\star,0}$, with $\alpha_{i}^{\star,0}, \gamma_{j}^{\star,0}, \rho_{t}^{\star,0} \sim \iid \N(0, 1 / 24)$. We set $\beta_{y}^{0} = 0.5$ and $\beta_{x}^{0} = 1$, and generate samples with $N_{1} \in \{60, 120, 240\}$, $N_{2} = N_{1} / 2$, and $T(N_{1}) = N_{1} / 5$. This design ensures that $N_{1}$, $N_{2}$, and $T$ grow at constant relative rates, consistent with our asymptotic analysis. All results are based on $10{,}000$ simulated samples for each $N_{1}$.

Table \ref{tab:simulation_results_overlapping} reports the results for DGP I. For ABC, we consider the bandwidths $h \in \{0, 1, 2\}$. A bandwidth of $h = 0$ corrects only for the incidental parameter bias, whereas $h > 0$ additionally corrects for the Nickell-type bias.
\begin{table}[!t]
\centering
\caption{DGP I: Finite Sample Properties of Estimators for Model Parameters}
\label{tab:simulation_results_overlapping}
\begin{threeparttable}
\begin{tabular}{lrrrrrr}
  \toprule
Estimator & Bias (in \%) & Bias / SD  &  Coverage  & Bias (in \%) & Bias / SD  &  Coverage  \\ 
  \midrule
  & \multicolumn{6}{c}{($N_{1} = 60, N_{2} = 30 , T = 12$)}\\
  & \multicolumn{3}{c}{$\beta_x^{0}$}  & \multicolumn{3}{c}{$\beta_y^{0}$}\\
  \cmidrule(lr){2-4}\cmidrule(lr){5-7} 
   FE               & 21.902 & 10.028 & 0.000 & -41.031 & -6.937 & 0.000 \\ 
   ABC ($h = 0$)    & -1.560 & -1.008 & 0.832 & -52.592 & -10.999 & 0.000 \\ 
   ABC ($h = 1$)    & -0.590 & -0.376 & 0.933 & -3.662 & -0.760 & 0.913 \\ 
   ABC ($h = 2$)    & -0.145 & -0.091 & 0.944 & -2.835 & -0.550 & 0.921 \\ 
   JK (MVU)         & -14.936 & -7.285 & 0.000 & 15.031 & 2.411 & 0.210 \\ 
   JK ($t_{3}$)     & -14.936 & -7.285 & 0.050 & 15.031 & 2.411 & 0.671 \\ 
   \midrule
   & \multicolumn{6}{c}{($N_{1} = 120, N_{2} = 60 , T = 24$)}\\
   & \multicolumn{3}{c}{$\beta_x^{0}$}  & \multicolumn{3}{c}{$\beta_y^{0}$}\\
   \cmidrule(lr){2-4}\cmidrule(lr){5-7} 
   FE               & 9.128 & 14.979 & 0.000 & -19.048 & -10.482 & 0.000 \\ 
   ABC ($h = 0$)    & -0.336 & -0.632 & 0.897 & -26.064 & -15.725 & 0.000 \\ 
   ABC ($h = 1$)    & -0.113 & -0.212 & 0.940 & -2.284 & -1.376 & 0.741 \\ 
   ABC ($h = 2$)    & -0.007 & -0.013 & 0.946 & -0.407 & -0.238 & 0.946 \\ 
   JK (MVU)         & -2.462 & -4.359 & 0.005 & 2.496 & 1.354 & 0.680 \\ 
   JK ($t_{3}$)     & -2.462 & -4.359 & 0.270 & 2.496 & 1.354 & 0.844 \\ 
   \midrule
   & \multicolumn{6}{c}{($N_{1} = 240, N_{2} = 120 , T = 48$)}\\
   & \multicolumn{3}{c}{$\beta_x^{0}$}  & \multicolumn{3}{c}{$\beta_y^{0}$}\\
   \cmidrule(lr){2-4}\cmidrule(lr){5-7} 
   FE               & 4.228 & 22.096 & 0.000 & -9.237 & -15.286 & 0.000 \\ 
   ABC ($h = 0$)    & -0.084 & -0.465 & 0.925 & -12.982 & -22.434 & 0.000 \\ 
   ABC ($h = 1$)    & -0.031 & -0.171 & 0.948 & -1.361 & -2.352 & 0.362 \\ 
   ABC ($h = 2$)    & -0.005 & -0.027 & 0.950 & -0.156 & -0.266 & 0.944 \\ 
   JK (MVU)         & -0.497 & -2.689 & 0.220 & 0.513 & 0.846 & 0.849 \\ 
   JK ($t_{3}$)     & -0.497 & -2.689 & 0.597 & 0.513 & 0.846 & 0.908 \\
   \bottomrule
\end{tabular}
\begin{tablenotes}
    \footnotesize
    \item\textbf{Notes:} $h$ refers to bandwidth parameter of ABC; SD and Coverage indicate standard deviation and coverage rates of confidence intervals with 95\% nominal level, respectively; results based on $10{,}000$ simulated samples for each $(N_{1}, N_{2}, T)$.
\end{tablenotes}
\end{threeparttable}
\end{table}

We first examine the results for the strictly exogenous regressor in the left panel of Table \ref{tab:simulation_results_overlapping}. At the smallest sample size ($N_1 = 60, N_2 = 30, T = 12$), the uncorrected estimator has a relative bias of $21.902\%$. As theory predicts, this bias decreases with the sample size: it is of order $N_{1}^{-1} + N_{2}^{-1} + T^{-1}$ and falls to $4.228\%$ at the largest sample size ($N_1 = 240, N_2 = 120, T = 48$). It nonetheless remains large relative to the dispersion of the estimator. In fact, the bias-to-standard-deviation ratio increases with the sample size, so the problem worsens in relative terms. Coverage rates are accordingly zero throughout. As our asymptotic theory predicts, the uncorrected estimator exhibits a more severe bias problem than the estimators studied in, for example, \textcite{fw2016} and \textcite{wz2021}.

The left panel also reports results for the debiased estimators. Since $x_{ijt}$ is strictly exogenous, correcting only for the incidental parameter bias (ABC with $h = 0$) suffices in principle. We nevertheless report results for $h > 0$ as well, to reflect the more realistic scenario in which the researcher does not know whether the regressors are strictly exogenous. ABC with $h = 0$ outperforms the uncorrected estimator on every metric in every sample. At the smallest sample size, the bias falls from $21.902\%$ to $-1.560\%$ and coverage improves from zero to $83.2\%$. ABC with $h > 0$ also performs reliably for strictly exogenous regressors. Notably, it outperforms $h = 0$ at every sample size. At ($N_{1} = 60, N_{2} = 30, T = 12$), for example, coverage improves from $83.2\%$ to $93.3\%$. This might be because, in finite samples, the incidental parameter bias and the Nickell-type bias are not as cleanly separated as asymptotic theory suggests, due to correlation between the strictly exogenous and the predetermined regressor. This provides further motivation for the recommendation to use ABC with $h > 0$ in practice.

The two jackknife approaches share the same point estimate and differ only in how the confidence intervals are constructed. The jackknife substantially reduces bias, but is not as effective as the best-performing analytical correction ($h = 2$). Bias-to-standard-deviation ratios decrease and coverage rates increase with the sample size. The $t_{3}$-statistic approach clearly outperforms MVU.

The right panel of Table \ref{tab:simulation_results_overlapping} presents the results for the predetermined regressor. The qualitative pattern is the same as for the strictly exogenous regressor, but the biases of the uncorrected estimator are considerably larger because of the additional Nickell-type bias. Analytical corrections with $h > 0$ and the jackknife approaches substantially reduce these biases and improve coverage. Interestingly, the jackknife approaches deliver much better coverage for the predetermined regressor than for the strictly exogenous one. ABC with $h = 0$, which corrects for the incidental parameter bias but not the Nickell-type bias, yields unreliable inference, similar to no correction at all.

We turn next to DGP II, with results in Table \ref{tab:simulation_results_nonoverlapping}. As established by our theory, the uncorrected estimator with non-interacted fixed effects is consistent and has a correctly centered asymptotic distribution, even for predetermined regressors, so no bias correction is needed. Both panels confirm this: biases relative to standard deviation are negligible and coverage rates are close to the nominal $95\%$ level.
\begin{table}[!t]
\centering
\caption{DGP II: Finite Sample Properties of Uncorrected Estimator for Model Parameters}
\label{tab:simulation_results_nonoverlapping}
\begin{threeparttable}
\begin{tabular}{lrrrrrr}
  \toprule
$(N_{1}, N_{2}, T)$ & Bias (in \%) & Bias / SD  &  Coverage  & Bias (in \%) & Bias / SD  &  Coverage  \\ 
  \midrule
  & \multicolumn{3}{c}{$\beta_x^{0}$}  & \multicolumn{3}{c}{$\beta_y^{0}$}\\
  \cmidrule(lr){2-4}\cmidrule(lr){5-7} 
   (60, 30, 12)     & 0.594 & 0.412 & 0.924 & -1.728 & -0.375 & 0.929 \\ 
   (120, 60, 24)    & 0.139 & 0.279 & 0.944 & -0.492 & -0.305 & 0.941 \\ 
   (240, 120, 48)   & 0.035 & 0.197 & 0.945 & -0.118 & -0.209 & 0.944 \\ 
   \bottomrule
\end{tabular}
\begin{tablenotes}
    \footnotesize
    \item\textbf{Notes:} SD and Coverage indicate standard deviation and coverage rates of confidence intervals with 95\% nominal level, respectively; results based on $10{,}000$ simulated samples for each $(N_{1}, N_{2}, T)$.
\end{tablenotes}
\end{threeparttable}
\end{table}

Overall, the simulation experiments support the validity of our asymptotic results in samples of sufficient size. Based on the results for DGP I, we recommend that practitioners use an analytical correction with $h > 0$ whenever it is unclear whether the regressors are strictly exogenous. Practitioners who prefer a jackknife approach should use the $t_{3}$-statistic rather than the MVU.

\section{Empirical Application}
\label{sec:empirical_application}

We illustrate our inferential procedure using bilateral trade data on plastic articles (HS6 code 392690), a broad category of standardized plastic components. We adapt the dynamic network formation model of \textcite{g2016} to a \textit{directed} network setting. This product category is well-suited for two reasons. First, it has the highest country participation in the data, as nearly every country in the world produces or consumes products in this category. Second, and more importantly, plastic articles of this type can in principle be produced by any country, implying that the probability of a trade link forming is bounded away from zero for all country pairs. This satisfies Assumption \ref{assumption:general} (v), which requires $0 < \mathbb{P}\big(y_{ijt} = 1 \mid \mathcal{X}_{ij}^{t,\ast}\big) < 1$ for all $i, j, t, N_{1}, T$, i.e., the network is dense.

\paragraph{Dynamic Model of Network Formation.} \textcite{g2016} proposes a dynamic model of network formation for \textit{undirected} network panel data in which the probability of a pair $(i,j)$ forming a link at time $t$ increases if (i) the pair was already directly connected in the previous period (\textit{state dependence}), (ii) the pair shared many links in common in the previous period (a taste for \textit{transitivity}), and (iii) the pair shares unobserved time-invariant characteristics (\textit{homophily}). Channels (i) and (ii) are observed and enter the utility function directly through the lagged link indicator $y_{ij(t - 1)}$ and the lagged count of two-paths $r_{ij(t - 1)}$, respectively.

Trade networks are inherently directed: exporting to a country is structurally distinct from importing from it, so the undirected framework of \textcite{g2016} is not directly applicable. Directed network data has two consequences for the model. First, the single lagged two-path count $r_{ij(t - 1)}$ of \textcite{g2016} decomposes into four distinct triadic variables:
\begin{align*}
    &r_{1, ij(t-1)} \coloneqq \sum_{k \neq i, j} y_{ik(t-1)} y_{kj(t-1)} \, , \qquad r_{2, ij(t-1)} \coloneqq \sum_{k \neq i, j} y_{ki(t-1)} y_{jk(t-1)} \, , \\
    &r_{3, ij(t-1)} \coloneqq \sum_{k \neq i, j} y_{ki(t-1)} y_{kj(t-1)} \, , \qquad r_{4, ij(t-1)} \coloneqq \sum_{k \neq i, j} y_{ik(t-1)} y_{jk(t-1)} \, .
\end{align*}
In trade networks, these triadic variables can be interpreted as \textit{supply chain transitivity}, \textit{cyclic closure}, \textit{shared upstream suppliers}, and \textit{shared export markets}, respectively. Second, the directed setting introduces a natural analogue of undirected link persistence: the lagged reverse-direction link $y_{ji(t-1)}$, which we refer to as \textit{dynamic reciprocity}. If country $j$ exported to country $i$ in the previous period, existing bilateral infrastructure, such as shipping routes, lowers the cost of establishing a link in the reverse direction. 

Based on these considerations, we model whether exporter $i$ and importer $j$ have a trading relationship at time $t$ as:
\begin{equation}
    \label{eq:dynamic_network_model}
    \EX{y_{ijt} \mid \mathcal{X}_{ij}^{t,\ast}} = F_{\epsilon}\big(\beta_{1}^{0} \, y_{ij(t-1)} + \beta_{2}^{0} \, y_{ji(t-1)} + \beta_{3}^{0} \, r_{1,ij(t-1)} + \beta_{4}^{0} \, r_{3,ij(t-1)} + \alpha_{it}^{0} + \gamma_{jt}^{0} + \rho_{ij}^{0}\big) \, ,
\end{equation}
where $\alpha_{it}^{0}$ and $\gamma_{jt}^{0}$ capture time-varying out-degree and in-degree unobserved heterogeneity, $\rho_{ij}^{0}$ is pair-specific unobserved heterogeneity, and $F_{\epsilon}(\pi)$ denotes the standard normal CDF. We include $r_{1, ij(t-1)}$ and $r_{3,ij(t-1)}$ as the two triadic variables with the clearest economic motivation in a trade setting: supply chain transitivity, where an indirect export chain $i \to k \to j$ raises the probability of a direct link $i \to j$, and shared upstream suppliers, where two countries sourcing from a common supplier ($k \to i$  and $k \to  j$ ) are more likely to establish direct bilateral trade. We exclude the cyclic closure term $r_{2, ij(t-1)}$, which has no compelling trade interpretation for a homogeneous good, and the shared-export-market term $r_{4, ij(t-1)}$, which is theoretically ambiguous due to competing effects.\footnote{When we include $r_{2,ij(t-1)}$ and $r_{4,ij(t-1)}$ in our empirical specification, we find insignificant coefficients close to zero.} We account for pairwise clustering (see Remark \ref{remark:reciprocity_directed_network}).

\paragraph{Data.} We use the \textit{BACI international trade database}, which provides harmonized bilateral trade flows at the HS6 level.\footnote{BACI reconciles import and export declarations reported to UN Comtrade. The reconciliation procedure effectively discards flows that are very small or inconsistently reported. See \textcite{baci_database} for details.} Our sample covers the period 1995--2019 ($T = 25$), includes $N_{1} = 233$ countries, and defines $y_{ijt} = 1$ if country $i$ reports positive exports to country $j$ in year $t$, and zero otherwise. The panel is unbalanced.

The network becomes markedly denser over the sample period: in 1995, 5,796 of 45,156 directed country pairs traded (12.8\%), rising to 11,461 of 49,506 pairs (23.2\%) in 2006 and 14,823 of 50,850 pairs (29.2\%) in 2019. This near-doubling of network density over 25 years reflects the broad globalization of trade in this product category. It also motivates the dynamic specification in \eqref{eq:dynamic_network_model}: the gradual densification suggests that new trade relationships do not form in isolation, but build on existing ones through state dependence and triadic closure, as new exporters enter markets where indirect trade chains or shared suppliers are already in place.

\paragraph{Results.} We estimate \eqref{eq:dynamic_network_model} by probit maximum likelihood. Table \ref{tab:empirical_results} reports the estimates, comparing the uncorrected estimator against the debiased estimator for bandwidths $h \in \{1, \dots , 4\}$.\footnote{Obtaining uncorrected estimates takes roughly 20 seconds on a standard laptop computer using \textit{alpaca}. Debiasing the estimates for $h = 4$ takes an additional 7 seconds.}
\begin{table}[!htbp]
\centering
\caption{Probit Coefficients: Dynamic Formation of a Directed Trade Network}
\label{tab:empirical_results}
\begin{threeparttable}
\begin{tabular}{llllll}
\toprule
\multicolumn{6}{c}{Dependent variable: trade relationship $i \to j$} \\
\midrule
 & Uncorrected & \multicolumn{4}{c}{Debiased} \\
\cmidrule(lr){3-6}
 &  & $h=1$ & $h=2$ & $h=3$ & $h=4$ \\
\midrule
State dependence & $0.337^{\ast\ast\ast}$ & $0.485^{\ast\ast\ast}$ & $0.500^{\ast\ast\ast}$ & $0.498^{\ast\ast\ast}$ & $0.490^{\ast\ast\ast}$ \\ 
   & (0.007) & (0.007) & (0.007) & (0.007) & (0.007) \\[0.5em]
Dynamic reciprocity & $0.088^{\ast\ast\ast}$ & $0.079^{\ast\ast\ast}$ & $0.083^{\ast\ast\ast}$ & $0.084^{\ast\ast\ast}$ & $0.085^{\ast\ast\ast}$ \\ 
  & (0.010) & (0.010) & (0.010) & (0.010) & (0.010) \\[0.5em]
Supply chain transitivity & $0.019^{\ast\ast\ast}$ & $0.017^{\ast\ast\ast}$ & $0.017^{\ast\ast\ast}$ & $0.017^{\ast\ast\ast}$ & $0.017^{\ast\ast\ast}$ \\ 
  & (0.001) & (0.001) & (0.001) & (0.001) & (0.001) \\[0.5em] 
Shared upstream suppliers & $0.015^{\ast\ast\ast}$ & $0.013^{\ast\ast\ast}$ & $0.013^{\ast\ast\ast}$ & $0.013^{\ast\ast\ast}$ & $0.013^{\ast\ast\ast}$ \\ 
  & (0.001) & (0.001) & (0.001) & (0.001) & (0.001) \\ 
\bottomrule
\end{tabular}
\begin{tablenotes}
\footnotesize
\item \textbf{Notes:} $h$ refers to bandwidth parameter of debiased estimator; clustered standard errors in parentheses (see Remark \ref{remark:reciprocity_directed_network}); significance levels: 
$^{\ast\ast\ast}$ $p<0.01$, $^{\ast\ast}$ $p<0.05$, $^{\ast}$ $p<0.1$. 
\end{tablenotes}
\end{threeparttable}
\end{table}
The debiased estimates are stable across all values of $h$, which serves as a sensitivity check on the bandwidth choice. All estimates are positive and statistically significant under both estimators. The estimates of $\beta_{1}^{0}$ indicate strong persistence in bilateral trade relationships, consistent with sunk costs at the firm level generating state dependence \parencite{hsw2020}. The uncorrected estimate of $\beta_{1}^{0}$ is substantially smaller than the debiased estimates across all bandwidths, consistent with a downward Nickell-type bias. The estimates of $\beta_{2}^{0}$ confirm that an existing import relationship from $j$ to $i$ raises the propensity of a subsequent export link from $i$ to $j$, consistent with bilateral trade infrastructure lowering costs in both directions. The estimates of $\beta_{3}^{0}$ and $\beta_{4}^{0}$ provide evidence for both triadic channels: supply chain transitivity and shared upstream suppliers each raise the probability of a direct trade link forming, in line with global value chain mechanisms. The comparable magnitudes of the estimates for $\beta_{3}^{0}$ and $\beta_{4}^{0}$ suggest that both channels contribute similarly to network densification. For $\beta_{2}^{0}$, $\beta_{3}^{0}$, and $\beta_{4}^{0}$, the difference between the uncorrected and debiased estimates is less pronounced but still relatively large compared to the very small standard errors. 

\section{Concluding Remarks}
\label{sec:concluding_remarks}

We provide the first extensive theoretical analysis of fixed effects M-estimators for three-dimensional panel data across different data structures and specifications of multi-way unobserved heterogeneity. Our analysis reveals a sharp dichotomy: non-interacted specifications are free of asymptotic bias, whereas interacted specifications suffer from a severe inference problem manifested in degenerate asymptotic distributions, except for OLS and (pseudo-)Poisson maximum likelihood estimators with strictly exogenous regressors. The choice of heterogeneity specification and estimator therefore has first-order consequences for inference. We resolve the inference problem with analytical debiased estimators that have nondegenerate, correctly centered asymptotic distributions. Simulation evidence confirms good finite-sample performance, and our empirical application shows that the correction is economically meaningful.

We see several promising extensions for future work. First, for some models, in particular binary choice models, the extension of our analysis to average partial effects would be valuable, as these provide directly interpretable quantities. A treatment of average partial effects, however, cannot be handled by a common set of assumptions across several fixed effects specifications as we used for $\beta^{0}$ (see Assumption \ref{assumption:general}). \textcite{fw2016} show that the distribution of the unobserved effects is important for the properties of estimators for average partial effects, and that different sampling assumptions lead to different properties. Extending our analysis to average partial effects would therefore require different sampling assumptions for each fixed effects specification, and an analysis of how these assumptions affect the properties of the resulting estimators. Our general expansions developed in the appendix  provide a starting point to derive the corresponding theory. Second, we conjecture that for some specifications, it is possible to relax the asymptotics, e.g., to derive results under fixed-$T$ asymptotics. Third, recall that \textcite{wz2021} build on the two-dimensional panel results of \textcite{fw2016} to derive the properties of the three-way fixed effects (pseudo-)Poisson maximum likelihood estimator in the three-dimensional panel setting. Analogously, we conjecture that our three-dimensional panel results will be helpful in deriving the properties of (pseudo-)Poisson maximum likelihood estimators for four-dimensional panel data models with  four-way, triple interacted unobserved heterogeneity under a strict exogeneity assumption and fixed-$T$ asymptotics. Such an extension would allow researchers to study bilateral trade flows at the product level using exporter-importer-product-time panel data.

\clearpage
\printbibliography
\clearpage

\appendix
\numberwithin{equation}{section}

\begin{center}
    \LARGE\bfseries
    Appendix
\end{center}

\section{Notation and Definitions}
\label{appendix:notation_definitions}

Before presenting the proofs, we provide additional definitions and comment on the notation
used throughout the appendix.

\paragraph{Unobserved Effects Models.} The asymptotic analysis focuses on the bipartite panel ($s = 1$) with $n = N_{1} N_{2} T$ and $L = N_{1}T + N_{2}T + N_{1} N_{2}$. Differences across data structures are immaterial. The linear index is $\pi_{ijt}(\beta, \mu_{ijt}(\phi)) = (X \beta)_{ijt} + (D \phi)_{ijt}$, with $X$ a dense $n \times K$ matrix and $D$ a sparse $n \times L$ matrix of dummies. We write $\pi(\beta, \phi) \coloneqq X \beta + D \phi$. $K$ is fixed, $L$ grows with at least one panel dimension. $D$ has up to $\mathcal{M} \in \{1, 2, 3\}$ blocks $D = (D_{1}, \ldots, D_{\mathcal{M}})$ with $L_{m}$ columns each and $L = \sum_{m} L_{m}$. The block design and the constraint matrix $V$ depend on the linear-index specification.

\paragraph{Estimators.} The profile estimator is
\begin{equation}
    \label{eq:uncorrected_estimator_profile}
    \hat{\beta} = \underset{\beta \in \mathbb{R}^{K}}{\argmin} \, \mathcal{L}_{n}(\beta, \hat{\phi}(\beta)) \, , \quad
    \hat{\phi}(\beta) = \underset{\phi \in \mathbb{R}^{L}}{\argmin} \, \mathcal{L}_{n}(\beta, \phi) \, .
\end{equation}
 
\paragraph{Notation.} $C \in (0, \infty)$ is a generic finite constant. Vector $p$-norms are standard, with $\norm{v}_{p} \leq \dim(v)^{1/p - 1/p^{\prime}} \norm{v}_{p^{\prime}}$ for $1 \leq p \leq p^{\prime} < \infty$. Matrix norms are induced by vector $p$-norms ($p = 2$: spectral; $p = 1, \infty$: column/row sum). $I_m$, $\iota_m$, $e_m$ denote the identity, ones-vector, and $m$-th standard basis vector, and
\begin{equation*}
    \dop{\mathrm{r}} \psi(\beta, \phi) \coloneqq \frac{\partial^{\mathrm{r}} \psi(y, \pi(\beta, \phi))}{(\partial \pi)^{\mathrm{r}}} \, , \qquad \nabla^{\mathrm{r}} \psi(\beta, \phi) \coloneqq \diag(\dop{\mathrm{r}} \psi(\beta, \phi)) \, .
\end{equation*}
Define
\begin{align*}
    u(\beta, \phi) \coloneqq& \, \frac{D^{\prime} \dop{1} \psi(\beta, \phi)}{w_{n}} \, , &  H(\beta, \phi) \coloneqq& \, \frac{D^{\prime} \nabla^{2} \psi(\beta, \phi) D + V V^{\prime}}{w_{n}} \, ,\\
    Q(\beta, \phi) \coloneqq& \, \frac{D (H(\beta, \phi))^{- 1} D^{\prime}}{w_{n}} \, , & P(\beta, \phi) \coloneqq& \, Q(\beta, \phi) \, \nabla^{2} \psi(\beta, \phi) \, , \\
    M(\beta, \phi) \coloneqq& \, I_{n} - P(\beta, \phi) \, , & W(\beta, \phi) \coloneqq& \, \frac{(M(\beta, \phi) X)^{\prime} \nabla^{2} \psi(\beta, \phi) X}{n} \, .
\end{align*}
Arguments are omitted at $(\beta^{0}, \phi^{0})$. We write $\overline{H} \coloneqq \EX{H}$, $\widetilde{H} \coloneqq H - \overline{H}$, and $\overline{Q} \coloneqq w_{n}^{-1} D \overline{H}^{-1} D^{\prime}$.
 
For $\mathcal{M} > 1$ we use the decomposition $\overline{H} = \overline{F} + \overline{G}$ with $\overline{F} \coloneqq \bdiag(\overline{F}_{1}, \ldots, \overline{F}_{\mathcal{M}}) = w_{n}^{-1} \diag(\overline{f})$ and $F_{m}(\beta, \phi) \coloneqq w_{n}^{-1} D_{m}^{\prime} \nabla^{2} \psi(\beta, \phi) D_{m}$. Let $v^{\circ - 1}$ denote the elementwise inverse, and set $\overline{\mathcal{Q}} \coloneqq w_{n}^{-1} D \overline{F}^{-1} D^{\prime}$, $\overline{\mathcal{Q}}_{m} \coloneqq w_{n}^{-1} D_{m} \overline{F}_{m}^{-1} D_{m}^{\prime}$, so that $\overline{\mathcal{Q}} \dop{1} \psi = \sum_{m} \overline{\mathcal{Q}}_{m} \dop{1} \psi$ and, for $\mathcal{M} > 1$,
\begin{equation*}
    D \overline{F}^{- 1} \big(\overline{G} - w_{n}^{- 1} V V^{\prime}\big) \overline{F}^{- 1} u = \sum_{m \neq m^{\prime}} \overline{\mathcal{Q}}_{m} \overline{\nabla^{2} \psi} \, \overline{\mathcal{Q}}_{m^{\prime}} \dop{1} \psi \, .
\end{equation*}
Finally, $\mathfrak{D}_{\pi}^{\mathrm{r}} \coloneqq \nabla^{\mathrm{r}} \psi \ddot{X}$ with $\ddot{X} = X - \mathfrak{X}$, $\mathfrak{X} = \overline{Q} \, \overline{\nabla^{2} \psi X} = D \Xi^{0}$, $\Xi^{0} = (\xi_{1}^{0}, \ldots, \xi_{K}^{0})$. For $\varepsilon, \eta \geq 0$, $\mathfrak{B}(\varepsilon) \coloneqq \{\beta \colon \norm{\beta - \beta^{0}}_{2} \leq \varepsilon\}$ and $\mathfrak{P}(\eta, p) \coloneqq \{\phi \colon \norm{\phi - \phi^{0}}_{p} \leq \eta\}$.

\section{Statements Required for the Proofs of Theorems \ref{theorem:asymptotic_distributions_interacted}--\ref{theorem:asymptotic_distributions_noninteracted}}
\label{appendix:statements}

This section collects the auxiliary results invoked in the proofs of Theorems \ref{theorem:asymptotic_distributions_interacted}, \ref{theorem:asymptotic_distributions_interacted_debiased}, and \ref{theorem:asymptotic_distributions_noninteracted}. Statements appear here. Proofs (except for Lemma \ref{lemma:clt_mds}) are given in the Online Appendix.

\begin{lemma}[Central limit theorem for martingale difference sequences]
	\label{lemma:clt_mds}
	Consider the scalar process $\{\vartheta_{mt} \colon m \in \{1, \ldots, \abs{\mathcal{D}}\}, t \in \{1, \ldots, T\}\}$. Let $\{\{\vartheta_{m1}, \ldots, \vartheta_{mT}\} \colon m \in \{1, \ldots, \abs{\mathcal{D}}\}\}$ be independent across $m$, and be a martingale difference sequence for each $m, \abs{\mathcal{D}}, T$. Let $\EX{\abs{\vartheta_{mt}}^{2 + \nu}}$ be uniformly bounded across $m, t, \abs{\mathcal{D}}, T$ for some $\nu > 0$. Let $\bar{\sigma} > 0$ for all sufficiently large $n = \abs{\mathcal{D}} T$, and let $n^{- 1} \sum_{m = 1}^{\abs{\mathcal{D}}} \sum_{t = 1}^{T} \vartheta_{mt}^{2} - \bar{\sigma}^{2} \xrightarrow{p} 0$ as $n \rightarrow \infty$. Then, $\tfrac{1}{\bar{\sigma} \sqrt{n}} \sum_{m = 1}^{\abs{\mathcal{D}}} \sum_{t = 1}^{T} \vartheta_{mt} \xrightarrow{d} \N(0, 1)$.
\end{lemma}

\paragraph{Proof of Lemma \ref{lemma:clt_mds}.} The lemma is a reformulation of Lemma S.1 of \textcite{fw2016} (see their Supplement S.6.1; derived from Corollary 5.26 of \textcite{w2001}).

\subsection{Asymptotic Expansions for Interacted Specification}

\begin{lemma}[Implied Regularity Conditions 1]
	\label{lemma:regularity_conditions1_interacted}
	Let Assumption \ref{assumption:general} hold. Furthermore, let $w_{n} = T$, $q > 5$, $\varepsilon = o(T^{- 1 / 2})$, and $\eta = \mathcal{O}(T^{- 1 / 10})$ for some $\varepsilon > 0$ and $\eta > 0$. Then,
	\begin{enumerate}[(i)]
		\item for $p \geq 1$, $\norm{D}_{p} = \mathcal{O}(T^{1 / p})$, $\norm{D^\prime}_{p} = \mathcal{O}(T^{1 - 1 / p})$, $\norm{V}_{p} = \mathcal{O}(T^{1 / p})$, $\norm{V^{\prime}}_{p} = \mathcal{O}(T^{1 - 1 / p})$;
		\item $\overline{H} > 0$ a.\,s., $\norm{\overline{H}^{- 1}}_{\infty} = \mathcal{O}(1)$ a.\,s., $\norm{\overline{Q}}_{\infty} = \mathcal{O}(1)$ a.\,s., $\overline{F} > 0$ a.\,s., $\norm{\overline{F}^{- 1}}_{\infty} = \mathcal{O}(1)$ a.\,s., $\norm{\overline{\mathcal{Q}}}_{\infty} = \mathcal{O}(1)$ a.\,s., $\norm{\overline{G}}_{\infty} = \mathcal{O}(1)$ a.\,s.;
		\item for $1 \leq p \leq 20 + \nu$, $\nu > 0$, and $1 \leq \mathrm{r} \leq 5$, $\sup_{ijt} \, \EX{\max_{k} \abs{(\mathfrak{D}_{\pi}^{\mathrm{r}})_{ijt} e_{k}}^{p}}$ is a.\,s. uniformly bounded over $N_{1}, N_{2}, T$;
		\item for $p \geq 1$ and $1 \leq \mathrm{r} \leq \mathrm{r}^{\prime} \leq 5$, $\max_{k} \norm{\overline{\dop{\mathrm{r}} \psi \odot X} e_{k}}_{p} = \mathcal{O}(T^{3 / p})$ a.\,s., $\norm{\overline{\dop{\mathrm{r}} \psi}}_{p} = \mathcal{O}(T^{3 / p})$ a.\,s., $\max_{k} \norm{\overline{\mathfrak{D}_{\pi}^{\mathrm{r}}} e_{k}}_{p} = \mathcal{O}(T^{3 / p})$ a.\,s., $\max_{k} \norm{D^{\prime} \overline{\nabla^{\mathrm{r}} \psi X} e_{k}}_{p} = \mathcal{O}(T^{1 + 2 / p})$ a.\,s., $\max_{k} \norm{\mathfrak{X} e_{k}}_{p} = \mathcal{O}(T^{3 / p})$ a.\,s., $\norm{D^{\prime} \overline{\nabla^{\mathrm{r}} \psi} D}_{\infty} = \mathcal{O}(T)$ a.\,s., $\norm{D^{\prime} \overline{\nabla^{\mathrm{r}} \psi} D}_{\infty} = \mathcal{O}(T)$ a.\,s., $\norm{D^{\prime} \overline{\nabla^{\mathrm{r}} \psi \nabla^{\mathrm{r}^{\prime}} \psi} D}_{\infty} = \mathcal{O}(T)$ a.\,s., $\max_{k} \norm{D^{\prime} \diag(\overline{\mathfrak{D}_{\pi}^{\mathrm{r}}} e_{k}) D}_{\infty} = \mathcal{O}(T)$ a.\,s., \linebreak $\max_{k, k^{\prime}} \norm{D^{\prime} \diag(\overline{\mathfrak{D}_{\pi}^{\mathrm{r}} e_{k} \odot \mathfrak{D}_{\pi}^{\mathrm{r}^{\prime}}} e_{k^{\prime}}) D}_{\infty} = \mathcal{O}(T)$ a.\,s.;
		\item for $1 \leq p \leq 20$ and $1 \leq \mathrm{r} \leq 5$, $\max_{k} \norm{D^{\prime} \widetilde{\nabla^{\mathrm{r}} \psi X} e_{k}}_{p} = \mathcal{O}_{P}(T^{1 / 2 + 2 / p})$, $\norm{D^{\prime} \widetilde{\dop{\mathrm{r}} \psi}}_{p} = \mathcal{O}_{P}(T^{1 / 2 + 2 / p})$, $\max_{k} \norm{D^{\prime} \widetilde{\mathfrak{D}_{\pi}^{\mathrm{r}}} e_{k}}_{p} = \mathcal{O}_{P}(T^{1 / 2 + 2 / p})$, $\norm{\overline{Q} \dop{1} \psi}_{p} = \mathcal{O}_{P}(T^{- 1 / 2 + 3 / p})$, \linebreak $\max_{k} \norm{\overline{\mathcal{Q}} \widetilde{\mathfrak{D}_{\pi}^{\mathrm{r}}} e_{k}}_{p} = \mathcal{O}_{P}(T^{- 1 / 2 + 3 / p})$, $\norm{\overline{\mathcal{Q}} \dop{1} \psi}_{p} = \mathcal{O}_{P}(T^{- 1 / 2 + 3 / p})$, \linebreak $\max_{k} \norm{V^{\prime} \overline{F}^{- 1} D^{\prime} \widetilde{\nabla^{\mathrm{r}} \psi X} e_{k}}_{p} = \mathcal{O}_{P}(T^{1 + 1 / p})$, $\norm{V^{\prime} \overline{F}^{- 1} D^{\prime} \dop{1} \psi}_{p} = \mathcal{O}_{P}(T^{1 + 1 / p})$, \linebreak $\max_{k} \norm{V^{\prime} \overline{F}^{- 1} D^{\prime} \widetilde{\mathfrak{D}_{\pi}^{\mathrm{r}}} e_{k}}_{p} = \mathcal{O}_{P}(T^{1 + 1 / p})$, $\norm{\overline{G} \, \overline{F}^{- 1} D^{\prime} \dop{1} \psi}_{p} = \mathcal{O}_{P}(T^{2 / p})$, \linebreak $\max_{k} \norm{\overline{G} \, \overline{F}^{- 1} D^{\prime} \widetilde{\mathfrak{D}_{\pi}^{\mathrm{r}}} e_{k}}_{p} = \mathcal{O}_{P}(T^{2 / p})$, $\max_{k} \norm{V^{\prime} \overline{F}^{- 1} D^{\prime} \diag(\overline{\mathfrak{D}_{\pi}^{\mathrm{r}}} e_{k}) \overline{\mathcal{Q}} \dop{1} \psi}_{p} = \mathcal{O}_{P}(T^{1 + 1 / p})$, $\max_{k} \norm{\overline{G} \, \overline{F}^{- 1} D^{\prime} \diag(\overline{\mathfrak{D}_{\pi}^{\mathrm{r}}} e_{k}) \overline{\mathcal{Q}} \dop{1} \psi}_{p} = \mathcal{O}_{P}(T^{2 / p})$; 
		\item for $p > 2$ and $1 \leq \mathrm{r} \leq \mathrm{r}^{\prime} \leq 5$, $\max_{k} \norm{D^{\prime} \diag(\widetilde{\mathfrak{D}_{\pi}^{\mathrm{r}}} e_{k}) D}_{2} = \mathcal{O}_{P}(T^{3 / 5})$, $\norm{D^{\prime} \widetilde{\nabla^{\mathrm{r}} \psi} D}_{2} = \mathcal{O}_{P}(T^{3 / 5})$, $\max_{k} \norm{D^{\prime} \diag(\widetilde{\mathfrak{D}_{\pi}^{\mathrm{r}}} e_{k}) D}_{p} = o_{P}(T^{11 / 10 - 1 / p})$, $\norm{D^{\prime} \widetilde{\nabla^{\mathrm{r}} \psi} D}_{p} = o_{P}(T^{11 / 10 - 1 / p})$, $\max_{k, k^{\prime}} \norm{D^{\prime} \diag(\widetilde{\mathfrak{D}_{\pi}^{\mathrm{r}} e_{k} \odot \mathfrak{D}_{\pi}^{\mathrm{r}^{\prime}}} e_{k^{\prime}}) D}_{2} = \mathcal{O}_{P}(T^{7 / 10})$, $\norm{D^{\prime} \widetilde{\nabla^{\mathrm{r}} \psi \nabla^{\mathrm{r}^{\prime}} \psi} D}_{2} = \mathcal{O}_{P}(T^{7 / 10})$, \linebreak $\max_{k, k^{\prime}} \norm{D^{\prime} \diag(\widetilde{\mathfrak{D}_{\pi}^{\mathrm{r}} e_{k} \mathfrak{D}_{\pi}^{\mathrm{r}^{\prime}}} e_{k^{\prime}}) D}_{p} = \mathcal{O}_{P}(T^{6 / 5 - 1 / p})$, $\norm{D^{\prime} \widetilde{\nabla^{\mathrm{r}} \psi \nabla^{\mathrm{r}^{\prime}} \psi} D}_{p} = o_{P}(T^{6 / 5 - 1 / p})$, $\max_{k} \norm{\diag(\mathfrak{D}_{\pi}^{\mathrm{r}} e_{k}) D}_{2} = \mathcal{O}_{P}(T^{1 / 2})$, $\norm{\nabla^{\mathrm{r}} \psi D}_{2} = \mathcal{O}_{P}(T^{1 / 2})$, $\max_{k} \norm{\diag(\mathfrak{D}_{\pi}^{\mathrm{r}} e_{k}) D}_{p} = o_{P}(T^{3 / 20 + 7 / (10 p)})$, $\norm{\nabla^{\mathrm{r}} \psi D}_{p} = o_{P}(T^{3 / 20 + 7 / (10 p)})$;
		\item for $1 \leq \mathrm{r} \leq \mathrm{r}^{\prime} \leq 5$, $\sup_{(\beta, \phi) \in \mathfrak{B}(\varepsilon) \times \mathfrak{P}(\eta, q)} \, \max_{k} \norm{D^{\prime} \abs{\diag(\dop{\mathrm{r}} \psi(\beta, \phi) \odot X e_{k})} D}_{\infty} = o_{P}(T^{11 / 10})$, $\sup_{(\beta, \phi) \in \mathfrak{B}(\varepsilon) \times \mathfrak{P}(\eta, q)} \, \norm{D^{\prime} \abs{\nabla^{\mathrm{r}} \psi(\beta, \phi)} D}_{\infty} = o_{P}(T^{11 / 10})$, \linebreak $\sup_{(\beta, \phi) \in \mathfrak{B}(\varepsilon) \times \mathfrak{P}(\eta, q)} \, \norm{D^{\prime} \abs{\nabla^{\mathrm{r}} \psi(\beta, \phi)} \abs{\nabla^{\mathrm{r}^{\prime}} \psi(\beta, \phi)} D}_{\infty} = o_{P}(T^{6 / 5})$; 
		\item for $2 \leq p \leq 10$, $H(\beta, \phi) > 0$ wpa1 $\forall \, (\beta, \phi) \in \mathfrak{B}(\varepsilon) \times \mathfrak{P}(\eta, q)$, \linebreak $\sup_{(\beta, \phi) \in \mathfrak{B}(\varepsilon) \times \mathfrak{P}(\eta, q)} \, \norm{(H(\beta, \phi))^{- 1}}_{p} = \mathcal{O}_{P}(1)$, $\sup_{(\beta, \phi) \in \mathfrak{B}(\varepsilon) \times \mathfrak{P}(\eta, q)} \, \norm{Q(\beta, \phi)}_{p} = \mathcal{O}_{P}(1)$;
		\item for $2 < p \leq 10$, $\norm{H^{- 1} - \overline{H}^{- 1}}_{2} = \mathcal{O}_{P}(T^{- 2 / 5})$, $\norm{H^{- 1} - \overline{H}^{- 1} + \overline{H}^{- 1} \widetilde{H} \, \overline{H}^{- 1}}_{2} = \mathcal{O}_{P}(T^{- 4 / 5})$, $\norm{H^{- 1} - \overline{H}^{- 1} + \overline{H}^{- 1} \widetilde{H} \, \overline{H}^{- 1} - \overline{H}^{- 1} \widetilde{H} \, \overline{H}^{- 1} \widetilde{H} \, \overline{H}^{- 1}}_{2} = \mathcal{O}_{P}(T^{- 6 / 5})$, $\norm{Q - \overline{Q}}_{2} = \mathcal{O}_{P}(T^{- 2 / 5})$, $\norm{H^{- 1} - \overline{H}^{- 1}}_{p} = o_{P}(T^{1 / 10 - 1 / p})$, $\norm{H^{- 1} - \overline{H}^{- 1} + \overline{H}^{- 1} \widetilde{H} \, \overline{H}^{- 1}}_{p} = o_{P}(T^{1 / 5 - 2 / p})$, $\norm{H^{- 1} - \overline{H}^{- 1} + \overline{H}^{- 1} \widetilde{H} \, \overline{H}^{- 1} - \overline{H}^{- 1} \widetilde{H} \, \overline{H}^{- 1} \widetilde{H} \, \overline{H}^{- 1}}_{p} = \mathcal{O}_{P}(T^{3 / 10 - 3 / p})$, $\norm{Q - \overline{Q}}_{p} = \mathcal{O}_{P}(T^{1 / 10 - 1 / p})$;
		\item for $1 \leq p \leq 20 + \nu$, $\nu > 0$, and $1 \leq \mathrm{r} \leq 5$, $\sup_{(\beta, \phi) \in \mathfrak{B}(\varepsilon) \times \mathfrak{P}(\eta, q)} \max_{k} \norm{\dop{\mathrm{r}} \psi(\beta, \phi) \odot X e_{k}}_{p} = \mathcal{O}_{P}(T^{3 / p})$, $\sup_{(\beta, \phi) \in \mathfrak{B}(\varepsilon) \times \mathfrak{P}(\eta, q)} \, \norm{\dop{\mathrm{r}} \psi(\beta, \phi)}_{p} = \mathcal{O}_{P}(T^{3 / p})$, $\max_{k} \norm{X e_{k}}_{p} = \mathcal{O}_{P}(T^{3 / p})$, $\max_{k} \norm{\mathfrak{D}_{\pi}^{\mathrm{r}} e_{k}}_{p} = \mathcal{O}_{P}(T^{3 / p})$, $\sup_{(\beta, \phi) \in \mathfrak{B}(\varepsilon) \times \mathfrak{P}(\eta, q)} \max_{k} \norm{D^{\prime} \nabla^{\mathrm{r}} \psi(\beta, \phi) X e_{k}}_{p} = \mathcal{O}_{P}(T^{1 + 2 / p})$;
		\item for $1 \leq p \leq 10$, $\sup_{(\beta, \phi) \in \mathfrak{B}(\varepsilon) \times \mathfrak{P}(\eta, q)} \max_{k} \norm{P(\beta, \phi) X e_{k}}_{p} = \mathcal{O}_{P}(T^{3 / p})$, \linebreak $\sup_{(\beta, \phi) \in \mathfrak{B}(\varepsilon) \times \mathfrak{P}(\eta, q)} \max_{k} \norm{M(\beta, \phi) X e_{k}}_{p} = \mathcal{O}_{P}(T^{3 / p})$, \linebreak $\sup_{(\beta, \phi) \in \mathfrak{B}(\varepsilon) \times \mathfrak{P}(\eta, q)} \, \norm{Q(\beta, \phi) \dop{1} \psi}_{p} = \mathcal{O}_{P}(T^{- 1 / 2 + 3 / p})$
		\item for $1 \leq p \leq 2$ and $1 \leq \mathrm{r} \leq 5$, $\max_{k} \norm{\nabla^{\mathrm{r}} \psi M X e_{k} - \mathfrak{D}_{\pi}^{\mathrm{r}} e_{k}}_{p} = \mathcal{O}_{P}(T^{- 1 / 2 + 3 / p})$, \linebreak $\max_{k} \norm{\nabla^{\mathrm{r}} \psi M X e_{k} - \mathfrak{D}_{\pi}^{\mathrm{r}} e_{k} + \nabla^{\mathrm{r}} \psi \overline{\mathcal{Q}} \widetilde{\mathfrak{D}_{\pi}^{2}} e_{k}}_{p} = \mathcal{O}_{P}(T^{- 9 / 10 + 3 / p})$, $\max_{k} \norm{\nabla^{\mathrm{r}} \psi M X e_{k} - \mathfrak{D}_{\pi}^{\mathrm{r}} e_{k} + \nabla^{\mathrm{r}} \psi \overline{Q} \widetilde{\mathfrak{D}_{\pi}^{2}} e_{k} - \nabla^{\mathrm{r}} \psi \overline{Q} \widetilde{\nabla^{2} \psi} \overline{\mathcal{Q}} \widetilde{\mathfrak{D}_{\pi}^{2}} e_{k}}_{p} = \mathcal{O}_{P}(T^{- 6 / 5 + 3 / p})$, \linebreak $\max_{k} \norm{\mathfrak{X} e_{k} - P X e_{k}}_{p} = \mathcal{O}_{P}(T^{- 1 / 2 + 3 / p})$, $\max_{k} \norm{\mathfrak{X} e_{k} - P X e_{k} + \overline{\mathcal{Q}} \widetilde{\mathfrak{D}_{\pi}^{2}} e_{k}}_{p} = \mathcal{O}_{P}(T^{- 9 / 10 + 3 / p})$, $\max_{k} \norm{\mathfrak{X} e_{k} - P X e_{k} + \overline{Q} \widetilde{\mathfrak{D}_{\pi}^{2}} e_{k} - \overline{Q} \widetilde{\nabla^{2} \psi} \overline{\mathcal{Q}} \widetilde{\mathfrak{D}_{\pi}^{2}} e_{k}}_{p} = \mathcal{O}_{P}(T^{- 6 / 5 + 3 / p})$, \linebreak $\norm{\nabla^{\mathrm{r}} \psi Q \dop{1} \psi - \nabla^{\mathrm{r}} \psi \overline{\mathcal{Q}} \dop{1} \psi}_{p} = \mathcal{O}_{P}(T^{- 9 / 10 + 3 / p})$, \linebreak $\norm{\nabla^{\mathrm{r}} \psi Q \dop{1} \psi - \nabla^{\mathrm{r}} \psi \overline{Q} d^{1} + \nabla^{\mathrm{r}} \psi \overline{Q} \widetilde{\nabla^{2} \psi} \overline{\mathcal{Q}} \dop{1} \psi}_{p} = \mathcal{O}_{P}(T^{- 13 / 10 + 3 / p})$, $\norm{Q \dop{1} \psi - \overline{\mathcal{Q}} \dop{1} \psi}_{p} = \mathcal{O}_{P}(T^{- 9 / 10 + 3 / p})$, $\norm{Q \dop{1} \psi - \overline{Q} d^{1} + \overline{Q} \widetilde{\nabla^{2} \psi} \overline{\mathcal{Q}} \dop{1} \psi}_{p} = \mathcal{O}_{P}(T^{- 13 / 10 + 3 / p})$;
		\item for all $\beta \in \mathfrak{B}(\varepsilon)$, the domain $u(\beta, \mathfrak{P}(\eta, q))$ includes $\mathbf{0}_{L}$ wpa1;
		\item $\overline{W} > 0$ a.\,s., $\norm{\overline{W}^{- 1}}_{2} = \mathcal{O}_{P}(1)$, $\norm{U}_{2} = \mathcal{O}_{P}(T^{1 / 2})$, $\norm{\overline{B}_{\alpha}}_{2} = \mathcal{O}_{P}(1)$, $\norm{\overline{B}_{\gamma}}_{2} = \mathcal{O}_{P}(1)$, $\norm{\overline{B}_{\rho}}_{2} = \mathcal{O}_{P}(1)$;
		\item $\norm{W - \overline{W}}_{2} = o_{P}(T^{- 1 / 2})$;
		\item $W(\beta, \phi) > 0$ wpa1 $\forall \, (\beta, \phi) \in \mathfrak{B}(\varepsilon) \times \mathfrak{P}(\eta, q)$.
	\end{enumerate}
\end{lemma}

\begin{theorem}[Asymptotic Expansion]
	\label{theorem:asymptotic_expansion_profile_score_interacted}
	\label{theorem:asymptotic_expansion_lincom_ip_function_interacted}
	Let Lemma \ref{lemma:regularity_conditions1_interacted} hold. In addition, assume that $\beta \in \mathfrak{B}(\varepsilon)$ with $\varepsilon > 0$ and $\varepsilon = o(T^{- 1 / 2})$.
	\begin{enumerate}[(i)]
		\item For any $\beta \in \mathfrak{B}(\varepsilon)$,
		\begin{equation*}
			\frac{\partial \mathcal{L}_{n}(\beta, \hat{\phi}(\beta))}{\partial \beta} = \left(\frac{\sqrt{N_{1} N_{2}}}{\sqrt{T}}\right) U + N_{1} N_{2} \, \overline{W} (\beta - \beta^{0}) + o_{P}\big(T^{\frac{1}{2}}\big) + o_{P}\big(T^{\frac{3}{2}} \bignorm{\beta - \beta^{0}}_{2}\big) \, ,
		\end{equation*}
		where
		\begin{align*}
			& U \coloneqq U_{1} + \sqrt{N_{1}} \left(\frac{\sqrt{T}}{\sqrt{N_{2}}}\right) \big(- \overline{U}_{2, 1} + \overline{U}_{3, 1}\big) + \sqrt{N_{2}} \left(\frac{\sqrt{T}}{\sqrt{N_{1}}}\right) \big(- \overline{U}_{2, 2} + \overline{U}_{3, 2}\big) \, + \\
			& \quad \sqrt{T} \left(\frac{\sqrt{N_{1}}}{\sqrt{T}}\right) \left(\frac{\sqrt{N_{2}}}{\sqrt{T}}\right) \big(- \overline{U}_{2, 3} + \overline{U}_{3, 3}\big)
		\end{align*}
		with
		\begin{align*}
			&U_{1} \coloneqq \frac{1}{\sqrt{N_{1} N_{2} T}} \sum_{i = 1}^{N_{1}} \sum_{j = 1}^{N_{2}} \sum_{t = 1}^{T} (\mathfrak{D}_{\pi}^{1})_{ijt} \, , \quad \overline{U}_{2, 1} \coloneqq \frac{1}{N_{1} T} \sum_{i = 1}^{N_{1}} \sum_{t = 1}^{T} \frac{\sum_{j = 1}^{N_{2}} \EX{(\mathfrak{D}_{\pi}^{2})_{ijt} (\dop{1} \psi)_{ijt}}}{\sum_{j = 1}^{N_{2}} \EX{(\dop{2} \psi)_{ijt}}} \, , \\
			&\overline{U}_{2, 2} \coloneqq \frac{1}{N_{2} T} \sum_{j = 1}^{N_{2}} \sum_{t = 1}^{T} \frac{\sum_{i = 1}^{N_{1}} \EX{(\mathfrak{D}_{\pi}^{2})_{ijt} (\dop{1} \psi)_{ijt}}}{\sum_{i = 1}^{N_{1}} \EX{(\dop{2} \psi)_{ijt}}} \, , \\
			&\overline{U}_{2, 3} \coloneqq \frac{1}{N_{1} N_{2}} \sum_{i = 1}^{N_{1}} \sum_{j = 1}^{N_{2}} \frac{\sum_{t = 1}^{T} \sum_{t^{\prime} = 1}^{T} \EX{(\mathfrak{D}_{\pi}^{2})_{ijt^{\prime}} (\dop{1} \psi)_{ijt}}}{\sum_{t = 1}^{T} \EX{(\dop{2} \psi)_{ijt}}} \, , \\
			&\overline{U}_{3, 1} \coloneqq \frac{1}{2 \, N_{1} T} \sum_{i = 1}^{N_{1}} \sum_{t = 1}^{T} \frac{\big\{\sum_{j = 1}^{N_{2}} \EX{(\mathfrak{D}_{\pi}^{3})_{ijt}}\big\} \sum_{j = 1}^{N_{2}} \EX{\big\{(\dop{1} \psi)_{ijt}\big\}^{2}}}{\big\{\sum_{j = 1}^{N_{2}} \EX{(\dop{2} \psi)_{ijt}}\big\}^{2}} \, , \\
			&\overline{U}_{3, 2} \coloneqq \frac{1}{2 \, N_{2} T} \sum_{j = 1}^{N_{2}} \sum_{t = 1}^{T} \frac{\big\{\sum_{i = 1}^{N_{1}} \EX{(\mathfrak{D}_{\pi}^{3})_{ijt}}\big\} \sum_{i = 1}^{N_{1}} \EX{\big\{(\dop{1} \psi)_{ijt}\big\}^{2}}}{\big\{\sum_{i = 1}^{N_{1}} \EX{(\dop{2} \psi)_{ijt}}\big\}^{2}} \, , \\
			&\overline{U}_{3, 3} \coloneqq \frac{1}{2 \, N_{1} N_{2}} \sum_{i = 1}^{N_{1}} \sum_{j = 1}^{N_{2}} \frac{\big\{\sum_{t = 1}^{T} \EX{(\mathfrak{D}_{\pi}^{3})_{ijt}}\big\} \EX{\big\{\sum_{t = 1}^{T} (\dop{1} \psi)_{ijt}\big\}^{2}}}{\big\{\sum_{t = 1}^{T} \EX{(\dop{2} \psi)_{ijt}}\big\}^{2}} \, , \\
			&\overline{W} \coloneqq \frac{1}{N_{1} N_{2} T} \sum_{i = 1}^{N_{1}} \sum_{j = 1}^{N_{2}} \sum_{t = 1}^{T} \EX{(\mathfrak{D}_{\pi}^{2})_{ijt} (x_{ijt} - \mathfrak{x}_{ijt})^{\prime}} \, .
		\end{align*}
		\item Let $a \in \Real^{L}$ with $\norm{a}_{p^{\prime}} = \mathcal{O}_{P}(1)$, $10 / 3 \leq p \leq 10$, and $p^{\prime} = p / (p - 1)$. Then, for any $\beta \in \mathfrak{B}(\varepsilon)$,
		\begin{equation*}
			a^{\prime} \big(\hat{\phi}(\beta) - \phi^{0}\big) = a^{\prime} \mathbb{U} +  a^{\prime} \overline{\mathbb{W}} \big(\beta - \beta^{0}\big) + o_{P}\big(T^{- \frac{3}{5} + \frac{2}{p}}\big) + o_{P}\big(T^{\frac{2}{p}} \bignorm{\beta - \beta^{0}}_{2}\big) \, ,
		\end{equation*}
		where $\mathbb{U} \coloneqq - H^{- 1} D^{\prime} \dop{1} \psi / T$ and $\overline{\mathbb{W}} \coloneqq - \, \Xi^{0}$.
	\end{enumerate}
\end{theorem}

\begin{theorem}[Consistency of $\hat{\beta}$ and of $\hat{\phi}$]
	\label{theorem:consistency_interacted}
	Let Lemma \ref{lemma:regularity_conditions1_interacted} hold. Then, $\norm{\hat{\beta} - \beta^{0}}_{2} = \mathcal{O}_{P}(T^{- 1})$. In addition, for $1 \leq p \leq 10$, $\norm{\hat{\phi} - \phi^{0}}_{p} = \mathcal{O}_{P}(T^{- 1 / 2 + 2 / p})$ and $\norm{\hat{\phi} - \phi^{0} - \mathbb{U}}_{p} = o_{P}(T^{- 3 / 5 + 2 / p})$.
\end{theorem}

\begin{corollary}[Asymptotic Expansion of $\hat{\beta}$]
	\label{corollary:asymptotic_expansion_score_interacted}
	Let Lemma \ref{lemma:regularity_conditions1_interacted} hold. Then, $\sqrt{N_{1} N_{2} T} \, (\hat{\beta} - \beta^{0}) = - \overline{W}^{- 1} U + o_{P}(1)$.
\end{corollary}

\begin{lemma}[Implied Regularity Conditions 2]
	\label{lemma:regularity_conditions2_interacted}
	Let Assumption \ref{assumption:general} hold. Furthermore, let $w_{n} = T$, $q > 5$, $\varepsilon = o(T^{- 1 / 2})$, and $\eta = \mathcal{O}(T^{- 1 / 10})$ for some $\varepsilon > 0$ and $\eta > 0$. Then,
	\begin{enumerate}[(i)]
		\item for $1 \leq p \leq 2$ and $1 \leq \mathrm{r} \leq 3$, $\max_{k} \bignorm{\big(\widehat{\mathfrak{D}_{\pi}^{\mathrm{r}}} - \mathfrak{D}_{\pi}^{\mathrm{r}}\big) e_{k}}_{p} = \mathcal{O}_{P}(T^{- 1 / 2 + 3 / p})$, $\max_{k} \norm{\big(\widehat{\mathfrak{D}_{\pi}^{\mathrm{r}}} - \mathfrak{D}_{\pi}^{\mathrm{r}} + \nabla^{\mathrm{r}} \psi \overline{\mathcal{Q}} \widetilde{\mathfrak{D}_{\pi}^{2}} + \diag(\overline{\mathcal{Q}} \dop{1} \psi) \mathfrak{D}_{\pi}^{\mathrm{r} + 1} - \nabla^{\mathrm{r}} \psi \overline{\mathcal{Q}} \diag(\overline{\mathcal{Q}} \dop{1} \psi) \overline{\mathfrak{D}_{\pi}^{3}}\big) e_{k}}_{p} = o_{P}(T^{- 3 / 5 + 3 / p})$, $\norm{\widehat{\dop{\mathrm{r}} \psi} - \dop{\mathrm{r}} \psi}_{p} = \mathcal{O}_{P}(T^{- 1 / 2 + 3 / p})$, $\norm{\widehat{\dop{\mathrm{r}} \psi} - \dop{\mathrm{r}} \psi + \nabla^{\mathrm{r} + 1} \psi \overline{\mathcal{Q}} \dop{1} \psi}_{p} = o_{P}(T^{- 3 / 5 + 3 / p})$;
		\item $\norm{\widetilde{F}}_{\infty} = \mathcal{O}_{P}(T^{- 2 / 5})$;
		\item $F(\beta, \phi) > 0$ wpa1 $\forall \, (\beta, \phi) \in \mathfrak{B}(\varepsilon) \times \mathfrak{P}(\eta, q)$, $\sup_{(\beta, \phi) \in \mathfrak{B}(\varepsilon) \times \mathfrak{P}(\eta, q)} \, \bignorm{(F(\beta, \phi))^{- 1}}_{\infty} = \mathcal{O}_{P}(1)$;
		\item for $1 \leq p \leq 10$, $\norm{\hat{f}^{\circ - 1} - \bar{f}^{\circ - 1}}_{p} = \mathcal{O}_{P}(T^{- 3 / 2 + 2 / p})$, $\norm{\hat{f}^{\circ - 1} - \bar{f}^{\circ - 1} - \bar{f}^{\circ - 2} \odot D^{\prime} \overline{\nabla^{3} \psi} \, \overline{\mathcal{Q}} \dop{1} \psi + \bar{f}^{\circ - 2} \odot \tilde{f}}_{p} = o_{P}(T^{- 3 / 2 + 2 / p})$.
	\end{enumerate}
\end{lemma}

\subsection{Asymptotic Expansions for Non-interacted Specification}

\begin{lemma}[Implied Regularity Conditions 1]
	\label{lemma:regularity_conditions1_noninteracted}
	Let Assumption \ref{assumption:general} hold. Furthermore, let $w_{n} = T^{2}$, $q > 20 / 19$, $\varepsilon = o(T^{- 1})$, and $\eta = \mathcal{O}(T^{- 1 / 20})$ for some $\varepsilon > 0$ and $\eta > 0$. Then,
	\begin{enumerate}[(i)]
		\item for $p \geq 1$, $\norm{D}_{p} = \mathcal{O}(T^{2 / p})$, $\norm{D^\prime}_{p} = \mathcal{O}(T^{2 - 2 / p})$;
		\item $\overline{H} > 0$ a.\,s., $\norm{\overline{H}^{- 1}}_{\infty} = \mathcal{O}(1)$ a.\,s., $\norm{\overline{Q}}_{\infty} = \mathcal{O}(1)$ a.\,s.;
		\item for $1 \leq p \leq 20 + \nu$, $\nu > 0$, and $1 \leq \mathrm{r} \leq 5$, $\sup_{ijt} \, \EX{\max_{k} \abs{(\mathfrak{D}_{\pi}^{\mathrm{r}})_{ijt} e_{k}}^{p}}$ is a.\,s. uniformly bounded over $N_{1}, N_{2}, T$;
        \item for $p \geq 1$, $\max_{k} \norm{\xi_{k}^{0}}_{p} = \mathcal{O}(T^{1 / p})$ a.\,s.;
		\item for $1 \leq p \leq 20$ and $1 \leq \mathrm{r} \leq 5$, $\max_{k} \norm{D^{\prime} \widetilde{\nabla^{\mathrm{r}} \psi X} e_{k}}_{p} = \mathcal{O}_{P}(T^{1 + 1 / p})$, $\norm{D^{\prime} \widetilde{\dop{\mathrm{r}} \psi}}_{p} = \mathcal{O}_{P}(T^{1 + 1 / p})$;
		\item for $p > 2$ and $1 \leq \mathrm{r} \leq 5$, $\norm{D^{\prime} \widetilde{\nabla^{\mathrm{r}} \psi} D}_{2} = \mathcal{O}_{P}(T^{21 / 20})$, $\norm{D^{\prime} \widetilde{\nabla^{\mathrm{r}} \psi} D}_{p} = o_{P}(T^{41 / 20 - 2 / p})$;
		\item for $1 \leq \mathrm{r} \leq 5$, $\sup_{(\beta, \phi) \in \mathfrak{B}(\varepsilon) \times \mathfrak{P}(\eta, q)} \, \max_{k} \norm{D^{\prime} \abs{\diag(\dop{\mathrm{r}} \psi(\beta, \phi) \odot X e_{k})} D}_{\infty} = o_{P}(T^{41 / 20})$, $\sup_{(\beta, \phi) \in \mathfrak{B}(\varepsilon) \times \mathfrak{P}(\eta, q)} \, \norm{D^{\prime} \abs{\nabla^{\mathrm{r}} \psi(\beta, \phi)} D}_{\infty} = o_{P}(T^{41 / 20})$;
		\item for $2 \leq p \leq 40$, $H(\beta, \phi) > 0 \quad \text{wpa1} \quad \forall \, (\beta, \phi) \in \mathfrak{B}(\varepsilon) \times \mathfrak{P}(\eta, q)$ wpa1 $\forall \, (\beta, \phi) \in \mathfrak{B}(\varepsilon) \times \mathfrak{P}(\eta, q)$, $\sup_{(\beta, \phi) \in \mathfrak{B}(\varepsilon) \times \mathfrak{P}(\eta, q)} \, \norm{(H(\beta, \phi))^{- 1}}_{p} = \mathcal{O}_{P}(1)$, $\sup_{(\beta, \phi) \in \mathfrak{B}(\varepsilon) \times \mathfrak{P}(\eta, q)} \, \norm{Q(\beta, \phi)}_{p} = \mathcal{O}_{P}(1)$;
		\item for $2 < p \leq 40$, $\norm{H^{- 1} - \overline{H}^{- 1}}_{2} = \mathcal{O}_{P}(T^{- 19 / 20})$, $\norm{H^{- 1} - \overline{H}^{- 1}}_{p} = o_{P}(T^{1 / 20 - 2 / p})$;
		\item for $1 \leq p \leq 20 + \nu$, $\nu > 0$, and $1 \leq \mathrm{r} \leq 5$, $\sup_{(\beta, \phi) \in \mathfrak{B}(\varepsilon) \times \mathfrak{P}(\eta, q)} \max_{k} \norm{\dop{\mathrm{r}} \psi(\beta, \phi) \odot X e_{k}}_{p} = \mathcal{O}_{P}(T^{3 / p})$, $\sup_{(\beta, \phi) \in \mathfrak{B}(\varepsilon) \times \mathfrak{P}(\eta, q)} \, \norm{\dop{\mathrm{r}} \psi(\beta, \phi)}_{p} = \mathcal{O}_{P}(T^{3 / p})$, $\max_{k} \norm{X e_{k}}_{p} = \mathcal{O}_{P}(T^{3 / p})$, $\max_{k} \norm{\mathfrak{D}_{\pi}^{\mathrm{r}} e_{k}}_{p} = \mathcal{O}_{P}(T^{3 / p})$, $\sup_{(\beta, \phi) \in \mathfrak{B}(\varepsilon) \times \mathfrak{P}(\eta, q)} \max_{k} \norm{D^{\prime} \nabla^{\mathrm{r}} \psi(\beta, \phi) X e_{k}}_{p} = \mathcal{O}_{P}(T^{1 + 2 / p})$;
		\item for $1 \leq p \leq 20$, $p^{\prime} = p + \nu$, and $\nu > 0$, $\sup_{(\beta, \phi) \in \mathfrak{B}(\varepsilon) \times \mathfrak{P}(\eta, q)} \max_{k} \norm{M(\beta, \phi) X e_{k}}_{p^{\prime}} = \mathcal{O}_{P}(T^{3 / p^{\prime}})$, $\sup_{(\beta, \phi) \in \mathfrak{B}(\varepsilon) \times \mathfrak{P}(\eta, q)} \, \norm{Q(\beta, \phi) \dop{1} \psi}_{p} = \mathcal{O}_{P}(T^{- 1 + 3 / p})$;
		\item for $1 \leq p \leq 2$ and $1 \leq \mathrm{r} \leq 2$, $\max_{k} \norm{\nabla^{\mathrm{r}} \psi M X e_{k} - \mathfrak{D}_{\pi}^{\mathrm{r}} e_{k}}_{p} = o_{P}(T^{3 / p})$;
		\item for all $\beta \in \mathfrak{B}(\varepsilon)$, the domain $u(\beta, \mathfrak{P}(\eta, q))$ includes $\mathbf{0}_{L}$ wpa1,
		\item $\overline{W} > 0$ a.\,s., $\norm{\overline{W}^{- 1}}_{2} = \mathcal{O}_{P}(1)$, $\norm{U}_{2} = \mathcal{O}_{P}(1)$;
		\item $\norm{W - \overline{W}}_{2} = o_{P}(1)$;
		\item $W(\beta, \phi) > 0$ wpa1 $\forall \, (\beta, \phi) \in \mathfrak{B}(\varepsilon) \times \mathfrak{P}(\eta, q)$.
	\end{enumerate}
\end{lemma}

\begin{theorem}[Asymptotic Expansion]
	\label{theorem:asymptotic_expansion_profile_score_noninteracted}
	\label{theorem:asymptotic_expansion_lincom_ip_function_noninteracted}
	Let Lemma \ref{lemma:regularity_conditions1_noninteracted} hold. In addition, assume that $\beta \in \mathfrak{B}(\varepsilon)$ with $\varepsilon > 0$ and $\varepsilon = o(T^{- 1})$.
	\begin{enumerate}[(i)]
		\item For any $\beta \in \mathfrak{B}(\varepsilon)$,
		\begin{equation*}
			\frac{\partial \mathcal{L}_{n}(\beta, \hat{\phi}(\beta))}{\partial \beta} = \sqrt{\frac{N_{1} N_{2}}{T^{3}}} \, U + \bigg(\frac{N_{1} N_{2}}{T}\bigg) \, \overline{W} (\beta - \beta^{0}) + o_{P}\big(T^{- \frac{1}{2}}\big) + o_{P}\big(T \bignorm{\beta - \beta^{0}}_{2}\big) \, ,
		\end{equation*}
		where $U \coloneqq (N_{1} N_{2} T)^{- 1 / 2} \sum_{i = 1}^{N_{1}} \sum_{j = 1}^{N_{2}} \sum_{t = 1}^{T} (\mathfrak{D}_{\pi}^{1})_{ijt}$.
		\item Let $a \in \Real^{L}$ with $\norm{a}_{p^{\prime}} = \mathcal{O}_{P}(1)$, $1 \leq p \leq 20$, and $p^{\prime} = p / (p - 1)$. Then, for any $\beta \in \mathfrak{B}(\varepsilon)$,
		\begin{equation*}
			a^{\prime} \big(\hat{\phi}(\beta) - \phi^{0}\big) = a^{\prime} \mathbb{U} +  a^{\prime} \overline{\mathbb{W}} \big(\beta - \beta^{0}\big) + o_{P}\big(T^{- \frac{9}{5} + \frac{1}{p}}\big) + o_{P}\big(T^{\frac{1}{p}} \bignorm{\beta - \beta^{0}}_{2}\big) \, ,
		\end{equation*}
		where $\mathbb{U} \coloneqq - H^{- 1} D^{\prime} \dop{1} \psi / T^{2}$ and $\overline{\mathbb{W}} \coloneqq - \, \Xi^{0}$.
	\end{enumerate}
\end{theorem}

\begin{theorem}[Consistency of $\hat{\beta}$ and of $\hat{\phi}$]
	\label{theorem:consistency_noninteracted}
	Let Lemma \ref{lemma:regularity_conditions1_noninteracted} hold. Then, $\norm{\hat{\beta} - \beta^{0}}_{2} = \mathcal{O}_{P}(T^{- 3 / 2})$. In addition, for $1 \leq p \leq 20$, $\norm{\hat{\phi} - \phi^{0}}_{p} = \mathcal{O}_{P}(T^{- 1 + 1 / p})$.
\end{theorem}

\begin{corollary}[Asymptotic Expansion of $\hat{\beta}$]
	\label{corollary:asymptotic_expansion_score_noninteracted}
	Let Lemma \ref{lemma:regularity_conditions1_noninteracted} hold. Then, $\sqrt{N_{1} N_{2} T} \, (\hat{\beta} - \beta^{0}) = - \overline{W}^{- 1} U + o_{P}(1)$.
\end{corollary}

\begin{lemma}[Implied Regularity Conditions 2]
	\label{lemma:regularity_conditions2_noninteracted}
	Let Assumption \ref{assumption:general} hold. Furthermore, let $w_{n} = T^{2}$, $q > 20 / 19$, $\varepsilon = o(T^{- 1})$, and $\eta = \mathcal{O}(T^{- 1 / 20})$ for some $\varepsilon > 0$ and $\eta > 0$. Then, for a positive $p$ with $1 \leq p \leq 2$ and a positive integer $\mathrm{r}$ with $1 \leq \mathrm{r} \leq 2$, $\max_{k} \norm{(\widehat{\mathfrak{D}_{\pi}^{\mathrm{r}}} - \mathfrak{D}_{\pi}^{\mathrm{r}}) e_{k}}_{p} = o_{P}(T^{3 / p})$.
\end{lemma}

\section{Proofs of Theorems \ref{theorem:asymptotic_distributions_interacted}--\ref{theorem:asymptotic_distributions_noninteracted}}
\label{appendix:main_results}

\subsection{Proof of Theorem \ref{theorem:asymptotic_distributions_interacted}}
\label{appendix:main_results_distributions_interacted}

By Corollary \ref{corollary:asymptotic_expansion_score_interacted},
\begin{equation*}
	\sqrt{N_{1} N_{2} T} \, (\hat{\beta} - \beta^{0}) = - \overline{W}^{- 1} U_{1} - \overline{W}^{- 1} (U - U_{1}) + o_{P}(1) \, .
\end{equation*}

$\{(\mathfrak{D}_{\pi}^{1})_{ijt}\}$ is a martingale difference sequence (MDS) in $t$ for each $(i,j)$ and independent across $(i,j)$ given $\Phi$, so $\EX{U_{1}} = \mathbf{0}_{K}$ and $\text{Var}(U_{1}) = \overline{\Sigma}$. By Lemma \ref{lemma:clt_mds} and the Cram\'er-Wold device, $U_{1} \xrightarrow{d} \N(0, \overline{\Sigma}_{\infty})$. Hence, $- \overline{W}^{- 1} U_{1} \xrightarrow{d} \N(0, \overline{W}_{\infty}^{- 1} \overline{\Sigma}_{\infty} \overline{W}_{\infty}^{- 1})$.

The MDS property yields $\sum_{t^{\prime} = 1}^{T} \EX{(\mathfrak{D}_{\pi}^{2})_{ijt^{\prime}} (\dop{1} \psi)_{ijt}} = \sum_{t^{\prime} = t}^{T} \EX{(\mathfrak{D}_{\pi}^{2})_{ijt^{\prime}} (\dop{1} \psi)_{ijt}}$ and \linebreak $\EX{\big\{\sum_{t = 1}^{T} (\dop{1} \psi)_{ijt}\big\}^{2}} = \sum_{t = 1}^{T} \EX{\big\{(\dop{1} \psi)_{ijt}\big\}^{2}}$. Setting $\overline{B}_{\alpha} \coloneqq - \overline{U}_{2, 1} + \overline{U}_{3, 1}$, $\overline{B}_{\gamma} \coloneqq - \overline{U}_{2, 2} + \overline{U}_{3, 2}$, $\overline{B}_{\rho} \coloneqq - \overline{U}_{2, 3} + \overline{U}_{3, 3}$,
\begin{equation*}
	U - U_{1} = \sqrt{N_{1}} \tfrac{\sqrt{T}}{\sqrt{N_{2}}} \overline{B}_{\alpha} + \sqrt{N_{2}} \tfrac{\sqrt{T}}{\sqrt{N_{1}}} \overline{B}_{\gamma} + \sqrt{T} \tfrac{\sqrt{N_{1} N_{2}}}{T} \overline{B}_{\rho} + o_{P}(1) \, .
\end{equation*}

$\overline{W}^{- 1} (U - U_{1}) = \mathcal{O}_{P}(\sqrt{T})$. Rescaling, $\sqrt{N_{1} N_{2}} \, (\hat{\beta} - \beta^{0}) = - T^{-1/2} \, \overline{W}^{- 1} U_{1} + \overline{b} + o_{P}(1)$, where $\overline{b} \coloneqq - \overline{W}^{- 1} \big( \tfrac{\sqrt{N_{1}}}{\sqrt{N_{2}}} \overline{B}_{\alpha} + \tfrac{\sqrt{N_{2}}}{\sqrt{N_{1}}} \overline{B}_{\gamma} + \tfrac{\sqrt{N_{1} N_{2}}}{T} \overline{B}_{\rho} \big)$. By Slutsky's Theorem, $\sqrt{N_{1} N_{2}} \, (\hat{\beta} - \beta^{0}) \xrightarrow{d} \delta_{\overline{b}_{\infty}}$.\hfill\qedsymbol

\paragraph{Theorem \ref{theorem:asymptotic_distributions_interacted} - Expressions for $s = 2$ and $s = 3$.} The $s = 3$ expressions are obtained from the $s = 2$ expressions by: (i) replacing $j \neq i$ with $j > i$; (ii) multiplying every prefactor by $2$; (iii) dropping $\overline{B}_{2, \gamma, \infty}$.
\begin{align*}
	\overline{B}_{2, \alpha, \infty} \coloneqq& \, \overline{\mathbb{E}}\bigg[ - \frac{1}{N_{1} T} \sum_{i, t} \frac{\sum_{j \neq i} \EX{\ddot{x}_{ijt} (\dop{2} \psi)_{ijt} (\dop{1} \psi)_{ijt}}}{\sum_{j \neq i} \EX{(\dop{2}_{\mathcal{X}} \psi)_{ijt}}} \\
	& \, + \frac{1}{2 N_{1} T} \sum_{i, t} \frac{\big\{\sum_{j \neq i} \EX{\ddot{x}_{ijt} (\dop{3}_{\mathcal{X}} \psi)_{ijt}}\big\} \sum_{j \neq i} \EX{((\dop{1} \psi)_{ijt})^{2}}}{\big\{\sum_{j \neq i} \EX{(\dop{2}_{\mathcal{X}} \psi)_{ijt}}\big\}^{2}} \bigg] \, .
\end{align*}
$\overline{B}_{2, \gamma, \infty}$ is obtained from $\overline{B}_{2, \alpha, \infty}$ by interchanging $i \leftrightarrow j$ throughout (with $j = 1, \ldots, N_{1}$, $i \neq j$). Further,
\begin{align*}
	\overline{B}_{2, \rho, \infty} \coloneqq& \, \overline{\mathbb{E}}\bigg[ - \frac{1}{N_{1} (N_{1} - 1)} \sum_{i, j \neq i} \frac{\sum_{t \leq t^{\prime}} \EX{\ddot{x}_{ijt^{\prime}} (\dop{2} \psi)_{ijt^{\prime}} (\dop{1} \psi)_{ijt}}}{\sum_{t} \EX{(\dop{2}_{\mathcal{X}} \psi)_{ijt}}} \\
	& \, + \frac{1}{2 N_{1} (N_{1} - 1)} \sum_{i, j \neq i} \frac{\big\{\sum_{t} \EX{\ddot{x}_{ijt} (\dop{3}_{\mathcal{X}} \psi)_{ijt}}\big\} \big\{\sum_{t} \EX{((\dop{1} \psi)_{ijt})^{2}}\big\}}{\big\{\sum_{t} \EX{(\dop{2}_{\mathcal{X}} \psi)_{ijt}}\big\}^{2}} \bigg] \, ,  \\
	\overline{W}_{2, \infty} \coloneqq& \, \LEX{\frac{1}{N_{1} (N_{1} - 1) T} \sum_{i, j \neq i, t} \EX{(\dop{2}_{\mathcal{X}} \psi)_{ijt} \, \ddot{x}_{ijt} \, \ddot{x}_{ijt}^{\prime}}} \, , \\
	\overline{\Sigma}_{2, \infty} \coloneqq& \, \LEX{\frac{1}{N_{1} (N_{1} - 1) T} \sum_{i, j \neq i, t} \EX{((\dop{1} \psi)_{ijt})^{2} \, \ddot{x}_{ijt} \, \ddot{x}_{ijt}^{\prime}}} \, .
\end{align*}

\subsection{Proof of Theorem \ref{theorem:asymptotic_distributions_interacted_debiased}}
\label{appendix:main_results_distributions_interacted_debiased}

The proof of Theorem \ref{theorem:asymptotic_distributions_interacted_debiased} uses the following auxiliary result. Proof given in Online Appendix.

\begin{theorem}[Consistency of Estimators of Bias and Variance Components]
	\label{theorem:consistency_bias_variance_interacted}
	Let Lemmas \ref{lemma:regularity_conditions1_interacted} and \ref{lemma:regularity_conditions2_interacted} hold. Then, 
	\begin{align*}
		&\bignorm{\widehat{B}_{\alpha} - \overline{B}_{\alpha}}_{2} = o_{P}\big(T^{- \frac{1}{2}}\big) \, , \quad \bignorm{\widehat{B}_{\gamma} - \overline{B}_{\gamma}}_{2} = o_{P}\big(T^{- \frac{1}{2}}\big) \, , \quad \bignorm{\widehat{B}_{\rho} - \overline{B}_{\rho}}_{2} = o_{P}\big(T^{- \frac{1}{2}}\big) \, , \\
		&\norm{\widehat{W} - \overline{W}}_{2} = o_{P}\big(T^{- \frac{1}{2}}\big) \, , \quad \bignorm{\widehat{W}^{- 1} - \overline{W}^{- 1}}_{2} = o_{P}\big(T^{- \frac{1}{2}}\big) \, , \quad \norm{\widehat{\Sigma} - \overline{\Sigma}}_{2} = o_{P}(1) \, .
	\end{align*}
\end{theorem}

\noindent From the proof of Theorem \ref{theorem:asymptotic_distributions_interacted}, $\sqrt{N_{1} N_{2} T} (\hat{\beta} - \beta^{0}) = - \overline{W}^{- 1} U_{1} + \sqrt{T} \, \overline{b} + o_{P}(1)$. Decompose
\begin{equation*}
	\widehat{W}^{- 1} \Big( \sqrt{N_{1}} \tfrac{\sqrt{T}}{\sqrt{N_{2}}} \widehat{B}_{\alpha} + \sqrt{N_{2}} \tfrac{\sqrt{T}}{\sqrt{N_{1}}} \widehat{B}_{\gamma} + \sqrt{T} \tfrac{\sqrt{N_{1} N_{2}}}{T} \widehat{B}_{\rho} \Big) + \sqrt{T} \, \overline{b} \eqqcolon \mathfrak{E}_{1} + \cdots + \mathfrak{E}_{9} \, ,
\end{equation*}
where $\mathfrak{E}_{1}$, $\mathfrak{E}_{2}$, and $\mathfrak{E}_{3}$ involve $(\widehat{W}^{- 1} - \overline{W}^{- 1})(\widehat{B}_{\cdot} - \overline{B}_{\cdot})$, $\mathfrak{E}_{4}$, $\mathfrak{E}_{5}$, and $\mathfrak{E}_{6}$ involve $(\widehat{W}^{- 1} - \overline{W}^{- 1}) \overline{B}_{\cdot}$, $\mathfrak{E}_{7}$, $\mathfrak{E}_{8}$, and $\mathfrak{E}_{9}$ involve $\overline{W}^{- 1}(\widehat{B}_{\cdot} - \overline{B}_{\cdot})$, and $\cdot$ is a placeholder. By Assumption \ref{assumption:general}, Lemma \ref{lemma:regularity_conditions1_interacted}, and Theorem \ref{theorem:consistency_bias_variance_interacted},
\begin{equation*}
	\norm{\mathfrak{E}_{1}}_{2} \leq \sqrt{T} \tfrac{\sqrt{N_{1}}}{\sqrt{T}} \tfrac{\sqrt{T}}{\sqrt{N_{2}}} \bignorm{\widehat{W}^{- 1} - \overline{W}^{- 1}}_{2} \bignorm{\widehat{B}_{\alpha} - \overline{B}_{\alpha}}_{2} = o_{P}(1) \, ,
\end{equation*}
and $\mathfrak{E}_{4}, \mathfrak{E}_{7}$ admit identical bounds. The $\gamma$- and $\rho$-analogues are analogous. Thus $\sqrt{N_{1} N_{2} T} \, (\tilde{\beta} - \beta^{0}) = - \overline{W}^{- 1} U_{1} + o_{P}(1)$ with $\tilde{\beta} = \hat{\beta} + \widehat{W}^{- 1}(N_{2}^{- 1} \widehat{B}_{\alpha} + N_{1}^{- 1} \widehat{B}_{\gamma} + T^{- 1} \widehat{B}_{\rho})$. The remaining claims follow from Theorem \ref{theorem:consistency_bias_variance_interacted} and $- \overline{W}^{- 1} U_{1} \xrightarrow{d} \N(0, \overline{W}_{\infty}^{- 1} \overline{\Sigma}_{\infty} \overline{W}_{\infty}^{- 1})$ from the proof of Theorem \ref{theorem:asymptotic_distributions_interacted}.\hfill\qedsymbol

\paragraph{Theorem \ref{theorem:asymptotic_distributions_interacted_debiased} - Expressions for $s = 2$ and $s = 3$.} The $s = 3$ estimators are obtained from the $s = 2$ estimators by: (i) replacing $j \neq i$ with $j > i$; (ii) multiplying every prefactor by $2$; (iii) dropping $\widehat{B}_{2, \gamma}$.
\begin{align*}
	\widehat{B}_{2, \alpha} \coloneqq& \, - \frac{1}{N_{1} T} \sum_{i, t} \frac{\sum_{j \neq i} \hat{\ddot{x}}_{ijt} (\widehat{\dop{2} \psi})_{ijt} (\widehat{\dop{1} \psi})_{ijt}}{\sum_{j \neq i} (\widehat{\dop{2}_{\mathcal{X}} \psi})_{ijt}} \\
	& \, + \frac{1}{2 N_{1} T} \sum_{i, t} \frac{\big\{\sum_{j \neq i} \hat{\ddot{x}}_{ijt} (\widehat{\dop{3}_{\mathcal{X}} \psi})_{ijt}\big\} \sum_{j \neq i} ((\widehat{\dop{1} \psi})_{ijt})^{2}}{\big\{\sum_{j \neq i} (\widehat{\dop{2}_{\mathcal{X}} \psi})_{ijt}\big\}^{2}} \, .
\end{align*}
$\widehat{B}_{2, \gamma}$ is obtained from $\widehat{B}_{2, \alpha}$ by interchanging $i \leftrightarrow j$. Further,
\begin{align*}
	\widehat{B}_{2, \rho} \coloneqq& \, - \frac{1}{N_{1} (N_{1} - 1)} \sum_{i, j \neq i} \frac{\sum_{m = 0}^{h} T / (T - m) \sum_{t = m + 1}^{T} \hat{\ddot{x}}_{ijt} (\widehat{\dop{2} \psi})_{ijt} (\widehat{\dop{1} \psi})_{ij(t - q)}}{\sum_{t} (\widehat{\dop{2}_{\mathcal{X}} \psi})_{ijt}} \\
	& \, + \frac{1}{2 N_{1} (N_{1} - 1)} \sum_{i, j \neq i} \frac{\big\{\sum_{t} \hat{\ddot{x}}_{ijt} (\widehat{\dop{3}_{\mathcal{X}} \psi})_{ijt}\big\} \big\{\sum_{t} ((\widehat{\dop{1} \psi})_{ijt})^{2}\big\}}{\big\{\sum_{t} (\widehat{\dop{2}_{\mathcal{X}} \psi})_{ijt}\big\}^{2}} \, , \\
	\widehat{W}_{2} \coloneqq& \, \frac{1}{N_{1} (N_{1} - 1) T} \sum_{i, j \neq i, t} (\widehat{\dop{2}_{\mathcal{X}} \psi})_{ijt} \, \hat{\ddot{x}}_{ijt} \, \hat{\ddot{x}}_{ijt}^{\prime} \, , \\
	\widehat{\Sigma}_{2} \coloneqq& \, \frac{1}{N_{1} (N_{1} - 1) T} \sum_{i, j \neq i, t} ((\widehat{\dop{1} \psi})_{ijt})^{2} \, \hat{\ddot{x}}_{ijt} \, \hat{\ddot{x}}_{ijt}^{\prime} \, .
\end{align*}

\subsection{Proof of Theorem \ref{theorem:asymptotic_distributions_noninteracted}}
\label{appendix:main_results_distributions_noninteracted}

The proof of Theorem \ref{theorem:asymptotic_distributions_noninteracted} uses the following auxiliary result. Proof given in Online Appendix.

\begin{theorem}[Consistency of estimators of variance components]
	\label{theorem:consistency_variance_noninteracted}
	Let Lemmas \ref{lemma:regularity_conditions1_noninteracted} and \ref{lemma:regularity_conditions2_noninteracted} hold. Then, $\norm{\widehat{W} - \overline{W}}_{2} = o_{P}(1)$, $\norm{\widehat{W}^{- 1} - \overline{W}^{- 1}}_{2} = o_{P}(1)$, and $\norm{\widehat{\Sigma} - \overline{\Sigma}}_{2} = o_{P}(1)$.
\end{theorem}

\noindent By Corollary \ref{corollary:asymptotic_expansion_score_noninteracted}, $\sqrt{N_{1} N_{2} T} (\hat{\beta} - \beta^{0}) = - \overline{W}^{- 1} U + o_{P}(1)$. As in the proof of Theorem \ref{theorem:asymptotic_distributions_interacted}, $U \xrightarrow{d} \N(0, \overline{\Sigma}_{\infty})$, so $- \overline{W}^{- 1} U \xrightarrow{d} \N(0, \overline{W}_{\infty}^{- 1} \overline{\Sigma}_{\infty} \overline{W}_{\infty}^{- 1})$. Theorem \ref{theorem:consistency_variance_noninteracted} gives $\widehat{\Sigma} \xrightarrow{p} \overline{\Sigma}_{\infty}$ and $\widehat{W}^{- 1} \xrightarrow{p} \overline{W}^{- 1}_{\infty}$.\hfill\qedsymbol

\section{Jackknife Inference}
\label{supplement:jackknife_inference}

We outline jackknife inference using the framework of \textcite{h2026_jackknife} as an alternative to analytical bias correction. We focus on $s = 1$ and specification (3.b). $s \in \{2, 3\}$ and other specifications follow analogously. The setting matches Example 3 of \textcite{h2026_jackknife} with $N_{3} = T$.

\begin{assumption}[Homogeneity]
	\label{assumption:homogeneity}
	For each $s \in \{1, 2, 3\}$, the sequence $\{(y_{ijt}, x_{ijt}, \alpha_{it}^{0}, \gamma_{jt}^{0}, \rho_{ij}^{0})\}$ is identically distributed across $(i, j)$ and strictly stationary across $t$, for each $\lvert \mathcal{D}_{s} \rvert, T$.
\end{assumption}

Theorem \ref{theorem:asymptotic_distributions_interacted} gives
\begin{equation*}
	\sqrt{N_{1} N_{2} T}(\hat{\beta} - \beta^{0}) = \overline{W}_{\infty}^{- 1} \partial_{\beta} \mathcal{L}_{n}(\beta^{0}, \phi^{0}) - \overline{W}_{\infty}^{- 1} \overline{B}_{\infty} + o_{P}(1) \, ,
\end{equation*}
with $\partial_{\beta} \mathcal{L}_{n}(\beta^{0}, \phi^{0}) \xrightarrow{d} \mathcal{N}(0, \overline{\Sigma}_{\infty})$ and $\overline{B}_{\infty} = \overline{B}_{\alpha, \infty} \sqrt{N_{1} T / N_{2}} + \overline{B}_{\gamma, \infty} \sqrt{N_{2} T / N_{1}} + \overline{B}_{\rho, \infty} \sqrt{N_{1} N_{2} / T}$. For a $K$-vector $a$, $\varphi \coloneqq a^{\prime} \beta^{0}$, and $\hat{\varphi}^{(0)} \coloneqq a^{\prime} \hat{\beta}$,
\begin{equation*}
	r_{N}(\hat{\varphi}^{(0)} - \varphi) = z_{N} + \mu_{1, N} + \mu_{2, N} + \mu_{3, N} + o_{P}(1) \, ,
\end{equation*}
with $r_{N} \coloneqq \sqrt{N_{1} N_{2} T}$, $z_{N} \coloneqq a^{\prime} \overline{W}_{\infty}^{- 1} \partial_{\beta} \mathcal{L}_{n}(\beta^{0}, \phi^{0}) \xrightarrow{d} \mathcal{N}(0, \sigma^{2})$, $\mu_{1, N} \coloneqq - a^{\prime} \overline{W}_{\infty}^{- 1} \overline{B}_{\alpha, \infty} \sqrt{N_{1} T / N_{2}}$, $\mu_{2, N} \coloneqq - a^{\prime} \overline{W}_{\infty}^{- 1} \overline{B}_{\gamma, \infty} \sqrt{N_{2} T / N_{1}}$, $\mu_{3, N} \coloneqq - a^{\prime} \overline{W}_{\infty}^{- 1} \overline{B}_{\rho, \infty} \sqrt{N_{1} N_{2} / T}$. Assumption AD$^{\dagger}$ holds with $R = 3$. The divergence of $\mu_{r, N}$ under Assumption \ref{assumption:general} is permitted by Remark 10 of \textcite{h2026_jackknife}. Assumptions \ref{assumption:general} and \ref{assumption:homogeneity} suffice for Assumption JK$^{\dagger}$.

With $N_{1} = 2 N_{1, 2}$, $N_{2} = 2 N_{2, 2}$, $T = 2 T_{2}$, define six subsamples by halving each dimension in turn:
\begin{align*}
	S_{1} \coloneqq& \, \{1, \ldots, N_{1, 2}\} \times \{1, \ldots, N_{2}\} \times \{1, \ldots, T\} \, , \\
	S_{3} \coloneqq& \, \{1, \ldots, N_{1}\} \times \{1, \ldots, N_{2, 2}\} \times \{1, \ldots, T\} \, , \\
	S_{5} \coloneqq& \, \{1, \ldots, N_{1}\} \times \{1, \ldots, N_{2}\} \times \{1, \ldots, T_{2}\} \, ,
\end{align*}
where $S_{2}$, $S_{4}$, and $S_{6}$ denote the corresponding complementary subsamples, e.g., $S_{1} \cup S_{2} = S_{0}$, where $S_{0}$ denotes the full sample. Set $\hat{\varphi}^{(j)} \coloneqq a^{\prime} \hat{\beta}^{(j)}$ on $S_{j}$ and $\hat{\boldsymbol{\varphi}} \coloneqq (\hat{\varphi}^{(0)}, \ldots, \hat{\varphi}^{(6)})^{\top}$.

Assumption JK$^{\dagger}$ holds with $m = 7$ and
\begin{equation*}
	A = \begin{pmatrix}
		1 & 1 & 1 \\
        1 & 2 & 1 \\
		1 & 2 & 1 \\
        2 & 1 & 1 \\
		2 & 1 & 1 \\
		1 & 1 & 2 \\
		1 & 1 & 2
	\end{pmatrix} \, ,
	\quad
	C = \begin{pmatrix}
		1 & 1 & 1 & 1 & 1 & 1 & 1 \\
		1 & 2 & 0 & 1 & 1 & 1 & 1 \\
		1 & 0 & 2 & 1 & 1 & 1 & 1 \\
		1 & 1 & 1 & 2 & 0 & 1 & 1 \\
		1 & 1 & 1 & 0 & 2 & 1 & 1 \\
		1 & 1 & 1 & 1 & 1 & 2 & 0 \\
		1 & 1 & 1 & 1 & 1 & 0 & 2
	\end{pmatrix} \, .
\end{equation*}
The columns of $A$ correspond to $\mu_{1, N}$, $\mu_{2, N}$, and $\mu_{3, N}$, i.e., to the biases induced by $\alpha_{it}^{0}$, $\gamma_{jt}^{0}$, and $\rho_{ij}^{0}$. Each $\alpha_{it}^{0}$ is estimated from $N_{2}$ observations, each $\gamma_{jt}^{0}$ from $N_{1}$, and each $\rho_{ij}^{0}$ from $T$. Row $j$ of $A$ collects the inflation factors for subsample $S_{j}$. For example, halving $N_{1}$ doubles $\mu_{2, N}$, halving $N_{2}$ doubles $\mu_{1, N}$, and halving $T$ doubles $\mu_{3, N}$, leaving the other two bias terms unchanged in each case. The MVUJ weights solving $\min_{v \in \mathcal{V}} v^{\top} C v$ subject to $v^{\prime} A = \mathbf{0}_{3}^{\prime}$, $v^{\prime} \iota_{7} = 1$ are $v^{\ast} = \linebreak (4, -\tfrac{1}{2}, -\tfrac{1}{2}, -\tfrac{1}{2}, -\tfrac{1}{2}, -\tfrac{1}{2}, -\tfrac{1}{2})^{\prime}$, yielding $\tilde{\varphi} \coloneqq v^{\ast \prime} \hat{\boldsymbol{\varphi}} = 4 \hat{\varphi}^{(0)} - \tfrac{1}{2} \sum_{j = 1}^{6} \hat{\varphi}^{(j)}$. Since $m - \text{rank}(A, \iota_{7}) = 3$, up to $q = 3$ variance-weight vectors are available. A convenient choice is
\begin{equation*}
	u_{1}^{\ast} = (0, -\tfrac{1}{2}, \tfrac{1}{2}, 0, 0, 0, 0)^{\prime} \, , \quad
	u_{2}^{\ast} = (0, 0, 0, -\tfrac{1}{2}, \tfrac{1}{2}, 0, 0)^{\prime} \, , \quad
	u_{3}^{\ast} = (0, 0, 0, 0, 0, -\tfrac{1}{2}, \tfrac{1}{2})^{\prime} \, .
\end{equation*}
For $q \in \{1, 2, 3\}$, $\tilde{\sigma}_{q}^{2} = (4 q)^{- 1} \sum_{l = 1}^{q} (\hat{\varphi}^{(2 l)} - \hat{\varphi}^{(2 l - 1)})^{2}$, and Theorem 2.4 of \textcite{h2026_jackknife} gives the following.

\begin{proposition}[Jackknife inference]
	\label{proposition:jackknife_inference}
	Let Assumptions \ref{assumption:general} and \ref{assumption:homogeneity} hold. Then for $q \in \{1, 2, 3\}$,
	\begin{equation*}
		J_{q} \coloneqq \frac{\tilde{\varphi} - \varphi}{\tilde{\sigma}_{q}} \xrightarrow{d} t_{q} \, .
	\end{equation*}
\end{proposition}

\clearpage


\setcounter{page}{1}
\setcounter{section}{0}
\setcounter{theorem}{0}
\setcounter{lemma}{0}
\setcounter{corollary}{0}
\renewcommand{\thesection}{S.\arabic{section}}
\renewcommand{\thetheorem}{S.\arabic{theorem}}
\renewcommand{\thelemma}{S.\arabic{lemma}}
\renewcommand{\thecorollary}{S.\arabic{corollary}}

\begin{center}
	\LARGE\bfseries
	Online Appendix
\end{center}

\section{Proof of Asymptotic Expansion Theorems}
\label{supplement:proof_of_asymptotic_expansions}

\subsection{Partial Derivatives of Objective Function}
\label{supplement:partial_derivatives}

Given the criterion function $\psi(y, \pi)$ and the linear index $\pi(\beta, \phi) = X \beta + D \phi$, the first- and second-order derivatives of $\mathcal{L}_{n}(\beta, \phi)$ involving the constraint term $VV^{\prime}$ are
\begin{align*}
	&\frac{\partial \mathcal{L}_{n}(\beta, \phi)}{\partial \beta} = \frac{X^{\prime} \dop{1} \psi(\beta, \phi)}{w_{n}} \, , \quad \frac{\partial \mathcal{L}_{n}(\beta, \phi)}{\partial \phi} = u(\beta, \phi) + \frac{V V^{\prime} \phi}{w_{n}} \, , \\
	&\frac{\partial^{2} \mathcal{L}_{n}(\beta, \phi)}{\partial \phi \partial \phi^{\prime}} = H(\beta, \phi) \, .
\end{align*}
All remaining derivatives take the unified form
\begin{equation*}
	\frac{\partial^{\mathrm{r}+2} \mathcal{L}_{n}(\beta, \phi)}{\partial \theta_{1} \cdots \partial \theta_{\mathrm{r}} \, \partial \omega \, \partial \omega^{\prime}} = \frac{A^{\prime} \, \nabla^{\mathrm{r}+2} \psi(\beta, \phi) \, \diag(v_{1} \odot \cdots \odot v_{\mathrm{r}}) \, B}{w_{n}} \, ,
\end{equation*}
where $\mathrm{r} \geq 0$, $\omega, \omega^{\prime} \in \{\beta, \phi\}$ with $A = X$ if $\omega = \beta$ and $A = D$ if $\omega = \phi$ (and similarly $B$ for $\omega^{\prime}$). Each scalar differentiator $\theta_{m} \in \{\beta_{k}, \phi_{l}\}$ contributes $v_{m} = X e_{k}$ or $v_{m} = D e_{l}$, and for $\mathrm{r} = 0$ the $\diag$ factor is absent. This covers the $\partial^2\mathcal{L}_n/\partial\beta\partial\beta^{\prime}$ and $\partial^2\mathcal{L}_n/\partial\beta\partial\phi^{\prime}$ blocks ($\mathrm{r}=0$) as well as all third-, fourth-, and fifth-order mixed derivatives.

\subsection{Taylor Expansions of Legendre Transforms}
\label{supplement:legendre_taylor}

Let $\mathfrak{B}(\varepsilon) \times \mathfrak{P}(\eta, q)$ denote the neighborhood of the true parameters $(\beta^{0}, \phi^{0})$ for some $\varepsilon > 0$, $\eta > 0$, and $q \geq 1$. Following \textcite{fw2016}, we apply the Legendre transform to $\mathcal{L}_{n}(\beta, \phi)$ to expand $\partial \mathcal{L}_{n}(\beta, \hat{\phi}(\beta)) / \partial \beta$ for any $\beta \in \mathfrak{B}(\varepsilon)$.

\begin{lemma}[Taylor Expansions of Legendre Transforms]
	\label{lemma:taylor_expansions}
	\label{theorem:taylor_expansions}
	Let $\varepsilon, \eta > 0$, $q \geq 1$. Assume that (i) $H(\beta, \phi)$ is invertible on $\mathfrak{B}(\varepsilon) \times \mathfrak{P}(\eta, q)$; (ii) $\mathcal{L}_{n}$ is five times continuously differentiable there; (iii) $\mathbf{0}_{L} \in u(\beta, \mathfrak{P}(\eta, q))$ for all $\beta \in \mathfrak{B}(\varepsilon)$; (iv) $V V^{\prime} \phi^{0} = \mathbf{0}_{L}$. Then, for all $\beta \in \mathfrak{B}(\varepsilon)$,
	\begin{align*}
		&\frac{\partial \mathcal{L}_{n}(\beta, \hat{\phi}(\beta))}{\partial \beta} = \mathcal{T}^{(1)} + \mathcal{T}^{(2)}(\beta) + \mathcal{T}^{(3)} + \mathcal{T}^{(4)}(\beta) + \mathcal{T}^{(5)}(\beta) + \mathcal{T}^{(6)} \, + \\
		& \quad \sum_{r = 1}^{2} \mathcal{T}_{r}^{(7)}(\beta) + \sum_{r = 1}^{3} \mathcal{T}_{r}^{(8)}(\beta) + \sum_{r = 1}^{3} \mathcal{T}_{r}^{(9)}(\beta) + \sum_{r = 1}^{2} \mathcal{T}_{r}^{(10)} + \sum_{r = 1}^{8} \mathcal{T}_{r}^{(11)}
	\end{align*}
	and
	\begin{equation*}
		\hat{\phi}(\beta) = \phi^{0} + \mathcal{T}^{(12)}(\beta) + \mathcal{T}^{(13)} + \mathcal{T}^{(14)}(\beta) + \mathcal{T}^{(15)}(\beta) + \mathcal{T}^{(16)} + \sum_{r = 1}^{2} \mathcal{T}_{r}^{(17)} \, ,
	\end{equation*}
	where
	\begin{align*}
		&\mathcal{T}^{(1)} \coloneqq \frac{X^{\prime} \dop{1} \psi}{w_{n}} \, , \quad \mathcal{T}^{(2)}(\beta) \coloneqq \frac{X^{\prime} \nabla^{2} \psi M X (\beta - \beta^{0})}{w_{n}} = \frac{(M X)^{\prime} \nabla^{2} \psi M X (\beta - \beta^{0})}{w_{n}} \, , \\
		&\mathcal{T}^{(3)} \coloneqq - \frac{X^{\prime} \nabla^{2} \psi Q \dop{1} \psi}{w_{n}} \, , \quad \mathcal{T}^{(4)}(\beta) \coloneqq \frac{(M X)^{\prime} \nabla^{3} \psi \diag(M X (\beta - \beta^{0})) M X (\beta - \beta^{0})}{2 w_{n}} \, , \\
		&\mathcal{T}^{(5)}(\beta) \coloneqq - \frac{(M X)^{\prime} \nabla^{3} \psi \diag(Q \dop{1} \psi) M X (\beta - \beta^{0})}{w_{n}} \, , \\
		&\mathcal{T}^{(6)} \coloneqq \frac{(M X)^{\prime} \nabla^{3} \psi \diag(Q \dop{1} \psi) Q \dop{1} \psi}{2 w_{n}} \, , \\
		&\mathcal{T}_{1}^{(7)}(\beta) \coloneqq - \frac{(\check{M} X)^{\prime} \check{\nabla^{3} \psi} \diag(\check{M} X (\beta - \beta^{0})) \check{Q} \diag(\check{M} X (\beta - \beta^{0})) \check{\nabla^{3} \psi} \check{M} X (\beta - \beta^{0})}{6 w_{n}}  \, , \\
		&\mathcal{T}_{2}^{(7)}(\beta) \coloneqq - \frac{(\check{M} X)^{\prime} \check{\nabla^{4} \psi} \diag(\check{M} X (\beta - \beta^{0})) \diag(\check{M} X (\beta - \beta^{0})) \check{M} X (\beta - \beta^{0})}{6 w_{n}} \, , \\
		&\mathcal{T}_{1}^{(8)}(\beta) \coloneqq \frac{(\check{M} X)^{\prime} \check{\nabla^{3} \psi} \diag(\check{M} X (\beta - \beta^{0})) \check{Q} \diag(\check{M} X (\beta - \beta^{0})) \check{\nabla^{3} \psi} \check{Q} \dop{1} \psi}{w_{n}} \, , \\
		&\mathcal{T}_{2}^{(8)}(\beta) \coloneqq \frac{(\check{M} X)^{\prime} \check{\nabla^{3} \psi} \diag(\check{Q} \dop{1} \psi) \check{Q} \diag(\check{M} X (\beta - \beta^{0})) \check{\nabla^{3} \psi} \check{M} X (\beta - \beta^{0})}{2 w_{n}} \, , \\
		&\mathcal{T}_{3}^{(8)}(\beta) \coloneqq - \frac{(\check{M} X)^{\prime} \check{\nabla^{4} \psi} \diag(\check{M} X (\beta - \beta^{0})) \diag(\check{M} X (\beta - \beta^{0})) \check{Q} \dop{1} \psi}{2 w_{n}} \, , \\
		&\mathcal{T}_{1}^{(9)}(\beta) \coloneqq - \frac{(\check{M} X)^{\prime} \check{\nabla^{3} \psi} \diag(\check{Q} \dop{1} \psi) \check{Q} \diag(\check{Q} \dop{1} \psi) \check{\nabla^{3} \psi} \check{M} X (\beta - \beta^{0})}{w_{n}} \, , \\
		&\mathcal{T}_{2}^{(9)}(\beta) \coloneqq - \frac{(\check{M} X)^{\prime} \check{\nabla^{3} \psi} \diag(\check{M} X (\beta - \beta^{0})) \check{Q} \diag(\check{Q} \dop{1} \psi) \check{\nabla^{3} \psi} \check{Q} \dop{1} \psi}{2 w_{n}} \, , \\
		&\mathcal{T}_{3}^{(9)}(\beta) \coloneqq \frac{(\check{M} X)^{\prime} \check{\nabla^{4} \psi} \diag(\check{M} X (\beta - \beta^{0})) \diag(\check{Q} \dop{1} \psi) \check{Q} \dop{1} \psi}{2 w_{n}} \, , \\
		&\mathcal{T}_{1}^{(10)} \coloneqq \frac{(M X)^{\prime} \nabla^{3} \psi \diag(Q \dop{1} \psi) Q \diag(Q \dop{1} \psi) \nabla^{3} \psi Q \dop{1} \psi}{2 w_{n}} \, ,  \\
		&\mathcal{T}_{2}^{(10)} \coloneqq - \frac{(M X)^{\prime} \nabla^{4} \psi \diag(Q \dop{1} \psi) \diag(Q \dop{1} \psi) Q \dop{1} \psi}{6 w_{n}} \, , \\
		&\mathcal{T}_{1}^{(11)} \coloneqq - \frac{(\check{M} X)^{\prime} \check{\nabla^{3} \psi}  \diag(\check{Q} \dop{1} \psi) \check{Q} \diag(D u) \diag(\check{Q} \dop{1} \psi) \check{Q} \dop{1} \psi}{24 w_{n}} \, , \\
		&\mathcal{T}_{2}^{(11)} \coloneqq \frac{(\check{M} X)^{\prime} \check{\nabla^{3} \psi} \diag(\check{Q} \dop{1} \psi) \check{Q} \diag(\check{Q} \dop{1} \psi) \check{Q} \diag(\check{Q} \dop{1} \psi) \check{\nabla^{3} \psi} \check{Q} \dop{1} \psi}{24 w_{n}} \, , \\
		&\mathcal{T}_{3}^{(11)} \coloneqq \frac{(\check{M} X)^{\prime} \check{\nabla^{3} \psi} \diag(\check{Q} \dop{1} \psi) \check{Q} \check{\nabla^{3} \psi} \diag(\check{Q} \dop{1} \psi) \check{Q} \diag(\check{Q} \dop{1} \psi) \check{\nabla^{3} \psi} \check{Q} \dop{1} \psi}{3 w_{n}} \, , \\
		&\mathcal{T}_{4}^{(11)} \coloneqq - \frac{(\check{M} X)^{\prime} \check{\nabla^{4} \psi} \diag(\check{Q} \dop{1} \psi) \diag(\check{Q} \dop{1} \psi) \check{Q} \diag(\check{Q} \dop{1} \psi) \check{\nabla^{3} \psi} \check{Q} \dop{1} \psi}{6 w_{n}} \, , \\
		&\mathcal{T}_{5}^{(11)} \coloneqq - \frac{(\check{M} X)^{\prime} \check{\nabla^{3} \psi} \diag(\check{Q} \dop{1} \psi) \check{Q} \diag(\check{Q} \dop{1} \psi) \diag(\check{Q} \dop{1} \psi) \check{\nabla^{4} \psi} \check{Q} \dop{1} \psi}{12 w_{n}} \, , \\
		&\mathcal{T}_{6}^{(11)} \coloneqq - \frac{(\check{M} X)^{\prime} \check{\nabla^{4} \psi} \diag(D u) \diag(\check{Q} \dop{1} \psi) \check{Q} \diag(\check{Q} \dop{1} \psi) \check{\nabla^{3} \psi} \check{Q} \dop{1} \psi}{24 w_{n}} \, , \\
		&\mathcal{T}_{7}^{(11)} \coloneqq \frac{(\check{M} X)^{\prime} \check{\nabla^{4} \psi} \diag(\check{Q} \diag(\check{Q} \dop{1} \psi) \check{\nabla^{3} \psi} \check{Q} \dop{1} \psi) \check{Q} \diag(\check{Q} \dop{1} \psi)  \check{\nabla^{3} \psi}  \check{Q} \dop{1} \psi}{24 w_{n}} \, , \\
		&\mathcal{T}_{8}^{(11)} \coloneqq \frac{(\check{M} X)^{\prime} \check{\nabla^{5} \psi} \diag(\check{Q} \dop{1} \psi) \diag(\check{Q} \dop{1} \psi) \diag(\check{Q} \dop{1} \psi) \check{Q} \dop{1} \psi}{24 w_{n}} \, , \\
		&\mathcal{T}^{(12)}(\beta) \coloneqq - \frac{H^{- 1} D^{\prime} \nabla^{2} \psi X (\beta - \beta^{0})}{w_{n}} \, , \quad \mathcal{T}^{(13)} \coloneqq - \frac{H^{- 1} D^{\prime} \dop{1} \psi}{w_{n}}= - H^{- 1} u \, , \\
		&\mathcal{T}^{(14)}(\beta) \coloneqq - \frac{\check{H}^{- 1} D^{\prime} \check{\nabla^{3} \psi} \diag(\check{M} X (\beta - \beta^{0})) \check{M} X (\beta - \beta^{0})}{2 \, w_{n}} \, , \\
		&\mathcal{T}^{(15)}(\beta) \coloneqq \frac{\check{H}^{- 1} D^{\prime} \check{\nabla^{3} \psi} \diag(\check{Q} \dop{1} \psi) \check{M} X (\beta - \beta^{0})}{w_{n}} \, , \\
		&\mathcal{T}^{(16)} \coloneqq - \frac{H^{- 1} D^{\prime} \nabla^{3} \psi \diag(Q \dop{1} \psi) Q \dop{1} \psi}{2 \, w_{n}} \, , \\
		&\mathcal{T}_{1}^{(17)} \coloneqq - \frac{\check{H}^{- 1} D^{\prime} \check{\nabla^{3} \psi} \diag(\check{Q} \dop{1} \psi) \check{Q} \diag(\check{Q} \dop{1} \psi) \check{\nabla^{3} \psi} \check{Q} \dop{1} \psi}{2 \, w_{n}} \, , \\
		&\mathcal{T}_{2}^{(17)} \coloneqq \frac{\check{H}^{- 1} D^{\prime} \check{\nabla^{4} \psi} \diag(\check{Q} \dop{1} \psi) \diag(\check{Q} \dop{1} \psi) \check{Q} \dop{1} \psi}{6 \, w_{n}} \, ,
	\end{align*}
	with a check mark indicating evaluation at intermediate values $\check{\beta}$ and $\check{\phi}$, which may differ across elements; $\check{\beta}$ lies on the segment between $\beta^{0}$ and $\beta$, and $\check{\upsilon} \coloneqq u(\check{\beta}, \check{\phi})$ lies on the segment between $u = u(\beta^{0}, \phi^{0})$ and $0$, with $(\check{\phi}, \check{\upsilon})$ in one-to-one correspondence.
\end{lemma}

\paragraph{Proof of Lemma \ref{lemma:taylor_expansions}.} Assumption (i) implies $\mathcal{L}_{n}(\beta, \phi)$ is strictly convex in $\phi$ for all $\beta \in \mathfrak{B}(\varepsilon)$, so the Legendre transform
\begin{align}
	\mathcal{L}_{n}^{\ast}(\beta, \upsilon) =& \, \max_{\phi \in \mathfrak{P}(\eta, q)} (\upsilon^{\prime} \phi - \mathcal{L}_{n}(\beta, \phi)) \, , \label{eq:legendre_transform} \\
	\phi^{\ast}(\beta, \upsilon) =& \; \underset{\phi \in \mathfrak{P}(\eta, q)}{\argmax} (\upsilon^{\prime} \phi - \mathcal{L}_{n}(\beta, \phi)) \label{eq:legendre_transform_argmax}
\end{align}
is well-defined on $\Theta^{\ast} \coloneqq \{(\beta, \upsilon) \in \Real^{K + L} \colon (\beta, \phi^{\ast}(\beta, \upsilon)) \in \mathfrak{B}(\varepsilon) \times \mathfrak{P}(\eta, q)\}$, and $\phi \mapsto u(\beta, \phi)$ is a bijection for each $\beta \in \mathfrak{B}(\varepsilon)$. Differentiating the first-order condition $\upsilon - \partial_{\phi} \mathcal{L}_{n}(\beta, \phi^{\ast}(\beta, \upsilon)) = \mathbf{0}_{L}$ with respect to $\beta^{\prime}$ and $\upsilon^{\prime}$ and rearranging gives
\begin{equation}
	\label{eq:pdes}
	\frac{\partial \phi^{\ast}(\beta, \upsilon)}{\partial \beta^{\prime}} = - \underbrace{(H(\beta, \upsilon))^{- 1} J(\beta, \upsilon)}_{\eqqcolon S(\beta, \upsilon)} \, , \qquad \frac{\partial \phi^{\ast}(\beta, \upsilon)}{\partial \upsilon^{\prime}} = (H(\beta, \upsilon))^{- 1} \, , 
\end{equation}
where
\begin{equation*}
	H(\beta, \upsilon) \coloneqq \frac{\partial^{2} \mathcal{L}_{n}(\beta, \phi^{\ast}(\beta, \upsilon))}{\partial \phi \partial \phi^{\prime}} \, , \qquad  J(\beta, \upsilon) \coloneqq \frac{\partial^{2} \mathcal{L}_{n}(\beta, \phi^{\ast}(\beta, \upsilon))}{\partial \phi \partial \beta^{\prime}} \, .
\end{equation*}
Assumption (iii) ensures $\mathbf{0}_{L} \in u(\beta, \mathfrak{P}(\eta, q))$ for all $\beta \in \mathfrak{B}(\varepsilon)$, so evaluating \eqref{eq:legendre_transform} at $\upsilon = \mathbf{0}_{L}$ gives $\mathcal{L}_{n}^{\ast}(\beta, \mathbf{0}_{L}) = \mathcal{L}_{n}(\beta, \hat{\phi}(\beta))$.

From \eqref{eq:legendre_transform} and \eqref{eq:legendre_transform_argmax},
\begin{equation}
	\label{eq:legendre_transform2}
	\mathcal{L}_{n}^{\ast}(\beta, \upsilon) = \upsilon^{\prime} \phi^{\ast}(\beta, \upsilon) - \mathcal{L}_{n}(\beta, \phi^{\ast}(\beta, \upsilon)) \, .
\end{equation}
Since $\partial_{\beta} \mathcal{L}_{n}(\beta, \hat{\phi}(\beta)) = -\partial_{\beta} \mathcal{L}_{n}^{\ast}(\beta, \upsilon)|_{\upsilon = 0}$ and $\hat{\phi}(\beta) = \partial_{\upsilon} \mathcal{L}_{n}^{\ast}(\beta, \upsilon)|_{\upsilon = 0}$, following \textcite{fwct2014} we apply Taylor's theorem element-wise. Assumption (ii) ensures that, for example, $\partial^{5} \mathcal{L}_{n}^{\ast}/(\partial \beta_{k} \partial \upsilon_{l} \partial \upsilon_{l^{\prime}} \partial \upsilon \partial \upsilon^{\prime})$ exists. For each $k \in \{1, \ldots, K\}$, expanding $\partial_{\beta_{k}} \mathcal{L}_{n}^{\ast}(\beta, \upsilon)$ around $(\beta^{0}, u)$ gives
\begin{align}
	\label{eq:taylor_expansion_legendre_scores_beta}
	&\frac{\partial \mathcal{L}_{n}(\beta, \hat{\phi}(\beta))}{\partial \beta_{k}} = - \frac{\partial \mathcal{L}_{n}^{\ast}(\beta, \upsilon)}{\partial \beta_{k}}\biggr\rvert_{\substack{\beta \, = \, \beta^{0} \\ \upsilon \, = \, u^{\phantom{0}}}} - \frac{\partial^{2} \mathcal{L}_{n}^{\ast}(\beta, \upsilon)}{\partial \beta_{k} \partial \beta^{\prime}}\biggr\rvert_{\substack{\beta \, = \, \beta^{0} \\ \upsilon \, = \, u^{\phantom{0}}}} (\beta - \beta^{0}) + \frac{\partial^{2} \mathcal{L}_{n}^{\ast}(\beta, \upsilon)}{\partial \beta_{k} \partial \upsilon^{\prime}}\biggr\rvert_{\substack{\beta \, = \, \beta^{0} \\ \upsilon \, = \, u^{\phantom{0}}}} u \, - \nonumber \\
	& \quad \frac{1}{2} (\beta - \beta^{0})^{\prime} \frac{\partial^{3} \mathcal{L}_{n}^{\ast}(\beta, \upsilon)}{\partial \beta_{k} \partial \beta \partial \beta^{\prime}}\biggr\rvert_{\substack{\beta \, = \, \beta^{0} \\ \upsilon \, = \, u^{\phantom{0}}}} (\beta - \beta^{0}) + (\beta - \beta^{0})^{\prime} \frac{\partial^{3} \mathcal{L}_{n}^{\ast}(\beta, \upsilon)}{\partial \beta_{k} \partial \beta \partial \upsilon^{\prime}}\biggr\rvert_{\substack{\beta \, = \, \beta^{0} \\ \upsilon \, = \, u^{\phantom{0}}}} u \, - \nonumber \\
	& \quad \frac{1}{2} u^{\prime} \frac{\partial^{3} \mathcal{L}_{n}^{\ast}(\beta, \upsilon)}{\partial \beta_{k} \partial \upsilon \partial \upsilon^{\prime}}\biggr\rvert_{\substack{\beta \, = \, \beta^{0} \\ \upsilon \, = \, u^{\phantom{0}}}} u - \frac{1}{6} (\beta - \beta^{0})^{\prime} \sum_{k^{\prime} = 1}^{K} \frac{\partial^{4} \mathcal{L}_{n}^{\ast}(\beta, \upsilon)}{\partial \beta_{k} \partial \beta_{k^{\prime}} \partial \beta \partial \beta^{\prime}}\biggr\rvert_{\substack{\beta \, = \, \check{\beta} \\ \upsilon \, = \, \check{\upsilon}}} (\beta_{k^{\prime}} - \beta_{k^{\prime}}^{0}) (\beta - \beta^{0}) \, + \nonumber \\
	& \quad \frac{1}{2} (\beta - \beta^{0})^{\prime} \sum_{k^{\prime} = 1}^{K} \frac{\partial^{4} \mathcal{L}_{n}^{\ast}(\beta, \upsilon)}{\partial \beta_{k} \partial \beta_{k^{\prime}} \partial \beta \partial \upsilon^{\prime}}\biggr\rvert_{\substack{\beta \, = \, \check{\beta} \\ \upsilon \, = \, \check{\upsilon}}} (\beta_{k^{\prime}} - \beta_{k^{\prime}}^{0}) u \, - \nonumber \\
	& \quad \frac{1}{2} u^{\prime} \sum_{k^{\prime} = 1}^{K} \frac{\partial^{4} \mathcal{L}_{n}^{\ast}(\beta, \upsilon)}{\partial \beta_{k} \partial \beta_{k^{\prime}} \partial \upsilon \partial \upsilon^{\prime}}\biggr\rvert_{\substack{\beta \, = \, \check{\beta} \\ \upsilon \, = \, \check{\upsilon}}} (\beta_{k^{\prime}} - \beta_{k^{\prime}}^{0}) u + \frac{1}{6} u^{\prime} \sum_{l = 1}^{L} \frac{\partial^{4} \mathcal{L}_{n}^{\ast}(\beta, \upsilon)}{\partial \beta_{k} \partial \upsilon_{l} \partial \upsilon \partial \upsilon^{\prime}}\biggr\rvert_{\substack{\beta \, = \, \beta^{0} \\ \upsilon \, = \, u^{\phantom{0}}}} \upsilon_{l}^{0} u \, - \nonumber \\
	& \quad \frac{1}{24} u^{\prime} \sum_{l = 1}^{L} \sum_{l^{\prime} = 1}^{L} \frac{\partial^{5} \mathcal{L}_{n}^{\ast}(\beta, \upsilon)}{\partial \beta_{k} \partial \upsilon_{l} \partial \upsilon_{l^{\prime}} \partial \upsilon \partial \upsilon^{\prime}}\biggr\rvert_{\substack{\beta \, = \, \check{\beta} \\ \upsilon \, = \, \check{\upsilon}}} \upsilon_{l}^{0} \upsilon_{l^{\prime}}^{0}  u  
\end{align}
and for each $l \in \{1, \ldots, L\}$, expanding $\partial_{\upsilon_{l}} \mathcal{L}_{n}^{\ast}(\beta, \upsilon)$ gives
\begin{align}
	\label{eq:taylor_expansion_legendre_scores_upsilon}
	&e_{l}^{\prime} \hat{\phi}(\beta) = \frac{\partial \mathcal{L}_{n}^{\ast}(\beta, \upsilon)}{\partial \upsilon_{l}}\biggr\rvert_{\substack{\beta \, = \, \beta^{0} \\ \upsilon \, = \, u^{\phantom{0}}}} + \frac{\partial^{2} \mathcal{L}_{n}^{\ast}(\beta, \upsilon)}{\partial \upsilon_{l} \partial \beta^{\prime}}\biggr\rvert_{\substack{\beta \, = \, \beta^{0} \\ \upsilon \, = \, u^{\phantom{0}}}} (\beta - \beta^{0}) - \frac{\partial^{2} \mathcal{L}_{n}^{\ast}(\beta, \upsilon)}{\partial \upsilon_{l} \partial \upsilon^{\prime}}\biggr\rvert_{\substack{\beta \, = \, \beta^{0} \\ \upsilon \, = \, u^{\phantom{0}}}} u \, + \nonumber \\
	& \quad \frac{1}{2} (\beta - \beta^{0})^{\prime} \frac{\partial^{3} \mathcal{L}_{n}^{\ast}(\beta, \upsilon)}{\partial \upsilon_{l} \partial \beta \partial \beta^{\prime}}\biggr\rvert_{\substack{\beta \, = \, \check{\beta} \\ \upsilon \, = \, \check{\upsilon}}} (\beta - \beta^{0}) - (\beta - \beta^{0})^{\prime} \frac{\partial^{3} \mathcal{L}_{n}^{\ast}(\beta, \upsilon)}{\partial \upsilon_{l} \partial \beta \partial \upsilon^{\prime}}\biggr\rvert_{\substack{\beta \, = \, \check{\beta} \\ \upsilon \, = \, \check{\upsilon}}} u \, + \nonumber \\
	& \quad \frac{1}{2} u^{\prime} \frac{\partial^{3} \mathcal{L}_{n}^{\ast}(\beta, \upsilon)}{\partial \upsilon_{l} \partial \upsilon \partial \upsilon^{\prime}}\biggr\rvert_{\substack{\beta \, = \, \beta^{0} \\ \upsilon \, = \, u^{\phantom{0}}}} u - \frac{1}{6} u^{\prime} \sum_{l^{\prime} = 1}^{L} \frac{\partial^{4} \mathcal{L}_{n}^{\ast}(\beta, \upsilon)}{\partial \upsilon_{l} \partial \upsilon_{l^{\prime}} \partial \upsilon \partial \upsilon^{\prime}}\biggr\rvert_{\substack{\beta \, = \, \check{\beta} \\ \upsilon \, = \, \check{\upsilon}}} \upsilon_{l^{\prime}}^{0} u \, .
\end{align}
Differentiating \eqref{eq:legendre_transform2} gives
\begin{align}
	&\frac{\partial \mathcal{L}_{n}^{\ast}(\beta, \upsilon)}{\partial \beta_{k}} = - \, \frac{\partial \mathcal{L}_{n}(\beta, \phi^{\ast}(\beta, \upsilon))}{\partial \beta_{k}} \, , \quad \frac{\partial \mathcal{L}_{n}^{\ast}(\beta, \upsilon)}{\partial \upsilon} = \phi^{\ast}(\beta, \upsilon) \, , \label{eq:derivatives_legendre} \\
	&\frac{\partial^{2} \mathcal{L}_{n}^{\ast}(\beta, \upsilon)}{\partial \beta_{k} \partial \beta^{\prime}} = - \, \frac{\partial^{2} \mathcal{L}_{n}(\beta, \phi^{\ast}(\beta, \upsilon))}{\partial \beta_{k} \partial \beta^{\prime}} + (S(\beta, \upsilon) e_{k})^{\prime} J(\beta, \upsilon) \, , \nonumber \\
	&\frac{\partial^{2} \mathcal{L}_{n}^{\ast}(\beta, \upsilon)}{\partial \beta_{k} \partial \upsilon^{\prime}} = - (S(\beta, \upsilon) e_{k})^{\prime} \, , \quad  \frac{\partial^{2} \mathcal{L}_{n}^{\ast}(\beta, \upsilon)}{\partial \upsilon \partial \upsilon^{\prime}} = (H(\beta, \upsilon))^{- 1} \, , \nonumber \\
	&\frac{\partial^{3} \mathcal{L}_{n}^{\ast}(\beta, \upsilon)}{\partial \beta_{k} \partial \beta \partial \beta^{\prime}} = - A_{k}^{(1)}(\beta, \upsilon) + A_{k}^{(2)}(\beta, \upsilon) S(\beta, \upsilon) + (S(\beta, \upsilon))^{\prime} (A_{k}^{(2)}(\beta, \upsilon))^{\prime} -  \nonumber \\
	&\quad (S(\beta, \upsilon))^{\prime} A_{k}^{(3)}(\beta, \upsilon) S(\beta, \upsilon) \, , \nonumber \\
	&\frac{\partial^{3} \mathcal{L}_{n}^{\ast}(\beta, \upsilon)}{\partial \beta_{k} \partial \beta \partial \upsilon^{\prime}} = - A_{k}^{(2)}(\beta, \upsilon) (H(\beta, \upsilon))^{- 1} + (S(\beta, \upsilon))^{\prime} A_{k}^{(3)}(\beta, \upsilon) (H(\beta, \upsilon))^{- 1} \, , \nonumber \\ 
	&\frac{\partial^{3} \mathcal{L}_{n}^{\ast}(\beta, \upsilon)}{\partial \beta_{k} \partial \upsilon \partial \upsilon^{\prime}} = - (H(\beta, \upsilon))^{- 1} A_{k}^{(3)}(\beta, \upsilon) (H(\beta, \upsilon))^{- 1} \, , \nonumber \\
	&\frac{\partial^{3} \mathcal{L}_{n}^{\ast}(\beta, \upsilon)}{\partial \upsilon_{l} \partial \upsilon \partial \upsilon^{\prime}} = - (H(\beta, \upsilon))^{- 1} B_{l}^{(2)}(\beta, \upsilon) (H(\beta, \upsilon))^{- 1} \, , \nonumber \\
	&\frac{\partial^{4} \mathcal{L}_{n}^{\ast}(\beta, \upsilon)}{\partial \beta_{k} \partial \beta_{k^{\prime}} \partial \beta \partial \beta^{\prime}} = - G_{kk^{\prime}}^{(1)}(\beta, \upsilon) + G_{kk^{\prime}}^{(2)}(\beta, \upsilon) S(\beta, \upsilon) + (S(\beta, \upsilon))^{\prime} (G_{kk^{\prime}}^{(2)}(\beta, \upsilon))^{\prime} \, - \nonumber \\
	&\quad (S(\beta, \upsilon))^{\prime} G_{kk^{\prime}}^{(3)}(\beta, \upsilon) S(\beta, \upsilon) + \big(A_{k}^{(2)}(\beta, \upsilon) - (S(\beta, \upsilon))^{\prime} A_{k}^{(3)}(\beta, \upsilon) \big) V_{k^{\prime}}(\beta, \upsilon) \, + \nonumber \\
	&\quad (V_{k^{\prime}}(\beta, \upsilon))^{\prime} \big((A_{k}^{(2)}(\beta, \upsilon))^{\prime} - A_{k}^{(3)}(\beta, \upsilon) S(\beta, \upsilon)\big)  \, , \nonumber \\
	&\frac{\partial^{4} \mathcal{L}_{n}^{\ast}(\beta, \upsilon)}{\partial \beta_{k} \partial \beta_{k^{\prime}} \partial \beta \partial \upsilon^{\prime}} = - \big(G_{kk^{\prime}}^{(2)}(\beta, \upsilon) - (S(\beta, \upsilon))^{\prime} G_{kk^{\prime}}^{(3)}(\beta, \upsilon)\big) (H(\beta, \upsilon))^{- 1} \, + \nonumber \\
	&\quad (V_{k}(\beta, \upsilon))^{\prime} A_{k^{\prime}}^{(3)}(\beta, \upsilon) (H(\beta, \upsilon))^{- 1} + (V_{k^{\prime}}(\beta, \upsilon))^{\prime} A_{k}^{(3)}(\beta, \upsilon) (H(\beta, \upsilon))^{- 1} \, , \nonumber \\
	&\frac{\partial^{4} \mathcal{L}_{n}^{\ast}(\beta, \upsilon)}{\partial \beta_{k} \partial \beta_{k^{\prime}} \partial \upsilon \partial \upsilon^{\prime}} = - (H(\beta, \upsilon))^{- 1} G_{kk^{\prime}}^{(3)}(\beta, \upsilon) (H(\beta, \upsilon))^{- 1} + \nonumber \\
	&\quad (H(\beta, \upsilon))^{- 1} A_{k}^{(3)}(\beta, \upsilon) (H(\beta, \upsilon))^{- 1} A_{k^{\prime}}^{(3)}(\beta, \upsilon) (H(\beta, \upsilon))^{- 1} \, + \nonumber \\
	&\quad (H(\beta, \upsilon))^{- 1} A_{k^{\prime}}^{(3)}(\beta, \upsilon) (H(\beta, \upsilon))^{- 1} A_{k}^{(3)}(\beta, \upsilon) (H(\beta, \upsilon))^{- 1} \, , \nonumber \\
	&\frac{\partial^{4} \mathcal{L}_{n}^{\ast}(\beta, \upsilon)}{\partial \beta_{k} \partial \upsilon_{l} \partial \upsilon \partial \upsilon^{\prime}} = - (H(\beta, \upsilon))^{- 1} \Big( \sum_{l^{\prime} = 1}^{L} E_{kl^{\prime}}^{(3)}(\beta, \upsilon) e_{l^{\prime}}^{\prime} (H(\beta, \upsilon))^{- 1} e_{l} \Big) (H(\beta, \upsilon))^{- 1} + \nonumber \\
	&\quad (H(\beta, \upsilon))^{- 1} C_{kl}(\beta, \upsilon) (H(\beta, \upsilon))^{- 1} + \nonumber \\
	&\quad (H(\beta, \upsilon))^{- 1} A_{k}^{(3)}(\beta, \upsilon) (H(\beta, \upsilon))^{- 1} B_{l}^{(2)}(\beta, \upsilon) (H(\beta, \upsilon))^{- 1} \, + \nonumber \\
	&\quad (H(\beta, \upsilon))^{- 1} B_{l}^{(2)}(\beta, \upsilon) (H(\beta, \upsilon))^{- 1} A_{k}^{(3)}(\beta, \upsilon) (H(\beta, \upsilon))^{- 1} \, , \nonumber \\
	&\frac{\partial^{4} \mathcal{L}_{n}^{\ast}(\beta, \upsilon)}{\partial \upsilon_{l} \partial \upsilon_{l^{\prime}} \partial \upsilon \partial \upsilon^{\prime}} = (H(\beta, \upsilon))^{- 1} B_{l}^{(2)}(\beta, \upsilon) (H(\beta, \upsilon))^{- 1} B_{l^{\prime}}^{(2)}(\beta, \upsilon) (H(\beta, \upsilon))^{- 1} \, + \nonumber \\
	&\quad (H(\beta, \upsilon))^{- 1} B_{l^{\prime}}^{(2)}(\beta, \upsilon) (H(\beta, \upsilon))^{- 1} B_{l}^{(2)}(\beta, \upsilon) (H(\beta, \upsilon))^{- 1} \, + \nonumber \\
	&\quad (H(\beta, \upsilon))^{- 1} \big(P_{ll^{\prime}}^{(1)}(\beta, \upsilon) - P_{ll^{\prime}}^{(2)}(\beta, \upsilon)\big) (H(\beta, \upsilon))^{- 1} \nonumber \, , \\
	&\frac{\partial^{5} \mathcal{L}_{n}^{\ast}(\beta, \upsilon)}{\partial \beta_{k} \partial \upsilon_{l} \partial \upsilon_{l^{\prime}} \partial \upsilon \partial \upsilon^{\prime}} = \nonumber \\
	&\quad - (H(\beta, \upsilon))^{- 1} B_{l}^{(2)}(\beta, \upsilon) (H(\beta, \upsilon))^{- 1} B_{l^{\prime}}^{(2)}(\beta, \upsilon) (H(\beta, \upsilon))^{- 1} A_{k}^{(3)}(\beta, \upsilon) (H(\beta, \upsilon))^{- 1} \, + \nonumber\\
	&\quad (H(\beta, \upsilon))^{- 1} B_{l}^{(2)}(\beta, \upsilon) (H(\beta, \upsilon))^{- 1} \Big( \sum_{l^{\prime\prime} = 1}^{L} E_{kl^{\prime\prime}}^{(3)}(\beta, \upsilon) e_{l^{\prime\prime}}^{\prime} (H(\beta, \upsilon))^{- 1} e_{l^{\prime}} \Big) (H(\beta, \upsilon))^{- 1} \, - \nonumber \\
	&\quad (H(\beta, \upsilon))^{- 1} B_{l}^{(2)}(\beta, \upsilon) (H(\beta, \upsilon))^{- 1} C_{kl^{\prime}}(\beta, \upsilon) (H(\beta, \upsilon))^{- 1} \, - \nonumber \\
	&\quad (H(\beta, \upsilon))^{- 1} B_{l}^{(2)}(\beta, \upsilon) (H(\beta, \upsilon))^{- 1} A_{k}^{(3)}(\beta, \upsilon) (H(\beta, \upsilon))^{- 1} B_{l^{\prime}}^{(2)}(\beta, \upsilon) (H(\beta, \upsilon))^{- 1} \, + \nonumber \\
	&\quad (H(\beta, \upsilon))^{- 1} F_{ll^{\prime}}^{(2)}(\beta, \upsilon) (H(\beta, \upsilon))^{- 1} A_{k}^{(3)}(\beta, \upsilon) (H(\beta, \upsilon))^{- 1} \, - \nonumber \\
	&\quad (H(\beta, \upsilon))^{- 1} B_{l^{\prime}}^{(2)}(\beta, \upsilon) (H(\beta, \upsilon))^{- 1} B_{l}^{(2)}(\beta, \upsilon) (H(\beta, \upsilon))^{- 1} A_{k}^{(3)}(\beta, \upsilon) (H(\beta, \upsilon))^{- 1} \, +\nonumber \\
	&\quad (H(\beta, \upsilon))^{- 1} B_{l^{\prime}}^{(2)}(\beta, \upsilon) (H(\beta, \upsilon))^{- 1} \Big( \sum_{l^{\prime\prime} = 1}^{L} E_{kl^{\prime\prime}}^{(3)}(\beta, \upsilon) e_{l^{\prime\prime}}^{\prime} (H(\beta, \upsilon))^{- 1} e_{l} \Big) (H(\beta, \upsilon))^{- 1} \, - \nonumber \\
	&\quad (H(\beta, \upsilon))^{- 1} \Big( \sum_{l^{\prime\prime} = 1}^{L} \sum_{l^{\prime\prime\prime} = 1}^{L}  R_{kl^{\prime\prime}l^{\prime\prime\prime}}(\beta, \upsilon) e_{l^{\prime\prime\prime}}^{\prime} (H(\beta, \upsilon))^{- 1} e_{l^{\prime}} e_{l^{\prime\prime}}^{\prime} (H(\beta, \upsilon))^{- 1} e_{l} \Big) (H(\beta, \upsilon))^{- 1} \, + \nonumber \\
	&\quad (H(\beta, \upsilon))^{- 1} \Big( \sum_{l^{\prime\prime} = 1}^{L} \sum_{l^{\prime\prime\prime} = 1}^{L} \frac{\partial^{4} \mathcal{L}_{n}(\beta, \phi^{\ast}(\beta, \upsilon))}{\partial \phi_{l^{\prime\prime}} \partial \phi_{l^{\prime\prime\prime}} \partial \phi \partial \phi^{\prime}} e_{l^{\prime\prime\prime}}^{\prime} U_{l^{\prime}}(\beta, \upsilon) e_{k} e_{l^{\prime\prime}}^{\prime} (H(\beta, \upsilon))^{- 1} e_{l} \Big) (H(\beta, \upsilon))^{- 1} \, + \nonumber \\
	&\quad (H(\beta, \upsilon))^{- 1} \Big( \sum_{l^{\prime\prime} = 1}^{L} E_{kl^{\prime\prime}}^{(3)}(\beta, \upsilon) e_{l^{\prime\prime}}^{\prime} (H(\beta, \upsilon))^{- 1} B_{l^{\prime}}^{(2)}(\beta, \upsilon) (H(\beta, \upsilon))^{- 1} e_{l} \Big) (H(\beta, \upsilon))^{- 1} \, + \nonumber \\
	&\quad (H(\beta, \upsilon))^{- 1} \Big( \sum_{l^{\prime\prime} = 1}^{L} E_{kl^{\prime\prime}}^{(3)}(\beta, \upsilon) e_{l^{\prime\prime}}^{\prime} (H(\beta, \upsilon))^{- 1} e_{l} \Big) (H(\beta, \upsilon))^{- 1} B_{l^{\prime}}^{(2)}(\beta, \upsilon) (H(\beta, \upsilon))^{- 1} \, - \nonumber \\
	&\quad (H(\beta, \upsilon))^{- 1} B_{l^{\prime}}^{(2)}(\beta, \upsilon) (H(\beta, \upsilon))^{- 1} C_{kl}(\beta, \upsilon) (H(\beta, \upsilon))^{- 1} \, + \nonumber \\
	&\quad (H(\beta, \upsilon))^{- 1} \Big( \sum_{l^{\prime\prime} = 1}^{L} \sum_{l^{\prime\prime\prime} = 1}^{L} \frac{\partial^{4} \mathcal{L}_{n}(\beta, \phi^{\ast}(\beta, \upsilon))}{\partial \phi_{l^{\prime\prime}} \partial \phi_{l^{\prime\prime\prime}} \partial \phi \partial \phi^{\prime}} e_{l^{\prime\prime\prime}}^{\prime} (H(\beta, \upsilon))^{- 1} e_{l^{\prime}} e_{l^{\prime\prime}}^{\prime} U_{l}(\beta, \upsilon) e_{k} \Big) (H(\beta, \upsilon))^{- 1} \, + \nonumber \\
	&\quad (H(\beta, \upsilon))^{- 1} \Big( \sum_{l^{\prime\prime} = 1}^{L} \frac{\partial^{3} \mathcal{L}_{n}(\beta, \phi^{\ast}(\beta, \upsilon))}{\partial \phi_{l^{\prime\prime}} \partial \phi \partial \phi^{\prime}} e_{l^{\prime\prime}}^{\prime} W_{ll^{\prime}}(\beta, \upsilon) e_{k} \Big) (H(\beta, \upsilon))^{- 1} \, - \nonumber \\
	&\quad (H(\beta, \upsilon))^{- 1} C_{kl}(\beta, \upsilon) (H(\beta, \upsilon))^{- 1} B_{l^{\prime}}^{(2)}(\beta, \upsilon) (H(\beta, \upsilon))^{- 1} \, , \nonumber
\end{align}
where
\begin{align*}
	U_{l}(\beta, \upsilon) \coloneqq& \, (H(\beta, \upsilon))^{- 1} \big(B_{l}^{(1)}(\beta, \upsilon) - B_{l}^{(2)}(\beta, \upsilon) S(\beta, \upsilon)\big) \, , \\
	V_{k}(\beta, \upsilon) \coloneqq& \, (H(\beta, \upsilon))^{- 1} \big((A_{k}^{(2)}(\beta, \upsilon))^{\prime} - A_{k}^{(3)}(\beta, \upsilon) S(\beta, \upsilon)\big) \, , \\
	W_{ll^{\prime}}(\beta, \upsilon) \coloneqq& \, (H(\beta, \upsilon))^{- 1} \big(F_{ll^{\prime}}^{(1)}(\beta, \upsilon) - F_{ll^{\prime}}^{(2)}(\beta, \upsilon) S(\beta, \upsilon)\big) - \\
	& \, (H(\beta, \upsilon))^{- 1} \big(B_{l^{\prime}}^{(2)}(\beta, \upsilon) U_{l}(\beta, \upsilon) + B_{l}^{(2)}(\beta, \upsilon) U_{l^{\prime}}(\beta, \upsilon)\big) \, , \\
	A_{k}^{(1)}(\beta, \upsilon) \coloneqq& \, \frac{\partial^{3} \mathcal{L}_{n}(\beta, \phi^{\ast}(\beta, \upsilon))}{\partial \beta_{k} \partial \beta \partial \beta^{\prime}} - \sum_{l = 1}^{L} \frac{\partial^{3} \mathcal{L}_{n}(\beta, \phi^{\ast}(\beta, \upsilon))}{\partial \phi_{l} \partial \beta \partial \beta^{\prime}} e_{l}^{\prime} S(\beta, \upsilon) e_{k} \, , \\
	A_{k}^{(2)}(\beta, \upsilon) \coloneqq& \, \frac{\partial^{3} \mathcal{L}_{n}(\beta, \phi^{\ast}(\beta, \upsilon))}{\partial \beta_{k} \partial \beta \partial \phi^{\prime}} - \sum_{l = 1}^{L} \frac{\partial^{3} \mathcal{L}_{n}(\beta, \phi^{\ast}(\beta, \upsilon))}{\partial \phi_{l} \partial \beta \partial \phi^{\prime}} e_{l}^{\prime} S(\beta, \upsilon) e_{k} \, , \\
	A_{k}^{(3)}(\beta, \upsilon) \coloneqq& \, \frac{\partial^{3} \mathcal{L}_{n}(\beta, \phi^{\ast}(\beta, \upsilon))}{\partial \beta_{k} \partial \phi \partial \phi^{\prime}} - \sum_{l = 1}^{L} \frac{\partial^{3} \mathcal{L}_{n}(\beta, \phi^{\ast}(\beta, \upsilon))}{\partial \phi_{l} \partial \phi \partial \phi^{\prime}} e_{l}^{\prime} S(\beta, \upsilon) e_{k} \, , \\
	B_{l}^{(1)}(\beta, \upsilon) \coloneqq& \, \sum_{l^{\prime} = 1}^{L} \frac{\partial^{3} \mathcal{L}_{n}(\beta, \phi^{\ast}(\beta, \upsilon))}{\partial \phi_{l^{\prime}} \partial \phi \partial \beta^{\prime}} e_{l^{\prime}}^{\prime} (H(\beta, \upsilon))^{- 1} e_{l} \, , \\
	B_{l}^{(2)}(\beta, \upsilon) \coloneqq& \, \sum_{l^{\prime} = 1}^{L} \frac{\partial^{3} \mathcal{L}_{n}(\beta, \phi^{\ast}(\beta, \upsilon))}{\partial \phi_{l^{\prime}} \partial \phi \partial \phi^{\prime}} e_{l^{\prime}}^{\prime} (H(\beta, \upsilon))^{- 1} e_{l} \, , \\
	C_{kl}(\beta, \upsilon) \coloneqq& \, \sum_{l^{\prime} = 1}^{L} \frac{\partial^{3} \mathcal{L}_{n}(\beta, \phi^{\ast}(\beta, \upsilon))}{\partial \phi_{l^{\prime}} \partial \phi \partial \phi^{\prime}} e_{l^{\prime}}^{\prime} U_{l}(\beta, \upsilon) e_{k} \, , \\
	D_{kk^{\prime}}^{(1)}(\beta, \upsilon) \coloneqq& \, \frac{\partial^{4} \mathcal{L}_{n}(\beta, \phi^{\ast}(\beta, \upsilon))}{\partial \beta_{k} \partial \beta_{k^{\prime}} \partial \beta \partial \beta^{\prime}} - \sum_{l = 1}^{L} \frac{\partial^{4} \mathcal{L}_{n}(\beta, \phi^{\ast}(\beta, \upsilon))}{\partial \beta_{k} \partial \phi_{l} \partial \beta \partial \beta^{\prime}} e_{l}^{\prime} S(\beta, \upsilon) e_{k^{\prime}}  \, , \\
	D_{kk^{\prime}}^{(2)}(\beta, \upsilon) \coloneqq& \, \frac{\partial^{4} \mathcal{L}_{n}(\beta, \phi^{\ast}(\beta, \upsilon))}{\partial \beta_{k} \partial \beta_{k^{\prime}} \partial \beta \partial \phi^{\prime}} - \sum_{l = 1}^{L} \frac{\partial^{4} \mathcal{L}_{n}(\beta, \phi^{\ast}(\beta, \upsilon))}{\partial \beta_{k} \partial \phi_{l} \partial \beta \partial \phi^{\prime}} e_{l}^{\prime} S(\beta, \upsilon) e_{k^{\prime}}  \, , \\
	D_{kk^{\prime}}^{(3)}(\beta, \upsilon) \coloneqq& \, \frac{\partial^{4} \mathcal{L}_{n}(\beta, \phi^{\ast}(\beta, \upsilon))}{\partial \beta_{k} \partial \beta_{k^{\prime}} \partial \phi \partial \phi^{\prime}} - \sum_{l = 1}^{L} \frac{\partial^{4} \mathcal{L}_{n}(\beta, \phi^{\ast}(\beta, \upsilon))}{\partial \beta_{k} \partial \phi_{l} \partial \phi \partial \phi^{\prime}} e_{l}^{\prime} S(\beta, \upsilon) e_{k^{\prime}}  \, , \\
	E_{kl}^{(1)}(\beta, \upsilon) \coloneqq& \, \frac{\partial^{4} \mathcal{L}_{n}(\beta, \phi^{\ast}(\beta, \upsilon))}{\partial \beta_{k} \partial \phi_{l} \partial \beta \partial \beta^{\prime}} - \sum_{l^{\prime} = 1}^{L} \frac{\partial^{4} \mathcal{L}_{n}(\beta, \phi^{\ast}(\beta, \upsilon))}{\partial \phi_{l} \partial \phi_{l^{\prime}} \partial \beta \partial \beta^{\prime}} e_{l^{\prime}}^{\prime} S(\beta, \upsilon) e_{k}  \, , \\
	E_{kl}^{(2)}(\beta, \upsilon) \coloneqq& \, \frac{\partial^{4} \mathcal{L}_{n}(\beta, \phi^{\ast}(\beta, \upsilon))}{\partial \beta_{k} \partial \phi_{l} \partial \beta \partial \phi^{\prime}} - \sum_{l^{\prime} = 1}^{L} \frac{\partial^{4} \mathcal{L}_{n}(\beta, \phi^{\ast}(\beta, \upsilon))}{\partial \phi_{l} \partial \phi_{l^{\prime}} \partial \beta \partial \phi^{\prime}} e_{l^{\prime}}^{\prime} S(\beta, \upsilon) e_{k}  \, , \\
	E_{kl}^{(3)}(\beta, \upsilon) \coloneqq& \, \frac{\partial^{4} \mathcal{L}_{n}(\beta, \phi^{\ast}(\beta, \upsilon))}{\partial \beta_{k} \partial \phi_{l} \partial \phi \partial \phi^{\prime}} - \sum_{l^{\prime} = 1}^{L} \frac{\partial^{4} \mathcal{L}_{n}(\beta, \phi^{\ast}(\beta, \upsilon))}{\partial \phi_{l} \partial \phi_{l^{\prime}} \partial \phi \partial \phi^{\prime}} e_{l^{\prime}}^{\prime} S(\beta, \upsilon) e_{k}  \, , \\
	F_{ll^{\prime}}^{(1)}(\beta, \upsilon) \coloneqq& \, \sum_{l^{\prime\prime} = 1}^{L} \frac{\partial^{4} \mathcal{L}_{n}(\beta, \phi^{\ast}(\beta, \upsilon))}{\partial \phi_{l^{\prime}} \phi_{l^{\prime\prime}} \partial \phi \partial \beta^{\prime}} e_{l^{\prime\prime}}^{\prime} (H(\beta, \upsilon))^{- 1} (I - B_{l^{\prime}}^{(2)}(\beta, \upsilon) (H(\beta, \upsilon))^{- 1}) e_{l} \, , \\
	F_{ll^{\prime}}^{(2)}(\beta, \upsilon) \coloneqq& \, \sum_{l^{\prime\prime} = 1}^{L} \frac{\partial^{4} \mathcal{L}_{n}(\beta, \phi^{\ast}(\beta, \upsilon))}{\partial \phi_{l^{\prime}} \phi_{l^{\prime\prime}} \partial \phi \partial \phi^{\prime}} e_{l^{\prime\prime}}^{\prime} (H(\beta, \upsilon))^{- 1} (I - B_{l^{\prime}}^{(2)}(\beta, \upsilon) (H(\beta, \upsilon))^{- 1}) e_{l} \, , \\
	G_{kk^{\prime}}^{(1)}(\beta, \upsilon) \coloneqq& \, D_{kk^{\prime}}^{(1)}(\beta, \upsilon) - \sum_{l = 1}^{L} E_{k^{\prime}l}^{(1)}(\beta, \upsilon) e_{l}^{\prime} S(\beta, \upsilon) e_{k} - \sum_{l = 1}^{L} \frac{\partial^{3} \mathcal{L}_{n}(\beta, \phi^{\ast}(\beta, \upsilon))}{\partial \phi_{l} \partial \beta \partial \beta^{\prime}} e_{l}^{\prime} V_{k^{\prime}}(\beta, \upsilon) e_{k} \, ,\\
	G_{kk^{\prime}}^{(2)}(\beta, \upsilon) \coloneqq& \, D_{kk^{\prime}}^{(2)}(\beta, \upsilon) - \sum_{l = 1}^{L} E_{k^{\prime}l}^{(2)}(\beta, \upsilon) e_{l}^{\prime} S(\beta, \upsilon) e_{k} - \sum_{l = 1}^{L} \frac{\partial^{3} \mathcal{L}_{n}(\beta, \phi^{\ast}(\beta, \upsilon))}{\partial \phi_{l} \partial \beta \partial \phi^{\prime}} e_{l}^{\prime} V_{k^{\prime}}(\beta, \upsilon) e_{k} \, ,\\
	G_{kk^{\prime}}^{(3)}(\beta, \upsilon) \coloneqq& \, D_{kk^{\prime}}^{(3)}(\beta, \upsilon) - \sum_{l = 1}^{L} E_{k^{\prime}l}^{(3)}(\beta, \upsilon) e_{l}^{\prime} S(\beta, \upsilon) e_{k} - \sum_{l = 1}^{L} \frac{\partial^{3} \mathcal{L}_{n}(\beta, \phi^{\ast}(\beta, \upsilon))}{\partial \phi_{l} \partial \phi \partial \phi^{\prime}} e_{l}^{\prime} V_{k^{\prime}}(\beta, \upsilon) e_{k} \, ,\\
	P_{ll^{\prime}}^{(1)}(\beta, \upsilon) \coloneqq& \, \sum_{l^{\prime\prime} = 1}^{L} \frac{\partial^{3} \mathcal{L}_{n}(\beta, \phi^{\ast}(\beta, \upsilon))}{\partial \phi_{l^{\prime\prime}} \partial \phi \partial \phi^{\prime}} e_{l^{\prime\prime}}^{\prime} (H(\beta, \upsilon))^{- 1} B_{l}^{(2)}(\beta, \upsilon) (H(\beta, \upsilon))^{- 1} e_{l^{\prime}} \, ,\\
	P_{ll^{\prime}}^{(2)}(\beta, \upsilon) \coloneqq& \, \sum_{l^{\prime\prime} = 1}^{L} \sum_{l^{\prime\prime\prime} = 1}^{L} \frac{\partial^{4} \mathcal{L}_{n}(\beta, \phi^{\ast}(\beta, \upsilon))}{\partial \phi_{l^{\prime\prime}} \partial \phi_{l^{\prime\prime\prime}} \partial \phi \partial \phi^{\prime}} e_{l^{\prime\prime}}^{\prime} (H(\beta, \upsilon))^{- 1} e_{l} e_{l^{\prime\prime\prime}}^{\prime} (H(\beta, \upsilon))^{- 1} e_{l^{\prime}} \, , \\
	R_{kll^{\prime}}(\beta, \upsilon) \coloneqq& \, \frac{\partial^{5} \mathcal{L}_{n}(\beta, \phi^{\ast}(\beta, \upsilon))}{\partial \beta_{k} \partial \phi_{l} \partial \phi_{l^{\prime}} \partial \phi \partial \phi^{\prime}} - \sum_{l^{\prime\prime} = 1}^{L} \frac{\partial^{5} \mathcal{L}_{n}(\beta, \phi^{\ast}(\beta, \upsilon))}{\partial \phi_{l} \partial \phi_{l^{\prime}} \partial \phi_{l^{\prime\prime}} \partial \phi \partial \phi^{\prime}} e_{l^{\prime\prime}}^{\prime} S(\beta, \upsilon) e_{k} \, .
\end{align*}
Stacking \eqref{eq:taylor_expansion_legendre_scores_beta} over $k$ and \eqref{eq:taylor_expansion_legendre_scores_upsilon} over $l$, and substituting \eqref{eq:derivatives_legendre}, assumption (iv), and the expressions in Section \ref{supplement:partial_derivatives} yields the stated results.\hfill\qedsymbol

\subsection{Proofs for Interacted Specification}
\label{supplement:proof_of_asymptotic_expansions_interacted}

\paragraph{Proof of Theorem \ref{theorem:asymptotic_expansion_profile_score_interacted} (i).} The conditions of Theorem \ref{theorem:taylor_expansions} hold wpa1. Thus,
\begin{align*}
	\frac{\partial \mathcal{L}_{n}(\beta, \hat{\phi}(\beta))}{\partial \beta} =& \, \mathcal{T}^{(1)} + \mathcal{T}^{(2)}(\beta) + \mathcal{T}^{(3)} + \mathcal{T}^{(5)}(\beta) + \mathcal{T}^{(6)} + \sum_{r = 1}^{2} \mathcal{T}_{r}^{(10)} + \mathcal{R}^{(1)}(\beta) \, ,
\end{align*}
where $\mathcal{R}^{(1)}(\beta) \coloneqq \mathcal{T}^{(4)}(\beta) + \sum_{r = 1}^{2} \mathcal{T}_{r}^{(7)}(\beta) + \sum_{r = 1}^{3} \mathcal{T}_{r}^{(8)}(\beta) + \sum_{r = 1}^{3} \mathcal{T}_{r}^{(9)}(\beta) + \sum_{r = 1}^{8} \mathcal{T}_{r}^{(11)}$. Throughout, $\beta \in \mathfrak{B}(\varepsilon)$ with $\varepsilon = o(T^{- 1 / 2})$.

\vspace{0.5em}
\noindent\# \underline{Part 1.} By the triangle inequality,
\begin{equation*}
	\bignorm{\mathcal{R}^{(1)}(\beta)}_{2} \leq \bignorm{\mathcal{T}^{(4)}(\beta)}_{2} + \sum_{r = 1}^{2} \bignorm{\mathcal{T}_{r}^{(7)}(\beta)}_{2} + \sum_{r = 1}^{3} \bignorm{\mathcal{T}_{r}^{(8)}(\beta)}_{2} + \sum_{r = 1}^{3} \bignorm{\mathcal{T}_{r}^{(9)}(\beta)}_{2} + \sum_{r = 1}^{8} \bignorm{\mathcal{T}_{r}^{(11)}}_{2} \, .
\end{equation*}
By H\"older's inequality,
\begin{align*}
	&\bignorm{\mathcal{T}^{(4)}(\beta)}_{2} \leq C \, T^{- 1} \max_{k} \bignorm{M X e_{k}}_{4}^{3} \, \bignorm{\dop{3} \psi}_{4} \, \bignorm{\beta - \beta^{0}}_{2}^{2} = o_{P}\big(T^{\frac{3}{2}} \bignorm{\beta - \beta^{0}}_{2}\big) \, , \\
	&\bignorm{\mathcal{T}_{1}^{(7)}(\beta)}_{2} \leq C \, T^{- 1} \bignorm{\check{Q}}_{2} \max_{k} \bignorm{\check{M} X e_{k}}_{6}^{4} \, \bignorm{\check{\dop{3} \psi}}_{6}^{2} \, \bignorm{\beta - \beta^{0}}_{2}^{3} = o_{P}\big(T^{\frac{3}{2}} \bignorm{\beta - \beta^{0}}_{2}\big) \, , \\
	&\bignorm{\mathcal{T}_{1}^{(11)}}_{2} \leq C \, T^{- 2} \bignorm{\check{Q}}_{2} \, \max_{k} \bignorm{\check{M} X e_{k}}_{6} \, \bignorm{\check{\dop{3} \psi}}_{6} \, \bignorm{\check{Q} \dop{1} \psi}_{6}^{3} \, \norm{D}_{6} \, \bignorm{D^{\prime} \dop{1} \psi}_{6} = o_{P}\big(T^{\frac{1}{2}}\big) \, .
\end{align*}
The remaining terms satisfy analogous bounds. Hence, $\norm{\mathcal{R}^{(1)}(\beta)}_{2} = o_{P}(T^{1 / 2}) + o_{P}(T^{3 / 2} \norm{\beta - \beta^{0}}_{2})$ for all $\beta \in \mathfrak{B}(\varepsilon)$.

\vspace{0.5em}
\noindent\# \underline{Part 2.} Recall $\mathcal{T}^{(2)}(\beta) = N_{1} N_{2} W (\beta - \beta^{0})$. Then,
\begin{align*}
	\mathcal{T}^{(2)}(\beta) =& \, N_{1} N_{2} \, \overline{W} (\beta - \beta^{0}) + T^{2} \, \bigg(\frac{N_{1} N_{2}}{T^{2}}\bigg) \, \big(W - \overline{W}\big) (\beta - \beta^{0}) \\
	=& \, N_{1} N_{2} \, \overline{W} (\beta - \beta^{0}) + o_{P}\big(T^{\frac{3}{2}} \bignorm{\beta - \beta^{0}}_{2}\big) \, .
\end{align*}
Decomposing $\mathcal{T}^{(3)}$ using Lemma \ref{lemma:inverse_approximation},
\begin{align*}
	&\mathcal{T}^{(3)} = - \frac{\mathfrak{X}^{\prime} \dop{1} \psi}{T} - \frac{(\widetilde{\mathfrak{D}_{\pi}^{2}})^{\prime} \sum_{m = 1}^{3} \overline{\mathcal{Q}}_{m} \dop{1} \psi}{T} + \frac{(\widetilde{\mathfrak{D}_{\pi}^{2}})^{\prime} \sum_{m = 1}^{3} \sum_{m^{\prime} \neq m}^{3} \overline{\mathcal{Q}}_{m} \overline{\nabla^{2} \psi} \, \overline{\mathcal{Q}}_{m^{\prime}} \dop{1} \psi}{T} \, + \\
	&\qquad \frac{(\widetilde{\mathfrak{D}_{\pi}^{2}})^{\prime} \sum_{m = 1}^{3} \sum_{m^{\prime} = 1}^{3} \overline{\mathcal{Q}}_{m} \widetilde{\nabla^{2} \psi} \, \overline{\mathcal{Q}}_{m^{\prime}} \dop{1} \psi}{T} + \frac{(\widetilde{\mathfrak{D}_{\pi}^{2}})^{\prime} D \overline{F}^{- 1} V V^{\prime} \, \overline{F}^{- 1} D^{\prime} \dop{1} \psi}{T^{3}} \, - \\
	&\qquad \frac{(\widetilde{\mathfrak{D}_{\pi}^{2}})^{\prime} D \overline{F}^{- 1} \overline{G} \, \overline{H}^{- 1} \overline{G} \, \overline{F}^{- 1} D^{\prime} \dop{1} \psi}{T^{2}} -  \frac{(\widetilde{\mathfrak{D}_{\pi}^{2}})^{\prime} D \overline{F}^{- 1} \overline{G} \, \overline{H}^{- 1} \widetilde{H} \, \overline{H}^{- 1} D^{\prime} \dop{1} \psi}{T^{2}} \, - \\
	&\qquad \frac{(\widetilde{\mathfrak{D}_{\pi}^{2}})^{\prime} D \overline{F}^{- 1} \widetilde{H} \, \overline{H}^{- 1} \overline{G} \, \overline{F}^{- 1} D^{\prime} \dop{1} \psi}{T^{2}} - \frac{(\widetilde{\nabla^{2} \psi X})^{\prime} \overline{Q} \, \widetilde{\nabla^{2} \psi} \, \overline{Q} \, \widetilde{\nabla^{2} \psi} \, \overline{Q} \dop{1} \psi}{T} \, - \\
	&\qquad \frac{X^{\prime} \nabla^{2} \psi D (H^{- 1} - \overline{H}^{- 1} + \overline{H}^{- 1} \widetilde{H} \, \overline{H}^{- 1} - \overline{H}^{- 1} \widetilde{H} \, \overline{H}^{- 1} \widetilde{H} \, \overline{H}^{- 1}) D^{\prime} \dop{1} \psi}{T^{2}} \\
	&\qquad\eqqcolon \mathfrak{T}_{1}^{(3)} + \ldots + \mathfrak{T}_{10}^{(3)} \, .
\end{align*}
By H\"older's inequality,
\begin{equation*}
	\bignorm{\mathfrak{T}_{5}^{(3)}}_{2} \leq C \, T^{- 3} \max_{k} \bignorm{V^{\prime} \overline{F}^{- 1} D^{\prime} \widetilde{\mathfrak{D}_{\pi}^{2}} e_{k}}_{2} \, \bignorm{V^{\prime} \overline{F}^{- 1} D^{\prime} \dop{1} \psi}_{2} = o_{P}\big(T^{\frac{1}{2}}\big) \, .
\end{equation*}
The bounds $\norm{\mathfrak{T}_{r}^{(3)}}_{2} = o_{P}(T^{1 / 2})$ for $r = 6, \ldots, 10$ follow analogously. Combining $\mathcal{T}^{(1)}$ and $\mathcal{T}^{(3)}$,
\begin{align*}
	\mathcal{T}^{(1)} + \mathcal{T}^{(3)} =& \, \left(\frac{\sqrt{N_{1} N_{2}}}{\sqrt{T}}\right) \Bigg\{\underbrace{\frac{(X - \mathfrak{X})^{\prime} \dop{1} \psi}{\sqrt{N_{1} N_{2} T}}}_{= \, U_{1}}\Bigg\} + \mathfrak{T}_{2}^{(3)} + \mathfrak{T}_{3}^{(3)} + \mathfrak{T}_{4}^{(3)} + \, o_{P}\big(T^{\frac{1}{2}}\big) \, .
\end{align*}
Decomposing $\mathcal{T}^{(5)}(\beta)$,
\begin{align*}
	&\mathcal{T}^{(5)}(\beta) = - \Bigg\{\frac{(X - \mathfrak{X})^{\prime} \sum_{m = 1}^{3} \diag(\overline{\mathcal{Q}}_{m} \dop{1} \psi) \mathfrak{D}_{\pi}^{3}}{T} + \frac{(X - \mathfrak{X})^{\prime} \diag((Q \dop{1} \psi - \overline{\mathcal{Q}} \dop{1} \psi)) \mathfrak{D}_{\pi}^{3}}{T} \, + \\
	&\qquad \frac{(X - \mathfrak{X})^{\prime} \diag(Q \dop{1} \psi) (\nabla^{3} \psi M X - \mathfrak{D}_{\pi}^{3})}{T} + \frac{(\nabla^{3} \psi M X - \mathfrak{D}_{\pi}^{3})^{\prime} \diag(Q \dop{1} \psi) M X}{T} \bigg\} (\beta - \beta^{0}) \\
	&\qquad \eqqcolon \mathfrak{T}_{1}^{(5)}(\beta) + \ldots + \mathfrak{T}_{4}^{(5)}(\beta) \, .
\end{align*}
By H\"older's inequality, $\norm{\mathfrak{T}_{r}^{(5)}(\beta)}_{2} = o_{P}(T^{3 / 2} \norm{\beta - \beta^{0}}_{2})$ for $r = 2, 3, 4$. Hence $\mathcal{T}^{(5)}(\beta) = \mathfrak{T}_{1}^{(5)}(\beta) + o_{P}(T^{3 / 2} \norm{\beta - \beta^{0}}_{2})$. Decomposing $\mathcal{T}^{(6)}$ using Lemma \ref{lemma:inverse_approximation},
\begin{align*}
	&\mathcal{T}^{(6)} = \frac{(\mathfrak{D}_{\pi}^{3})^{\prime} \sum_{m = 1}^{3} \sum_{m^{\prime} = 1}^{3} \diag(\overline{\mathcal{Q}}_{m} \dop{1} \psi) \overline{\mathcal{Q}}_{m^{\prime}} \dop{1} \psi}{2 \, T} \, - \\
	&\qquad \frac{(\overline{\mathfrak{D}_{\pi}^{3}})^{\prime} \sum_{m = 1}^{3} \sum_{m^{\prime} = 1}^{3} \sum_{m^{\prime\prime} \neq m^{\prime}}^{3} \diag(\overline{\mathcal{Q}}_{m} \dop{1} \psi) \overline{\mathcal{Q}}_{m^{\prime}}  \overline{\nabla^{2} \psi} \, \overline{\mathcal{Q}}_{m^{\prime\prime}} \dop{1} \psi}{T} \, - \\
	&\qquad \frac{(\overline{\mathfrak{D}_{\pi}^{3}})^{\prime} \sum_{m = 1}^{3} \sum_{m^{\prime} = 1}^{3} \sum_{m^{\prime\prime} = 1}^{3} \diag(\overline{\mathcal{Q}}_{m} \dop{1} \psi) \overline{\mathcal{Q}}_{m^{\prime}}  \widetilde{\nabla^{2} \psi} \, \overline{\mathcal{Q}}_{m^{\prime\prime}} \dop{1} \psi}{T} \, - \\
	&\qquad \frac{(\widetilde{\mathfrak{D}_{\pi}^{2}})^{\prime} \sum_{m = 1}^{3} \sum_{m^{\prime} = 1}^{3} \sum_{m^{\prime\prime} = 1}^{3} \overline{\mathcal{Q}}_{m} \nabla^{3} \psi \diag(\overline{\mathcal{Q}}_{m^{\prime}} \dop{1} \psi) \overline{\mathcal{Q}}_{m^{\prime\prime}} \dop{1} \psi}{2 \, T} \, - \\
	&\qquad \frac{(\overline{\mathfrak{D}_{\pi}^{3}})^{\prime} \diag(\overline{\mathcal{Q}} \dop{1} \psi) D \overline{F}^{- 1} V V^{\prime} \overline{F}^{- 1} D^{\prime} \dop{1} \psi}{T^{3}} - \frac{(\widetilde{\mathfrak{D}_{\pi}^{3}})^{\prime} \diag(\overline{\mathcal{Q}} \dop{1} \psi) D \overline{F}^{- 1} \overline{G} \, \overline{F}^{- 1} D^{\prime} \dop{1} \psi}{T^{2}} \, + \\
	&\qquad \frac{(\mathfrak{D}_{\pi}^{3})^{\prime} \diag(D \overline{H}^{- 1} \overline{G} \, \overline{F}^{- 1} D^{\prime} \dop{1} \psi) D \overline{F}^{- 1} \overline{G} \, \overline{F}^{- 1} D^{\prime} \dop{1} \psi}{2 \, T^{3}} \, + \\
	&\qquad \frac{(\overline{\mathfrak{D}_{\pi}^{3}})^{\prime} \diag(\overline{\mathcal{Q}} \dop{1} \psi) D \overline{F}^{- 1} \overline{G} \, \overline{H}^{- 1} \overline{G} \, \overline{F}^{- 1} D^{\prime} \dop{1} \psi}{T^{2}} \, + \\
	&\qquad \frac{(\widetilde{\mathfrak{D}_{\pi}^{3}})^{\prime} \diag(\overline{\mathcal{Q}} \dop{1} \psi) D \overline{F}^{- 1} \overline{G} \, \overline{H}^{- 1} \overline{G} \, \overline{F}^{- 1} D^{\prime} \dop{1} \psi}{T^{2}} \, - \\
	&\qquad \frac{(\mathfrak{D}_{\pi}^{3})^{\prime} \diag(D \overline{H}^{- 1} \overline{G} \, \overline{F}^{- 1} D^{\prime} \dop{1} \psi) D \overline{F}^{- 1} \overline{G} \, \overline{H}^{- 1} \overline{G} \, \overline{F}^{- 1} D^{\prime} \dop{1} \psi}{2 \, T^{3}} \, + \\
	&\qquad \frac{(\mathfrak{D}_{\pi}^{3})^{\prime} \diag(D \overline{H}^{- 1} \overline{G} \, \overline{F}^{- 1} D^{\prime} \dop{1} \psi) \overline{Q} \widetilde{\nabla^{2} \psi} \overline{\mathcal{Q}} \dop{1} \psi}{2 \, T^{2}} \, + \\
	&\qquad \frac{(\mathfrak{D}_{\pi}^{3})^{\prime} \diag(\overline{Q} \dop{1} \psi) (Q \dop{1} \psi - \overline{Q} \dop{1} \psi + \overline{Q} \widetilde{\nabla^{2} \psi} \overline{\mathcal{Q}} \dop{1} \psi)}{2 \, T} \, - \\
	&\qquad \frac{(\overline{\mathfrak{D}_{\pi}^{3}})^{\prime} \diag(\overline{\mathcal{Q}} \dop{1} \psi) D \overline{F}^{- 1} \overline{G} \, \overline{H}^{- 1} D^{\prime} \widetilde{\nabla^{2} \psi} \overline{\mathcal{Q}} \dop{1} \psi}{T^{2}} - \frac{(\widetilde{\mathfrak{D}_{\pi}^{3}})^{\prime} \diag(\overline{\mathcal{Q}} \dop{1} \psi) \overline{Q} \widetilde{\nabla^{2} \psi} \overline{\mathcal{Q}} \dop{1} \psi}{T} \, - \\
	&\qquad \frac{(\mathfrak{D}_{\pi}^{3})^{\prime} \diag((Q \dop{1} \psi - \overline{\mathcal{Q}} \dop{1} \psi)) \overline{Q} \widetilde{\nabla^{2} \psi} \overline{\mathcal{Q}} \dop{1} \psi}{2 \, T} \, + \\
	&\qquad \frac{(\mathfrak{D}_{\pi}^{3})^{\prime} \diag(Q \dop{1} \psi) (Q \dop{1} \psi - \overline{Q} \dop{1} \psi + \overline{Q} \widetilde{\nabla^{2} \psi} \overline{\mathcal{Q}} \dop{1} \psi)}{2 \, T} \, - \\
	&\qquad \frac{(\nabla^{3} \psi \overline{\mathcal{Q}} \widetilde{\mathfrak{D}_{\pi}^{2}})^{\prime} \diag(\overline{\mathcal{Q}} \dop{1} \psi) (Q \dop{1} \psi - \overline{\mathcal{Q}} \dop{1} \psi)}{2 \, T} \, - \\
	&\qquad \frac{(\nabla^{3} \psi \overline{\mathcal{Q}} \widetilde{\mathfrak{D}_{\pi}^{2}})^{\prime} \diag(Q \dop{1} \psi) (Q \dop{1} \psi - \overline{\mathcal{Q}} \dop{1} \psi)}{2 \, T} \, + \\
	&\qquad \frac{(\nabla^{3} \psi M X - \mathfrak{D}_{\pi}^{3} + \nabla^{3} \psi \overline{\mathcal{Q}} \widetilde{\mathfrak{D}_{\pi}^{2}})^{\prime} \diag(Q \dop{1} \psi) Q \dop{1} \psi}{2 \, T} \\
	&\qquad \eqqcolon \mathfrak{T}_{1}^{(6)} + \ldots + \mathfrak{T}_{19}^{(6)} \, .
\end{align*}
By H\"older's inequality,
\begin{equation*}
	\bignorm{\mathfrak{T}_{5}^{(6)}}_{2} \leq C \, T^{- 3} \, \max_{k} \bignorm{V^{\prime} \overline{F}^{- 1} D^{\prime} \diag(\overline{\mathfrak{D}_{\pi}^{3}} e_{k}) \overline{\mathcal{Q}} \dop{1} \psi}_{2} \, \bignorm{V^{\prime} \overline{F}^{- 1} D^{\prime} \dop{1} \psi}_{2} = o_{P}\big(T^{\frac{1}{2}}\big) \, .
\end{equation*}
The bounds $\norm{\mathfrak{T}_{r}^{(6)}}_{2} = o_{P}(T^{1 / 2})$ for $r = 6, \ldots, 19$ follow analogously. Hence, $\mathcal{T}^{(6)} = \linebreak \mathfrak{T}_{1}^{(6)} + \mathfrak{T}_{2}^{(6)} + \mathfrak{T}_{3}^{(6)} + \mathfrak{T}_{4}^{(6)} + o_{P}(T^{1 / 2})$. Expanding $\mathcal{T}_{1}^{(10)}$,
\begin{align*}
	&\mathcal{T}_{1}^{(10)} = \frac{(\overline{\mathfrak{D}_{\pi}^{3}})^{\prime} \sum_{m = 1}^{3} \sum_{m^{\prime} = 1}^{3} \sum_{m^{\prime\prime} = 1}^{3} \sum_{m^{\prime\prime\prime} = 1}^{3} \diag(\overline{\mathcal{Q}}_{m} \dop{1} \psi) \overline{\mathcal{Q}}_{m^{\prime}} \nabla^{3} \psi \diag(\overline{\mathcal{Q}}_{m^{\prime\prime}} \dop{1} \psi) \overline{\mathcal{Q}}_{m^{\prime\prime\prime}} \dop{1} \psi}{2 \, T} \, - \\
	&\qquad \frac{(\overline{\mathfrak{D}_{\pi}^{3}})^{\prime} \diag(\overline{\mathcal{Q}} \dop{1} \psi) D \overline{F}^{- 1} \overline{G} \, \overline{H}^{- 1} D^{\prime} \nabla^{3} \psi \diag(\overline{\mathcal{Q}} \dop{1} \psi) \overline{\mathcal{Q}} \dop{1} \psi}{2 \, T^{2}} \, + \\
	&\qquad \frac{(\overline{\mathfrak{D}_{\pi}^{3}})^{\prime} \diag(\overline{\mathcal{Q}} \dop{1} \psi) \overline{Q} \nabla^{3} \psi \diag(\overline{\mathcal{Q}} \dop{1} \psi) (Q \dop{1} \psi - \overline{\mathcal{Q}} \dop{1} \psi)}{2 \, T} \, + \\
	&\qquad \frac{(\overline{\mathfrak{D}_{\pi}^{3}})^{\prime} \diag(\overline{\mathcal{Q}} \dop{1} \psi) \overline{Q} \nabla^{3} \psi \diag(Q \dop{1} \psi) (Q \dop{1} \psi - \overline{\mathcal{Q}} \dop{1} \psi)}{2 \, T} \, + \\
	&\qquad \frac{(\overline{\mathfrak{D}_{\pi}^{3}})^{\prime} \diag(\overline{\mathcal{Q}} \dop{1} \psi) (Q - \overline{Q}) \nabla^{3} \psi \diag(Q \dop{1} \psi) Q \dop{1} \psi}{2 \, T} \, + \\
	&\qquad \frac{(\overline{\mathfrak{D}_{\pi}^{3}})^{\prime} \diag((Q \dop{1} \psi - \overline{\mathcal{Q}} \dop{1} \psi)) Q \nabla^{3} \psi \diag(Q \dop{1} \psi) Q \dop{1} \psi}{2 \, T} \, + \\
	&\qquad \frac{(\widetilde{\mathfrak{D}_{\pi}^{3}})^{\prime} \diag(Q \dop{1} \psi) Q \nabla^{3} \psi \diag(Q \dop{1} \psi) Q \dop{1} \psi}{2 \, T} \, + \\
	&\qquad \frac{(\nabla^{3} \psi M X - \mathfrak{D}_{\pi}^{3})^{\prime} \diag(Q \dop{1} \psi) Q \nabla^{3} \psi \diag(Q \dop{1} \psi) Q \dop{1} \psi}{2 \, T} \\
	&\qquad \eqqcolon \mathfrak{T}_{1, 1}^{(10)} + \ldots + \mathfrak{T}_{1, 8}^{(10)} \, .
\end{align*}
By H\"older's inequality, $\norm{\mathfrak{T}_{1, r}^{(10)}}_{2} = o_{P}(T^{1 / 2})$ for $r = 2, \ldots, 8$. Hence, $\mathcal{T}_{1}^{(10)} = \mathfrak{T}_{1, 1}^{(10)} + o_{P}(T^{1 / 2})$. Expanding $\mathcal{T}_{2}^{(10)}$,
\begin{align*}
	&\mathcal{T}_{2}^{(10)} = - \frac{(\mathfrak{D}_{\pi}^{4})^{\prime} \sum_{m = 1}^{3} \sum_{m^{\prime} = 1}^{3} \sum_{m^{\prime\prime} = 1}^{3} \diag(\overline{\mathcal{Q}}_{m} \dop{1} \psi) \diag(\overline{\mathcal{Q}}_{m^{\prime}} \dop{1} \psi) \overline{\mathcal{Q}}_{m^{\prime\prime}} \dop{1} \psi}{6 \, T} \, - \\
	&\qquad \frac{(\mathfrak{D}_{\pi}^{4})^{\prime} \diag(\overline{\mathcal{Q}} \dop{1} \psi) \diag(\overline{\mathcal{Q}} \dop{1} \psi) (Q \dop{1} \psi - \overline{\mathcal{Q}} \dop{1} \psi)}{6 \, T} \, - \\
	&\qquad \frac{(\mathfrak{D}_{\pi}^{4})^{\prime} \diag(\overline{\mathcal{Q}} \dop{1} \psi) \diag(Q \dop{1} \psi) (Q \dop{1} \psi - \overline{\mathcal{Q}} \dop{1} \psi)}{6 \, T} \, - \\
	&\qquad \frac{(\mathfrak{D}_{\pi}^{4})^{\prime} \diag(Q \dop{1} \psi) \diag(Q \dop{1} \psi) (Q \dop{1} \psi - \overline{\mathcal{Q}} \dop{1} \psi)}{6 \, T} \, - \\
	&\qquad \frac{(\nabla^{4} \psi M X - \mathfrak{D}_{\pi}^{4})^{\prime} \diag(Q \dop{1} \psi) \diag(Q \dop{1} \psi) Q \dop{1} \psi}{6 \, T} \\
	&\qquad \eqqcolon \mathfrak{T}_{2, 1}^{(10)} + \ldots + \mathfrak{T}_{2, 5}^{(10)} \, .
\end{align*}
By H\"older's inequality, $\norm{\mathfrak{T}_{2, r}^{(10)}}_{2} = o_{P}(T^{1 / 2})$ for $r = 2, \ldots, 5$. Hence, $\mathcal{T}_{2}^{(10)} = \mathfrak{T}_{2, 1}^{(10)} + o_{P}(T^{1 / 2})$.

\vspace{0.5em}
\noindent\# \underline{Part 3.} It remains to analyze $\sum_{r = 2}^{4} \mathfrak{T}_{r}^{(3)} + \mathfrak{T}_{1}^{(5)}(\beta) + \sum_{r = 1}^{4} \mathfrak{T}_{r}^{(6)} + \mathfrak{T}_{1, 1}^{(10)} + \mathfrak{T}_{2, 1}^{(10)}$. We establish four argument patterns at their first occurrence and subsequently refer back to them.

Decomposing $\mathfrak{T}_{2}^{(3)}$ into three terms,
\begin{align*}
	&\mathfrak{T}_{2}^{(3)} = - \underbrace{\frac{1}{T} \sum_{i = 1}^{N_{1}} \sum_{t = 1}^{T} \frac{\big\{\sum_{j = 1}^{N_{2}} (\widetilde{\mathfrak{D}_{\pi}^{2}} )_{ijt}\big\} \big\{\sum_{j = 1}^{N_{2}} (\dop{1} \psi)_{ijt}\big\}}{\sum_{j = 1}^{N_{2}} (\overline{\dop{2} \psi})_{ijt}}}_{\eqqcolon \mathfrak{T}_{2, 1}^{(3)}} - \ldots \, - \\
	& \quad \underbrace{\frac{1}{T} \sum_{i = 1}^{N_{1}} \sum_{j = 1}^{N_{2}} \frac{\big\{\sum_{t = 1}^{T} (\widetilde{\mathfrak{D}_{\pi}^{2}})_{ijt}\big\} \big\{\sum_{t = 1}^{T} (\dop{1} \psi)_{ijt}\big\}}{\sum_{t = 1}^{T} (\overline{\dop{2} \psi})_{ijt}}}_{\eqqcolon \mathfrak{T}_{2, 3}^{(3)}} \, .
\end{align*}
Decomposing the $k$-th element of $\mathfrak{T}_{2, 1}^{(3)}$,
\begin{align*}
	\mathfrak{T}_{2, 1, k}^{(3)} =& \, \frac{1}{T} \sum_{i = 1}^{N_{1}} \sum_{t = 1}^{T} \underbrace{\frac{\big\{\sum_{j = 1}^{N_{2}} (\widetilde{\mathfrak{D}_{\pi}^{2}} e_{k})_{ijt}\big\} \big\{\sum_{j = 1}^{N_{2}} (\dop{1} \psi)_{ijt}\big\}}{\sum_{j = 1}^{N_{2}} (\overline{\dop{2} \psi})_{ijt}}}_{\eqqcolon (\vartheta_{1, k})_{it}} \\
	=& \, \frac{1}{T} \sum_{i = 1}^{N_{1}} \sum_{t = 1}^{T} (\overline{\vartheta}_{1, k})_{it} + \frac{1}{T} \sum_{i = 1}^{N_{1}} \sum_{t = 1}^{T} (\widetilde{\vartheta}_{1, k})_{it} \, .
\end{align*}
By the Cauchy-Schwarz inequality, Jensen's inequality, and Lemma \ref{lemma:moment_bounds_mixing},
\begin{align*}
	&\sup_{it} \EX{\max_{k} \abs{(\vartheta_{1, k})_{it}}^{10}} \leq C \, \left\{\sup_{it} \EX{\bigg(\frac{1}{\sqrt{N_{2}}} \sum_{j = 1}^{N_{2}} \max_{k} (\widetilde{\mathfrak{D}_{\pi}^{2}} e_{k})_{ijt}\bigg)^{20}} \sup_{it} \EX{\bigg(\frac{1}{\sqrt{N_{2}}} \sum_{j = 1}^{N_{2}} (\dop{1} \psi)_{ijt}\bigg)^{20}}\right\}^{\frac{1}{2}} \\
	&\quad \leq C \quad \text{a.\,s.}
\end{align*}
Hence, by Lemma \ref{lemma:moment_bounds_mixing},
\begin{equation*}
	\EX{\max_{k} \bigg(\frac{1}{T} \sum_{i = 1}^{N_{1}} \sum_{t = 1}^{T} (\widetilde{\vartheta}_{1, k})_{it}\bigg)^{2}} \leq \bigg(\frac{N_{1}}{T}\bigg) \, \sup_{i} \EX{\bigg(\frac{1}{\sqrt{T}} \sum_{t = 1}^{T} \max_{k} (\widetilde{\vartheta}_{1, k})_{it}\bigg)^{2}} = \mathcal{O}(1) \quad \text{a.\,s.} \, ,
\end{equation*}
so $\max_{k} T^{- 1} \sum_{i, t} (\widetilde{\vartheta}_{1, k})_{it} = o_{P}(T^{1 / 2})$ by Markov's inequality. By conditional independence across both cross-sectional dimensions, $T^{- 1} \sum_{i, t} (\overline{\vartheta}_{1, k})_{it} = N_{1} \, \overline{U}_{2, 1} e_{k}$. Thus, $\mathfrak{T}_{2, 1}^{(3)} = N_{1} \, \overline{U}_{2, 1} + o_{P}(T^{1 / 2})$. We refer to this argument as \textit{Pattern 1}. By Pattern 1, $\mathfrak{T}_{2, 2}^{(3)} = N_{2} \, \overline{U}_{2, 2} + o_{P}(T^{1 / 2})$ and $\mathfrak{T}_{2, 3}^{(3)} = (N_{1} N_{2} / T) \, \overline{U}_{2, 3} + o_{P}(T^{1 / 2})$. Thus, $\mathfrak{T}_{2}^{(3)} = - N_{1} \, \overline{U}_{2, 1} - N_{2} \, \overline{U}_{2, 2} - (N_{1} N_{2} / T) \, \overline{U}_{2, 3} + o_{P}(T^{1 / 2})$. Decomposing $\mathfrak{T}_{3}^{(3)}$ into six terms,
\begin{align*}
	&\mathfrak{T}_{3}^{(3)} = \underbrace{\frac{1}{T} \sum_{i = 1}^{N_{1}} \sum_{j = 1}^{N_{2}} \sum_{t = 1}^{T} \frac{\big\{\sum_{j^{\prime} = 1}^{N_{2}} (\widetilde{\mathfrak{D}_{\pi}^{2}})_{ij^{\prime}t}\big\} (\overline{\dop{2} \psi})_{ijt} \big\{\sum_{i^{\prime} = 1}^{N_{1}} (\dop{1} \psi)_{i^{\prime}jt}\big\}}{\big\{\sum_{j^{\prime} = 1}^{N_{2}} (\overline{\dop{2} \psi})_{ij^{\prime}t}\big\} \big\{\sum_{i^{\prime} = 1}^{N_{1}} (\overline{\dop{2} \psi})_{i^{\prime}jt}\big\}}}_{\eqqcolon \mathfrak{T}_{3, 1}^{(3)}} + \ldots \, + \\
	&\quad \underbrace{\frac{1}{T} \sum_{i = 1}^{N_{1}} \sum_{j = 1}^{N_{2}} \sum_{t = 1}^{T} \frac{\big\{\sum_{t^{\prime} = 1}^{T} (\widetilde{\mathfrak{D}_{\pi}^{2}})_{ijt^{\prime}}\big\} (\overline{\dop{2} \psi})_{ijt} \big\{\sum_{i^{\prime} = 1}^{N_{1}} (\dop{1} \psi)_{i^{\prime}jt}\big\}}{\big\{\sum_{t^{\prime} = 1}^{T} (\overline{\dop{2} \psi})_{ijt^{\prime}}\big\} \big\{\sum_{i^{\prime} = 1}^{N_{1}} (\overline{\dop{2} \psi})_{i^{\prime}jt}\big\}}}_{\eqqcolon \mathfrak{T}_{3, 6}^{(3)}} \, .
\end{align*}
The $k$-th element of $\mathfrak{T}_{3, 1}^{(3)}$ is
\begin{align*}
	\mathfrak{T}_{3, 1, k}^{(3)} =& \, \frac{1}{\sqrt{N_{1} N_{2}} T} \sum_{i = 1}^{N_{1}} \sum_{j = 1}^{N_{2}} \sum_{t = 1}^{T} \underbrace{\frac{N_{1} N_{2} (\overline{\dop{2} \psi})_{ijt}}{\big\{\sum_{i^{\prime} = 1}^{N_{1}} (\overline{\dop{2} \psi})_{i^{\prime}jt}\big\} \big\{\sum_{j^{\prime} = 1}^{N_{2}} (\overline{\dop{2} \psi})_{ij^{\prime}t}\big\}}}_{\eqqcolon (\overline{\zeta}_{1})_{ijt}} \\
	&\qquad \bigg\{\frac{1}{\sqrt{N_{1}}} \sum_{i^{\prime} = 1}^{N_{1}} (\dop{1} \psi)_{i^{\prime}jt}\bigg\} \bigg\{\frac{1}{\sqrt{N_{2}}} \sum_{j^{\prime} = 1}^{N_{2}} (\widetilde{\mathfrak{D}_{\pi}^{2}} e_{k})_{ij^{\prime}t}\bigg\} \, ,
\end{align*}
with
\begin{align*}
	&\Gamma_{1, k, ijt, i^{\prime} j^{\prime} t^{\prime}} \coloneqq \bigg\{\frac{1}{\sqrt{N_{1}}} \sum_{i^{\prime\prime} = 1}^{N_{1}} (\dop{1} \psi)_{i^{\prime\prime}jt}\bigg\} \bigg\{\frac{1}{\sqrt{N_{2}}} \sum_{j^{\prime\prime} = 1}^{N_{2}} (\widetilde{\mathfrak{D}_{\pi}^{2}} e_{k})_{ij^{\prime\prime}t}\bigg\} \bigg\{\frac{1}{\sqrt{N_{1}}} \sum_{i^{\prime\prime\prime} = 1}^{N_{1}} (\dop{1} \psi)_{i^{\prime\prime\prime}j^{\prime}t^{\prime}}\bigg\} \\
	&\qquad \bigg\{\frac{1}{\sqrt{N_{2}}} \sum_{j^{\prime\prime\prime} = 1}^{N_{2}} (\widetilde{\mathfrak{D}_{\pi}^{2}} e_{k})_{i^{\prime}j^{\prime\prime\prime}t^{\prime}}\bigg\} \, .
\end{align*}
By H\"older's inequality, Jensen's inequality, and the Lyapunov inequality, $\sup_{ijt} (\overline{\zeta}_{1})_{ijt} \leq C$ a.\,s.\ and $\sup_{ijtt^{\prime}} \mathbb{E}[\max_{k} \Gamma_{1, k, ijt, ijt^{\prime}}] \leq C$ a.\,s. Hence,
\begin{align*}
	\EX{\max_{k} \Big(\mathfrak{T}_{3, 1, k}^{(3)}\Big)^{2}} \leq& \, C \big(\sup_{ijt} (\overline{\zeta}_{1})_{ijt}\big)^{2} \sup_{ijtt^{\prime}} \EX{\max_{k} \Gamma_{1, k, ijt, ijt^{\prime}}} \leq C \quad \text{a.\,s.} \, ,
\end{align*}
so $\norm{\mathfrak{T}_{3, 1}^{(3)}}_{2} = o_{P}(T^{1 / 2})$ by Markov's inequality. We refer to this argument as \textit{Pattern 2}. By Pattern 2, $\norm{\mathfrak{T}_{3, r}^{(3)}}_{2} = o_{P}(T^{1 / 2})$ for $r = 2, \ldots, 6$. Thus, $\norm{\mathfrak{T}_{3}^{(3)}}_{2} = o_{P}(T^{1 / 2})$. Decomposing $\mathfrak{T}_{4}^{(3)}$ into nine terms,
\begin{align*}
	&\mathfrak{T}_{4}^{(3)} = \underbrace{\frac{1}{T} \sum_{i = 1}^{N_{1}} \sum_{t = 1}^{T} \frac{\big\{\sum_{j = 1}^{N_{2}} (\widetilde{\mathfrak{D}_{\pi}^{2}})_{ijt}\big\} \big\{\sum_{j = 1}^{N_{2}} (\widetilde{\dop{2} \psi})_{ijt}\big\} \big\{\sum_{j = 1}^{N_{2}} (\dop{1} \psi)_{ijt}\big\}}{\big\{\sum_{j = 1}^{N_{2}} (\overline{\dop{2} \psi})_{ijt}\big\}^{2}}}_{\eqqcolon \mathfrak{T}_{4, 1}^{(3)}} + \ldots \, + \\
	&\quad \underbrace{\frac{1}{T} \sum_{i = 1}^{N_{1}} \sum_{j = 1}^{N_{2}} \frac{\big\{\sum_{t = 1}^{T} (\widetilde{\mathfrak{D}_{\pi}^{2}})_{ijt}\big\} \big\{\sum_{t = 1}^{T} (\widetilde{\dop{2} \psi})_{ijt}\big\} \big\{\sum_{t = 1}^{T} (\dop{1} \psi)_{ijt}\big\}}{\big\{\sum_{t = 1}^{T} (\overline{\dop{2} \psi})_{ijt}\big\}^{2}}}_{\eqqcolon \mathfrak{T}_{4, 9}^{(3)}}
\end{align*}
Decomposing the $k$-th element of $\mathfrak{T}_{4, 1}^{(3)}$,
\begin{align*}
	\mathfrak{T}_{4, 1, k}^{(3)} =& \, \frac{1}{T} \sum_{i = 1}^{N_{1}} \sum_{t = 1}^{T} \underbrace{\frac{\big\{\sum_{j = 1}^{N_{2}} (\widetilde{\mathfrak{D}_{\pi}^{2}} e_{k})_{ijt}\big\} \big\{\sum_{j = 1}^{N_{2}} (\widetilde{\dop{2} \psi})_{ijt}\big\} \big\{\sum_{j = 1}^{N_{2}} (\dop{1} \psi)_{ijt}\big\}}{\big\{\sum_{j = 1}^{N_{2}} (\overline{\dop{2} \psi})_{ijt}\big\}^{2}}}_{\eqqcolon (\vartheta_{2, k})_{it}} \\
	=& \, \frac{1}{T} \sum_{i = 1}^{N_{1}} \sum_{t = 1}^{T} (\overline{\vartheta}_{2, k})_{it} + \frac{1}{T} \sum_{i = 1}^{N_{1}} \sum_{t = 1}^{T} (\widetilde{\vartheta}_{2, k})_{it} \, .
\end{align*}
By H\"older's inequality, Jensen's inequality, and the Lyapunov inequality, \linebreak $\sup_{it} \EX{\max_{k} \abs{\sqrt{N_{2}} \, (\vartheta_{2, k})_{it}}^{20 / 3}} \leq C$ a.\,s. Hence, by Lemma \ref{lemma:moment_bounds_mixing},
\begin{align*}
	&\EX{\max_{k} \bigg\{\frac{1}{T} \sum_{i = 1}^{N_{1}} \sum_{t = 1}^{T} (\widetilde{\vartheta}_{2, k})_{it}\bigg\}^{2}} \leq \bigg(\frac{1}{T}\bigg) \bigg(\frac{N_{1}}{T}\bigg) \bigg(\frac{T}{N_{2}}\bigg) \sup_{i} \EX{\bigg\{\frac{1}{\sqrt{T}} \sum_{t = 1}^{T} \sqrt{N_{2}} \max_{k} (\widetilde{\vartheta}_{2, k})_{it}\bigg\}^{2}} \\
	&\qquad = \mathcal{O}\big(T^{- 1}\big) \quad \text{a.\,s.} \, ,
\end{align*}
so $\max_{k} T^{- 1} \sum_{i, t} (\widetilde{\vartheta}_{2, k})_{it} = o_{P}(T^{1 / 2})$ by Markov's inequality. By conditional independence, Jensen's inequality, H\"older's inequality, and the Lyapunov inequality,
\begin{align*}
	&\max_{k} \biggabs{\frac{1}{T} \sum_{i = 1}^{N_{1}} \sum_{t = 1}^{T} (\overline{\vartheta}_{2, k})_{it}} \leq \frac{C}{N_{2}^{2} T} \sum_{i = 1}^{N_{1}} \sum_{j = 1}^{N_{2}} \sum_{t = 1}^{T} \max_{k} \biggabs{\EX{(\widetilde{\mathfrak{D}_{\pi}^{2}} e_{k})_{ijt} (\widetilde{\dop{2} \psi})_{ijt} (\dop{1} \psi)_{ijt}}} \\
	&\qquad = \mathcal{O}(1) \quad \text{a.\,s.} \, .
\end{align*}
Hence $\norm{\mathfrak{T}_{4, 1}^{(3)}}_{2} = o_{P}(T^{1 / 2})$. We refer to this argument as \textit{Pattern 3}. By Pattern 3, $\norm{\mathfrak{T}_{4, 4}^{(3)}}_{2} = o_{P}(T^{1 / 2})$. For the $k$-th element of $\mathfrak{T}_{4, 9}^{(3)}$,
\begin{align*}
	\mathfrak{T}_{4, 9, k}^{(3)} =& \, \bigg(\frac{N_{1} N_{2}}{T}\bigg) \, \frac{1}{N_{1} N_{2}} \sum_{i = 1}^{N_{1}} \sum_{j = 1}^{N_{2}} \underbrace{\frac{\sum_{t = 1}^{T} \sum_{t^{\prime} = 1}^{T} \sum_{t^{\prime\prime} = 1}^{T} \EX{(\widetilde{\mathfrak{D}_{\pi}^{2}} e_{k})_{ijt} (\widetilde{\dop{2} \psi})_{ijt} (\dop{1} \psi)_{ijt}}}{\big\{\sum_{t = 1}^{T} (\overline{\dop{2} \psi})_{ijt}\big\}^{2}}}_{\eqqcolon (\overline{\vartheta}_{3, k})_{ij}} + \, o_{P}\big(T^{\frac{1}{2}}\big) \, .
\end{align*}
By Lemma \ref{lemma:asymptotic_bound_triple_sum}, $\sup_{ij} \max_{k} \abs{(\overline{\vartheta}_{3, k})_{ij}} = \mathcal{O}_{P}(T^{- 1})$. Hence, $\norm{\mathfrak{T}_{4, 9}^{(3)}}_{2} = o_{P}(T^{1 / 2})$. We refer to this argument as \textit{Pattern 4}. By Pattern 2, $\norm{\mathfrak{T}_{4, r}^{(3)}}_{2} = o_{P}(T^{1 / 2})$ for $r \in \{2, 3, 5, 6, 7, 8\}$. Hence, $\mathfrak{T}_{4}^{(3)} = o_{P}(T^{1 / 2})$. Thus, $\sum_{r = 2}^{4} \mathfrak{T}_{r}^{(3)} = - N_{1} \, \overline{U}_{2, 1} - N_{2} \, \overline{U}_{2, 2} - (N_{1} N_{2} / T) \, \overline{U}_{2, 3} + o_{P}(T^{1 / 2})$.

Decomposing $\mathfrak{T}_{1}^{(5)}(\beta)$ into three terms,
\begin{align*}
	&\mathfrak{T}_{1}^{(5)}(\beta) = - \underbrace{\frac{1}{T} \sum_{i = 1}^{N_{1}} \sum_{t = 1}^{T} \frac{\big\{\sum_{j = 1}^{N_{2}} (X - \mathfrak{X})_{ijt} (\mathfrak{D}_{\pi}^{3})_{ijt}\big\} \big\{\sum_{j = 1}^{N_{2}} (\dop{1} \psi)_{ijt}\big\}}{\sum_{j = 1}^{N_{2}} (\overline{\dop{2} \psi})_{ijt}}}_{\eqqcolon \mathfrak{T}_{1, 1}^{(5)}} (\beta - \beta^{0}) - \ldots \, -  \\
	& \quad \underbrace{\frac{1}{T} \sum_{i = 1}^{N_{1}} \sum_{j = 1}^{N_{2}} \frac{\big\{\sum_{t = 1}^{T} (X - \mathfrak{X})_{ijt} (\mathfrak{D}_{\pi}^{3})_{ijt}\big\} \big\{\sum_{t = 1}^{T} (\dop{1} \psi)_{ijt}\big\}}{\sum_{t = 1}^{T} (\overline{\dop{2} \psi})_{ijt}}}_{\eqqcolon \mathfrak{T}_{1, 3}^{(5)}} (\beta - \beta^{0}) \, .
\end{align*}
Decomposing the $(k, k^{\prime})$ element of $\mathfrak{T}_{1, 1}^{(5)}$,
\begin{align*}
	\mathfrak{T}_{1, 1, k, k^{\prime}}^{(5)} =& \, \frac{1}{T} \sum_{i = 1}^{N_{1}} \sum_{t = 1}^{T} \underbrace{\frac{\big\{\sum_{j = 1}^{N_{2}} ((X - \mathfrak{X}) e_{k})_{ijt} (\mathfrak{D}_{\pi}^{3} e_{k^{\prime}})_{ijt}\big\} \big\{\sum_{j = 1}^{N_{2}} (\dop{1} \psi)_{ijt}\big\}}{\sum_{j = 1}^{N_{2}} (\overline{\dop{2} \psi})_{ijt}}}_{\eqqcolon (\vartheta_{6, k, k^{\prime}})_{it}} \\
	=& \, \frac{1}{T} \sum_{i, t} (\overline{\vartheta}_{4, k, k^{\prime}})_{it} + \frac{1}{T} \sum_{i, t} (\widetilde{\vartheta}_{4, k, k^{\prime}})_{it} \, .
\end{align*}
By an argument analogous to Pattern 3, both the variance and mean terms are $\mathcal{O}(T)$ a.\,s., giving $\norm{\mathfrak{T}_{1, 1}^{(5)}}_{2} = \mathcal{O}_{P}(T)$ and hence $\norm{\mathfrak{T}_{1, 1}^{(5)} (\beta - \beta^{0})}_{2} = o_{P}(T^{3 / 2} \norm{\beta - \beta^{0}}_{2})$. By analogous arguments, $\norm{\mathfrak{T}_{1, 2}^{(5)} (\beta - \beta^{0})}_{2} = o_{P}(T^{3 / 2} \norm{\beta - \beta^{0}}_{2})$ and $\norm{\mathfrak{T}_{1, 3}^{(5)} (\beta - \beta^{0})}_{2} = o_{P}(T^{3 / 2} \norm{\beta - \beta^{0}}_{2})$, the latter using Lemma \ref{lemma:covariance_inequality_mixing}. Hence, $\norm{\mathfrak{T}_{1}^{(5)}(\beta)}_{2} = o_{P}(T^{3 / 2} \norm{\beta - \beta^{0}}_{2})$.

Decomposing $\mathfrak{T}_{1}^{(6)}$,
\begin{equation*}
	\mathfrak{T}_{1}^{(6)} = \frac{(\overline{\mathfrak{D}_{\pi}^{3}})^{\prime} \diag(\overline{\mathcal{Q}} \dop{1} \psi) \overline{\mathcal{Q}} \dop{1} \psi}{2 \, T} + \frac{(\widetilde{\mathfrak{D}_{\pi}^{3}})^{\prime} \diag(\overline{\mathcal{Q}} \dop{1} \psi) \overline{\mathcal{Q}} \dop{1} \psi}{2 \, T} \eqqcolon \mathfrak{T}_{1, 1}^{(6)} + \mathfrak{T}_{1, 2}^{(6)} \, .
\end{equation*}
Decomposing $\mathfrak{T}_{1, 1}^{(6)}$ into six terms,
\begin{align*}
	&\mathfrak{T}_{1, 1}^{(6)} = \underbrace{\frac{1}{2 \, T} \sum_{i = 1}^{N_{1}} \sum_{t = 1}^{T} \frac{\big\{\sum_{j = 1}^{N_{2}} (\overline{\mathfrak{D}_{\pi}^{3}})_{ijt}\big\} \big\{\sum_{j = 1}^{N_{2}} (\dop{1} \psi)_{ijt}\big\}^{2}}{\big\{\sum_{j = 1}^{N_{2}} (\overline{\dop{2} \psi})_{ijt}\big\}^{2}}}_{\eqqcolon \mathfrak{T}_{1, 1, 1}^{(6)}} + \ldots \, + \\
	& \quad \underbrace{\frac{1}{2 \, T} \sum_{i = 1}^{N_{1}} \sum_{j = 1}^{N_{2}} \frac{\big\{\sum_{t = 1}^{T} (\overline{\mathfrak{D}_{\pi}^{3}})_{ijt}\big\} \big\{\sum_{t = 1}^{T} (\dop{1} \psi)_{ijt}\big\}^{2}}{\big\{\sum_{t = 1}^{T} (\overline{\dop{2} \psi})_{ijt}\big\}^{2}}}_{\eqqcolon \mathfrak{T}_{1, 1, 6}^{(6)}}
\end{align*}
By Pattern 1: $\mathfrak{T}_{1, 1, 1}^{(6)} = N_{1} \, \overline{U}_{3, 1} + o_{P}(T^{1 / 2})$, $\mathfrak{T}_{1, 1, 4}^{(6)} = N_{2} \, \overline{U}_{3, 2} + o_{P}(T^{1 / 2})$, and $\mathfrak{T}_{1, 1, 6}^{(6)} = (N_{1} N_{2} / T) \, \overline{U}_{3, 3} + o_{P}(T^{1 / 2})$. By Pattern 2: $\norm{\mathfrak{T}_{1, 1, r}^{(6)}}_{2} = o_{P}(T^{1 / 2})$ for $r \in \{2, 3, 5\}$. Hence, $\mathfrak{T}_{1, 1}^{(6)} = N_{1} \, \overline{U}_{3, 1} + N_{2} \, \overline{U}_{3, 2} + (N_{1} N_{2} / T) \, \overline{U}_{3, 3} + o_{P}(T^{1 / 2})$. Decomposing $\mathfrak{T}_{1, 2}^{(6)}$ into six terms,
\begin{align*}
	&\mathfrak{T}_{1, 2}^{(6)} = \underbrace{\frac{1}{2 \, T} \sum_{i = 1}^{N_{1}} \sum_{t = 1}^{T} \frac{\big\{\sum_{j = 1}^{N_{2}} (\widetilde{\mathfrak{D}_{\pi}^{3}})_{ijt}\big\} \big\{\sum_{j = 1}^{N_{2}} (\dop{1} \psi)_{ijt}\big\}^{2}}{\big\{\sum_{j = 1}^{N_{2}} (\overline{\dop{2} \psi})_{ijt}\big\}^{2}}}_{\eqqcolon \mathfrak{T}_{1, 2, 1}^{(6)}} + \ldots \, + \\
	& \quad \underbrace{\frac{1}{2 \, T} \sum_{i = 1}^{N_{1}} \sum_{j = 1}^{N_{2}} \frac{\big\{\sum_{t = 1}^{T} (\widetilde{\mathfrak{D}_{\pi}^{3}})_{ijt}\big\} \big\{\sum_{t = 1}^{T} (\dop{1} \psi)_{ijt}\big\}^{2}}{\big\{\sum_{t = 1}^{T} (\overline{\dop{2} \psi})_{ijt}\big\}^{2}}}_{\eqqcolon \mathfrak{T}_{1, 2, 6}^{(6)}}
\end{align*}
By Pattern 3: $\norm{\mathfrak{T}_{1, 2, 1}^{(6)}}_{2} = o_{P}(T^{1 / 2})$ and $\norm{\mathfrak{T}_{1, 2, 4}^{(6)}}_{2} = o_{P}(T^{1 / 2})$. By Pattern 4: $\norm{\mathfrak{T}_{1, 2, 6}^{(6)}}_{2} = o_{P}(T^{1 / 2})$. By Pattern 2: $\norm{\mathfrak{T}_{1, 2, r}^{(6)}}_{2} = o_{P}(T^{1 / 2})$ for $r \in \{2, 3, 5\}$. Hence, $\norm{\mathfrak{T}_{1, 2}^{(6)}}_{2} = o_{P}(T^{1 / 2})$, and thus $\mathfrak{T}_{1}^{(6)} = N_{1} \, \overline{U}_{3, 1} + N_{2} \, \overline{U}_{3, 2} + (N_{1} N_{2} / T) \, \overline{U}_{3, 3} + o_{P}(T^{1 / 2})$. Decomposing $\mathfrak{T}_{2}^{(6)}$ into 18 terms,
\begin{align*}
	&\mathfrak{T}_{2}^{(6)} = - \underbrace{\frac{1}{T} \sum_{i = 1}^{N_{1}} \sum_{j = 1}^{N_{2}} \sum_{t = 1}^{T} \frac{(\overline{\dop{2} \psi})_{ijt} \big\{\sum_{i^{\prime} = 1}^{N_{1}} (\dop{1} \psi)_{i^{\prime}jt}\big\} \big\{\sum_{j^{\prime} = 1}^{N_{2}} (\overline{\mathfrak{D}_{\pi}^{3}})_{ij^{\prime}t}\big\} \big\{\sum_{j^{\prime} = 1}^{N_{2}} (\dop{1} \psi)_{ij^{\prime}t}\big\}}{\big\{\sum_{i^{\prime} = 1}^{N_{1}} (\overline{\dop{2} \psi})_{i^{\prime}jt}\big\} \big\{\sum_{j^{\prime} = 1}^{N_{2}} (\overline{\dop{2} \psi})_{ij^{\prime}t}\big\}^{2}}}_{\eqqcolon \mathfrak{T}_{2, 1}^{(6)}} - \ldots \, - \\
	& \quad \underbrace{\frac{1}{T} \sum_{i = 1}^{N_{1}} \sum_{j = 1}^{N_{2}} \sum_{t = 1}^{T} \frac{(\overline{\dop{2} \psi})_{ijt} \big\{\sum_{i^{\prime} = 1}^{N_{1}} (\dop{1} \psi)_{i^{\prime}jt}\big\} \big\{\sum_{t^{\prime} = 1}^{T} (\overline{\mathfrak{D}_{\pi}^{3}})_{ijt^{\prime}}\big\} \big\{\sum_{t^{\prime} = 1}^{T} (\dop{1} \psi)_{ijt^{\prime}}\big\}}{\big\{\sum_{i^{\prime} = 1}^{N_{1}} (\overline{\dop{2} \psi})_{i^{\prime}jt}\big\} \big\{\sum_{t^{\prime} = 1}^{T} (\overline{\dop{2} \psi})_{ijt^{\prime}}\big\}^{2}}}_{\eqqcolon \mathfrak{T}_{2, 18}^{(6)}}
\end{align*}
By Pattern 2, $\norm{\mathfrak{T}_{2, r}^{(6)}}_{2} = o_{P}(T^{1 / 2})$ for $r \in \{1, \ldots, 18\}$. Hence, $\norm{\mathfrak{T}_{2}^{(6)}}_{2} = o_{P}(T^{1 / 2})$. Decomposing $\mathfrak{T}_{3}^{(6)}$ into 27 terms,
\begin{align*}
	&\mathfrak{T}_{3}^{(6)} = - \underbrace{\frac{1}{T} \sum_{i = 1}^{N_{1}} \sum_{t = 1}^{T} \frac{\big\{\sum_{j = 1}^{N_{2}} (\overline{\mathfrak{D}_{\pi}^{3}})_{ijt}\big\} \big\{\sum_{j = 1}^{N_{2}} (\dop{1} \psi)_{ijt}\big\}^{2} \big\{\sum_{j = 1}^{N_{2}} (\widetilde{\dop{2} \psi})_{ijt}\big\}}{\big\{\sum_{j = 1}^{N_{2}} (\overline{\dop{2} \psi})_{ijt}\big\}^{3}}}_{\eqqcolon \mathfrak{T}_{3, 1}^{(6)}} - \ldots \, - \\
	& \quad \underbrace{\frac{1}{T} \sum_{i = 1}^{N_{1}} \sum_{j = 1}^{N_{2}} \frac{\big\{\sum_{t = 1}^{T} (\overline{\mathfrak{D}_{\pi}^{3}})_{ijt}\big\} \big\{\sum_{t = 1}^{T} (\dop{1} \psi)_{ijt}\big\}^{2} \big\{\sum_{t = 1}^{T} (\widetilde{\dop{2} \psi})_{ijt}\big\}}{\big\{\sum_{t = 1}^{T} (\overline{\dop{2} \psi})_{ijt}\big\}^{3}}}_{\eqqcolon \mathfrak{T}_{3, 27}^{(6)}} \, .
\end{align*}
By Pattern 3: $\norm{\mathfrak{T}_{3, 1}^{(6)}}_{2} = o_{P}(T^{1 / 2})$ and $\norm{\mathfrak{T}_{3, 14}^{(6)}}_{2} = o_{P}(T^{1 / 2})$. By Pattern 4: $\norm{\mathfrak{T}_{3, 27}^{(6)}}_{2} = o_{P}(T^{1 / 2})$. By Pattern 2: $\norm{\mathfrak{T}_{3, r}^{(6)}}_{2} = o_{P}(T^{1 / 2})$ for $r \in \{1, \ldots, 27\} \setminus \{1, 14, 27\}$. Hence, $\norm{\mathfrak{T}_{3}^{(6)}}_{2} = o_{P}(T^{1 / 2})$. Decomposing $\mathfrak{T}_{4}^{(6)}$,
\begin{equation*}
	\mathfrak{T}_{4}^{(6)} = - \frac{(\overline{\mathcal{Q}} \widetilde{\mathfrak{D}_{\pi}^{2}})^{\prime} \overline{\nabla^{3} \psi} \diag(\overline{\mathcal{Q}} \dop{1} \psi) \overline{\mathcal{Q}} \dop{1} \psi}{2 \, T} - \frac{(\overline{\mathcal{Q}} \widetilde{\mathfrak{D}_{\pi}^{2}})^{\prime} \widetilde{\nabla^{3} \psi} \diag(\overline{\mathcal{Q}} \dop{1} \psi) \overline{\mathcal{Q}} \dop{1} \psi}{2 \, T} \eqqcolon \mathfrak{T}_{4, 1}^{(6)} + \mathfrak{T}_{4, 2}^{(6)} \, .
\end{equation*}
Decomposing $\mathfrak{T}_{4, 1}^{(6)}$ into 27 terms,
\begin{align*}
	&\mathfrak{T}_{4, 1}^{(6)} = - \underbrace{\frac{1}{2 \, T} \sum_{i = 1}^{N_{1}} \sum_{t = 1}^{T} \frac{\big\{\sum_{j = 1}^{N_{2}} (\widetilde{\mathfrak{D}_{\pi}^{2}})_{ijt}\big\} \big\{\sum_{j = 1}^{N_{2}} (\dop{1} \psi)_{ijt}\big\}^{2} \big\{\sum_{j = 1}^{N_{2}} (\overline{\dop{3} \psi})_{ijt}\big\}}{\big\{\sum_{j = 1}^{N_{2}} (\overline{\dop{2} \psi})_{ijt}\big\}^{3}}}_{\eqqcolon \mathfrak{T}_{4, 1, 1}^{(6)}} - \ldots \, - \\
	& \quad \underbrace{\frac{1}{2 \, T} \sum_{i = 1}^{N_{1}} \sum_{j = 1}^{N_{2}} \frac{\big\{\sum_{t = 1}^{T} (\widetilde{\mathfrak{D}_{\pi}^{2}})_{ijt}\big\} \big\{\sum_{t = 1}^{T} (\dop{1} \psi)_{ijt}\big\}^{2} \big\{\sum_{t = 1}^{T} (\overline{\dop{3} \psi})_{ijt}\big\}}{\big\{\sum_{t = 1}^{T} (\overline{\dop{2} \psi})_{ijt}\big\}^{3}}}_{\eqqcolon \mathfrak{T}_{4, 1, 27}^{(6)}} \, .
\end{align*}
By Pattern 3: $\norm{\mathfrak{T}_{4, 1, 1}^{(6)}}_{2} = o_{P}(T^{1 / 2})$ and $\norm{\mathfrak{T}_{4, 1, 14}^{(6)}}_{2} = o_{P}(T^{1 / 2})$. By Pattern 4: $\norm{\mathfrak{T}_{4, 1, 27}^{(6)}}_{2} = o_{P}(T^{1 / 2})$. By Pattern 2: $\norm{\mathfrak{T}_{4, 1, r}^{(6)}}_{2} = o_{P}(T^{1 / 2})$ for $r \in \{1, \ldots, 27\} \setminus \{1, 14, 27\}$. Hence, $\norm{\mathfrak{T}_{4, 1}^{(6)}}_{2} = o_{P}(T^{1 / 2})$. Decomposing $\mathfrak{T}_{4, 2}^{(6)}$ into 27 terms,
\begin{align*}
	&\mathfrak{T}_{4, 2}^{(6)} = - \underbrace{\frac{1}{2 \, T} \sum_{i = 1}^{N_{1}} \sum_{t = 1}^{T} \frac{\big\{\sum_{j = 1}^{N_{2}} (\widetilde{\mathfrak{D}_{\pi}^{2}})_{ijt}\big\} \big\{\sum_{j = 1}^{N_{2}} (\dop{1} \psi)_{ijt}\big\}^{2} \big\{\sum_{j = 1}^{N_{2}} (\widetilde{\dop{3} \psi})_{ijt}\big\}}{\big\{\sum_{j = 1}^{N_{2}} (\overline{\dop{2} \psi})_{ijt}\big\}^{3}}}_{\eqqcolon \mathfrak{T}_{4, 2, 1}^{(6)}} - \ldots \, - \\
	& \quad \underbrace{\frac{1}{2 \, T} \sum_{i = 1}^{N_{1}} \sum_{j = 1}^{N_{2}} \frac{\big\{\sum_{t = 1}^{T} (\widetilde{\mathfrak{D}_{\pi}^{2}})_{ijt}\big\} \big\{\sum_{t = 1}^{T} (\dop{1} \psi)_{ijt}\big\}^{2} \big\{\sum_{t = 1}^{T} (\widetilde{\dop{3} \psi})_{ijt}\big\}}{\big\{\sum_{t = 1}^{T} (\overline{\dop{2} \psi})_{ijt}\big\}^{3}}}_{\eqqcolon \mathfrak{T}_{4, 2, 27}^{(6)}} \, .
\end{align*}
By Pattern 3: $\norm{\mathfrak{T}_{4, 2, r}^{(6)}}_{2} = o_{P}(T^{1 / 2})$ for $r \in \{1, 14, 27\}$. By Pattern 2: $\norm{\mathfrak{T}_{4, 2, r}^{(6)}}_{2} = o_{P}(T^{1 / 2})$ for $r \in \{1, \ldots, 27\} \setminus \{1, 14, 27\}$. Hence, $\norm{\mathfrak{T}_{4, 2}^{(6)}}_{2} = o_{P}(T^{1 / 2})$, and thus $\norm{\mathfrak{T}_{4}^{(6)}}_{2} = o_{P}(T^{1 / 2})$. Combining, $\sum_{r = 1}^{4} \mathfrak{T}_{r}^{(6)} = N_{1} \, \overline{U}_{3, 1} + N_{2} \, \overline{U}_{3, 2} + (N_{1} N_{2} / T) \, \overline{U}_{3, 3} + o_{P}(T^{1 / 2})$.

Decomposing $\mathfrak{T}_{1, 1}^{(10)}$,
\begin{align*}
	&\mathfrak{T}_{1, 1}^{(10)} = \frac{(\overline{\mathfrak{D}_{\pi}^{3}})^{\prime} \diag(\overline{\mathcal{Q}} \dop{1} \psi) \overline{\mathcal{Q}} \, \overline{\nabla^{3} \psi} \diag(\overline{\mathcal{Q}} \dop{1} \psi) \overline{\mathcal{Q}} \dop{1} \psi}{2 \, T} \, + \\
	&\quad \frac{(\overline{\mathfrak{D}_{\pi}^{3}})^{\prime} \diag(\overline{\mathcal{Q}} \dop{1} \psi) \overline{\mathcal{Q}} \, \widetilde{\nabla^{3} \psi} \diag(\overline{\mathcal{Q}} \dop{1} \psi) \overline{\mathcal{Q}} \dop{1} \psi}{2 \, T} \eqqcolon \mathfrak{T}_{1, 1, 1}^{(10)} + \mathfrak{T}_{1, 1, 2}^{(10)} \, .
\end{align*}
Decomposing $\mathfrak{T}_{1, 1, 1}^{(10)}$ into 81 terms,
\begin{align*}
	&\mathfrak{T}_{1, 1, 1}^{(10)} = \underbrace{\frac{1}{2 \, T} \sum_{i = 1}^{N_{1}} \sum_{t = 1}^{T} \frac{\big\{\sum_{j = 1}^{N_{2}} (\overline{\mathfrak{D}_{\pi}^{3}})_{ijt}\big\} \big\{\sum_{j = 1}^{N_{2}} (\dop{1} \psi)_{ijt}\big\}^{3} \big\{\sum_{j = 1}^{N_{2}} (\overline{\dop{3} \psi})_{ijt}\big\}}{\big\{\sum_{j = 1}^{N_{2}} (\overline{\dop{2} \psi})_{ijt}\big\}^{4}}}_{\eqqcolon \mathfrak{T}_{1, 1, 1, 1}^{(10)}} + \ldots \, + \\
	& \quad \underbrace{\frac{1}{2 \, T} \sum_{i = 1}^{N_{1}} \sum_{j = 1}^{N_{2}} \frac{\big\{\sum_{t = 1}^{T} (\overline{\mathfrak{D}_{\pi}^{3}})_{ijt}\big\} \big\{\sum_{t = 1}^{T} (\dop{1} \psi)_{ijt}\big\}^{3} \big\{\sum_{t = 1}^{T} (\overline{\dop{3} \psi})_{ijt}\big\}}{\big\{\sum_{t = 1}^{T} (\overline{\dop{2} \psi})_{ijt}\big\}^{4}}}_{\eqqcolon \mathfrak{T}_{1, 1, 1, 81}^{(10)}} \, .
\end{align*}
By Pattern 3: $\norm{\mathfrak{T}_{1, 1, 1, 1}^{(10)}}_{2} = o_{P}(T^{1 / 2})$ and $\norm{\mathfrak{T}_{1, 1, 1, 41}^{(10)}}_{2} = o_{P}(T^{1 / 2})$. By Pattern 4: $\norm{\mathfrak{T}_{1, 1, 1, 81}^{(10)}}_{2} = o_{P}(T^{1 / 2})$. By Pattern 2: $\norm{\mathfrak{T}_{1, 1, 1, r}^{(10)}}_{2} = o_{P}(T^{1 / 2})$ for $r \in \{1, \ldots, 81\} \setminus \{1, 41, 81\}$. Hence, $\norm{\mathfrak{T}_{1, 1, 1}^{(10)}}_{2} = o_{P}(T^{1 / 2})$. By Pattern 3, $\norm{\mathfrak{T}_{1, 1, 2}^{(10)}}_{2} = o_{P}(T^{1 / 2})$, and thus $\norm{\mathfrak{T}_{1, 1}^{(10)}}_{2} = o_{P}(T^{1 / 2})$. Analogously, $\norm{\mathfrak{T}_{2, 1}^{(10)}}_{2} = o_{P}(T^{1 / 2})$.\hfill\qedsymbol

\paragraph{Proof of Theorem \ref{theorem:asymptotic_expansion_profile_score_interacted} (ii).} The conditions of \ref{lemma:taylor_expansions} hold wpa1. Thus, $a^{\prime} \big(\hat{\phi}(\beta) - \phi^{0}\big) = a^{\prime} \mathcal{T}^{(12)}(\beta) + a^{\prime} \mathcal{T}^{(13)} + a^{\prime} \mathcal{R}^{(2)}(\beta)$, where $\mathcal{R}^{(2)}(\beta) \coloneqq \mathcal{T}^{(14)}(\beta) + \mathcal{T}^{(15)}(\beta) + \mathcal{T}^{(16)} + \sum_{r = 1}^{2} \mathcal{T}_{r}^{(17)}$. Throughout, $\beta \in \mathfrak{B}(\varepsilon)$ with $\varepsilon = o(T^{- 1 / 2})$. By H\"older's inequality and the triangle inequality,
\begin{equation*}
	\bigabs{a^{\prime} \mathcal{R}^{(2)}(\beta)} \leq \norm{a}_{p^{\prime}} \, \Big\{ \bignorm{\mathcal{T}^{(14)}(\beta)}_{p} + \bignorm{\mathcal{T}^{(15)}(\beta)}_{p} + \bignorm{\mathcal{T}^{(16)}}_{p} + \sum_{r = 1}^{2} \bignorm{\mathcal{T}_{r}^{(17)}}_{p} \Big\} \, .
\end{equation*}
By Lemma \ref{lemma:matrix_norm_inequalties} and H\"older's inequality,
\begin{align*}
	&\sup_{\beta} \bignorm{\mathcal{T}^{(14)}(\beta)}_{p} \leq C \, T^{- 1} \, \norm{D^{\prime}}_{p} \bignorm{\check{H}^{- 1}}_{p} \max_{k} \bignorm{\check{\nabla^{3} \psi} X e_{k}}_{p} \max_{k} \bignorm{\check{M} X e_{k}}_{10} \, \bignorm{\beta - \beta^{0}}_{2} \, \varepsilon \, + \\
	&\qquad C \, T^{- 2} \, \norm{D^{\prime}}_{p} \bignorm{\check{H}^{- 1}}_{p}^{2} \, \bignorm{D^{\prime} \abs{\check{\nabla^{3} \psi}} D}_{\infty} \max_{k} \bignorm{\check{\nabla^{2} \psi} X e_{k}}_{p} \max_{k} \bignorm{\check{M} X e_{k}}_{10} \, \bignorm{\beta - \beta^{0}}_{2} \, \varepsilon \\
	&\qquad = o_{P}\big(T^{\frac{2}{p}} \bignorm{\beta - \beta^{0}}_{2}\big) \, , \\
	&\bignorm{\mathcal{T}^{(16)}}_{p} \leq C \, T^{- 2} \, \bignorm{H^{- 1}}_{p}^{2} \, \bignorm{D^{\prime} \abs{\nabla^{3} \psi} D}_{\infty} \, \bignorm{D^{\prime} \dop{1} \psi}_{p} \, \bignorm{Q \dop{1} \psi}_{10} = o_{P}\big(T^{- \frac{3}{5} + \frac{2}{p}}\big) \, .
\end{align*}
$\mathcal{T}^{(15)}(\beta)$ satisfies an analogous bound to $\mathcal{T}^{(14)}(\beta)$ and is $o_P(T^{2/p}\norm{\beta - \beta^0}_2)$. $\mathcal{T}_1^{(17)}$ and $\mathcal{T}_2^{(17)}$ satisfy analogous bounds to $\mathcal{T}^{(16)}$ and are $o_P(T^{-3/5+2/p})$. Hence, $\norm{\mathcal{R}^{(2)}(\beta)}_{p} = o_{P}(T^{- 3 / 5 + 2 / p}) + o_{P}(T^{2 / p} \norm{\beta - \beta^{0}}_{2})$, for all $\beta \in \mathfrak{B}(\varepsilon)$. Decomposing $\mathcal{T}^{(12)}(\beta)$,
\begin{equation*}
	\mathcal{T}^{(12)}(\beta) = \overline{\mathbb{W}} \big(\beta - \beta^{0}\big) - \underbrace{\frac{\overline{H}^{- 1} \, D^{\prime} \widetilde{\nabla^{2} \psi X} (\beta - \beta^{0})}{T}}_{\eqqcolon \mathfrak{T}_{1}^{(12)}(\beta)} - \underbrace{\frac{(H^{- 1} - \overline{H}^{- 1}) \, D^{\prime} \nabla^{2} \psi X (\beta - \beta^{0})}{T}}_{\eqqcolon \mathfrak{T}_{2}^{(12)}(\beta)}
\end{equation*}
By H\"older's inequality,
\begin{align*}
	\bignorm{\mathfrak{T}_{1}^{(12)}(\beta)}_{p} \leq& \, C \, T^{- 1} \,\bignorm{\overline{H}^{- 1}}_{\infty} \max_{k} \bignorm{D^{\prime} \widetilde{\nabla^{2} \psi X} e_{k}}_{p} \, \bignorm{\beta - \beta^{0}}_{2} = o_{P}\big(T^{\frac{2}{p}} \bignorm{\beta - \beta^{0}}_{2}\big) \, , \\
	\bignorm{\mathfrak{T}_{2}^{(12)}(\beta)}_{p} \leq& \, C \, T^{- 1} \, \norm{D^{\prime}}_{p} \bignorm{H^{- 1} - \overline{H}^{- 1}}_{p} \max_{k} \bignorm{\nabla^{2} \psi X e_{k}}_{p} \, \bignorm{\beta - \beta^{0}}_{2} = o_{P}\big(T^{\frac{2}{p}} \bignorm{\beta - \beta^{0}}_{2}\big) \, .
\end{align*}
Thus, $\mathcal{T}^{(12)}(\beta) = \overline{\mathbb{W}} (\beta - \beta^{0}) + o_{P}(T^{2/p} \|\beta - \beta^{0}\|_2)$. Combining all terms and using $\norm{a}_{p^{\prime}} = \mathcal{O}_{P}(1)$ yields the stated result.\hfill\qedsymbol

\paragraph{Proof of Theorem \ref{theorem:consistency_interacted}.} Let $r_{\beta} = 2 \, (N_{1} N_{2} T)^{- 1 / 2} \, \overline{W}^{- 1} \abs{U} = \mathcal{O}_{P}(T^{- 1})$ be a $K$-dimensional vector.

\vspace{0.5em}
\noindent\# \underline{Part 1.} Suppose that $K = 1$ and define $\underline{b} \coloneqq \beta^{0} - r_{\beta}$ and $\overline{b} \coloneqq \beta^{0} + r_{\beta}$. For $\beta \in [\underline{b}, \overline{b}]$, we have $\abs{\beta - \beta^{0}} = \abs{r_{\beta}} = \mathcal{O}_{P}(T^{- 1}) = o_{P}(\varepsilon)$, so $\beta \in \mathfrak{B}(\varepsilon)$ wpa1. By Theorem \ref{theorem:asymptotic_expansion_profile_score_interacted} (i) and sufficiently large $n$,
\begin{align*}
	\frac{\partial \mathcal{L}_{n}(\beta, \hat{\phi}(\beta))}{\partial \beta}\biggr\rvert_{\substack{\beta \, = \, \underline{b}}} =& \, \left(\frac{\sqrt{N_{1} N_{2}}}{\sqrt{T}}\right) \underbrace{\big(U - 2 \, \abs{U}\big)}_{< \, 0} + \, o_{P}\big(T^{\frac{1}{2}}\big) \, , \\
	\frac{\partial \mathcal{L}_{n}(\beta, \hat{\phi}(\beta))}{\partial \beta}\biggr\rvert_{\substack{\beta \, = \, \overline{b}}} =& \, \left(\frac{\sqrt{N_{1} N_{2}}}{\sqrt{T}}\right) \underbrace{\big(U + 2 \, \abs{U}\big)}_{> 0} + \, o_{P}\big(T^{\frac{1}{2}}\big) \, .
\end{align*}
Since $\partial_{\beta} \mathcal{L}_{n}(\hat{\beta}, \hat{\phi}(\hat{\beta})) = 0$ and $\mathcal{L}_{n}(\beta, \hat{\phi}(\beta))$ is strictly convex on $\mathfrak{B}(\varepsilon)$, these sign conditions imply $\abs{\hat{\beta} - \beta^{0}} \leq r_{\beta}$, and hence, $\norm{\hat{\beta} - \beta^{0}} = \mathcal{O}_{P}(T^{- 1})$ for $K = 1$.

For $K > 1$, define $\underline{\beta} \coloneqq \beta^{0} - r_{\beta} \, (\hat{\beta} - \beta^{0}) / \norm{\hat{\beta} - \beta^{0}}_{2}$, $\overline{\beta} \coloneqq \beta^{0} + r_{\beta} \, (\hat{\beta} - \beta^{0}) / \norm{\hat{\beta} - \beta^{0}}_{2}$, and let $B$ be the line segment between them. Restricting $\beta \in B$ and reapplying the $K = 1$ argument yields $\norm{\hat{\beta} - \beta^{0}}_{2} \leq r_{\beta} = \mathcal{O}_{P}(T^{- 1})$.

\vspace{0.5em}
\noindent\# \underline{Part 2.} Define $p^{\prime} \coloneqq p / (p - 1)$. Then, by the definition of the $p$-norm, $\norm{\hat{\phi} - \phi^{0}}_{p} = \sup_{\norm{a}_{p^{\prime}} = 1} \abs{a^{\prime} (\hat{\phi} - \phi^{0})}$. By Lemma \ref{lemma:matrix_norm_inequalties}, $\norm{\mathbb{U}}_{10} = \mathcal{O}_{P}(T^{- 3 / 10})$ and $\norm{\overline{\mathbb{W}}}_{10} = \mathcal{O}_{P}(T^{1 / 5})$. By Theorem \ref{theorem:asymptotic_expansion_profile_score_interacted} (ii),
\begin{align*}
	&\bignorm{\hat{\phi} - \phi^{0}}_{10} \leq \norm{\mathbb{U}}_{10} +  \bignorm{\overline{\mathbb{W}}}_{10} \, \bignorm{\hat{\beta} - \beta^{0}}_{2} + o_{P}\big(T^{- \frac{2}{5}}\big) + o_{P}\big(T^{\frac{1}{5}} \bignorm{\hat{\beta} - \beta^{0}}_{2}\big) = \mathcal{O}_{P}\big(T^{- \frac{3}{10}}\big) \, , \\
	&\bignorm{\hat{\phi} - \phi^{0} - \mathbb{U}}_{10} \leq \bignorm{\overline{\mathbb{W}}}_{10} \, \bignorm{\hat{\beta} - \beta^{0}}_{2} + o_{P}\big(T^{- \frac{2}{5}}\big) + o_{P}\big(T^{\frac{1}{5}} \bignorm{\hat{\beta} - \beta^{0}}_{2}\big) = o_{P}\big(T^{- \frac{2}{5}}\big) \, .
\end{align*}
For $1 \leq p \leq 10$, the stated rates follow by H\"older's inequality and $L = \mathcal{O}(T^{2})$.\hfill\qedsymbol

\paragraph{Proof of Corollary \ref{corollary:asymptotic_expansion_score_interacted}.} By Theorem \ref{theorem:consistency_interacted}, $\norm{\hat{\beta} - \beta^{0}}_{2} = \mathcal{O}_{P}(T^{- 1}) = o_{P}(\varepsilon)$, and $\partial_{\beta} \mathcal{L}_{n}(\hat{\beta}, \hat{\phi}(\hat{\beta})) = \mathbf{0}_{K}$ by the definition of $\hat{\beta}$. Evaluating Theorem \ref{theorem:asymptotic_expansion_profile_score_interacted} (i) at $\hat{\beta}$ and rearranging,
\begin{equation*}
	N_{1} N_{2} \, \overline{W} (\hat{\beta} - \beta^{0}) = - \left(\frac{\sqrt{N_{1} N_{2}}}{\sqrt{T}}\right) U + o_{P}\big(T^{\frac{1}{2}}\big) + o_{P}\big(T^{\frac{3}{2}} \norm{\hat{\beta} - \beta^{0}}_{2}\big) \, .
\end{equation*}
The stated result follows upon dividing by $\sqrt{N_{1} N_{2}} / \sqrt{T}$ and using $T \norm{\hat{\beta} - \beta^{0}}_{2} = \mathcal{O}_{P}(1)$.\hfill\qedsymbol

\begin{corollary}[Consistency of $\hat{\pi}$,]
	\label{corollary:consistency_interacted}
	Let Lemma \ref{lemma:regularity_conditions1_interacted} hold. Then, for $1 \leq p \leq 10$, $\norm{\hat{\pi} - \pi^{0}}_{p} = \mathcal{O}_{P}(T^{- 1 / 2 + 3 / p})$ and $\norm{\hat{\pi} - \pi^{0} + Q \dop{1} \psi}_{p} = o_{P}(T^{- 3 / 5 + 3 / p})$.
\end{corollary}

\paragraph{Proof of Corollary \ref{corollary:consistency_interacted}.} Recall $\pi(\beta, \phi) = X \beta + D \phi$ and $D \mathbb{U} = - \, Q \dop{1} \psi$. Decomposing,
\begin{equation*}
	\hat{\pi} - \pi^{0} = X (\hat{\beta} - \beta^{0}) + D (\hat{\phi} - \phi^{0}) = X (\hat{\beta} - \beta^{0}) + D (\hat{\phi} - \phi^{0} - \mathbb{U}) - Q \dop{1} \psi \, .
\end{equation*}
By H\"older's inequality, Jensen's inequality, and Theorem \ref{theorem:consistency_interacted}, $\norm{X (\hat{\beta} - \beta^{0})}_{p} = \mathcal{O}_{P}(T^{- 1 + 3 / p})$, $\norm{D (\hat{\phi} - \phi^{0})}_{p} = \mathcal{O}_{P}(T^{- 1 / 2 + 3 / p})$, and $\norm{D (\hat{\phi} - \phi^{0} - \mathbb{U})}_{p} = o_{P}(T^{- 3 / 5 + 3 / p})$. The stated results follow by the triangle inequality.\hfill\qedsymbol

\paragraph{Proof of Theorem \ref{theorem:consistency_bias_variance_interacted}.} \noindent\# \underline{Part 1.} Let $\overline{B}_{\alpha, 1}$ denote the first term in $\overline{B}_{\alpha}$. The $k$-th element of $\overline{B}_{\alpha, 1}$ can be expressed as
\begin{equation*}
	\big(\overline{B}_{\alpha, 1}\big)_{k} = - \frac{1}{N_{1} T} (\bar{f}_{1}^{\circ - 1})^{\prime} \bar{g}_{1, 1, k} = - \frac{1}{N_{1} T} \sum_{i = 1}^{N_{1}} \sum_{t = 1}^{T} \frac{(\bar{g}_{1, 1, k})_{it}}{(\bar{f}_{1})_{it}}
\end{equation*}
where $(\bar{g}_{1, 1, k})_{it} \coloneqq \sum_{j = 1}^{N_{2}} \EX{(\mathfrak{D}_{\pi}^{2} e_{k})_{ijt} (\dop{1} \psi)_{ijt}}$. The corresponding estimator is 
$(\hat{g}_{1, 1, k})_{it} \coloneqq \sum_{j = 1}^{N_{2}} (\widehat{\mathfrak{D}_{\pi}^{2}} e_{k})_{ijt} (\widehat{\dop{1} \psi})_{ijt}$. Decomposing
\begin{align*}
	&\hat{g}_{1, 1, k} - \bar{g}_{1, 1, k} = D_{1}^{\prime} \big(\widehat{\mathfrak{D}_{\pi}^{2}} e_{k} \odot \widehat{\dop{1} \psi} - \overline{\mathfrak{D}_{\pi}^{2} e_{k} \odot \dop{1} \psi}\big) =\\
	&\qquad D_{1}^{\prime} \big( (\widehat{\mathfrak{D}_{\pi}^{2}} e_{k} - \mathfrak{D}_{\pi}^{2} e_{k}) \odot (\widehat{\dop{1} \psi} - \dop{1} \psi) \big) + D_{1}^{\prime} \big( (\widehat{\mathfrak{D}_{\pi}^{2}} e_{k} - \mathfrak{D}_{\pi}^{2} e_{k}) \odot \dop{1} \psi \big) \, +\\
	&\qquad D_{1}^{\prime} \big( \mathfrak{D}_{\pi}^{2} e_{k} \odot (\widehat{\dop{1} \psi} - \dop{1} \psi) \big) + D_{1}^{\prime} \big( \mathfrak{D}_{\pi}^{2} e_{k} \odot \dop{1} \psi - \overline{\mathfrak{D}_{\pi}^{2} e_{k} \odot \dop{1} \psi} \big) \, = \\
	&\qquad - D_{1}^{\prime} \diag(\mathfrak{D}_{\pi}^{2} e_{k}) \nabla^{2} \psi \overline{\mathcal{Q}} \dop{1} \psi - D_{1}^{\prime} \nabla^{1} \psi \nabla^{2} \psi \overline{\mathcal{Q}} \widetilde{\mathfrak{D}_{\pi}^{2}} e_{k} \, - \\
	&\qquad D_{1}^{\prime} \diag(\mathfrak{D}_{\pi}^{3} e_{k}) \overline{\mathcal{Q}} \dop{1} \psi + D_{1}^{\prime} \nabla^{2} \psi \overline{\mathcal{Q}} \diag(\overline{\mathfrak{D}_{\pi}^{3}} e_{k}) \overline{\mathcal{Q}} \dop{1} \psi \, + \\
	&\qquad D_{1}^{\prime} \big( \mathfrak{D}_{\pi}^{2} e_{k} \odot \dop{1} \psi - \overline{\mathfrak{D}_{\pi}^{2} e_{k} \odot \dop{1} \psi} \big) + D_{1}^{\prime} \big( (\widehat{\mathfrak{D}_{\pi}^{2}} e_{k} - \mathfrak{D}_{\pi}^{2} e_{k}) \odot (\widehat{\dop{1} \psi} - \dop{1} \psi) \big) \, + \\
	&\qquad D_{1}^{\prime} \big( (\widehat{\mathfrak{D}_{\pi}^{2}} e_{k} - \mathfrak{D}_{\pi}^{2} e_{k} + \nabla^{2} \psi \overline{\mathcal{Q}} \widetilde{\mathfrak{D}_{\pi}^{2}} e_{k} + \ldots) \odot \dop{1} \psi \big) \, + \\
	&\qquad D_{1}^{\prime} \big( \mathfrak{D}_{\pi}^{2} e_{k} \odot (\widehat{\dop{1} \psi} - \dop{1} \psi + \nabla^{2} \psi \overline{\mathcal{Q}} \dop{1} \psi) \big) \\
	&\qquad \eqqcolon \mathfrak{g}_{1, 1, k, 1} + \ldots + \mathfrak{g}_{1, 1, k, 8} \, .
\end{align*}
By H\"older's inequality,
\begin{equation*}
	\max_{k} \bignorm{\mathfrak{g}_{1, 1, k, 6}}_{1} \leq \max_{k} \bignorm{\widehat{\mathfrak{D}_{\pi}^{2}} e_{k} - \mathfrak{D}_{\pi}^{2} e_{k}}_{2} \, \bignorm{\widehat{\dop{1} \psi} - \dop{1} \psi}_{2} = o_{P}\big(T^{\frac{12}{5}}\big) \, .
\end{equation*}
Analogously, $\max_{k} \norm{\mathfrak{g}_{1, 1, k, r}}_{1} = o_{P}(T^{12/5})$ for $r \in \{7, 8\}$. By H\"older's inequality,
\begin{equation*}
	\max_{k} \bignorm{\mathfrak{g}_{1, 1, k, 1}}_{1} \leq \bignorm{\dop{2} \psi}_{20 + \nu} \max_{k} \bignorm{\mathfrak{D}_{\pi}^{2} e_{k}}_{20 + \nu} \, \bignorm{\overline{\mathcal{Q}} \dop{1} \psi}_{1} = o_{P}\big(T^{\frac{14}{5}}\big) \, .
\end{equation*}
Analogously, $\max_{k} \norm{\mathfrak{g}_{1, 1, k, r}}_{1} = o_{P}(T^{14/5})$ for $r \in \{2, 3, 4\}$. Let $\vartheta_{1, k} \coloneqq \mathfrak{D}_{\pi}^{2} e_{k} \odot \dop{1} \psi$. By Lemma \ref{lemma:moment_bounds_mixing},
\begin{equation*}
	\EX{\max_{k} \bignorm{\mathfrak{g}_{1, 1, k, 5}}_{10}^{10}} \leq T^{7} \, \bigg(\frac{N_{1}}{T}\bigg) \bigg(\frac{N_{2}}{T}\bigg)^{5} \sup_{it} \EX{ \bigg\{\frac{1}{\sqrt{N_{2}}}\sum_{j = 1}^{N_{2}} \max_{k} (\widetilde{\vartheta}_{1, k})_{ijt} \bigg\}^{10} } = \mathcal{O}\big(T^{7}\big) \quad \text{a.\,s.} \, ,
\end{equation*}
so $\max_{k} \norm{\mathfrak{g}_{1, 1, k, 5}}_{10} = \mathcal{O}_{P}(T^{7 / 10})$ by Markov's inequality, and thus \linebreak $\max_{k} \norm{\mathfrak{g}_{1, 1, k, 5}}_{p} \leq (N_{1} T)^{1 / p - 1 / 10} \max_{k} \norm{\mathfrak{g}_{1, 1, k, 5}}_{10} = \mathcal{O}_{P}(T^{1 / 2 + 2 / p})$ for $1 \leq p \leq 10$. Hence, $\max_{k} \norm{\hat{g}_{1, 1, k} - \bar{g}_{1, 1, k}}_{1} = o_{P}(T^{14 / 5})$ and $\max_{k} \norm{\hat{g}_{1, 1, k} - \bar{g}_{1, 1, k} - \mathfrak{g}_{1, 1, k, 1} - \ldots}_{1} = o_{P}(T^{12 / 5})$. Let $f_{1}^{\ast} \coloneqq D_{1}^{\prime} \overline{\nabla^{3} \psi} \, \overline{\mathcal{Q}} \dop{1} \psi$. Decomposing the estimator $(\hat{f}_{1}^{\circ - 1})^{\prime} \hat{g}_{1, 1, k}$ for each $k \in \{1, \ldots, K\}$,
\begin{align*}
	&(\hat{f}_{1}^{\circ - 1})^{\prime} \hat{g}_{1, 1, k} - (\bar{f}_{1}^{\circ - 1})^{\prime} \bar{g}_{1, 1, k} = \\
	& \qquad (\hat{f}_{1}^{\circ - 1} - \bar{f}_{1}^{\circ - 1})^{\prime} (\hat{g}_{1, 1, k} - \bar{g}_{1, 1, k}) + (\hat{f}_{1}^{\circ - 1} - \bar{f}_{1}^{\circ - 1} - \bar{f}_{1}^{\circ - 2} \odot f_{1}^{\ast} + \bar{f}_{1}^{\circ - 2} \odot \tilde{f}_{1})^{\prime} \bar{g}_{1, 1, k} \, + \\
	&\qquad (\bar{f}_{1}^{\circ - 1})^{\prime} \big(\hat{g}_{1, 1, k} - \bar{g}_{1, 1, k} - \mathfrak{g}_{1, 1, k, 1} - \ldots - \mathfrak{g}_{1, 1, k, 5}\big) + (\bar{f}_{1}^{\circ - 2} \odot f_{1}^{\ast})^{\prime} \bar{g}_{1, 1, k} \, - \\
	&\qquad (\bar{f}_{1}^{\circ - 2} \odot \tilde{f}_{1})^{\prime} \bar{g}_{1, 1, k} + (\bar{f}_{1}^{\circ - 1})^{\prime} \mathfrak{g}_{1, 1, k, 1} + \ldots + (\bar{f}_{1}^{\circ - 1})^{\prime} \mathfrak{g}_{1, 1, k, 5}  \\
	& \qquad \eqqcolon \mathfrak{E}_{1, 1, k, 1} + \ldots + \mathfrak{E}_{1, 1, k, 10} \, .
\end{align*}
By H\"older's inequality,
\begin{align*}
	\max_{k} \bigabs{\mathfrak{E}_{1, 1, k, 1}} \leq& \, \bignorm{\hat{f}_{1}^{\circ - 1} - \bar{f}_{1}^{\circ - 1}}_{10} \max_{k} \bignorm{\hat{g}_{1, 1, k} - \bar{g}_{1, 1, k}}_{1} = o_{P}\big(T^{\frac{3}{2}}\big) \, , \\
	\max_{k} \bigabs{\mathfrak{E}_{1, 1, k, 2}} \leq& \, \bignorm{\hat{f}_{1}^{\circ - 1} - \bar{f}_{1}^{\circ - 1} - \bar{f}_{1}^{\circ - 2} \odot f_{1}^{\ast} + \bar{f}_{1}^{\circ - 2} \odot \tilde{f}_{1}}_{1} \max_{k} \bignorm{\bar{g}_{1, 1, k}}_{\infty} = o_{P}\big(T^{\frac{3}{2}}\big) \, , \\
	\max_{k} \bigabs{\mathfrak{E}_{1, 1, k, 3}} \leq& \, \bignorm{\bar{f}_{1}^{\circ - 1}}_{\infty} \max_{k} \bignorm{\hat{g}_{1, 1, k} - \bar{g}_{1, 1, k} - \mathfrak{g}_{1, 1, k, 1} - \ldots - \mathfrak{g}_{1, 1, k, 5}}_{1} = o_{P}\big(T^{\frac{3}{2}}\big) \, .
\end{align*}
Decomposing $\mathfrak{E}_{1, 1, k, 4}$,
\begin{align*}
	\mathfrak{E}_{1, 1, k, 4} =& \, \big(\bar{f}_{1}^{\circ - 2} \odot D_{1}^{\prime} \overline{\nabla^{3} \psi} \, \overline{\mathcal{Q}}_{1} \dop{1} \psi\big)^{\prime} \bar{g}_{1, 1, k} + \ldots + \big(\bar{f}_{1}^{\circ - 2} \odot D_{1}^{\prime} \overline{\nabla^{3} \psi} \, \overline{\mathcal{Q}}_{3} \dop{1} \psi\big)^{\prime} \bar{g}_{1, 1, k} \\
	\eqqcolon& \, \mathfrak{E}_{1, 1, k, 4, 1} + \mathfrak{E}_{1, 1, k, 4, 2} + \mathfrak{E}_{1, 1, k, 4, 3} \, .
\end{align*}
Let $(\overline{\vartheta}_{2})_{ijt} \coloneqq N_{1} \, (\overline{\dop{3} \psi})_{ijt} / (\overline{f}_{2})_{jt}$. By Lemma \ref{lemma:moment_bounds_mixing} and the Lyapunov inequality,
\begin{align*}
	\EX{\bignorm{D_{1}^{\prime} \overline{\nabla^{3} \psi} \, \overline{\mathcal{Q}}_{2} \dop{1} \psi}_{20}^{20}} \leq& \, \bigg(\frac{T}{N_{1}}\bigg)^{9} \bigg(\frac{N_{2}}{T}\bigg)^{10} \, T^{2} \, \sup_{it} \, \EX{\bigg(\frac{1}{\sqrt{N_{1} N_{2}}} \sum_{i^{\prime} = 1}^{N_{1}} \sum_{j = 1}^{N_{2}} (\overline{\vartheta}_{2})_{ijt} (\dop{1} \psi)_{i^{\prime}jt} \bigg)^{20}} \\
	=& \, \mathcal{O}\big(T^{2}\big) \quad \text{a.\,s.} \, ,
\end{align*}
so $\norm{D_{1}^{\prime} \overline{\nabla^{3} \psi} \, \overline{\mathcal{Q}}_{2} \dop{1} \psi}_{20} = \mathcal{O}_{P}(T^{1 / 10})$ by Markov's inequality and thus $\norm{D_{1}^{\prime} \overline{\nabla^{3} \psi} \, \overline{\mathcal{Q}}_{2} \dop{1} \psi}_{p} = \mathcal{O}_{P}(T^{2 / p})$ for $1 \leq p \leq 20$. Analogously, $\norm{D_{1}^{\prime} \overline{\nabla^{3} \psi} \, \overline{\mathcal{Q}}_{3} \dop{1} \psi}_{p} = \mathcal{O}_{P}(T^{2 / p})$. Hence,
\begin{align*}
	\max_{k} \bigabs{\mathfrak{E}_{1, 1, k, 4, 2}} \leq& \, \bignorm{\bar{f}_{1}^{\circ - 1}}_{\infty}^{2} \max_{k} \bignorm{\bar{g}_{1, 1, k}}_{\infty} \norm{D_{1}^{\prime} \overline{\nabla^{3} \psi} \, \overline{\mathcal{Q}}_{2} \dop{1} \psi}_{1} = o_{P}\big(T^{\frac{3}{2}}\big) \, , \\
	\max_{k} \bigabs{\mathfrak{E}_{1, 1, k, 4, 3}} \leq& \, \bignorm{\bar{f}_{1}^{\circ - 1}}_{\infty}^{2} \max_{k} \bignorm{\bar{g}_{1, 1, k}}_{\infty} \norm{D_{1}^{\prime} \overline{\nabla^{3} \psi} \, \overline{\mathcal{Q}}_{3} \dop{1} \psi}_{1} = o_{P}\big(T^{\frac{3}{2}}\big) \, .
\end{align*}
Let $\overline{\vartheta}_{3} \coloneqq N_{2} \, (\bar{g}_{1, 1, k} \odot D_{1}^{\prime} \overline{\dop{3} \psi}) \odot \overline{f}_{1}^{\circ - 3}$. By similar arguments,
\begin{equation*}
	\EX{\max_{k} \bigabs{\mathfrak{E}_{1, 1, k, 4, 1}}^{2}} \leq C \, T \, \bigg(\frac{N_{1}}{T}\bigg)  \bigg(\frac{T}{N_{2}}\bigg) \sup_{ij} \EX{\bigg(\frac{1}{\sqrt{T}} \sum_{t = 1}^{T} (\overline{\vartheta}_{3})_{it} (\dop{1} \psi)_{ijt} \bigg)^{2}} = \mathcal{O}_{P}\big(T\big) \, ,
\end{equation*}
so $\max_{k} \abs{\mathfrak{E}_{1, 1, k, 4, 1}} = \mathcal{O}_{P}(T^{1 / 2}) = o_{P}(T^{3 / 2})$ by Markov's inequality. Hence, $\max_{k} \abs{\mathfrak{E}_{1, 1, k, 4}} = o_{P}(T^{3 / 2})$. By similar arguments, $\max_{k} \abs{\mathfrak{E}_{1, 1, k, r}} = o_{P}(T^{3 / 2})$ for all $r \in \{5, \ldots, 10\}$. Hence, combining all components,
\begin{equation*}
	\bignorm{\widehat{B}_{\alpha, 1} - \overline{B}_{\alpha, 1}}_{2} \leq C \, \bigg(\frac{T}{N_{1}}\bigg) \, T^{- 2} \max_{k} \bigabs{(\hat{f}_{1}^{\circ - 1})^{\prime} \hat{g}_{1, 1, k} - (\bar{f}_{1}^{\circ - 1})^{\prime} \bar{g}_{1, 1, k}} = o_{P}\big(T^{- \frac{1}{2}}\big) \, .
\end{equation*}
Analogously, $\norm{\widehat{B}_{\gamma, 1} - \overline{B}_{\gamma, 1}}_{2} = o_{P}(T^{- 1 / 2})$. By similar arguments, $\norm{\widehat{B}_{\alpha, 2} - \overline{B}_{\alpha, 2}}_{2} = o_{P}(T^{- 1 / 2})$, $\norm{\widehat{B}_{\gamma, 2} - \overline{B}_{\gamma, 2}}_{2} = o_{P}(T^{- 1 / 2})$, and $\norm{\widehat{B}_{\rho, 2} - \overline{B}_{\rho, 2}}_{2} = o_{P}(T^{- 1 / 2})$, where $\overline{B}_{\cdot, 2}$ denotes the second term in $\overline{B}_{\cdot}$ (with $\cdot$ as placeholder).

Let $\overline{B}_{\rho, 1}$ denote the first term in $\overline{B}_{\rho}$. The $k$-th element of $\overline{B}_{\rho, 1}$ can be expressed as
\begin{equation*}
	\big(\overline{B}_{\rho, 1}\big)_{k} = - \frac{1}{N_{1} N_{2}} (\bar{f}_{3}^{\circ - 1})^{\prime} \bar{g}_{3, 1, k} = - \frac{1}{N_{1} N_{2}} \sum_{i = 1}^{N_{1}} \sum_{j = 1}^{N_{2}} \frac{(\bar{g}_{3, 1, k})_{ij}}{(\bar{f}_{3})_{ij}}
\end{equation*}
where $(\bar{g}_{3, 1, k})_{ij} \coloneqq \sum_{t = 1}^{T} \sum_{t^{\prime} = t}^{T} \EX{(\mathfrak{D}_{\pi}^{2} e_{k})_{ijt^{\prime}} (\dop{1} \psi)_{ijt}}$. The corresponding (truncated) estimator is $(\hat{g}_{3, 1, k})_{ij} \coloneqq \sum_{t = 1}^{T} \sum_{t^{\prime} = t}^{(t + h) \wedge T} (\widehat{\mathfrak{D}_{\pi}^{2}} e_{k})_{ijt^{\prime}} (\widehat{\dop{1} \psi})_{ijt}$. Decomposing, for all $i, j, N_{1}, N_{2}$,
\begin{align*}
	&(\hat{g}_{3, 1, k} - \bar{g}_{3, 1, k})_{ij} = \sum_{t = 1}^{T} \sum_{t^{\prime} = t}^{(t + h) \wedge T} (\widehat{\mathfrak{D}_{\pi}^{2}} e_{k})_{ijt^{\prime}} (\widehat{\dop{1} \psi})_{ijt} - \sum_{t = 1}^{T} \sum_{t^{\prime} = t}^{T} \EX{(\mathfrak{D}_{\pi}^{2} e_{k})_{ijt^{\prime}} (\dop{1} \psi)_{ijt}} = \\
	&\qquad \sum_{t = 1}^{T} \sum_{t^{\prime} = t}^{(t + h) \wedge T} (\widehat{\mathfrak{D}_{\pi}^{2}} e_{k} - \mathfrak{D}_{\pi}^{2} e_{k})_{ijt^{\prime}} (\widehat{\dop{1} \psi} - \dop{1} \psi)_{ijt} \, + \\
	&\qquad \sum_{t = 1}^{T} \sum_{t^{\prime} = t}^{(t + h) \wedge T} (\widehat{\mathfrak{D}_{\pi}^{2}} e_{k} - \mathfrak{D}_{\pi}^{2} e_{k} + \nabla^{2} \psi \overline{\mathcal{Q}} \widetilde{\mathfrak{D}_{\pi}^{2}} e_{k} + \ldots)_{ijt^{\prime}} (\dop{1} \psi)_{ijt} \, + \\
	&\qquad \sum_{t = 1}^{T} \sum_{t^{\prime} = t}^{(t + h) \wedge T} (\mathfrak{D}_{\pi}^{2} e_{k})_{ijt^{\prime}} (\widehat{\dop{1} \psi} - \dop{1} \psi + \nabla^{2} \psi \overline{\mathcal{Q}} \dop{1} \psi)_{ijt} \, - \\
	&\qquad \sum_{t = 1}^{T} \sum_{t^{\prime} = t + h + 1}^{T} \EX{(\mathfrak{D}_{\pi}^{2} e_{k})_{ijt^{\prime}} (\dop{1} \psi)_{ijt}} \, - \\
	&\qquad \sum_{t = 1}^{T} \sum_{t^{\prime} = t}^{(t + h) \wedge T} \Big( (\mathfrak{D}_{\pi}^{2} e_{k})_{ijt^{\prime}} (\dop{1} \psi)_{ijt} - \EX{(\mathfrak{D}_{\pi}^{2} e_{k})_{ijt^{\prime}} (\dop{1} \psi)_{ijt}} \Big) \, - \\
	&\qquad \sum_{t = 1}^{T} \sum_{t^{\prime} = t}^{(t + h) \wedge T} (\nabla^{2} \psi \overline{\mathcal{Q}} \widetilde{\mathfrak{D}_{\pi}^{2}} e_{k})_{ijt^{\prime}} (\dop{1} \psi)_{ijt} \, - \\
	&\qquad \sum_{t = 1}^{T} \sum_{t^{\prime} = t}^{(t + h) \wedge T} (\diag(\mathfrak{D}_{\pi}^{3} e_{k}) \overline{\mathcal{Q}} \dop{1} \psi)_{ijt^{\prime}} (\dop{1} \psi)_{ijt} \, + \\
	&\qquad \sum_{t = 1}^{T} \sum_{t^{\prime} = t}^{(t + h) \wedge T} (\nabla^{2} \psi \overline{\mathcal{Q}} \diag(\mathfrak{D}_{\pi}^{3} e_{k}) \overline{\mathcal{Q}} \dop{1} \psi)_{ijt^{\prime}} (\dop{1} \psi)_{ijt} \, - \\
	&\qquad \sum_{t = 1}^{T} \sum_{t^{\prime} = t}^{(t + h) \wedge T} (\mathfrak{D}_{\pi}^{2} e_{k})_{ijt^{\prime}} (\nabla^{2} \psi \overline{\mathcal{Q}} \dop{1} \psi)_{ijt}  \\
	&\qquad \eqqcolon (\mathfrak{g}_{3, 1, k, 1})_{ij} + \ldots + (\mathfrak{g}_{3, 1, k, 9})_{ij} \, .
\end{align*}
By H\"older's inequality, Jensen's inequality, and arguments used before,
\begin{align*}
	\max_{k} \bignorm{\mathfrak{g}_{3, 1, k, 1}}_{1} \leq& \, (1 + h) \max_{k} \bignorm{\widehat{\mathfrak{D}_{\pi}^{2}} e_{k} - \mathfrak{D}_{\pi}^{2} e_{k}}_{2} \, \bignorm{\widehat{\dop{1} \psi} - \dop{1} \psi}_{2} = \mathcal{O}_{P}\big(T^{2} h\big) \, , \\
	\max_{k} \bignorm{\mathfrak{g}_{3, 1, k, 2}}_{1} \leq& \, (1 + h) \max_{k} \bignorm{\widehat{\mathfrak{D}_{\pi}^{2}} e_{k} - \mathfrak{D}_{\pi}^{2} e_{k} + \nabla^{2} \psi \overline{\mathcal{Q}} \widetilde{\mathfrak{D}_{\pi}^{2}} e_{k} + \ldots}_{2} \, \bignorm{\dop{1} \psi}_{2} = o_{P}\big(T^{\frac{24}{10}} h\big) \, , \\
	\max_{k} \bignorm{\mathfrak{g}_{3, 1, k, 3}}_{1} \leq& \, (1 + h) \max_{k} \bignorm{\mathfrak{D}_{\pi}^{2} e_{k}}_{2} \, \bignorm{\widehat{\dop{1} \psi} - \dop{1} \psi + \nabla^{2} \psi \overline{\mathcal{Q}} \dop{1} \psi}_{2} = o_{P}\big(T^{\frac{24}{10}} h\big) \, .
\end{align*}
By Lemma \ref{lemma:covariance_inequality_mixing} and the integral test for convergence,
\begin{align*}
	&\max_{k} \sup_{ijt} \sum_{t^{\prime} = t + h + 1}^{T} \Bigabs{\EX{(\mathfrak{D}_{\pi}^{2} e_{k})_{ijt^{\prime}} (\dop{1} \psi)_{ijt}}} \leq C \, \sum_{m = h + 1}^{\infty} m^{- 10} \leq C \, \int_{h}^{\infty} z^{- 10} dz = \mathcal{O}_{P}\big(h^{- 9}\big) \, ,
\end{align*}
so $\max_{k} \norm{\mathfrak{g}_{3, 1, k, 4}}_{1} = \mathcal{O}_{P}(T^{3} h^{- 9})$. Let $(\zeta_{1, k})_{ijtt^{\prime}} \coloneqq (\mathfrak{D}_{\pi}^{2} e_{k})_{ijt} (\dop{1} \psi)_{ijt^{\prime}}$. Using $\EX{(\widetilde{\zeta}_{1, k})_{ijtt^{\prime}}} = 0$ for all $i, j, t, t^{\prime}, N_{1}, N_{2}, T$,
\begin{align*}
	&\EX{\max_{k} \bignorm{\mathfrak{g}_{3, 1, k, 5}}_{10}^{10}} \leq \\
	&\quad (1 + h)^{10} \, T^{7} \, \bigg(\frac{N_{1}}{T}\bigg) \bigg(\frac{N_{2}}{T}\bigg) \sup_{ijm} \EX{ \bigg(\frac{1}{\sqrt{T - m}} \sum_{t = m + 1}^{T} \max_{k} (\widetilde{\zeta}_{1, k})_{ijt(t - m)} \bigg)^{10} } = \mathcal{O}_{P}\big(T^{7} h^{10}\big) \, ,
\end{align*}
so $\max_{k} \norm{\mathfrak{g}_{3, 1, k, 5}}_{10} = \mathcal{O}_{P}(T^{7 / 10} h)$ by Markov's inequality and $\max_{k} \norm{\mathfrak{g}_{3, 1, k, 5}}_{p} = \mathcal{O}_{P}(T^{1 / 2 + 2 / p} h)$ for $1 \leq p \leq 10$. By similar arguments, $\max_{k} \norm{\mathfrak{g}_{3, 1, k, r}}_{1} = \mathcal{O}_{P}(T^{5 / 2} h)$ for $r \in \{6, \ldots, 9\}$. Using $h \asymp T^{1 / 10}$,
\begin{align*}
	&\bignorm{\hat{g}_{3, 1, k} - \bar{g}_{3, 1, k}}_{1} = \mathcal{O}_{P}\big(T^{\frac{5}{2}} h\big) + \mathcal{O}_{P}\big(T^{3} h^{- 9}\big) = \mathcal{O}_{P}\big(T^{\frac{13}{5}}\big) \, , \\
	&\bignorm{\hat{g}_{3, 1, k} - \bar{g}_{3, 1, k} - \mathfrak{g}_{3, 1, k, 5} - \ldots - \mathfrak{g}_{3, 1, k, 9}}_{1} = o_{P}\big(T^{\frac{12}{5}} h\big) + \mathcal{O}_{P}\big(T^{3} h^{- 9}\big) = o_{P}\big(T^{\frac{5}{2}}\big) \, .
\end{align*}
Let $f_{3}^{\ast} \coloneqq D_{3}^{\prime} \overline{\nabla^{3} \psi} \, \overline{\mathcal{Q}} \dop{1} \psi$. Decomposing the estimator $(\hat{f}_{3}^{\circ - 1})^{\prime} \hat{g}_{3, 1, k}$ for each $k \in \{1, \ldots, K\}$,
\begin{align*}
	&(\hat{f}_{3}^{\circ - 1})^{\prime} \hat{g}_{3, 1, k} - (\bar{f}_{3}^{\circ - 1})^{\prime} \bar{g}_{3, 1, k} = \\
	&\qquad (\hat{f}_{3}^{\circ - 1} - \bar{f}_{3}^{\circ - 1})^{\prime} (\hat{g}_{3, 1, k} - \bar{g}_{3, 1, k}) + (\hat{f}_{3}^{\circ - 1} - \bar{f}_{3}^{\circ - 1} - \bar{f}_{3}^{\circ - 2} \odot f_{3}^{\ast} + \bar{f}_{3}^{\circ - 2} \odot \tilde{f}_{3})^{\prime} \bar{g}_{3, 1, k} \, + \\
	&\qquad (\bar{f}_{1}^{\circ - 1})^{\prime} \big(\hat{g}_{3, 1, k} - \bar{g}_{3, 1, k} - \mathfrak{g}_{3, 1, k, 5} - \ldots - \mathfrak{g}_{3, 1, k, 9}\big) + (\bar{f}_{3}^{\circ - 2} \odot f_{3}^{\ast})^{\prime} \bar{g}_{3, 1, k} \, - \\
	&\qquad (\bar{f}_{3}^{\circ - 2} \odot \tilde{f}_{3})^{\prime} \bar{g}_{3, 1, k} + (\bar{f}_{3}^{\circ - 1})^{\prime} \mathfrak{g}_{3, 1, k, 5} + \ldots + (\bar{f}_{3}^{\circ - 1})^{\prime} \mathfrak{g}_{3, 1, k, 9}  \\
	& \qquad \eqqcolon \mathfrak{E}_{3, 1, k, 1} + \ldots + \mathfrak{E}_{3, 1, k, 10} \, .
\end{align*}
By H\"older's inequality and Lemma \ref{lemma:covariance_inequality_mixing},
\begin{align*}
	\max_{k} \bigabs{\mathfrak{E}_{3, 1, k, 1}} \leq& \, \bignorm{\hat{f}_{3}^{\circ - 1} - \bar{f}_{3}^{\circ - 1}}_{10} \max_{k} \bignorm{\hat{g}_{3, 1, k} - \bar{g}_{3, 1, k}}_{1} = o_{P}\big(T^{\frac{3}{2}}\big) \, , \\
	\max_{k} \bigabs{\mathfrak{E}_{3, 1, k, 2}} \leq& \, \bignorm{\hat{f}_{3}^{\circ - 1} - \bar{f}_{3}^{\circ - 1} - \bar{f}_{3}^{\circ - 2} \odot f_{3}^{\ast} + \bar{f}_{3}^{\circ - 2} \odot \tilde{f}_{3}}_{1} \max_{k} \bignorm{\bar{g}_{3, 1, k}}_{\infty} = o_{P}\big(T^{\frac{3}{2}}\big) \, , \\
	\max_{k} \bigabs{\mathfrak{E}_{3, 1, k, 3}} \leq& \, \bignorm{\bar{f}_{3}^{\circ - 1}}_{\infty} \max_{k} \bignorm{\hat{g}_{3, 1, k} - \bar{g}_{3, 1, k} - \mathfrak{g}_{3, 1, k, 5} - \ldots - \mathfrak{g}_{3, 1, k, 9}}_{1} = o_{P}\big(T^{\frac{3}{2}}\big) \, .
\end{align*}
Decomposing $\mathfrak{E}_{3, 1, k, 4}$,
\begin{align*}
	\mathfrak{E}_{3, 1, k, 4} =& \, \big(\bar{f}_{3}^{\circ - 2} \odot D_{3}^{\prime} \overline{\nabla^{3} \psi} \, \overline{\mathcal{Q}}_{1} \dop{1} \psi\big)^{\prime} \bar{g}_{3, 1, k} + \ldots + \big(\bar{f}_{3}^{\circ - 2} \odot D_{3}^{\prime} \overline{\nabla^{3} \psi} \, \overline{\mathcal{Q}}_{3} \dop{1} \psi\big)^{\prime} \bar{g}_{3, 1, k} \\
	\eqqcolon& \, \mathfrak{E}_{3, 1, k, 4, 1} + \mathfrak{E}_{3, 1, k, 4, 2} + \mathfrak{E}_{3, 1, k, 4, 3} \, .
\end{align*}
By similar arguments as for $\mathfrak{E}_{1, 1, k, 4, 2}$ and $\mathfrak{E}_{1, 1, k, 4, 3}$, using $\norm{D_{3}^{\prime} \overline{\nabla^{3} \psi} \, \overline{\mathcal{Q}}_{1} \dop{1} \psi}_{p} = \mathcal{O}_{P}(T^{2 / p})$ and $\norm{D_{3}^{\prime} \overline{\nabla^{3} \psi} \, \overline{\mathcal{Q}}_{2} \dop{1} \psi}_{p} = \mathcal{O}_{P}(T^{2 / p})$ for $1 \leq p \leq 20$, we get $\max_{k} \abs{\mathfrak{E}_{3, 1, k, 4, 1}} = o_{P}(T^{3 / 2})$ and $\max_{k} \abs{\mathfrak{E}_{3, 1, k, 4, 2}} = o_{P}(T^{3 / 2})$. Let $\overline{\vartheta}_{4} \coloneqq T \, (\bar{g}_{3, 1, k} \odot D_{3}^{\prime} \overline{\dop{3} \psi}) \odot \overline{f}_{3}^{\circ - 3}$. By similar arguments as for $\mathfrak{E}_{1, 1, k, 4, 1}$,
\begin{equation*}
	\EX{\max_{k} \bigabs{\mathfrak{E}_{3, 1, k, 4, 3}}^{2}} \leq C \, T \, \bigg(\frac{N_{1}}{T}\bigg)  \bigg(\frac{N_{2}}{T}\bigg) \sup_{ij} \EX{\bigg(\frac{1}{\sqrt{T}} \sum_{t = 1}^{T} (\overline{\vartheta}_{4})_{ij} (\dop{1} \psi)_{ijt} \bigg)^{2}} = \mathcal{O}_{P}\big(T\big) \, ,
\end{equation*}
so $\max_{k} \abs{\mathfrak{E}_{3, 1, k, 4, 3}} = o_{P}(T^{3 / 2})$ by Markov's inequality. Hence, $\max_{k} \abs{\mathfrak{E}_{3, 1, k, 4}} = o_{P}(T^{3 / 2})$. By similar arguments, $\max_{k} \abs{\mathfrak{E}_{3, 1, k, 5}} = o_{P}(T^{3 / 2})$. Recall $(\zeta_{1, k})_{ijtt^{\prime}} = (\mathfrak{D}_{\pi}^{2} e_{k})_{ijt^{\prime}} (\dop{1} \psi)_{ijt}$ and let $\overline{\vartheta}_{5} \coloneqq T \, \overline{f}_{3}^{\circ - 1}$. Then
\begin{align*}
	\EX{\max_{k} \bigabs{\mathfrak{E}_{3, 1, k, 6}}^{2}} \leq& \, (1 + h)^{2} \, T \, \bigg(\frac{N_{1}}{T}\bigg) \bigg(\frac{N_{2}}{T}\bigg) \sup_{ijg} \EX{ \bigg(\frac{1}{\sqrt{T - g}} \sum_{t = g + 1}^{T} (\overline{\vartheta}_{5})_{ij} \max_{k} (\widetilde{\zeta}_{1, k})_{ijt(t - g)} \bigg)^{2} } \\
	=& \, \mathcal{O}_{P}\big(T h^{2}\big) \, ,
\end{align*}
so $\max_{k} \abs{\mathfrak{E}_{3, 1, k, 6}} = o_{P}(T^{3 / 2})$ by Markov's inequality. Analogously, $\max_{k} \abs{\mathfrak{E}_{3, 1, k, r}} = o_{P}(T^{3 / 2})$ for $r \in \{7, 8, 9, 10\}$. Hence, combining all intermediate results,
\begin{equation*}
	\bignorm{\widehat{B}_{\rho, 1} - \overline{B}_{\rho, 1}}_{2} \leq C \, \bigg(\frac{T^{2}}{N_{1} N_{2}}\bigg) \, T^{- 2} \max_{k} \bigabs{(\hat{f}_{3}^{\circ - 1})^{\prime} \hat{g}_{3, 1, k} - (\bar{f}_{3}^{\circ - 1})^{\prime} \bar{g}_{3, 1, k}} = o_{P}\big(T^{- \frac{1}{2}}\big) \, .
\end{equation*}

\vspace{0.5em}
\noindent\# \underline{Part 2.} The element at position $(k, k^{\prime})$ of $\overline{W}$ is
\begin{equation*}
	e_{k}^{\prime} \overline{W} e_{k^{\prime}} = \frac{\EX{(\mathfrak{D}_{\pi}^{2} e_{k})^{\prime} \ddot{X} e_{k^{\prime}}}}{N_{1} N_{2} T} = \frac{1}{N_{1} N_{2} T} \sum_{i = 1}^{N_{1}} \sum_{j = 1}^{N_{2}} \sum_{t = 1}^{T} \EX{(\mathfrak{D}_{\pi}^{2} e_{k})_{ijt} (\ddot{X} e_{k^{\prime}})_{ijt}}
\end{equation*}
with estimator $e_{k}^{\prime} \widehat{W} e_{k^{\prime}} = (N_{1} N_{2} T)^{- 1} (\widehat{\mathfrak{D}_{\pi}^{2}} e_{k})^{\prime} X e_{k^{\prime}}$. Since $\norm{W - \overline{W}}_{2} = o_{P}(T^{- 1 / 2})$, it suffices to show $\norm{\widehat{W} - W}_{2} = o_{P}(T^{- 1 / 2})$. Decomposing $e_{k}^{\prime}(\widehat{W} - W)e_{k^{\prime}}$,
\begin{align*}
	&e_{k}^{\prime} \big(\widehat{W} - W\big) e_{k^{\prime}} = \frac{\big(\widehat{\mathfrak{D}_{\pi}^{2}} e_{k} - \mathfrak{D}_{\pi}^{2} e_{k} + \nabla^{2} \psi \overline{\mathcal{Q}} \widetilde{\mathfrak{D}_{\pi}^{2}} e_{k} + \ldots\big)^{\prime} X e_{k^{\prime}}}{N_{1} N_{2} T} \, - \\
	&\qquad \frac{\big(\widetilde{\mathfrak{D}_{\pi}^{2}} e_{k}\big)^{\prime} \overline{\mathcal{Q}} \nabla^{2} \psi X e_{k^{\prime}}}{N_{1} N_{2} T} - \frac{(\dop{1} \psi)^{\prime} \overline{\mathcal{Q}} \diag(\mathfrak{D}_{\pi}^{3} e_{k}) X e_{k^{\prime}}}{N_{1} N_{2} T} + \frac{(\dop{1} \psi)^{\prime} \overline{\mathcal{Q}} \diag(\overline{\mathfrak{D}_{\pi}^{3}} e_{k}) \overline{\mathcal{Q}} \nabla^{2} \psi X e_{k^{\prime}}}{N_{1} N_{2} T} \\
	&\qquad \eqqcolon \mathfrak{E}_{4, k, k^{\prime}, 1} + \ldots + \mathfrak{E}_{4, k, k^{\prime}, 4} \, .
\end{align*}
By H\"older's inequality,
\begin{equation*}
	\max_{k, k^{\prime}} \bigabs{\mathfrak{E}_{4, k, k^{\prime}, 1}} \leq \frac{\max_{k} \norm{\widehat{\mathfrak{D}_{\pi}^{2}} e_{k} - \mathfrak{D}_{\pi}^{2} e_{k} + \ldots}_{2} \, \max_{k} \norm{X e_{k}}_{2}}{N_{1} N_{2} T} = o_{P}\big(T^{- \frac{1}{2}}\big) \, .
\end{equation*}
Decomposing $\mathfrak{E}_{4, k, k^{\prime}, 2}$,
\begin{equation*}
	\mathfrak{E}_{4, k, k^{\prime}, 2} = - \frac{\big(\widetilde{\mathfrak{D}_{\pi}^{2}} e_{k}\big)^{\prime} \mathfrak{X} e_{k^{\prime}}}{N_{1} N_{2} T} - \frac{\big(\widetilde{\mathfrak{D}_{\pi}^{2}} e_{k}\big)^{\prime} \overline{\mathcal{Q}} \widetilde{\nabla^{2} \psi X} e_{k^{\prime}}}{N_{1} N_{2} T} \eqqcolon \mathfrak{E}_{4, k, k^{\prime}, 2, 1} + \mathfrak{E}_{4, k, k^{\prime}, 2, 2} \, .
\end{equation*}
By the Cauchy-Schwarz inequality and Lemma \ref{lemma:matrix_norm_inequalties},
\begin{equation*}
	\max_{k, k^{\prime}} \bigabs{\mathfrak{E}_{4, k, k^{\prime}, 2, 2}} \leq (N_{1} N_{2} T^{2})^{- 1} \bignorm{\overline{F}^{- 1}}_{\infty} \max_{k} \bignorm{D^{\prime} \widetilde{\nabla^{2} \psi X} e_{k}}_{2} \max_{k} \bignorm{D^{\prime} \widetilde{\mathfrak{D}_{\pi}^{2}} e_{k}}_{2} = o_{P}\big(T^{- \frac{1}{2}}\big) \, .
\end{equation*}
By Lemma \ref{lemma:moment_bounds_mixing},
\begin{align*}
	&\EX{\max_{k, k^{\prime}} \bigabs{\mathfrak{E}_{4, k, k^{\prime}, 2, 1}}^{2}} \leq T^{- 3} \, \bigg(\frac{T}{N_{1}}\bigg) \bigg(\frac{T}{N_{2}}\bigg) \sup_{ij} \EX{\bigg(\frac{1}{\sqrt{T}} \sum_{t = 1}^{T} \max_{k, k^{\prime}} (\widetilde{\mathfrak{D}_{\pi}^{2}} e_{k})_{ijt} (\mathfrak{X} e_{k^{\prime}})_{ijt} \bigg)^{2}} \\
	& \quad = \mathcal{O}\big(T^{- 3}\big) \quad \text{a.\,s.} \, ,
\end{align*}
so $\max_{k, k^{\prime}} \abs{\mathfrak{E}_{4, k, k^{\prime}, 2, 1}} = o_{P}(T^{- 1 / 2})$ by Markov's inequality, and thus $\max_{k, k^{\prime}} \abs{\mathfrak{E}_{4, k, k^{\prime}, 2}} = o_{P}(T^{- 1 / 2})$. Analogously, $\max_{k, k^{\prime}} \abs{\mathfrak{E}_{4, k, k^{\prime}, r}} = o_{P}(T^{- 1 / 2})$ for $r \in \{3, 4\}$. Hence, $\norm{\widehat{W} - \overline{W}}_{2} = o_{P}(T^{- 1 / 2})$.

Decomposing $\widehat{W} = \overline{W} + (\widehat{W} - \overline{W})$ and applying Lemma \ref{lemma:inverse_neumann_series} wpa1, $\widehat{W}^{- 1} = \overline{W}^{- 1} \sum_{r = 0}^{\infty} \{- (\widehat{W} - \overline{W}) \overline{W}^{- 1}\}^{r}$, which implies
\begin{equation*}
	\bignorm{\widehat{W}^{- 1} - \overline{W}^{- 1}}_{2} \leq \frac{\norm{\overline{W}^{- 1}}_{2}^{2} \norm{\widehat{W} - \overline{W}}_{2}}{1 - \norm{\widehat{W} - \overline{W}}_{2} \, \norm{\overline{W}^{- 1}}_{2}} = o_{P}(T^{- 1 / 2}) \big(1 - o_{P}(1)\big)^{- 1} = o_{P}(T^{- 1 / 2}) \, .
\end{equation*}

\vspace{0.5em}
\noindent\# \underline{Part 3.} The element at position $(k, k^{\prime})$ of $\overline{\Sigma}$ is
\begin{equation*}
	e_{k}^{\prime} \overline{\Sigma} e_{k^{\prime}} = \frac{\EX{(\mathfrak{D}_{\pi}^{1} e_{k})^{\prime} \mathfrak{D}_{\pi}^{1} e_{k^{\prime}}}}{N_{1} N_{2} T}  = \frac{1}{N_{1} N_{2} T} \sum_{i = 1}^{N_{1}} \sum_{j = 1}^{N_{2}} \sum_{t = 1}^{T} \EX{(\mathfrak{D}_{\pi}^{1} e_{k})_{ijt} (\mathfrak{D}_{\pi}^{1} e_{k^{\prime}})_{ijt}}
\end{equation*}
with estimator $e_{k}^{\prime} \widehat{\Sigma} e_{k^{\prime}} = (N_{1} N_{2} T)^{- 1} (\widehat{\mathfrak{D}_{\pi}^{1}} e_{k})^{\prime} \widehat{\mathfrak{D}_{\pi}^{1}} e_{k^{\prime}}$. Decomposing $e_{k}^{\prime}(\widehat{\Sigma} - \overline{\Sigma})e_{k^{\prime}}$,
\begin{align*}
	e_{k}^{\prime} \big(\widehat{\Sigma} - \overline{\Sigma}\big) e_{k^{\prime}} =& \, \frac{\big(\widehat{\mathfrak{D}_{\pi}^{1}} e_{k} - \mathfrak{D}_{\pi}^{1} e_{k}\big)^{\prime} \big(\widehat{\mathfrak{D}_{\pi}^{1}} e_{k^{\prime}} - \mathfrak{D}_{\pi}^{1} e_{k^{\prime}}\big)}{N_{1} N_{2} T} \, + \frac{\big(\widehat{\mathfrak{D}_{\pi}^{1}} e_{k} - \mathfrak{D}_{\pi}^{1} e_{k}\big)^{\prime} \mathfrak{D}_{\pi}^{1} e_{k^{\prime}}}{N_{1} N_{2} T} \, + \\
	& \, \frac{\big(\mathfrak{D}_{\pi}^{1} e_{k}\big)^{\prime} \big(\widehat{\mathfrak{D}_{\pi}^{1}} e_{k^{\prime}} - \mathfrak{D}_{\pi}^{1} e_{k^{\prime}}\big)}{N_{1} N_{2} T} + \frac{\big(\mathfrak{D}_{\pi}^{1} e_{k}\big)^{\prime} \mathfrak{D}_{\pi}^{1} e_{k^{\prime}} - \EX{\big(\mathfrak{D}_{\pi}^{1} e_{k}\big)^{\prime} \mathfrak{D}_{\pi}^{1} e_{k^{\prime}}}}{N_{1} N_{2} T} \\
	&\eqqcolon \mathfrak{E}_{5, k, k^{\prime}, 1} + \ldots + \mathfrak{E}_{5, k, k^{\prime}, 4} \, .
\end{align*}
By the Cauchy-Schwarz inequality,
\begin{align*}
	\max_{k, k^{\prime}} \bigabs{\mathfrak{E}_{5, k, k^{\prime}, 1}} \leq& \, \frac{\max_{k} \norm{\widehat{\mathfrak{D}_{\pi}^{1}} e_{k} - \mathfrak{D}_{\pi}^{1} e_{k}}_{2}^{2}}{N_{1} N_{2} T} = o_{P}(1) \, , \\
	\max_{k, k^{\prime}} \bigabs{\mathfrak{E}_{5, k, k^{\prime}, 2}} \leq& \, \frac{\max_{k} \norm{\widehat{\mathfrak{D}_{\pi}^{1}} e_{k} - \mathfrak{D}_{\pi}^{1} e_{k}}_{2} \, \max_{k} \norm{\mathfrak{D}_{\pi}^{1} e_{k}}_{2}}{N_{1} N_{2} T} = o_{P}(1) \, , \\
	\max_{k, k^{\prime}} \bigabs{\mathfrak{E}_{5, k, k^{\prime}, 3}} \leq& \, \frac{\max_{k} \norm{\widehat{\mathfrak{D}_{\pi}^{1}} e_{k} - \mathfrak{D}_{\pi}^{1} e_{k}}_{2} \, \max_{k} \norm{\mathfrak{D}_{\pi}^{1} e_{k}}_{2}}{N_{1} N_{2} T} = o_{P}(1) \, .
\end{align*}
Let $\vartheta_{5, k, k^{\prime}} \coloneqq \mathfrak{D}_{\pi}^{1} e_{k} \odot \mathfrak{D}_{\pi}^{1} e_{k^{\prime}}$. By similar arguments,
\begin{equation*}
	\EX{\max_{k, k^{\prime}} \bigabs{\mathfrak{E}_{5, k, k^{\prime}, 4}}^{2}} \leq T^{- 3} \bigg(\frac{T^{2}}{N_{1} N_{2}}\bigg) \sup_{ij} \EX{\bigg(\frac{1}{\sqrt{T}} \sum_{t = 1}^{T} \max_{k, k^{\prime}} (\widetilde{\vartheta}_{5, k, k^{\prime}})_{ijt} \bigg)^{2}} = \mathcal{O}_{P}\big(T^{- 3}\big) \, ,
\end{equation*}
so $\max_{k, k^{\prime}} \abs{\mathfrak{E}_{5, k, k^{\prime}, 4}} = o_{P}(1)$ by Markov's inequality. Hence, $\norm{\widehat{\Sigma} - \overline{\Sigma}}_{2} = o_{P}(1)$.\hfill\qedsymbol

\subsection{Proofs for Non-interacted Specification}
\label{supplement:proof_of_asymptotic_expansions_noninteracted}

\paragraph{Proof of Theorem \ref{theorem:asymptotic_expansion_profile_score_noninteracted}.} The conditions of Theorem \ref{theorem:taylor_expansions} hold wpa1. Thus, $\partial \mathcal{L}_{n}(\beta, \hat{\phi}(\beta)) / \partial \beta = \mathcal{T}^{(1)} + \mathcal{T}^{(2)}(\beta) + \mathcal{T}^{(3)} + \mathcal{R}^{(1)}(\beta)$, where
\begin{align*}
	\mathcal{R}^{(1)}(\beta) \coloneqq& \, \mathcal{T}^{(4)}(\beta) + \mathcal{T}^{(5)}(\beta) + \mathcal{T}^{(6)} + \sum_{r = 1}^{2} \mathcal{T}_{r}^{(7)}(\beta) + \sum_{r = 1}^{3} \mathcal{T}_{r}^{(8)}(\beta) + \sum_{r = 1}^{3} \mathcal{T}_{r}^{(9)}(\beta) \, + \\
	&\quad  \sum_{r = 1}^{2} \mathcal{T}_{r}^{(10)} + \sum_{r = 1}^{8} \mathcal{T}_{r}^{(11)} \, .
\end{align*}
Throughout, $\beta \in \mathfrak{B}(\varepsilon)$ with $\varepsilon = o(T^{- 1})$.

\vspace{0.5em}
\noindent\# \underline{Part 1.} By the triangle inequality,
\begin{align*}
	\bignorm{\mathcal{R}^{(1)}(\beta)}_{2} \leq& \, \bignorm{\mathcal{T}^{(4)}(\beta)}_{2} + \bignorm{\mathcal{T}^{(5)}(\beta)}_{2} + \bignorm{\mathcal{T}^{(6)}}_{2} + \sum_{r = 1}^{2} \bignorm{\mathcal{T}_{r}^{(7)}(\beta)}_{2} + \sum_{r = 1}^{3} \bignorm{\mathcal{T}_{r}^{(8)}(\beta)}_{2} \, + \\
	&\quad \sum_{r = 1}^{3} \bignorm{\mathcal{T}_{r}^{(9)}(\beta)}_{2} + \sum_{r = 1}^{2} \bignorm{\mathcal{T}_{r}^{(10)}}_{2} + \sum_{r = 1}^{8} \bignorm{\mathcal{T}_{r}^{(11)}}_{2} \, .
\end{align*}
By the H\"older's inequality and $\sup_{\beta} \norm{\beta - \beta^{0}}_{2} \leq \varepsilon = o(T^{- 1})$,
\begin{align*}
	&\bignorm{\mathcal{T}^{(4)}(\beta)}_{2} \leq C \, T^{- 2} \max_{k} \bignorm{M X e_{k}}_{4}^{3} \, \bignorm{\dop{3} \psi}_{4} \, \bignorm{\beta - \beta^{0}}_{2}^{2} = o_{P}\big(T \bignorm{\beta - \beta^{0}}_{2}\big) \, , \\
	&\bignorm{\mathcal{T}^{(6)}}_{2} \leq C \, T^{- 2} \max_{k} \bignorm{M X e_{k}}_{4} \, \bignorm{\dop{3} \psi}_{4} \, \bignorm{Q \dop{1} \psi}_{4}^{2} = o_{P}\big(T^{- \frac{1}{2}}\big) \, .
\end{align*}
The remaining terms satisfy analogous bounds: $\mathcal{T}^{(5)}$, $\mathcal{T}_r^{(7)}$, $\mathcal{T}_r^{(8)}$, $\mathcal{T}_r^{(9)}$ are $o_P(T\|\beta - \beta^0\|_2)$, and $\mathcal{T}_r^{(10)}$, $\mathcal{T}_r^{(11)}$ are $o_P(T^{-1/2})$. Hence, $\norm{\mathcal{R}^{(1)}(\beta)}_{2} = o_{P}(T^{- 1 / 2}) + o_{P}(T \norm{\beta - \beta^{0}}_{2})$ for all $\beta \in \mathfrak{B}(\varepsilon)$.

\vspace{0.5em}
\noindent\# \underline{Part 2.} Recall $\mathcal{T}^{(2)}(\beta) = N_{1} N_{2} W (\beta - \beta^{0})$. Then,
\begin{align*}
	\mathcal{T}^{(2)}(\beta) =& \, \bigg(\frac{N_{1} N_{2}}{T}\bigg) \, \overline{W} (\beta - \beta^{0}) + T \, \bigg(\frac{N_{1} N_{2}}{T^{2}}\bigg) \, \big(W - \overline{W}\big) (\beta - \beta^{0}) \\
	=& \, \bigg(\frac{N_{1} N_{2}}{T}\bigg) \, \overline{W} (\beta - \beta^{0}) + o_{P}\big(T \bignorm{\beta - \beta^{0}}_{2}\big) \, .
\end{align*}
Decomposing $\mathcal{T}^{(3)}$,
\begin{align*}
	\mathcal{T}^{(3)} =& \, - \frac{\mathfrak{X}^{\prime} \dop{1} \psi}{T^{2}} - \frac{(\widetilde{\nabla^{2} \psi X})^{\prime} D \overline{H}^{- 1} D^{\prime} \dop{1} \psi}{T^{4}} - \frac{X^{\prime} \nabla^{2} \psi D (H^{- 1} - \overline{H}^{- 1}) D^{\prime} \dop{1} \psi}{T^{4}} \\
	\eqqcolon& \, \mathfrak{T}_{1}^{(3)} + \mathfrak{T}_{2}^{(3)} + \mathfrak{T}_{3}^{(3)} \, .
\end{align*}
By H\"older's inequality,
\begin{align*}
	\bignorm{\mathfrak{T}_{2}^{(3)}}_{2} \leq& \, C \, T^{- 4} \, \bignorm{\overline{H}^{- 1}}_{\infty} \, \max_{k} \bignorm{D^{\prime} \widetilde{\nabla^{2} \psi X} e_{k}}_{2} \, \bignorm{D^{\prime} \dop{1} \psi}_{2} = o_{P}\big(T^{- \frac{1}{2}}\big) \, , \\
	\bignorm{\mathfrak{T}_{3}^{(3)}}_{2} \leq& \, C \, T^{- 4} \, \bignorm{H^{- 1} - \overline{H}^{- 1}}_{2}  \max_{k} \bignorm{D^{\prime} \nabla^{2} \psi X e_{k}}_{2} \bignorm{D^{\prime} \dop{1} \psi}_{2} = o_{P}\big(T^{- \frac{1}{2}}\big) \, .
\end{align*}
Combining $\mathcal{T}^{(1)}$ and $\mathcal{T}^{(3)}$,
\begin{equation*}
	\mathcal{T}^{(1)} + \mathcal{T}^{(3)} =  \frac{X^{\prime} \dop{1} \psi}{T^{2}} + \mathfrak{T}_{1}^{(3)} + o_{P}\big(T^{- \frac{1}{2}}\big) = \sqrt{\frac{N_{1} N_{2}}{T^{3}}} \Bigg\{\underbrace{\frac{\ddot{X}^{\prime} \dop{1} \psi}{\sqrt{N_{1} N_{2} T}}}_{= \, U}\Bigg\} + \, o_{P}\big(T^{- \frac{1}{2}}\big) \, .
\end{equation*}
\hfill\qedsymbol

\paragraph{Proof of Theorem \ref{theorem:consistency_noninteracted}.} Let $r_{\beta} = 2 \, (N_{1} N_{2} T)^{- 1 / 2} \, \overline{W}^{- 1} \abs{U} = \mathcal{O}_{P}(T^{- 3 / 2})$ be a $K$-dimensional vector.

\vspace{0.5em}
\noindent\# \underline{Part 1.} Suppose that $K = 1$ and define $\underline{b} \coloneqq \beta^{0} - r_{\beta}$ and $\overline{b} \coloneqq \beta^{0} + r_{\beta}$. For $\beta \in [\underline{b}, \overline{b}]$, we have $\abs{\beta - \beta^{0}} = \abs{r_{\beta}} = \mathcal{O}_{P}(T^{- 3 / 2}) = o_{P}(\varepsilon)$, so $\beta \in \mathfrak{B}(\varepsilon)$ wpa1. By Theorem \ref{theorem:asymptotic_expansion_profile_score_noninteracted} and sufficiently large $n$,
\begin{align*}
	\frac{\partial \mathcal{L}_{n}(\beta, \hat{\phi}(\beta))}{\partial \beta}\biggr\rvert_{\substack{\beta \, = \, \underline{b}}} =& \, \sqrt{\frac{N_{1} N_{2}}{T^{3}}} \underbrace{\big(U - 2 \, \abs{U}\big)}_{< \, 0} + \, o_{P}\big(T^{- \frac{1}{2}}\big) \, , \\
	\frac{\partial \mathcal{L}_{n}(\beta, \hat{\phi}(\beta))}{\partial \beta}\biggr\rvert_{\substack{\beta \, = \, \overline{b}}} =& \, \sqrt{\frac{N_{1} N_{2}}{T^{3}}} \underbrace{\big(U + 2 \, \abs{U}\big)}_{> 0} + \, o_{P}\big(T^{- \frac{1}{2}}\big) \, .
\end{align*}
Since $\partial_{\beta} \mathcal{L}_{n}(\hat{\beta}, \hat{\phi}(\hat{\beta})) = 0$ and $\mathcal{L}_{n}(\beta, \hat{\phi}(\beta))$ is strictly convex on $\mathfrak{B}(\varepsilon)$, these sign conditions imply $\abs{\hat{\beta} - \beta^{0}} \leq r_{\beta}$, and hence $\norm{\hat{\beta} - \beta^{0}} = \mathcal{O}_{P}(T^{- 3 / 2})$ for $K = 1$.

For $K > 1$, define $\underline{\beta} \coloneqq \beta^{0} - r_{\beta} \, (\hat{\beta} - \beta^{0}) / \norm{\hat{\beta} - \beta^{0}}_{2}$, $\overline{\beta} \coloneqq \beta^{0} + r_{\beta} \, (\hat{\beta} - \beta^{0}) / \norm{\hat{\beta} - \beta^{0}}_{2}$, and let $B$ be the line segment between them. Restricting $\beta \in B$ and reapplying the $K = 1$ argument yields $\norm{\hat{\beta} - \beta^{0}}_{2} \leq r_{\beta} = \mathcal{O}_{P}(T^{- 3 / 2})$.

\vspace{0.5em}
\noindent\# \underline{Part 2.} Define $p^{\prime} \coloneqq p / (p - 1)$. By the definition of the $p$-norm, $\norm{\hat{\phi} - \phi^{0}}_{p} = \sup_{\norm{a}_{p^{\prime}} = 1} \abs{a^{\prime} (\hat{\phi} - \phi^{0})}$. By Lemma \ref{lemma:matrix_norm_inequalties}, $\norm{\mathbb{U}}_{20} \leq T^{- 2} \norm{H^{- 1}}_{20} \, \norm{D^{\prime} \dop{1} \psi}_{20} = \mathcal{O}_{P}(T^{- 19 / 20})$ and $\norm{\overline{\mathbb{W}}}_{20} \leq C \, \max_{k} \norm{\xi_{k}^{0}}_{20} = \mathcal{O}_{P}(T^{1 / 20})$. By Theorem \ref{theorem:asymptotic_expansion_lincom_ip_function_noninteracted} and $\norm{\hat{\beta} - \beta^{0}}_{2} = \mathcal{O}_{P}(T^{- 3 / 2})$ from Part 1,
\begin{equation*}
	\bignorm{\hat{\phi} - \phi^{0}}_{20} \leq \norm{\mathbb{U}}_{20} +  \bignorm{\overline{\mathbb{W}}}_{20} \, \bignorm{\hat{\beta} - \beta^{0}}_{2} + o_{P}\big(T^{- \frac{7}{4}}\big) + o_{P}\big(T^{\frac{1}{20}} \bignorm{\hat{\beta} - \beta^{0}}_{2}\big) = \mathcal{O}_{P}\big(T^{- \frac{19}{20}}\big) \, .
\end{equation*}
For $1 \leq p \leq 20$, the stated rates follow by H\"older's inequality and $L = \mathcal{O}(T^{2})$.\hfill\qedsymbol

\paragraph{Proof of Corollary \ref{corollary:asymptotic_expansion_score_noninteracted}.} By Theorem \ref{theorem:consistency_noninteracted}, $\norm{\hat{\beta} - \beta^{0}}_{2} = \mathcal{O}_{P}(T^{- 3 / 2}) = o_{P}(\varepsilon)$, and $\partial_{\beta} \mathcal{L}_{n}(\hat{\beta}, \hat{\phi}(\hat{\beta})) = \mathbf{0}_{K}$ by the definition of $\hat{\beta}$. Evaluating Theorem \ref{theorem:asymptotic_expansion_profile_score_noninteracted} at $\hat{\beta}$ and rearranging,
\begin{equation*}
	\bigg(\frac{N_{1} N_{2}}{T}\bigg) \, \overline{W} (\hat{\beta} - \beta^{0}) = - \sqrt{\frac{N_{1} N_{2}}{T^{3}}} \, U + o_{P}\big(T^{- \frac{1}{2}}\big) + o_{P}\big(T \norm{\hat{\beta} - \beta^{0}}_{2}\big) \, .
\end{equation*}
The stated result follows upon dividing by $\sqrt{N_{1} N_{2}} / \sqrt{T^{3}}$ and using $T^{3 / 2} \norm{\hat{\beta} - \beta^{0}}_{2} = \mathcal{O}_{P}(1)$.\hfill\qedsymbol

\begin{corollary}[Consistency of $\hat{\pi}$]
	\label{corollary:consistency_noninteracted}
	Let Lemma \ref{lemma:regularity_conditions1_noninteracted} hold. Then, for $1 \leq p \leq 20$, $\norm{\hat{\pi} - \pi^{0}}_{p} = \mathcal{O}_{P}(T^{- 1 + 3 / p})$.
\end{corollary}

\paragraph{Proof of Corollary \ref{corollary:consistency_noninteracted}.} Recall $\pi(\beta, \phi) = X \beta + D \phi$. Decomposing, $\hat{\pi} - \pi^{0} = X (\hat{\beta} - \beta^{0}) + D (\hat{\phi} - \phi^{0})$. By H\"older's inequality, Jensen's inequality, and Theorem \ref{theorem:consistency_noninteracted},
\begin{align*}
	\bignorm{X (\hat{\beta} - \beta^{0})}_{p} \leq& \, C \, \max_{k} \norm{X e_{k}}_{p} \, \bignorm{\hat{\beta} - \beta^{0}}_{2} = \mathcal{O}_{P}\big(T^{- 1 + \frac{3}{p}}\big) \, , \\
	\bignorm{D (\hat{\phi} - \phi^{0})}_{p} \leq& \, \norm{D}_{p} \, \bignorm{\hat{\phi} - \phi^{0}}_{p} = \mathcal{O}_{P}\big(T^{- 1 + \frac{3}{p}}\big) \, .
\end{align*}
The stated results follow by the triangle inequality.\hfill\qedsymbol

\paragraph{Proof of Theorem \ref{theorem:consistency_variance_noninteracted}.} The element at position $(k, k^{\prime})$ of $\overline{W}$ is
\begin{equation*}
	e_{k}^{\prime} \overline{W} e_{k^{\prime}} = \frac{\EX{(\mathfrak{D}_{\pi}^{2} e_{k})^{\prime} \ddot{X} e_{k^{\prime}}}}{N_{1} N_{2} T} = \frac{1}{N_{1} N_{2} T} \sum_{i = 1}^{N_{1}} \sum_{j = 1}^{N_{2}} \sum_{t = 1}^{T} \EX{(\mathfrak{D}_{\pi}^{2} e_{k})_{ijt} (\ddot{X} e_{k^{\prime}})_{ijt}}
\end{equation*}
with estimator $e_{k}^{\prime} \widehat{W} e_{k^{\prime}} = (N_{1} N_{2} T)^{- 1} (\widehat{\mathfrak{D}_{\pi}^{2}} e_{k})^{\prime} X e_{k^{\prime}}$. Since $\norm{W - \overline{W}}_{2} = o_{P}(1)$, it suffices to show $\norm{\widehat{W} - W}_{2} = o_{P}(1)$. Since $e_{k}^{\prime} \big(\widehat{W} - W\big) e_{k^{\prime}} = \big(\widehat{\mathfrak{D}_{\pi}^{2}} e_{k} - \mathfrak{D}_{\pi}^{2} e_{k}\big)^{\prime} X e_{k^{\prime}} / (N_{1} N_{2} T) \eqqcolon \mathfrak{E}_{1, k, k^{\prime}, 1}$, by H\"older's inequality,
\begin{equation*}
	\max_{k, k^{\prime}} \bigabs{\mathfrak{E}_{1, k, k^{\prime}, 1}} \leq \frac{\max_{k} \norm{\widehat{\mathfrak{D}_{\pi}^{2}} e_{k} - \mathfrak{D}_{\pi}^{2} e_{k}}_{2} \, \max_{k} \norm{X e_{k}}_{2}}{N_{1} N_{2} T} = o_{P}(1) \, ,
\end{equation*}
so $\norm{\widehat{W} - \overline{W}}_{2} = o_{P}(1)$. Applying Lemma  \ref{lemma:inverse_neumann_series} wpa1, $\widehat{W}^{- 1} = \overline{W}^{- 1} \sum_{r = 0}^{\infty} \{- (\widehat{W} - \overline{W}) \overline{W}^{- 1}\}^{r}$, which implies
\begin{equation*}
	\bignorm{\widehat{W}^{- 1} - \overline{W}^{- 1}}_{2} \leq \frac{\norm{\overline{W}^{- 1}}_{2}^{2} \norm{\widehat{W} - \overline{W}}_{2}}{1 - \norm{\widehat{W} - \overline{W}}_{2} \, \norm{\overline{W}^{- 1}}_{2}} = o_{P}(1) \big(1 - o_{P}(1)\big)^{- 1} = o_{P}(1) \, .
\end{equation*}

The element at position $(k, k^{\prime})$ of $\overline{\Sigma}$ is
\begin{equation*}
	e_{k}^{\prime} \overline{\Sigma} e_{k^{\prime}} = \frac{\EX{(\mathfrak{D}_{\pi}^{1} e_{k})^{\prime} \mathfrak{D}_{\pi}^{1} e_{k^{\prime}}}}{N_{1} N_{2} T}  = \frac{1}{N_{1} N_{2} T} \sum_{i = 1}^{N_{1}} \sum_{j = 1}^{N_{2}} \sum_{t = 1}^{T} \EX{(\mathfrak{D}_{\pi}^{1} e_{k})_{ijt} (\mathfrak{D}_{\pi}^{1} e_{k^{\prime}})_{ijt}}
\end{equation*}
with estimator $e_{k}^{\prime} \widehat{\Sigma} e_{k^{\prime}} = (N_{1} N_{2} T)^{- 1} (\widehat{\mathfrak{D}_{\pi}^{1}} e_{k})^{\prime} \widehat{\mathfrak{D}_{\pi}^{1}} e_{k^{\prime}}$. Decomposing $e_{k}^{\prime}(\widehat{\Sigma} - \overline{\Sigma})e_{k^{\prime}}$,
\begin{align*}
	e_{k}^{\prime} \big(\widehat{\Sigma} - \overline{\Sigma}\big) e_{k^{\prime}} =& \, \frac{\big(\widehat{\mathfrak{D}_{\pi}^{1}} e_{k} - \mathfrak{D}_{\pi}^{1} e_{k}\big)^{\prime} \big(\widehat{\mathfrak{D}_{\pi}^{1}} e_{k^{\prime}} - \mathfrak{D}_{\pi}^{1} e_{k^{\prime}}\big)}{N_{1} N_{2} T} \, + \frac{\big(\widehat{\mathfrak{D}_{\pi}^{1}} e_{k} - \mathfrak{D}_{\pi}^{1} e_{k}\big)^{\prime} \mathfrak{D}_{\pi}^{1} e_{k^{\prime}}}{N_{1} N_{2} T} \, + \\
	& \, \frac{\big(\mathfrak{D}_{\pi}^{1} e_{k}\big)^{\prime} \big(\widehat{\mathfrak{D}_{\pi}^{1}} e_{k^{\prime}} - \mathfrak{D}_{\pi}^{1} e_{k^{\prime}}\big)}{N_{1} N_{2} T} + \frac{\big(\mathfrak{D}_{\pi}^{1} e_{k}\big)^{\prime} \mathfrak{D}_{\pi}^{1} e_{k^{\prime}} - \EX{\big(\mathfrak{D}_{\pi}^{1} e_{k}\big)^{\prime} \mathfrak{D}_{\pi}^{1} e_{k^{\prime}}}}{N_{1} N_{2} T} \\
	&\eqqcolon \mathfrak{E}_{2, k, k^{\prime}, 1} + \ldots + \mathfrak{E}_{2, k, k^{\prime}, 4} \, .
\end{align*}
By the Cauchy-Schwarz inequality,
\begin{align*}
	\max_{k, k^{\prime}} \bigabs{\mathfrak{E}_{2, k, k^{\prime}, 1}} \leq& \, \frac{\max_{k} \norm{\widehat{\mathfrak{D}_{\pi}^{1}} e_{k} - \mathfrak{D}_{\pi}^{1} e_{k}}_{2}^{2}}{N_{1} N_{2} T} = o_{P}(1) \, , \\
	\max_{k, k^{\prime}} \bigabs{\mathfrak{E}_{2, k, k^{\prime}, 2}} \leq& \, \frac{\max_{k} \norm{\widehat{\mathfrak{D}_{\pi}^{1}} e_{k} - \mathfrak{D}_{\pi}^{1} e_{k}}_{2} \, \max_{k} \norm{\mathfrak{D}_{\pi}^{1} e_{k}}_{2}}{N_{1} N_{2} T} = o_{P}(1) \, , \\
	\max_{k, k^{\prime}} \bigabs{\mathfrak{E}_{2, k, k^{\prime}, 3}} \leq& \, \frac{\max_{k} \norm{\widehat{\mathfrak{D}_{\pi}^{1}} e_{k} - \mathfrak{D}_{\pi}^{1} e_{k}}_{2} \, \max_{k} \norm{\mathfrak{D}_{\pi}^{1} e_{k}}_{2}}{N_{1} N_{2} T} = o_{P}(1) \, .
\end{align*}
Let $\vartheta_{k, k^{\prime}} \coloneqq \mathfrak{D}_{\pi}^{1} e_{k} \odot \mathfrak{D}_{\pi}^{1} e_{k^{\prime}}$. By Lemma \ref{lemma:moment_bounds_mixing},
\begin{equation*}
	\EX{\max_{k, k^{\prime}} \bigabs{\mathfrak{E}_{2, k, k^{\prime}, 4}}^{2}} \leq T^{- 3} \bigg(\frac{T^{2}}{N_{1} N_{2}}\bigg) \sup_{ij} \EX{\bigg(\frac{1}{\sqrt{T}} \sum_{t = 1}^{T} \max_{k, k^{\prime}} (\widetilde{\vartheta}_{k, k^{\prime}})_{ijt} \bigg)^{2}} = \mathcal{O}_{P}\big(T^{- 3}\big) \, ,
\end{equation*}
so $\max_{k, k^{\prime}} \abs{\mathfrak{E}_{2, k, k^{\prime}, 4}} = o_{P}(1)$ by Markov's inequality. Hence $\norm{\widehat{\Sigma} - \overline{\Sigma}}_{2} = o_{P}(1)$.\hfill\qedsymbol

\section{Verifying the Implied Regularity Conditions}
\label{supplement:verifying_regularity_conditions}

\subsection{Interacted Specification}

Before verifying the implied regularity conditions for the interacted linear index specification with $\mathcal{M} = 3$, $\pi_{ijt}(\beta, \mu_{ijt}(\phi)) = (X \beta)_{ijt} + (D \phi)_{ijt} = x_{ijt}^{\prime} \beta + \alpha_{it} + \gamma_{jt} + \rho_{ij}$, we become more specific about the matrices $D$ and $V$. The sparse regressor matrix $D$ is
\begin{equation*}
	D = (D_{1}, D_{2}, D_{3}) = (I_{N_{1}} \otimes \iota_{N_{2}} \otimes I_{T}, \, \iota_{N_{1}} \otimes I_{N_{2}} \otimes I_{T}, \, I_{N_{1}} \otimes I_{N_{2}} \otimes \iota_{T}) \, .
\end{equation*}
We impose the following constraints on $\phi$,
\begin{equation*}
	V = 	
	\begin{pmatrix}
		\iota_{N_{1}} \otimes I_{T}   & I_{N_{1}} \otimes \iota_{T}     & 0	\\
		- \iota_{N_{2}} \otimes I_{T} & 0         & I_{N_{2}} \otimes \iota_{T} \\
		0      & - I_{N_{1}} \otimes \iota_{N_{2}}   & - \iota_{N_{1}} \otimes I_{N_{2}}
	\end{pmatrix} \eqqcolon 
	\begin{pmatrix}
		V_{1} & 	V_{2} & 	V_{3}\\
		V_{4} & 	V_{5} & 	V_{6}\\
		V_{7} & 	V_{8} & 	V_{9}
	\end{pmatrix} \, .
\end{equation*}

\paragraph{Proof of Lemma \ref{lemma:regularity_conditions1_interacted}.} \# (i) By the triangle inequality and $\norm{A \otimes B} = \norm{A} \norm{B}$, $\norm{D}_{1} = \mathcal{O}(T)$. Analogously, $\norm{D}_{\infty} = 3$. Hence, by Lemma \ref{lemma:matrix_norm_inequalties}, $\norm{D}_{p} \leq \norm{D}_{1}^{1/p} \norm{D}_{\infty}^{1 - 1/p} = \mathcal{O}(T^{1/p})$. By similar arguments, $\norm{V}_{1} = \mathcal{O}(T)$, $\norm{V}_{\infty} = 2$, $\norm{V}_{p} = \mathcal{O}(T^{1 / p})$, and $\norm{V^{\prime}}_{p} = \mathcal{O}(T^{1 - 1 / p})$.
\vspace{0.5em}

\noindent\# (ii) By the Courant–Fischer–Weyl min-max principle, $\lambda_{\min}(\overline{H}) =  \min_{\norm{v}_{2} = 1} v^{\prime} \big\{\frac{D^{\prime} \overline{\nabla^{2} \psi} D + V V^{\prime}}{T}\big\} v$. We distinguish between $c_{H} \geq 1$ and $c_{H} < 1$. For $c_{H} \geq 1$,
\begin{equation*}
	\lambda_{\min}\big(\overline{H}\big) \geq \min_{\norm{v}_{2} = 1} v^{\prime} \bigg\{\underbrace{\frac{D^{\prime} D + V V^{\prime}}{T}}_{\eqqcolon \, \mathbb{H}}\bigg\} v = \lambda_{\min}(\mathbb{H}) \, .
\end{equation*}
For $c_{H} < 1$, by Weyl's inequality (see \textcite{hj2012} Theorem 4.3.1) and $\lambda_{\min}(VV^{\prime}) = 0$, $\lambda_{\min}(\overline{H}) \geq \lambda_{\min}((D^{\prime} \overline{\nabla^{2} \psi} D + c_{H} \, V V^{\prime}) / T) \geq c_{H} \, \lambda_{\min}(\mathbb{H})$, so $\lambda_{\min}(\overline{H}) \geq \min(1, c_{H}) \lambda_{\min}(\mathbb{H})$ in both cases. The constraint structure gives $\mathbb{H}$ a $3\times3$ block-diagonal form,
$\mathbb{H} = \bdiag(\mathbb{H}_{1}, \mathbb{H}_{2}, \mathbb{H}_{3})$, where
\begin{align*}
	&\mathbb{H}_{1} \coloneqq I_{N_{1}} \otimes \frac{N_{2} \, I_{T} + \iota_{T} \iota_{T}^{\prime}}{T} + \frac{\iota_{N_{1}} \iota_{N_{1}}^{\prime}}{T} \otimes I_{T} \, , \quad
	\mathbb{H}_{2} \coloneqq I_{N_{2}} \otimes \frac{N_{1} \, I_{T} + \iota_{T} \iota_{T}^{\prime}}{T} + \frac{\iota_{N_{2}} \iota_{N_{2}}^{\prime}}{T} \otimes I_{T} \, , \\
	&\mathbb{H}_{3} \coloneqq I_{N_{1}} \otimes \frac{T \, I_{N_{2}} + \iota_{N_{2}} \iota_{N_{2}}^{\prime}}{T} + \frac{\iota_{N_{1}} \iota_{N_{1}}^{\prime}}{T} \otimes I_{N_{2}} \, .
\end{align*}
In particular,
\begin{align*}
	\mathbb{H}_{1} =& \, \frac{N_{2}}{T} \left\{ \underbrace{\begin{pmatrix}
			I_{T} + \frac{\iota_{T} \iota_{T}^{\prime}}{N_{2}} &  &  \\
			&\ddots &\\
			& & I_{T} + \frac{\iota_{T} \iota_{T}^{\prime}}{N_{2}}
	\end{pmatrix}}_{\eqqcolon \mathbb{F}_{1}} + 
	\underbrace{\frac{\iota_{N_{1}} \iota_{N_{1}}^{\prime} \otimes I_{T}}{N_{2}}}_{= \frac{V_{1} V_{1}^{\prime}}{N_{2}}} \right\} \, , \\
	\mathbb{H}_{2} =& \, \frac{N_{1}}{T} \left\{ \underbrace{\begin{pmatrix}
			I_{T} + \frac{\iota_{T} \iota_{T}^{\prime}}{N_{1}} &  &  \\
			&\ddots &\\
			& & I_{T} + \frac{\iota_{T} \iota_{T}^{\prime}}{N_{1}}
	\end{pmatrix}}_{\eqqcolon \mathbb{F}_{2}} + 
	\underbrace{\frac{\iota_{N_{2}} \iota_{N_{2}}^{\prime} \otimes I_{T}}{N_{1}}}_{= \frac{V_{4} V_{4}^{\prime}}{N_{1}}} \right\} \, , \\
	\mathbb{H}_{3} =& \left\{ \underbrace{\begin{pmatrix}
			I_{N_{2}} + \frac{\iota_{N_{2}} \iota_{N_{2}}^{\prime}}{T} &  &  \\
			&\ddots &\\
			& & I_{N_{2}} + \frac{\iota_{N_{2}} \iota_{N_{2}}^{\prime}}{T}
	\end{pmatrix}}_{\eqqcolon \mathbb{F}_{3}} + 
	\underbrace{\frac{\iota_{N_{1}} \iota_{N_{1}}^{\prime} \otimes I_{N_{2}}}{T}}_{= \frac{V_{9} V_{9}^{\prime}}{T}} \right\} \, .
\end{align*}
Invertibility of each block of $\mathbb{F}_1$, $\mathbb{F}_2$, $\mathbb{F}_3$ follows from the Sherman-Morrison formula: 
\begin{align*}
	\mathbb{F}_{1}^{- 1} =& \, I_{N_{1}} \otimes \underbrace{\left(I_{T} - \frac{\iota_{T} \iota_{T}^{\prime}}{N_{2} + T}\right)}_{\eqqcolon \mathbb{E}_{1}} \, , \quad \mathbb{F}_{2}^{- 1} = I_{N_{2}} \otimes \underbrace{\left(I_{T} - \frac{\iota_{T} \iota_{T}^{\prime}}{N_{1} + T}\right)}_{\eqqcolon \mathbb{E}_{2}} \, , \\
	\mathbb{F}_{3}^{- 1} =& \, I_{N_{1}} \otimes \underbrace{\left(I_{N_{2}} - \frac{\iota_{N_{2}} \iota_{N_{2}}^{\prime}}{N_{2} + T}\right)}_{\eqqcolon \mathbb{E}_{3}} \, .
\end{align*}
By the Woodbury identity,
\begin{align*}
	\mathbb{H}_{1}^{- 1} =& \, \left(\frac{T}{N_{2}}\right) \left\{\mathbb{F}_{1}^{- 1} - \mathbb{F}_{1}^{- 1} V_{1} \big(\underbrace{N_{2} \, I_{T} + V_{1}^{\prime} \mathbb{F}_{1}^{- 1} V_{1}}_{\eqqcolon \mathbb{G}_{1}}\big)^{- 1} V_{1}^{\prime} \mathbb{F}_{1}^{- 1}\right\} \, , \\
	\mathbb{H}_{2}^{- 1} =& \, \left(\frac{T}{N_{1}}\right) \left\{\mathbb{F}_{2}^{- 1} - \mathbb{F}_{2}^{- 1} V_{4} \big(\underbrace{N_{1} \, I_{T} + V_{4}^{\prime} \mathbb{F}_{2}^{- 1} V_{4}}_{\eqqcolon \mathbb{G}_{2}}\big)^{- 1} V_{4}^{\prime} \mathbb{F}_{2}^{- 1}\right\} \, , \\
	\mathbb{H}_{3}^{- 1} =& \, \mathbb{F}_{3}^{- 1} - \mathbb{F}_{3}^{- 1} V_{9} \big(\underbrace{T \, I_{N_{2}} + V_{9}^{\prime} \mathbb{F}_{3}^{- 1} V_{9}}_{\eqqcolon \mathbb{G}_{3}}\big)^{- 1} V_{9}^{\prime} \mathbb{F}_{3}^{- 1} \, ,
\end{align*}
where
\begin{align*}
	&\mathbb{G}_{1} =  \left\{(N_{1} + N_{2}) \, I_{T} - \left(\frac{N_{1}}{N_{2} + T}\right) \iota_{T} \iota_{T}^{\prime}\right\} \, , \quad \mathbb{G}_{2} =  \left\{(N_{1} + N_{2}) \, I_{T} - \left(\frac{N_{2}}{N_{1} + T}\right) \iota_{T} \iota_{T}^{\prime}\right\} \, , \\
	&\mathbb{G}_{3} =  \left\{(N_{1} + T) \, I_{N_{2}} - \left(\frac{N_{1}}{N_{2} + T}\right) \iota_{N_{2}} \iota_{N_{2}}^{\prime}\right\} \, .
\end{align*}
By the Sherman-Morrison formula,
\begin{equation*}
	\mathbb{H}^{- 1} = \begin{pmatrix}
		\mathbb{H}_{1}^{- 1} &  &  \\
		& \mathbb{H}_{2}^{- 1} &  \\
		&  & \mathbb{H}_{3}^{- 1} \\
	\end{pmatrix} \, , \quad \text{where} \quad
	\begin{cases}
		\mathbb{H}_{1}^{- 1} = \left(\frac{T}{N_{2}}\right) \left\{I_{N_{1}} \otimes \mathbb{E}_{1} - \iota_{N_{1}} \iota_{N_{1}}^{\prime} \otimes \mathbb{E}_{1} \mathbb{G}_{1}^{- 1} \mathbb{E}_{1} \right\} \, , \\
		\mathbb{H}_{2}^{- 1} = \left(\frac{T}{N_{1}}\right) \left\{I_{N_{2}} \otimes \mathbb{E}_{2} - \iota_{N_{2}} \iota_{N_{2}}^{\prime} \otimes \mathbb{E}_{2} \mathbb{G}_{2}^{- 1} \mathbb{E}_{2} \right\} \, , \\
		\mathbb{H}_{3}^{- 1} = I_{N_{1}} \otimes \mathbb{E}_{3} - \iota_{N_{1}} \iota_{N_{1}}^{\prime} \otimes \mathbb{E}_{3} \mathbb{G}_{3}^{- 1} \mathbb{E}_{3} \, .
	\end{cases}
\end{equation*}
By the definition of the spectral norm and Lemma \ref{lemma:matrix_norm_inequalties}, $\lambda_{\min}(\mathbb{H}) = (\norm{\mathbb{H}^{- 1}}_{2})^{- 1} \geq (\norm{\mathbb{H}^{- 1}}_{\infty})^{- 1} = \left\{\max_{m} \norm{\mathbb{H}_{m}^{- 1}}_{\infty} \right\}^{- 1}$. By $\norm{A \otimes B} = \norm{A}\norm{B}$, the triangle inequality, $\norm{\mathbb{E}_{m}}_{\infty} < 2$, $\norm{\mathbb{G}_{1}^{- 1}}_{\infty} < 1 / N_{2}$, $\norm{\mathbb{G}_{2}^{- 1}}_{\infty} < 1 / N_{1}$, and $\norm{\mathbb{G}_{3}^{- 1}}_{\infty} < 1 / T$, $\norm{\mathbb{H}_{1}^{- 1}}_{\infty} < (2 T) / N_{2} (1 + 2 N_{1} / N_{2})$, $\norm{\mathbb{H}_{2}^{- 1}}_{\infty} < (2 T) / N_{1} (1 + 2 N_{2} / N_{1})$, $\norm{\mathbb{H}_{3}^{- 1}}_{\infty} < 2 (1 + 2 N_{1} / T)$, so $\norm{\mathbb{H}^{- 1}}_{\infty} = \mathcal{O}(1)$ and $\overline{H} > 0$ a.\,s. By the definition of the operator norm, $\norm{\overline{H}^{- 1}}_{\infty} = \big\{\min_{\norm{v}_{\infty} = 1} \, \norm{\overline{H} \, v}_{\infty}\big\}^{- 1}$. For $c_{H} \geq 1$, $\norm{\overline{H}^{- 1}}_{\infty} \leq \norm{\mathbb{H}^{- 1}}_{\infty} = \mathcal{O}(1)$. For $c_{H} < 1$, $\norm{\overline{H}^{- 1}}_{\infty} \leq c_{H}^{-1} \norm{\mathbb{H}^{- 1}}_{\infty} = \mathcal{O}(1)$ a.\,s. Hence, $\overline{H} > 0$ and $\norm{\overline{H}^{- 1}}_{\infty} = \mathcal{O}(1)$ a.\,s. By (i) and Lemma \ref{lemma:matrix_norm_inequalties}, $\norm{\overline{Q}}_{\infty} \leq T^{- 1} \, \norm{D}_{1} \, \norm{D}_{\infty} \, \norm{\overline{H}^{-1}}_{\infty} = \mathcal{O}(1)$ a.\,s., $\overline{F} > 0$ a.\,s. and $\norm{\overline{F}_{1}^{- 1}}_{\infty} \leq c_{H}^{- 1} (T / N_{2})$, $\norm{\overline{F}_{2}^{- 1}}_{\infty} \leq c_{H}^{- 1} (T / N_{1})$, $\norm{\overline{F}_{3}^{- 1}}_{\infty} \leq c_{H}^{- 1}$ so $\norm{\overline{F}^{- 1}}_{\infty} = \mathcal{O}(1)$ a.\,s. Then, $\norm{\overline{\mathcal{Q}}}_{\infty} \leq T^{- 1} \, \norm{D}_{1} \, \norm{D}_{\infty} \, \norm{\overline{F}^{-1}}_{\infty} = \mathcal{O}(1)$ a.\,s. and, using $\norm{D_m}_1 \in \{N_1, N_2, T\}$, $\norm{D_m}_\infty = 1$ for $m \in \{1,2,3\}$, and (i), $\norm{\overline{G}}_{\infty} = \mathcal{O}(1)$ a.\,s.
\vspace{0.5em}

\noindent\# (iii) For $1 \leq \mathrm{r} \leq 5$, by the triangle inequality, Jensen's inequality, and $\norm{\overline{Q}}_{\infty}^{20 + \nu} \leq C$ a.\,s. by (ii), $\sup_{ijt} \EX{\max_{k} \abs{(\mathfrak{D}_{\pi}^{\mathrm{r}} e_{k})_{ijt}}^{20 + \nu}} \leq C$ a.\,s. Hence, by the Lyapunov inequality, $\sup_{ijt} \EX{\max_{k} \abs{(\mathfrak{D}_{\pi}^{\mathrm{r}} e_{k})_{ijt}}^{p}} \leq C$ a.\,s. for $1 \leq p \leq 20 + \nu$ and $1 \leq \mathrm{r} \leq 5$.
\vspace{0.5em}

\noindent\# (iv) For $p \geq 1$ and $1 \leq \mathrm{r} \leq 5$, by Jensen's inequality and the Lyapunov inequality, $\max_{k} \norm{\overline{\nabla^{\mathrm{r}} \psi X} e_{k}}_{p}^{p} = \mathcal{O}(T^{3})$ a.\,s., so $\max_{k} \norm{\overline{\nabla^{\mathrm{r}} \psi X} e_{k}}_{p} = \mathcal{O}(T^{3 / p})$ a.\,s. Similarly, using (iii), $\norm{\overline{\dop{\mathrm{r}} \psi}}_{p} = \mathcal{O}(T^{3 / p})$ a.\,s. and $\max_{k} \norm{\overline{\mathfrak{D}_{\pi}^{\mathrm{r}}} e_{k}}_{p} = \mathcal{O}(T^{3 / p})$ a.\,s. By Lemma \ref{lemma:matrix_norm_inequalties} and (i), $\max_{k} \norm{D^{\prime} \overline{\nabla^{\mathrm{r}} \psi X} e_{k}}_{p} = \mathcal{O}(T^{1 + 2 / p})$ a.\,s. Using $\norm{\overline{Q}}_{\infty} = \mathcal{O}(1)$ a.\,s. by (ii), $\max_{k} \norm{\mathfrak{X} e_{k}}_{p} \leq \norm{\overline{Q}}_{\infty} \max_{k} \norm{\overline{\nabla^{2} \psi X} e_{k}}_{p} = \mathcal{O}(T^{3 / p})$ a.\,s. By Jensen's inequality and the Lyapunov inequality, $\bignorm{D^{\prime} \overline{\nabla^{\mathrm{r}} \psi} D}_{\infty} = \mathcal{O}(T)$ a.\,s. By similar arguments and the Cauchy-Schwarz inequality, $\norm{D^{\prime} \overline{\nabla^{\mathrm{r}} \psi \nabla^{\mathrm{r}^{\prime}} \psi} D}_{\infty} = \mathcal{O}(T)$ a.\,s., $\max_{k} \norm{D^{\prime} \diag(\overline{\mathfrak{D}_{\pi}^{\mathrm{r}}} e_{k}) D}_{\infty} = \mathcal{O}(T)$ a.\,s., and \linebreak $\max_{k, k^{\prime}} \norm{D^{\prime} \diag(\overline{\mathfrak{D}_{\pi}^{\mathrm{r}} e_{k} \odot \mathfrak{D}_{\pi}^{\mathrm{r}^{\prime}}} e_{k^{\prime}}) D}_{\infty} = \mathcal{O}(T)$ a.\,s.
\vspace{0.5em}

\noindent\# (v) For $1 \leq \mathrm{r} \leq 5$, by the triangle inequality, $\norm{D^{\prime} \widetilde{\dop{\mathrm{r}} \psi}}_{20} \leq \norm{D_{1}^{\prime} \widetilde{\dop{\mathrm{r}} \psi}}_{20} + \norm{D_{2}^{\prime} \widetilde{\dop{\mathrm{r}} \psi}}_{20} + \norm{D_{3}^{\prime} \widetilde{\dop{\mathrm{r}} \psi}}_{20}$. By Lemma \ref{lemma:moment_bounds_mixing}, $\EX{\norm{D_{1}^{\prime} \widetilde{\dop{\mathrm{r}} \psi}}_{20}^{20}} = \mathcal{O}(T^{12})$ a.\,s. Analogously, $\EX{\norm{D_{2}^{\prime} \widetilde{\dop{\mathrm{r}} \psi}}_{20}^{20}} = \mathcal{O}(T^{12})$ and $\EX{\norm{D_{3}^{\prime} \widetilde{\dop{\mathrm{r}} \psi}}_{20}^{20}} = \mathcal{O}(T^{12})$ a.\,s. Hence, $\norm{D^{\prime} \widetilde{\dop{\mathrm{r}} \psi}}_{20} = \mathcal{O}_{P}(T^{3 / 5})$ by Markov's inequality, and for $1 \leq p \leq 20$, $\norm{D^{\prime} \widetilde{\dop{\mathrm{r}} \psi}}_{p} = \mathcal{O}_{P}(T^{1 / 2 + 2 / p})$. By similar arguments and (iii), $\max_{k} \norm{D^{\prime} \widetilde{\nabla^{\mathrm{r}} \psi X} e_{k}}_{p} = \mathcal{O}_{P}(T^{1 / 2 + 2 / p})$ and $\max_{k} \norm{D^{\prime} \widetilde{\mathfrak{D}_{\pi}^{\mathrm{r}}} e_{k}}_{p} = \mathcal{O}_{P}(T^{1 / 2 + 2 / p})$. By (i) and (ii),
\begin{align*}
	\bignorm{\overline{Q} \dop{1} \psi}_{p} \leq T^{- 1} \, \norm{D}_{p} \, \bignorm{\overline{H}^{- 1}}_{\infty} \, \bignorm{D^{\prime} \dop{1} \psi}_{p} = \mathcal{O}_{P}\big(T^{- \frac{1}{2} + \frac{3}{p}}\big) \, , \quad \bignorm{\overline{\mathcal{Q}} \dop{1} \psi}_{p} = \mathcal{O}_{P}\big(T^{- \frac{1}{2} + \frac{3}{p}}\big) \, .
\end{align*}
Analogously, $\max_{k} \norm{\overline{\mathcal{Q}} \widetilde{\mathfrak{D}_{\pi}^{\mathrm{r}}} e_{k}}_{p} = \mathcal{O}_{P}(T^{- 1 / 2 + 3 / p})$. By the triangle inequality, $\norm{V^{\prime} \overline{F}^{- 1} D^{\prime} \dop{1} \psi}_{20} \leq \norm{V_{1}^{\prime} \overline{F}_{1}^{- 1} D_{1}^{\prime} \dop{1} \psi}_{20} + \ldots + \norm{V_{9}^{\prime} \overline{F}_{3}^{- 1} D_{3}^{\prime} \dop{1} \psi}_{20}$. Let $\overline{\vartheta}_{1} \coloneqq T \, \overline{f}_{3}^{\circ - 1}$. By Lemma \ref{lemma:moment_bounds_mixing},
\begin{align*}
	&\EX{\bignorm{V_{9}^{\prime} \overline{F}_{3}^{- 1} D_{3}^{\prime} \dop{1} \psi}_{20}^{20}} \leq C \, \bigg(\frac{N_{1}}{T}\bigg)^{10} \bigg(\frac{N_{2}}{T}\bigg) \, T^{21} \sup_{j} \, \EX{\bigg(\frac{1}{\sqrt{N_{1} T}}\sum_{i = 1}^{N_{1}} \sum_{t = 1}^{T} (\overline{\vartheta}_{1})_{ij} (\dop{1} \psi)_{ijt} \bigg)^{20}} \\
	& \quad = \mathcal{O}\big(T^{21}\big) \quad \text{a.\,s.} \, ,
\end{align*}
so $\norm{V_{9}^{\prime} \overline{F}_{3}^{- 1} D_{3}^{\prime} \dop{1} \psi}_{20} = \mathcal{O}_{P}(T^{21 / 20})$ by Markov's inequality. The other five bounds follow analogously. Hence, $\norm{V^{\prime} \overline{F}^{- 1} D^{\prime} \dop{1} \psi}_{20} = \mathcal{O}_{P}(T^{21 / 20})$, and for $1 \leq p \leq 20$, $\norm{V^{\prime} \overline{F}^{- 1} D^{\prime} \dop{1} \psi}_{p} = \mathcal{O}_{P}(T^{1 + 1 / p})$. By similar arguments, $\max_{k} \norm{V^{\prime} \overline{F}^{- 1} D^{\prime} \widetilde{\nabla^{\mathrm{r}} \psi X e_{k}}}_{p} = \mathcal{O}_{P}(T^{1 + 1 / p})$ and \linebreak $\max_{k} \norm{V^{\prime} \overline{F}^{- 1} D^{\prime} \widetilde{\mathfrak{D}_{\pi}^{\mathrm{r}} e_{k}}}_{p} = \mathcal{O}_{P}(T^{1 + 1 / p})$ for $1 \leq \mathrm{r} \leq 5$. For $1 \leq \mathrm{r} \leq 5$, letting $\overline{\vartheta}_{2, \mathrm{r}, k} \coloneqq D_{3}^{\prime} \overline{\mathfrak{D}_{\pi}^{\mathrm{r}}} e_{k} \odot \overline{f}_{3}^{\circ - 1}$, $\EX{\max_{k} \norm{V_{9}^{\prime} \overline{F}_{3}^{- 1} D_{3}^{\prime} \diag(\overline{\mathfrak{D}_{\pi}^{\mathrm{r}}} e_{k}) \overline{\mathcal{Q}}_{3} \dop{1} \psi}_{20}^{20}} = \mathcal{O}(T^{21})$ a.\,s., so \linebreak $\max_{k} \norm{V_{9}^{\prime} \overline{F}_{3}^{- 1} D_{3}^{\prime} \diag(\overline{\mathfrak{D}_{\pi}^{\mathrm{r}}} e_{k}) \overline{\mathcal{Q}}_{3} \dop{1} \psi}_{20} = \mathcal{O}_{P}(T^{21 / 20})$ by Markov's inequality. The other 17 bounds follow analogously. Hence, for $1 \leq p \leq 20$, $\max_{k} \norm{V^{\prime} \overline{F}^{- 1} D^{\prime} \diag(\overline{\mathfrak{D}_{\pi}^{\mathrm{r}}} e_{k}) \overline{\mathcal{Q}} \dop{1} \psi}_{p} = \mathcal{O}_{P}(T^{1 + 1 / p})$. By the triangle inequality, $\norm{\overline{G} \, \overline{F}^{- 1} D^{\prime} \dop{1} \psi}_{20} \leq \norm{D_{1}^{\prime} \overline{\nabla^{2} \psi} \, \overline{\mathcal{Q}}_{2} \dop{1} \psi}_{20} + \ldots + T^{- 1} \norm{V}_{20} \, \norm{V^{\prime} \overline{F}^{- 1} D^{\prime} \dop{1} \psi}_{20}$. Let $(\overline{\vartheta}_{3})_{ijt} \coloneqq N_{1} \, (\overline{\dop{2} \psi})_{ijt} / (\overline{f}_{2})_{jt}$. By similar arguments, \linebreak $\EX{\norm{D_{1}^{\prime} \overline{\nabla^{2} \psi} \, \overline{\mathcal{Q}}_{2} \dop{1} \psi}_{20}^{20}} = \mathcal{O}(T^{2})$ a.\,s., so $\norm{D_{1}^{\prime} \overline{\nabla^{2} \psi} \, \overline{\mathcal{Q}}_{2} \dop{1} \psi}_{20} = \mathcal{O}_{P}(T^{1 / 10})$ by Markov's inequality. The other five bounds and $T^{- 1} \norm{V}_{20} \norm{V^{\prime} \overline{F}^{- 1} D^{\prime} \dop{1} \psi}_{20} = \mathcal{O}_{P}(T^{1 / 10})$ follow analogously. Hence, $\norm{\overline{G} \, \overline{F}^{- 1} D^{\prime} \dop{1} \psi}_{20} = \mathcal{O}_{P}(T^{1 / 10})$, and for $1 \leq p \leq 20$, $\norm{\overline{G} \, \overline{F}^{- 1} D^{\prime} \dop{1} \psi}_{p} = \mathcal{O}_{P}(T^{2 / p})$. By similar arguments, $\max_{k} \norm{\overline{G} \, \overline{F}^{- 1} D^{\prime} \widetilde{\mathfrak{D}_{\pi}^{\mathrm{r}}} e_{k}}_{p} = \mathcal{O}_{P}(T^{2 / p})$ and \linebreak $\max_{k} \norm{\overline{G} \, \overline{F}^{- 1} D^{\prime} \diag(\overline{\mathfrak{D}_{\pi}^{\mathrm{r}}} e_{k}) \overline{\mathcal{Q}} \dop{1} \psi}_{p} = \mathcal{O}_{P}(T^{2 / p})$ for $1 \leq \mathrm{r} \leq 5$.
\vspace{0.5em}

\noindent\# (vi) For $1 \leq \mathrm{r} \leq 5$, by the triangle inequality, $\bignorm{D^{\prime} \widetilde{\nabla^{\mathrm{r}} \psi} D}_{2} \leq \bignorm{D_{1}^{\prime} \widetilde{\nabla^{\mathrm{r}} \psi} D_{1}}_{2} + \ldots + 2 \, \bignorm{D_{3}^{\prime} \widetilde{\nabla^{\mathrm{r}} \psi} D_{2}}_{2}$. The first three are diagonal matrices. Using $\norm{D^{\prime} \widetilde{\dop{\mathrm{r}} \psi}}_{20} = \mathcal{O}_{P}(T^{3/5})$ by (v), $\norm{D_{1}^{\prime} \widetilde{\nabla^{\mathrm{r}} \psi} D_{1}}_{2} = \norm{D_{1}^{\prime} \widetilde{\dop{\mathrm{r}} \psi}}_{\infty} \leq \norm{D^{\prime} \widetilde{\dop{\mathrm{r}} \psi}}_{20} = \mathcal{O}_{P}(T^{3 / 5})$, and analogously for $D_2$ and $D_3$. The last three terms involve spectral norms of off-diagonal blocks of dimensions $N_{2} T \times N_{1} T$, $N_{1} N_{2} \times N_{1} T$, and $N_{1} N_{2} \times N_{2} T$. Each matrix has $N_{1} N_{2} T = n$ nonzero entries and can be permuted into a block-diagonal form. Let $\mathbb{P}_{1}$, $\mathbb{P}_{2}$, $\mathbb{P}_{3}$ be permutation matrices. By permutation invariance of the spectral norm and the triangle inequality, $\norm{D_{3}^{\prime} \widetilde{\nabla^{\mathrm{r}} \psi} D_{2}}_{2} \leq \norm{\mathbb{P}_{3} D_{3}^{\prime} \widetilde{\nabla^{\mathrm{r}} \psi} D_{2}}_{2} = \sup_{j} \norm{B_{j}}_{2}$, where
\begin{equation*}
	B_{j} \coloneqq \begin{pmatrix}
		\big(\widetilde{\dop{\mathrm{r}} \psi}\big)_{1j1} & \cdots & \big(\widetilde{\dop{\mathrm{r}} \psi}\big)_{1jT} \\
		\vdots&\ddots&\vdots \\
		\big(\widetilde{\dop{\mathrm{r}} \psi}\big)_{N_{1}j1} &\cdots & \big(\widetilde{\dop{\mathrm{r}} \psi}\big)_{N_{1}jT}
	\end{pmatrix}
\end{equation*}
is of dimension $N_{1} \times T$. By the union bound and Lemma \ref{lemma:asymptotic_bound_spectral_norm} with $p = 20 + \check{\nu}$ and $b = 1$,
\begin{align*}
	\EX{\bignorm{\mathbb{P}_{3} D_{3}^{\prime} \widetilde{\nabla^{\mathrm{r}} \psi} D_{2}}_{2}^{20 + \check{\nu}}} \leq \bigg(\frac{N_{2}}{T}\bigg) T \sup_{j} \EX{\bignorm{B_{j}}_{2}^{20 + \check{\nu}}} = \mathcal{O}\big(\big(\sqrt{\log(T)}\big)^{20 + \check{\nu}} \, T^{\frac{24 + \check{\nu}}{2}}\big) \quad \text{a.\,s.} \, ,
\end{align*}
so $\norm{D_{3}^{\prime} \widetilde{\nabla^{\mathrm{r}} \psi} D_{2}}_{2} = \mathcal{O}_{P}(\sqrt{\log(T)} \, T^{1 / 2 + 2 / (20 + \check{\nu})}) = \mathcal{O}_{P}(T^{3 / 5})$ by Markov's inequality. Analogously, $\norm{D_{2}^{\prime} \widetilde{\nabla^{\mathrm{r}} \psi} D_{1}}_{2} = \mathcal{O}_{P}(T^{3 / 5})$ and $\norm{D_{3}^{\prime} \widetilde{\nabla^{\mathrm{r}} \psi} D_{1}}_{2} = \mathcal{O}_{P}(T^{3 / 5})$, so $\norm{D^{\prime} \widetilde{\nabla^{\mathrm{r}} \psi} D}_{2} = \mathcal{O}_{P}(T^{3 / 5})$. By the triangle inequality, Jensen's inequality, Loeve's $c_{r}$ inequality, and the union bound, $\EX{\norm{D^{\prime} \widetilde{\nabla^{\mathrm{r}} \psi} D}_{\infty}^{20 + \nu}} = \mathcal{O}(T^{22 + \nu})$ a.\,s., so $\norm{D^{\prime} \widetilde{\nabla^{\mathrm{r}} \psi} D}_{\infty} = o_{P}(T^{11 / 10})$ by Markov's inequality. By Lemma \ref{lemma:matrix_norm_inequalties}, for $2 < p < \infty$, $\norm{D^{\prime} \widetilde{\nabla^{\mathrm{r}} \psi} D}_{p} \leq \norm{D^{\prime} \widetilde{\nabla^{\mathrm{r}} \psi} D}_{2}^{2 / p} \norm{D^{\prime} \widetilde{\nabla^{\mathrm{r}} \psi} D}_{\infty}^{1 - 2 / p} = o_{P}(T^{11 / 10 - 1 / p})$. By similar arguments and (iii), $\max_{k} \norm{D^{\prime} \diag(\widetilde{\mathfrak{D}_{\pi}^{\mathrm{r}}} e_{k}) D}_{2} = \mathcal{O}_{P}(T^{3 / 5})$ and $\max_{k} \norm{D^{\prime} \diag(\widetilde{\mathfrak{D}_{\pi}^{\mathrm{r}}} e_{k}) D}_{p} = \mathcal{O}_{P}(T^{11 / 10 - 1 / p})$ for $2 < p < \infty$. Using \linebreak $\sup_{ijt} \EX{\abs{(\dop{\mathrm{r}} \psi)_{ijt} (d^{\mathrm{r}^{\prime}} \psi)_{ijt}}^{10 + \nu / 2}} \leq C$ a.\,s., by the Cauchy-Schwarz and Jensen's inequalities, $\norm{D^{\prime} \widetilde{\nabla^{\mathrm{r}} \psi \nabla^{\mathrm{r}^{\prime}} \psi} D}_{2} = \mathcal{O}_{P}(T^{7 / 10})$ and $\norm{D^{\prime} \widetilde{\nabla^{\mathrm{r}} \psi \nabla^{\mathrm{r}^{\prime}} \psi} D}_{p} = \mathcal{O}_{P}(T^{6 / 5 - 1 / p})$ for $2 < p < \infty$ and $1 \leq \mathrm{r}, \mathrm{r}^{\prime} \leq 5$, and analogously for the $\mathfrak{D}_\pi^\mathrm{r}$ variants. By the property of the spectral norm, the Minkowski inequality, Lemma \ref{lemma:matrix_norm_inequalties}, and (iv), for $1 \leq \mathrm{r} \leq 5$, $\norm{\nabla^{\mathrm{r}} \psi D}_{2} \leq \norm{D^{\prime} \overline{\nabla^{\mathrm{r}} \psi \nabla^{\mathrm{r}} \psi} D}_{\infty}^{1 / 2} + \norm{D^{\prime} \widetilde{\nabla^{\mathrm{r}} \psi \nabla^{\mathrm{r}} \psi} D}_{2}^{1 / 2} = \mathcal{O}_{P}(T^{1 / 2})$. By similar arguments, $\max_{k} \norm{\diag(\mathfrak{D}_{\pi}^{\mathrm{r}} e_{k}) D}_{2} = \mathcal{O}_{P}(T^{1 / 2})$. By the union bound and Jensen's inequality, $\EX{\norm{\dop{\mathrm{r}} \psi}_{\infty}^{20 + \nu}} = \mathcal{O}(T^{3})$ a.\,s., so $\norm{\dop{\mathrm{r}} \psi}_{\infty} = o_{P}(T^{3 / 20})$ by Markov's inequality, and hence, $\norm{\nabla^{\mathrm{r}} \psi D}_{\infty} \leq 3 \norm{\dop{\mathrm{r}} \psi}_{\infty} = o_{P}(T^{3 / 20})$. By Lemma \ref{lemma:matrix_norm_inequalties}, for $2 < p < \infty$, $\norm{\nabla^{\mathrm{r}} \psi D}_{p} \leq \norm{\nabla^{\mathrm{r}} \psi D}_{2}^{2 / p} \norm{\nabla^{\mathrm{r}} \psi D}_{\infty}^{1 - 2 / p} = o_{P}(T^{3 / 20 + 7 / (10 p)})$. Analogously, $\max_{k} \norm{\diag(\mathfrak{D}_{\pi}^{\mathrm{r}} e_{k}) D}_{p} = o_{P}(T^{3 / 20 + 7 / (10 p)})$.
\vspace{0.5em}

\noindent\# (vii) For $1 \leq \mathrm{r} \leq 5$, by the triangle inequality, Jensen's inequality, Loeve's $c_{r}$ inequality, and the union bound, $\EX{\sup_{(\beta, \phi)} \, \norm{D^{\prime} \abs{\nabla^{\mathrm{r}} \psi(\beta, \phi)} D}_{\infty}^{20 + \nu}} = \mathcal{O}(T^{22 + \nu})$ a.\,s., so \linebreak $\sup_{(\beta, \phi)} \norm{D^{\prime} \abs{\nabla^{\mathrm{r}} \psi(\beta, \phi)} D}_{\infty} = o_{P}(T^{11 / 10})$ by Markov's inequality. By similar arguments and the Cauchy-Schwarz inequality, $\sup_{(\beta, \phi)} \max_{k} \norm{D^{\prime} \abs{\diag(\dop{\mathrm{r}} \psi(\beta, \phi) \odot X e_{k})} D}_{\infty} = o_{P}(T^{11 / 10})$ and $\sup_{(\beta, \phi)} \norm{D^{\prime} \abs{\nabla^{\mathrm{r}} \psi(\beta, \phi)} \abs{\nabla^{\mathrm{r}^{\prime}} \psi(\beta, \phi)} D}_{\infty} = o_{P}(T^{6 / 5})$ for $1 \leq \mathrm{r}, \mathrm{r}^{\prime} \leq 5$.
\vspace{0.5em}

\noindent\# (viii) For all $(\beta, \phi) \in \mathfrak{B}(\varepsilon) \times \mathfrak{P}(\eta, q)$, by a Taylor expansion of $H(\beta, \phi) = T^{- 1} D^{\prime} \nabla^{2} \psi(\beta, \phi) D$ around $(\beta^0, \phi^0)$, Lemma \ref{lemma:matrix_norm_inequalties}, the triangle inequality, and (vii), $\sup_{(\beta, \phi)} \, \norm{H(\beta, \phi) - H}_{p} \leq \sup_{(\beta, \phi)} \, \norm{H(\beta, \phi) - H}_{\infty} = o_{P}(1)$, using $\sup_{\beta} \norm{\beta - \beta^{0}}_{2} \leq \varepsilon$ and $\sup_{\phi} \norm{\phi - \phi^{0}}_{q} \leq \eta$. By (ii) and (vi), the conditions of Lemma \ref{lemma:invertibility} with $\bar{p} = 10$ hold wpa1. Thus, $H(\beta, \phi)$ is invertible for all $(\beta, \phi) \in \mathfrak{B}(\varepsilon) \times \mathfrak{P}(\eta, q)$ wpa1 and for $2 \leq p \leq 10$, $\sup_{(\beta, \phi)} \, \norm{(H(\beta, \phi))^{- 1}}_{p} = \mathcal{O}_{P}(1)$. By (i), $\sup_{(\beta, \phi)} \, \norm{Q(\beta, \phi)}_{p} \leq T^{- 1} \, \sup_{(\beta, \phi)} \, \norm{\big(H(\beta, \phi)\big)^{- 1}}_{p} \norm{D}_{p} \norm{D^{\prime}}_{p} = \mathcal{O}_{P}(1)$.
\vspace{0.5em}

\noindent\# (ix) Consider $H = \overline{H} + \widetilde{H}$. By (ii) and (vi), the conditions of Lemma \ref{lemma:inverse_neumann_series} hold wpa1. Let $E \coloneqq \sum_{r = 3}^{\infty} (- \widetilde{H} \, \overline{H}^{- 1})^{r}$. Then, $H^{- 1} = \overline{H}^{- 1} \sum_{r = 0}^{2} (- \widetilde{H} \, \overline{H}^{- 1})^{r} + \overline{H}^{- 1} E$, and for $2 \leq p \leq 10$, $\norm{E}_{p} \leq \norm{\widetilde{H}}_{p}^{3} \, \norm{\overline{H}^{- 1}}_{p}^{3} (1 - \norm{\widetilde{H}}_{p} \, \norm{\overline{H}^{- 1}}_{p})^{- 1}$. By the triangle inequality and (vi), $\norm{H^{- 1} - \overline{H}^{- 1}}_{2} = \mathcal{O}_{P}(T^{- 2 / 5})$, $\norm{H^{- 1} - \overline{H}^{- 1} + \overline{H}^{- 1} \widetilde{H} \, \overline{H}^{- 1}}_{2} = \mathcal{O}_{P}(T^{- 4 / 5})$, and $\norm{H^{- 1} - \overline{H}^{- 1} + \overline{H}^{- 1} \widetilde{H} \, \overline{H}^{- 1} - \overline{H}^{- 1} \widetilde{H} \, \overline{H}^{- 1} \widetilde{H} \, \overline{H}^{- 1}}_{2} = \mathcal{O}_{P}(T^{- 6 / 5})$. Analogously, $\norm{H^{- 1} - \overline{H}^{- 1}}_{p} = o_{P}(T^{1 / 10 - 1 / p})$, $\norm{H^{- 1} - \overline{H}^{- 1} + \overline{H}^{- 1} \widetilde{H} \, \overline{H}^{- 1}}_{p} = o_{P}(T^{1 / 5 - 2 / p})$, and $\norm{H^{- 1} - \overline{H}^{- 1} + \overline{H}^{- 1} \widetilde{H} \, \overline{H}^{- 1} - \overline{H}^{- 1} \widetilde{H} \, \overline{H}^{- 1} \widetilde{H} \, \overline{H}^{- 1}}_{p} = o_{P}(T^{3 / 10 - 3 / p})$. Then, using (i), $\norm{Q - \overline{Q}}_{2} = \mathcal{O}_{P}(T^{- 2 / 5})$ and $\norm{Q - \overline{Q}}_{p} = \mathcal{O}_{P}(T^{1 / 10 - 1 / p})$ for $2 < p \leq 10$.
\vspace{0.5em}

\noindent\# (x) For $1 \leq \mathrm{r} \leq 5$, by Jensen's inequality and the union bound, $\EX{\sup_{(\beta, \phi)} \norm{\dop{\mathrm{r}} \psi(\beta, \phi)}_{20 + \nu}^{20 + \nu}} = \mathcal{O}(T^{3})$ a.\,s., so $\sup_{(\beta, \phi)} \norm{\dop{\mathrm{r}} \psi(\beta, \phi)}_{20 + \nu} = \mathcal{O}_{P}(T^{3 / (20 + \nu)})$ by Markov's inequality, and for $1 \leq p \leq 20 + \nu$, $\sup_{(\beta, \phi)} \bignorm{\dop{\mathrm{r}} \psi(\beta, \phi)}_{p} = \mathcal{O}_{P}(T^{3 / p})$. By similar arguments and (iii), \linebreak $\sup_{(\beta, \phi)} \max_{k} \norm{\dop{\mathrm{r}} \psi(\beta, \phi) \odot X e_{k}}_{p} = \mathcal{O}_{P}(T^{3 / p})$, $\max_{k} \, \norm{X e_{k}}_{p} = \mathcal{O}_{P}(T^{3 / p})$, and $\max_{k} \, \norm{\mathfrak{D}_{\pi}^{\mathrm{r}} e_{k}}_{p} = \mathcal{O}_{P}(T^{3 / p})$ for $1 \leq p \leq 20 + \nu$. By (i), $\sup_{(\beta, \phi)} \max_{k} \norm{D^{\prime} \nabla^{\mathrm{r}} \psi(\beta, \phi) X e_{k}}_{p} = \mathcal{O}_{P}(T^{1 + 2 / p})$.
\vspace{0.5em}

\noindent\# (xi) By (viii) and (x), $\sup_{(\beta, \phi)} \max_{k} \, \norm{P(\beta, \phi) X e_{k}}_{10} = \mathcal{O}_{P}(T^{3 / 10})$. Hence, for $1 \leq p \leq 10$, $\sup_{(\beta, \phi)} \max_{k} \norm{P(\beta, \phi) X e_{k}}_{p} = \mathcal{O}_{P}(T^{3 / p})$ and $\sup_{(\beta, \phi)} \max_{k} \, \norm{M(\beta, \phi) X e_{k}}_{p} = \mathcal{O}_{P}(T^{3 / p})$. By (i), (v), and (viii), $\sup_{(\beta, \phi)} \norm{Q(\beta, \phi) \dop{1} \psi}_{10} = \mathcal{O}_{P}(T^{-1/5})$ and thus $\sup_{(\beta, \phi)} \norm{Q(\beta, \phi) \dop{1} \psi}_{p} = \mathcal{O}_{P}(T^{- 1 / 2 + 3 / p})$ for $1 \leq p \leq 10$.
\vspace{0.5em}

\noindent\# (xii) For $1 \leq \mathrm{r} \leq 5$ and each $k \in \{1, \ldots, K\}$, decomposing $\nabla^{\mathrm{r}} \psi M X e_{k}$ using Lemma \ref{lemma:inverse_approximation} with $\overline{H} = \overline{F} + \overline{G}$, $\nabla^{\mathrm{r}} \psi M X e_{k} - \mathfrak{D}_{\pi}^{\mathrm{r}} e_{k} = - \nabla^{\mathrm{r}} \psi \overline{Q} \widetilde{\mathfrak{D}_{\pi}^{2}} e_{k} + \ldots \eqqcolon \mathfrak{E}_{1, k} + \ldots + \mathfrak{E}_{5, k}$. By (ii), (v), (vi), (ix), and (x), $\max_{k} \norm{\mathfrak{E}_{1, k}}_{2} = \mathcal{O}_{P}(T)$ and $\max_{k} \norm{\mathfrak{E}_{r, k}}_{2} = \mathcal{O}_{P}(T^{3 / 5})$ for $r \in \{2, \ldots, 5\}$. Hence, $\max_{k} \norm{\nabla^{\mathrm{r}} \psi M X e_{k} - \mathfrak{D}_{\pi}^{\mathrm{r}} e_{k}}_{2} = \mathcal{O}_{P}(T)$, and for $1 \leq p \leq 2$, $\max_{k} \norm{\nabla^{\mathrm{r}} \psi M X e_{k} - \mathfrak{D}_{\pi}^{\mathrm{r}} e_{k}}_{p} = \mathcal{O}_{P}(T^{- 1 / 2 + 3 / p})$. Analogously, $\max_{k} \norm{\nabla^{\mathrm{r}} \psi M X e_{k} - \mathfrak{D}_{\pi}^{\mathrm{r}} e_{k} + \nabla^{\mathrm{r}} \psi \overline{\mathcal{Q}} \widetilde{\mathfrak{D}_{\pi}^{2}} e_{k}}_{p} = \mathcal{O}_{P}(T^{- 9 / 10 + 3 / p})$ and $\max_{k} \norm{\nabla^{\mathrm{r}} \psi M X e_{k} - \mathfrak{D}_{\pi}^{\mathrm{r}} e_{k} + \nabla^{\mathrm{r}} \psi \overline{Q} \widetilde{\mathfrak{D}_{\pi}^{2}} e_{k} - \nabla^{\mathrm{r}} \psi \overline{Q} \widetilde{\nabla^{\mathrm{r}} \psi} \overline{\mathcal{Q}} \widetilde{\mathfrak{D}_{\pi}^{2}} e_{k}}_{p} = \mathcal{O}_{P}(T^{- 6 / 5 + 3 / p})$, using $\mathfrak{E}_{1, k} = - \nabla^{\mathrm{r}} \psi \overline{\mathcal{Q}} \widetilde{\mathfrak{D}_{\pi}^{2}} e_{k} + T^{- 1} \nabla^{\mathrm{r}} \psi D \overline{H}^{- 1} \overline{G} \, \overline{F}^{- 1} D^{\prime} \widetilde{\mathfrak{D}_{\pi}^{2}} e_{k} \eqqcolon \mathfrak{E}_{1, k, 1} + \mathfrak{E}_{1, k, 2}$ by Lemma \ref{lemma:inverse_approximation}, and $\max_{k} \norm{\mathfrak{E}_{1, k, 2}}_{2} = \mathcal{O}_{P}(T^{1/2})$. By similar arguments, for $1 \leq p \leq 2$, $\max_{k} \norm{\mathfrak{X} e_{k} - P X e_{k}}_{p} = \mathcal{O}_{P}(T^{- 1 / 2 + 3 / p})$, $\max_{k} \norm{\mathfrak{X} e_{k} - P X e_{k} + \overline{\mathcal{Q}} \widetilde{\mathfrak{D}_{\pi}^{2}} e_{k}}_{p} = \mathcal{O}_{P}(T^{- 9 / 10 + 3 / p})$, and $\max_{k} \norm{\mathfrak{X} e_{k} - P X e_{k} + \overline{Q} \widetilde{\mathfrak{D}_{\pi}^{2}} e_{k} - \overline{Q} \widetilde{\nabla^{\mathrm{r}} \psi} \overline{\mathcal{Q}} \widetilde{\mathfrak{D}_{\pi}^{2}} e_{k}}_{p} = \mathcal{O}_{P}(T^{- 6 / 5 + 3 / p})$. For $1 \leq \mathrm{r} \leq 5$, decomposing $\nabla^{\mathrm{r}} \psi Q \dop{1} \psi$ using Lemma \ref{lemma:inverse_approximation}, $\nabla^{\mathrm{r}} \psi Q \dop{1} \psi = \nabla^{\mathrm{r}} \psi \overline{Q} \dop{1} \psi - \ldots \eqqcolon \mathfrak{E}_{6} + \ldots + \mathfrak{E}_{9}$. By (v) and (ix), $\norm{\mathfrak{E}_6}_{2} = \mathcal{O}_{P}(T)$ and $\norm{\mathfrak{E}_r}_{2} = \mathcal{O}_{P}(T^{3/5})$ for $r \in \{7, 8, 9\}$. By Lemma \ref{lemma:inverse_approximation}, $\mathfrak{E}_{6} = - \nabla^{\mathrm{r}} \psi \overline{\mathcal{Q}} \dop{1} \psi + T^{- 1} \nabla^{\mathrm{r}} \psi D \overline{H}^{- 1} \overline{G} \, \overline{F}^{- 1} D^{\prime} \dop{1} \psi \eqqcolon \mathfrak{E}_{6, 1} + \mathfrak{E}_{6, 2}$ with $\norm{\mathfrak{E}_{6, 2}}_{2} = \mathcal{O}_{P}(T^{1 / 2})$. Hence, $\norm{\nabla^{\mathrm{r}} \psi Q \dop{1} \psi - \nabla^{\mathrm{r}} \psi \overline{\mathcal{Q}} \dop{1} \psi}_{2} = \mathcal{O}_{P}(T^{3 / 5})$, and for $1 \leq p \leq 2$, $\norm{\nabla^{\mathrm{r}} \psi Q \dop{1} \psi - \nabla^{\mathrm{r}} \psi \overline{\mathcal{Q}} \dop{1} \psi}_{p} = \mathcal{O}_{P}\big(T^{- 9 / 10 + 3 / p}\big)$. Analogously, $\norm{\nabla^{\mathrm{r}} \psi Q \dop{1} \psi - \nabla^{\mathrm{r}} \psi \overline{Q} \dop{1} \psi + \nabla^{\mathrm{r}} \psi \overline{Q} \widetilde{\nabla^{\mathrm{r}} \psi} \overline{\mathcal{Q}} \dop{1} \psi}_{p} = \mathcal{O}_{P}(T^{- 6 / 5 + 3 / p})$, $\norm{Q \dop{1} \psi - \overline{\mathcal{Q}} \dop{1} \psi}_{p} = \mathcal{O}_{P}(T^{- 9 / 10 + 3 / p})$, and $\norm{Q \dop{1} \psi - \overline{Q} \dop{1} \psi + \overline{Q} \widetilde{\nabla^{\mathrm{r}} \psi} \overline{\mathcal{Q}} \dop{1} \psi}_{p} = \mathcal{O}_{P}(T^{- 6 / 5 + 3 / p})$ for $1 \leq p \leq 2$.
\vspace{0.5em}

\noindent\# (xiii) For all $\beta \in \mathfrak{B}(\varepsilon)$, let $\tilde{\phi}(\beta) \coloneqq \underset{\phi \in \mathfrak{P}(\eta, q)}{\argmin} \norm{u(\beta, \phi)}_{q}$, so $\norm{u(\beta, \tilde{\phi}(\beta))}_{q} \leq \norm{u(\beta, \phi^{0})}_{q}$. By the Legendre transformation (Section \ref{supplement:proof_of_asymptotic_expansions}) and a Taylor expansion of $\phi^{\ast}(\beta, u(\beta, \tilde{\phi}(\beta)))$ around $\upsilon = u(\beta, \phi^{0})$, $\tilde{\phi}(\beta) = \phi^{0} + (H(\beta, \phi^{\ast}(\beta, \check{\upsilon})))^{- 1} (u(\beta, \tilde{\phi}(\beta)) - u(\beta, \phi^{0}))$, where $\check{\upsilon}$ lies in the segment between $u(\beta, \tilde{\phi}(\beta))$ and $u(\beta, \phi^{0})$. By the triangle inequality, $\sup_{\beta} \norm{u(\beta, \tilde{\phi}(\beta)) - u(\beta, \phi^{0})}_{q} \leq 2 \, \sup_{\beta} \norm{u(\beta, \phi^{0})}_{q}$. By a Taylor expansion of $u(\beta, \phi^{0})$ around $\beta^{0}$, (v), (x), and $\sup_{\beta} \norm{\beta - \beta^{0}}_{2} \leq \varepsilon = o(T^{- 1 / 2})$, $\sup_{\beta} \bignorm{u(\beta, \phi^{0})}_{q} = \mathcal{O}_{P}(T^{- 1 / 2 + 2 / q})$. Hence, for $q > 5$, by (viii), $\sup_{\beta} \bignorm{\tilde{\phi}(\beta) - \phi^{0}}_{q} = o_{P}(\eta)$. So $\tilde{\phi}(\beta)$ is an interior solution wpa1. Since $\mathcal{L}_{n}(\beta, \phi)$ is strictly convex in $\phi$ for all $(\beta, \phi) \in \mathfrak{B}(\varepsilon) \times \mathfrak{P}(\eta, q)$ wpa1 by (viii), $u(\beta, \tilde{\phi}(\beta)) = \mathbf{0}_{L}$ wpa1, so $\mathbf{0}_{L} \in u(\beta, \mathfrak{P}(\eta, q))$ for all $\beta \in \mathfrak{B}(\varepsilon)$ wpa1.
\vspace{0.5em}

\noindent\# (xiv) By the Courant–Fischer–Weyl min-max principle, \linebreak $\lambda_{\min}(\overline{W}) = \underset{\{v \in \mathbb{R}^{K} \colon \norm{v}_{2} = 1\}}{\min} \tfrac{1}{N_{1} N_{2} T} \sum_{i = 1}^{N_{1}} \sum_{j = 1}^{N_{2}} \sum_{t = 1}^{T} \EX{(\dop{2} \psi)_{ijt} \{(\ddot{X} v)_{ijt} \}^{2}} \geq c_{W}$ a.\,s. Hence, $\overline{W} > 0$ a.\,s. and $\norm{\overline{W}^{- 1}}_{2} = \mathcal{O}_{P}(1)$. By the triangle inequality, $\norm{U}_{2} \leq K \, \max_{k} \bigabs{U_{1} e_{k}} + \ldots$. By Lemma \ref{lemma:moment_bounds_mixing}, $\max_{k} \abs{U_{1} e_{k}} = \mathcal{O}_{P}(1)$. By Jensen's inequality, the Cauchy-Schwarz inequality, the Lyapunov inequality, and (iii), $\max_{k} \abs{\overline{U}_{2, 1} e_{k}} = \mathcal{O}_{P}(1)$ and $\max_{k} \abs{\overline{U}_{2, 2} e_{k}} = \mathcal{O}_{P}(1)$. Using $\sum_{t = 1}^{T} (\mathfrak{D}_{\pi}^{2} e_{k})_{ijt} = \sum_{t = 1}^{T} (\widetilde{\mathfrak{D}_{\pi}^{2}} e_{k})_{ijt}$ implied by \eqref{eq:population_wls_program},
\begin{align*}
	&\max_{k} \bigabs{\overline{U}_{2, 3} e_{k}} \leq C \left(\sup_{ij} \EX{\bigg(\frac{1}{\sqrt{T}} \sum_{t = 1}^{T} \max_{k} (\widetilde{\mathfrak{D}_{\pi}^{2}} e_{k})_{ijt}\bigg)^{2}}\right)^{\frac{1}{2}} \left(\sup_{ij} \EX{\bigg(\frac{1}{\sqrt{T}} \sum_{t = 1}^{T} (\dop{1} \psi)_{ijt}\bigg)^{2}}\right)^{\frac{1}{2}} \\
	& \quad \leq C \quad \text{a.\,s.} \, ,
\end{align*}
so $\max_{k} \abs{U_{2, 3} e_{k}} = \mathcal{O}_{P}(1)$ by Markov's inequality. By similar arguments, $\max_{k} \abs{\overline{U}_{3, 1} e_{k}} = \mathcal{O}_{P}(1)$, $\max_{k} \abs{\overline{U}_{3, 2} e_{k}} = \mathcal{O}_{P}(1)$, and $\max_{k} \abs{U_{3, 3} e_{k}} = \mathcal{O}_{P}(1)$. Hence, $\norm{U}_{2} = \mathcal{O}_{P}(T^{1/2})$. By similar arguments, $\norm{\overline{B}_{\alpha}}_{2} = \mathcal{O}_{P}(1)$, $\norm{\overline{B}_{\gamma}}_{2} = \mathcal{O}_{P}(1)$, and $\norm{\overline{B}_{\rho}}_{2} = \mathcal{O}_{P}(1)$.
\vspace{0.5em}

\noindent\# (xv) Recall $W = (N_{1} N_{2} T)^{- 1} (M X)^{\prime} \nabla^{2} \psi M X$. Decomposing
\begin{equation*}
	W - \overline{W} = \frac{\widetilde{X^{\prime} \nabla^{2} \psi X}}{N_{1} N_{2} T} - \frac{\mathfrak{X}^{\prime} \widetilde{\nabla^{2} \psi X}}{N_{1} N_{2} T} - \frac{(\widetilde{\nabla^{2} \psi X})^{\prime} \mathfrak{X}}{N_{1} N_{2} T} + \frac{\mathfrak{X}^{\prime} \widetilde{\nabla^{2} \psi} \mathfrak{X}}{N_{1} N_{2} T} + \ldots \eqqcolon \mathfrak{E}_{1} + \ldots + \mathfrak{E}_{10} \, .
\end{equation*}
By (ii), (v), (x), and (xii), $\norm{\mathfrak{E}_{r}}_{2} = o_{P}(T^{- 1 / 2})$ for $r \in \{7, \ldots, 10\}$. By Lemma \ref{lemma:moment_bounds_mixing},
\begin{align*}
	&\EX{\max_{k, k^{\prime}} \bigabs{e_{k}^{\prime} \mathfrak{E}_{1} e_{k^{\prime}}}^{2}} \leq \bigg(\frac{T}{N_{1}}\bigg) \bigg(\frac{T}{N_{2}}\bigg) \, T^{- 3} \sup_{ij} \EX{\bigg(\frac{1}{\sqrt{T}} \sum_{t = 1}^{T} \max_{k, k^{\prime}} \widetilde{(\dop{2} \psi)_{ijt} (X e_{k})_{ijt} (X e_{k^{\prime}})_{ijt}}\bigg)^{2}} \\
	& \quad = \mathcal{O}\big(T^{- 3}\big) \quad \text{a.\,s.} \, ,
\end{align*}
so $\norm{\mathfrak{E}_{1}}_{2} = o_{P}(T^{- 1 / 2})$ by Markov's inequality. By Jensen's inequality, the Cauchy-Schwarz inequality, the Lyapunov inequality, and (v), $\norm{\mathfrak{E}_{r}}_{2} = o_{P}(T^{- 1 / 2})$ for $r \in \{2, 3, 4\}$. For each $k, k^{\prime} \in \{1, \ldots, K\}$,
\begin{align*}
	&e_{k}^{\prime} \mathfrak{E}_{5} e_{k^{\prime}} = \frac{1}{N_{1} N_{2} T} \sum_{i = 1}^{N_{1}} \sum_{t = 1}^{T} \frac{\big\{\sum_{j = 1}^{N_{2}} (\overline{\mathfrak{D}_{\pi}^{2}} e_{k})_{ijt}\big\} \big\{\sum_{j = 1}^{N_{2}} (\widetilde{\mathfrak{D}_{\pi}^{2}} e_{k^{\prime}})_{ijt}\big\}}{\sum_{j = 1}^{N_{2}} (\overline{\dop{2} \psi})_{ijt}} + \ldots \\
    &\quad \eqqcolon \mathfrak{E}_{5, 1, k, k^{\prime}} + \mathfrak{E}_{5, 2, k, k^{\prime}} + \mathfrak{E}_{5, 3, k, k^{\prime}} \, .
\end{align*}
Let $\overline{\vartheta}_{1, k} \coloneqq D_{1}^{\prime} \overline{\mathfrak{D}_{\pi}^{2}} e_{k} \odot \overline{f}_{1}^{\circ - 1}$. By Jensen's inequality, the Lyapunov inequality, (iii), and similar arguments, $\EX{\max_{k, k^{\prime}} \abs{\mathfrak{E}_{5, 1, k, k^{\prime}}}^{2}} = \mathcal{O}(T^{- 3})$ a.\,s., so $\max_{k, k^{\prime}}\abs{\mathfrak{E}_{5, 1, k, k^{\prime}}} = o_{P}(T^{- 1 / 2})$ by Markov's inequality. Analogously, $\max_{k, k^{\prime}}\abs{\mathfrak{E}_{5, 2, k, k^{\prime}}} = o_{P}(T^{- 1 / 2})$ and $\max_{k, k^{\prime}}\abs{\mathfrak{E}_{5, 3, k, k^{\prime}}} = o_{P}(T^{- 1 / 2})$. Hence, $\norm{\mathfrak{E}_{5}}_{2} = o_{P}(T^{- 1 / 2})$ and $\norm{\mathfrak{E}_{6}}_{2} = o_{P}(T^{- 1 / 2})$. By the triangle inequality, $\norm{W - \overline{W}} = o_{P}(T^{- 1 / 2})$.
\vspace{0.5em}

\noindent\# (xvi) For all $(\beta, \phi) \in \mathfrak{B}(\varepsilon) \times \mathfrak{P}(\eta, q)$, recall $W(\beta, \phi) = (N_{1} N_{2} T)^{- 1} (\nabla^{2} \psi(\beta, \phi) X)^{\prime} M(\beta, \phi) X$. By a Taylor expansion around $(\beta^{0}, \phi^{0})$ (see \textcite{fwct2014}),
\begin{equation*}
	e_{k}^{\prime} W(\beta, \phi) - W e_{k^{\prime}} = \frac{e_{k}^{\prime} X^{\prime} \check{\nabla^{3} \psi} \diag(X (\beta - \beta^{0})) \check{M} X e_{k^{\prime}}}{N_{1} N_{2} T} - \ldots \eqqcolon \mathfrak{E}_{1, k, k^{\prime}}(\beta) + \ldots + \mathfrak{E}_{4, k, k^{\prime}}(\phi) \, .
\end{equation*}
By H\"older's inequality, (vii), (x), (xi), and $\sup_{\beta} \norm{\beta - \beta^{0}}_{2} \leq \varepsilon$, $\sup_{\phi} \norm{\phi - \phi^{0}}_{q} \leq \eta$, $\sup_{\beta} \max_{k, k^{\prime}} \abs{\mathfrak{E}_{r, k, k^{\prime}}(\beta)} = o_{P}(1)$ for $r \in \{1, \ldots, 4\}$. Hence, $\sup_{(\beta, \phi)} \, \norm{W(\beta, \phi) - W}_{2} = o_{P}(1)$. By (xiv) and (xv), the conditions of Lemma \ref{lemma:invertibility} with $\bar{p} = 2$ hold wpa1. Thus, $W(\beta, \phi)$ is invertible for all $(\beta, \phi) \in \mathfrak{B}(\varepsilon) \times \mathfrak{P}(\eta, q)$ wpa1.\hfill\qedsymbol

\paragraph{Proof of Lemma \ref{lemma:regularity_conditions2_interacted}.} \# (i) By a Taylor expansion of $\widehat{\mathfrak{D}_{\pi}^{\mathrm{r}}} e_{k} = \nabla^{\mathrm{r}} \psi(\hat{\beta}, \hat{\phi}) M(\hat{\beta}, \hat{\phi}) X e_{k}$, $\mathrm{r} \in \{1, 2, 3\}$, around $(\beta^{0}, \phi^{0})$ for each $k \in \{1, \ldots, K\}$ (see \textcite{fwct2014}),
\begin{equation*}
	\widehat{\mathfrak{D}_{\pi}^{\mathrm{r}}} e_{k} = \nabla^{\mathrm{r}} \psi M X e_{k} + \nabla^{\mathrm{r} + 1} \psi \diag(M X e_{k}) (\hat{\pi} - \pi^{0}) - \nabla^{\mathrm{r}} \psi Q \nabla^{3} \psi \diag(M X e_{k}) (\hat{\pi} - \pi^{0}) + \ldots \, ,
\end{equation*}
where $\check{\beta}$, $\check{\phi}$ can differ for each element and each $k$. Decomposing further,
\begin{align*}
	&\widehat{\mathfrak{D}_{\pi}^{\mathrm{r}}} e_{k} - \mathfrak{D}_{\pi}^{\mathrm{r}} e_{k} = - \nabla^{\mathrm{r}} \psi \overline{\mathcal{Q}} \widetilde{\mathfrak{D}_{\pi}^{2}} e_{k} - \diag(\mathfrak{D}_{\pi}^{\mathrm{r} + 1} e_{k}) \overline{\mathcal{Q}} \dop{1} \psi + \nabla^{\mathrm{r}} \psi \overline{\mathcal{Q}} \diag(\overline{\mathfrak{D}_{\pi}^{3}} e_{k}) \overline{\mathcal{Q}} \dop{1} \psi + \ldots \\
	&\quad \eqqcolon \mathfrak{E}_{1, k} + \ldots + \mathfrak{E}_{17, k} \, .
\end{align*}
By H\"older's inequality and Corollary \ref{corollary:consistency_interacted}, $\max_{k} \norm{\mathfrak{E}_{r, k}}_{2} = \mathcal{O}_{P}(T)$ for $r \in \{1, 2, 3\}$ and $\max_{k} \norm{\mathfrak{E}_{r, k}}_{2} = o_{P}(T^{9 / 10})$ for $r \in \{4, \ldots, 17\}$. By the triangle inequality, for $1 \leq p \leq 2$ and $1 \leq \mathrm{r} \leq 3$, $\max_{k} \norm{(\widehat{\mathfrak{D}_{\pi}^{\mathrm{r}}} - \mathfrak{D}_{\pi}^{\mathrm{r}}) e_{k}}_{p} = \mathcal{O}_{P}(T^{- 1 / 2 + 3 / p})$ and $\max_{k} \norm{(\widehat{\mathfrak{D}_{\pi}^{\mathrm{r}}} - \mathfrak{D}_{\pi}^{\mathrm{r}} + \ldots) e_{k}}_{p} = o_{P}\big(T^{- \frac{3}{5} + \frac{3}{p}}\big)$. By a Taylor expansion of $\widehat{\dop{\mathrm{r}} \psi} = \dop{\mathrm{r}} \psi(\hat{\beta}, \hat{\phi})$, $\mathrm{r} \in \{1, 2, 3\}$, around $(\beta^{0}, \phi^{0})$ (see \textcite{fwct2014}), $\widehat{\dop{\mathrm{r}} \psi} = \dop{\mathrm{r}} \psi + \nabla^{\mathrm{r} + 1} \psi (\hat{\pi} - \pi^{0}) + \check{\nabla}^{\mathrm{r} + 2} \psi (\hat{\pi} - \pi^{0})^{\circ 2}$. Decomposing further, $\widehat{\dop{\mathrm{r}} \psi} - \dop{\mathrm{r}} \psi = - \nabla^{\mathrm{r} + 1} \psi \overline{\mathcal{Q}} \dop{1} \psi + \ldots \eqqcolon \mathfrak{E}_{1} + \ldots + \mathfrak{E}_{5}$. By H\"older's inequality and Corollary \ref{corollary:consistency_interacted}, $\norm{\mathfrak{E}_{1}}_{2} = \mathcal{O}_{P}(T)$ and $\norm{\mathfrak{E}_{r}}_{2} = o_{P}(T^{9 / 10})$ for $r \in \{2, \ldots, 5\}$. By the triangle inequality, for $1 \leq p \leq 2$ and $1 \leq \mathrm{r} \leq 3$, $\norm{\widehat{\dop{\mathrm{r}} \psi} - \dop{\mathrm{r}} \psi}_{p} = \mathcal{O}_{P}(T^{- 1 / 2 + 3 / p})$ and $\norm{\widehat{\dop{\mathrm{r}} \psi} - \dop{\mathrm{r}} \psi + \nabla^{\mathrm{r} + 1} \psi \overline{\mathcal{Q}} \dop{1} \psi}_{p} = o_{P}(T^{- 3 / 5 + 3 / p})$.
\vspace{0.5em}

\noindent\# (ii) Since $\widetilde{F}$ is diagonal and $\norm{D^{\prime} \widetilde{\dop{2} \psi}}_{20} = \mathcal{O}_{P}(T^{3 / 5})$, $\norm{\widetilde{F}}_{\infty} = T^{- 1} \norm{D^{\prime} \widetilde{\dop{2} \psi}}_{\infty} \leq T^{- 1} \norm{D^{\prime} \widetilde{\dop{2} \psi}}_{20} = \mathcal{O}_{P}(T^{- 2 / 5})$.
\vspace{0.5em}

\noindent\# (iii) By a Taylor expansion of $F(\beta, \phi)$ around $(\beta^{0}, \phi^{0})$, Lemma \ref{lemma:matrix_norm_inequalties}, the triangle inequality, and $\sup_{\beta} \norm{\beta - \beta^{0}}_{2} \leq \varepsilon$, $\sup_{\phi} \norm{\phi - \phi^{0}}_{q} \leq \eta$, $\sup_{(\beta, \phi)} \, \norm{F(\beta, \phi) - F}_{p} = o_{P}(1)$. By (vi), the conditions of Lemma \ref{lemma:invertibility} with $\bar{p} = \infty$ hold wpa1. Thus $F(\beta, \phi)$ is invertible for all $(\beta, \phi) \in \mathfrak{B}(\varepsilon) \times \mathfrak{P}(\eta, q)$ wpa1, and $\sup_{(\beta, \phi)} \, \norm{(F(\beta, \phi))^{- 1}}_{\infty} = \mathcal{O}_{P}(1)$.
\vspace{0.5em}

\noindent\# (iv) By Theorem \ref{theorem:consistency_interacted}, $\hat{\beta}$ and $\hat{\phi}$ are interior to $\mathfrak{B}(\varepsilon) \times \mathfrak{P}(\eta, q)$. Hence, by (iii), $(\hat{f})_{l} > 0$ wpa1 for all $l$, and $\norm{\hat{f}^{\circ - 1}}_{\infty} = T^{- 1} \norm{\widehat{F}^{- 1}}_{\infty} = \mathcal{O}_{P}(T^{- 1})$. Let $f^{\ast} \coloneqq D^{\prime} \overline{\nabla^{3} \psi} \, \overline{\mathcal{Q}} \dop{1} \psi$. Decomposing $\hat{f}^{\circ - 1} = \bar{f}^{\circ - 1} - \bar{f}^{\circ - 2} \odot (\hat{f} - f + f^{\ast} - f^{\ast} + \tilde{f}) + \hat{f}^{\circ - 1} \odot \bar{f}^{\circ - 2} \odot (\hat{f} - f + \tilde{f})^{\circ 2}$. By a Taylor expansion of $\hat{f} = D^{\prime} \widehat{\dop{2} \psi}$ around $(\beta^{0}, \phi^{0})$ (see \textcite{fwct2014}), $\hat{f} - f = - f^{\ast} + T^{- 1} D^{\prime} \overline{\nabla^{3} \psi} D \overline{H}^{- 1} \overline{G} \, \overline{F}^{- 1} D^{\prime} \dop{1} \psi - \ldots \eqqcolon -f^{\ast} + \mathfrak{E}_{1} + \ldots + \mathfrak{E}_{7}$. By H\"older's inequality, Theorem \ref{theorem:consistency_interacted}, and Corollary \ref{corollary:consistency_interacted}, $\norm{f^{\ast}}_{10} = \mathcal{O}_{P}(T^{7 / 10})$ and $\norm{\mathfrak{E}_{r}}_{10} = o_{P}(T^{7 / 10})$ for $r \in \{1, \ldots, 7\}$. By the triangle inequality, for $1 \leq p \leq 10$, $\bignorm{\hat{f} - f}_{p} = \mathcal{O}_{P}(T^{1 / 2 + 2 / p})$ and $\norm{\hat{f} - f + f^{\ast}}_{p} = o_{P}(T^{1 / 2 + 2 / p})$. By the triangle inequality and $\norm{\tilde{f}}_{p} = \norm{D^{\prime} \widetilde{\dop{2} \psi}}_{p} = \mathcal{O}_{P}(T^{1 / 2 + 2 / p})$, $\norm{\hat{f} - f + \tilde{f}}_{p} \leq \norm{\hat{f} - f}_{p} + \norm{\tilde{f}}_{p} = \mathcal{O}_{P}\big(T^{1 / 2 + 2 / p}\big)$. By H\"older's inequality, for $1 \leq p \leq 10$, $\norm{\hat{f}^{\circ - 1} - \bar{f}^{\circ - 1}}_{p} = \mathcal{O}_{P}(T^{- 3 / 2 + 2 / p})$ and $\norm{\hat{f}^{\circ - 1} - \bar{f}^{\circ - 1} - \ldots}_{p} = o_{P}(T^{- 3 / 2 + 2 / p})$.\hfill\qedsymbol

\subsection{Non-interacted Specification}

Before verifying the implied regularity conditions for the non-interacted linear index specification with $\mathcal{M} = 3$, $\pi_{ijt}(\beta, \mu_{ijt}(\phi)) = (X \beta)_{ijt} + (D \phi)_{ijt} = x_{ijt}^{\prime} \beta + \alpha_{i}^{\star} + \gamma_{j}^{\star} + \rho_{t}^{\star}$, the sparse regressor matrix is $D = (I_{N_{1}} \otimes \iota_{N_{2}} \otimes \iota_{T}, \, \iota_{N_{1}} \otimes I_{N_{2}} \otimes \iota_{T}, \, \iota_{N_{1}} \otimes \iota_{N_{2}} \otimes I_{T})$. We impose the following constraints on $\phi$,
\begin{equation*}
	V = 	
	\begin{pmatrix}
		\sqrt{T} \iota_{N_{1}}   &  \sqrt{N_{2}} \iota_{N_{1}}     & 0	\\
		-  \sqrt{T} \iota_{N_{2}}  & 0         &   \sqrt{N_{1}} \iota_{N_{2}} \\
		0      & -  \sqrt{N_{2}} \iota_{T}   & -  \sqrt{N_{1}} \iota_{T}
	\end{pmatrix} \eqqcolon 
	\begin{pmatrix}
		V_{1} & 	V_{2} & 	V_{3}\\
		V_{4} & 	V_{5} & 	V_{6}\\
		V_{7} & 	V_{8} & 	V_{9}
	\end{pmatrix} \, .
\end{equation*}

\paragraph{Proof of Lemma \ref{lemma:regularity_conditions1_noninteracted}.} \# (i) By the triangle inequality and $\norm{A \otimes B} = \norm{A} \norm{B}$, $\norm{D}_{1} \leq N_{2} T + N_{1} T + N_{1} N_{2} = \mathcal{O}(T^2)$ and $\norm{D}_{\infty} = 3$. Hence, by Lemma \ref{lemma:matrix_norm_inequalties}, $\norm{D}_{p} = \mathcal{O}(T^{2/p})$.
\vspace{0.5em}

\# (ii) By the Courant–Fischer–Weyl min-max principle, $\lambda_{\min}(\overline{H}) =  \min_{\norm{v}_{2} = 1} v^{\prime} ((D^{\prime} \overline{\nabla^{2} \psi} D + V V^{\prime}) / T^2) v$. By the same Weyl's inequality argument as in the interacted case (using $\lambda_{\min}(VV^{\prime})=0$), $\lambda_{\min}(\overline{H}) \geq \min(1, c_{H}) \, \lambda_{\min}(\mathbb{H})$ and $\mathbb{H} \coloneqq (D^{\prime} D + V V^{\prime}) / T^2$. The constraint structure gives $\mathbb{H}$ a $3\times3$ block-diagonal form with
\begin{align*}
	&\mathbb{H}_{1} \coloneqq \frac{N_{2} T}{T^{2}} I_{N_{1}}+\frac{N_{2} + T}{T^{2}} \iota_{N_{1}} \iota_{N_{1}}^{\prime} \, , \quad
	\mathbb{H}_{2} \coloneqq \frac{N_{1} T}{T^{2}} I_{N_{2}}+\frac{N_{1}+T}{T^{2}} \iota_{N_{2}} \iota_{N_{2}}^{\prime} \, , \\
	&\mathbb{H}_{3} \coloneqq \frac{N_{1}  N_{2}}{T^{2}} I_{T}+\frac{N_{1} + N_{2}}{T^{2}} \iota_{T} \iota_{T}^{\prime} \, .
\end{align*}
Invertibility of each block follows from the Sherman-Morrison formula. It follows that $\norm{\mathbb{H}^{-1}}_\infty = \mathcal{O}(1)$ and $\overline{H}>0$ a.\,s. For $c_H\geq1$, $\norm{\overline{H}^{-1}}_\infty\leq\norm{\mathbb{H}^{-1}}_\infty=\mathcal{O}(1)$. For $c_H<1$, $\norm{\overline{H}^{-1}}_\infty\leq c_H^{-1}\norm{\mathbb{H}^{-1}}_\infty=\mathcal{O}(1)$ a.\,s. By (i) and Lemma \ref{lemma:matrix_norm_inequalties}, $\norm{\overline{Q}}_{\infty} = \mathcal{O}(1)$ a.\,s.
\vspace{0.5em}

\noindent\# (iii) For $1 \leq \mathrm{r} \leq 5$, by the triangle inequality and ii), $\sup_{ijt} \EX{\max_{k} \abs{(\mathfrak{D}_{\pi}^{\mathrm{r}} e_{k})_{ijt}}^{20 + \nu}} \leq C$ a.\,s. Hence, by the Lyapunov inequality, $\sup_{ijt} \EX{\max_{k} \abs{(\mathfrak{D}_{\pi}^{\mathrm{r}} e_{k})_{ijt}}^{p}} \leq C$ a.\,s. for $1 \leq p \leq 20 + \nu$ and $1 \leq \mathrm{r} \leq 5$.
\vspace{0.5em}

\noindent\# (iv) For $p \geq 1$, by Jensen's inequality and the Lyapunov inequality, $\max_{k} \norm{\overline{\nabla^{2} \psi X} e_{k}}_{p} = \mathcal{O}(T^{3/p})$ a.\,s. By (i) and (ii), $\max_{k} \norm{\xi_{k}^{0}}_{p} = \mathcal{O}(T^{1 / p})$ a.\,s.
\hfill\qedsymbol
\vspace{0.5em}

\noindent\# v) For $1 \leq \mathrm{r} \leq 5$, by the triangle inequality, $\norm{D^{\prime} \widetilde{\dop{\mathrm{r}} \psi}}_{20} \leq \norm{D_{1}^{\prime} \widetilde{\dop{\mathrm{r}} \psi}}_{20} + \norm{D_{2}^{\prime} \widetilde{\dop{\mathrm{r}} \psi}}_{20} + \norm{D_{3}^{\prime} \widetilde{\dop{\mathrm{r}} \psi}}_{20}$. By Rosenthal's inequality and Lemma \ref{lemma:moment_bounds_mixing},
\begin{align*}
	\EX{\bignorm{D_{1}^{\prime} \widetilde{\dop{\mathrm{r}} \psi}}_{20}^{20}} &\leq \bigg(\frac{N_{1}}{T}\bigg) \bigg(\frac{N_{2}}{T}\bigg)^{10} T^{21} \sup_{i} \EX{\bigg( \frac{1}{\sqrt{N_{2} T}} \sum_{j, t} (\widetilde{\dop{\mathrm{r}} \psi})_{ijt} \bigg)^{20}} = \mathcal{O}(T^{21}) \quad \text{a.\,s.}
\end{align*}
Analogously, $\EX{\norm{D_{2}^{\prime} \widetilde{\dop{\mathrm{r}} \psi}}_{20}^{20}} = \mathcal{O}(T^{21})$ and $\EX{\norm{D_{3}^{\prime} \widetilde{\dop{\mathrm{r}} \psi}}_{20}^{20}} = \mathcal{O}(T^{21})$ a.\,s. Hence, $\norm{D^{\prime} \widetilde{\dop{\mathrm{r}} \psi}}_{20} = \mathcal{O}_{P}(T^{21/20})$ by Markov's inequality, and for $1 \leq p \leq 20$,
\begin{equation*}
	\bignorm{D^{\prime} \widetilde{\dop{\mathrm{r}} \psi}}_{p} \leq \bigg(\frac{N_{1}}{T} + \frac{N_{2}}{T}\bigg)^{\frac{1}{p} - \frac{1}{20}} T^{\frac{1}{p} - \frac{1}{20}} \bignorm{D^{\prime} \widetilde{\dop{\mathrm{r}} \psi}}_{20} = \mathcal{O}_{P}\big(T^{1 + \frac{1}{p}}\big) \, .
\end{equation*}
By similar arguments, $\max_{k} \norm{D^{\prime} \widetilde{\nabla^{\mathrm{r}} \psi X} e_{k}}_{p} = \mathcal{O}_{P}(T^{1 + 1/p})$.
\vspace{0.5em}

\noindent\# vi) For $1 \leq \mathrm{r} \leq 5$, by the triangle inequality,
\begin{align*}
	&\bignorm{D^{\prime} \widetilde{\nabla^{\mathrm{r}} \psi} D}_{2} \leq \bignorm{D_{1}^{\prime} \widetilde{\nabla^{\mathrm{r}} \psi} D_{1}}_{2} + \bignorm{D_{2}^{\prime} \widetilde{\nabla^{\mathrm{r}} \psi} D_{2}}_{2} + \bignorm{D_{3}^{\prime} \widetilde{\nabla^{\mathrm{r}} \psi} D_{3}}_{2} + \\
	&\quad 2 \, \bignorm{D_{1}^{\prime} \widetilde{\nabla^{\mathrm{r}} \psi} D_{2}}_{2} + 2 \, \bignorm{D_{1}^{\prime} \widetilde{\nabla^{\mathrm{r}} \psi} D_{3}}_{2} + 2 \, \bignorm{D_{2}^{\prime} \widetilde{\nabla^{\mathrm{r}} \psi} D_{3}}_{2} \, .
\end{align*}
The first three are diagonal, and using $\norm{D^{\prime} \widetilde{\dop{\mathrm{r}} \psi}}_{20} = \mathcal{O}_{P}(T^{21/20})$ by v), $\norm{D_{1}^{\prime} \widetilde{\nabla^{\mathrm{r}} \psi} D_{1}}_{2} = \norm{D_{1}^{\prime} \widetilde{\dop{\mathrm{r}} \psi}}_{\infty} \leq \norm{D^{\prime} \widetilde{\dop{\mathrm{r}} \psi}}_{20} = \mathcal{O}_{P}(T^{21 / 20})$, and analogously for $D_2$ and $D_3$. The off-diagonal blocks have explicit forms:
\begin{align*}
	D_{1}^{\prime} \widetilde{\nabla^{r} \psi} D_{2} =& \, \begin{pmatrix}
		\sum_{t} (\widetilde{\dop{\mathrm{r}} \psi})_{11t} & \cdots & \sum_{t} (\widetilde{\dop{\mathrm{r}} \psi})_{1N_{2}t} \\
		\vdots&\ddots&\vdots \\
		\sum_{t} (\widetilde{\dop{\mathrm{r}} \psi})_{N_{1}1t} & \cdots & \sum_{t} (\widetilde{\dop{\mathrm{r}} \psi})_{N_{1}N_{2}t}
	\end{pmatrix} \, , \\
	D_{1}^{\prime} \widetilde{\nabla^{\mathrm{r}} \psi} D_{3} =& \, \begin{pmatrix}
		\sum_{j} (\widetilde{\dop{\mathrm{r}} \psi})_{1j1} & \cdots & \sum_{j} (\widetilde{\dop{\mathrm{r}} \psi})_{1jT} \\
		\vdots&\ddots&\vdots \\
		\sum_{j} (\widetilde{\dop{\mathrm{r}} \psi})_{N_{1}j1} & \cdots & \sum_{j} (\widetilde{\dop{\mathrm{r}} \psi})_{N_{1}jT}
	\end{pmatrix} \, , \\
	D_{2}^{\prime} \widetilde{\nabla^{\mathrm{r}} \psi} D_{3} =& \, \begin{pmatrix}
		\sum_{i} (\widetilde{\dop{\mathrm{r}} \psi})_{i11} & \cdots & \sum_{i} (\widetilde{\dop{\mathrm{r}} \psi})_{i1T} \\
		\vdots&\ddots&\vdots \\
		\sum_{i} (\widetilde{\dop{\mathrm{r}} \psi})_{iN_{2}1} & \cdots & \sum_{i} (\widetilde{\dop{\mathrm{r}} \psi})_{iN_{2}T}
	\end{pmatrix} \, .
\end{align*}
Let $a_{jt}^{(2, 3)} \coloneqq N_{1}^{-1/2} \sum_{i} (\widetilde{\dop{\mathrm{r}} \psi})_{ijt}$. $\sup_{jt} \EX{|a_{jt}^{(2,3)}|^{20+\nu}} \leq C$ a.\,s. by Rosenthal's inequality. By Lemma \ref{lemma:asymptotic_bound_spectral_norm} with $p = 20 + \check{\nu}$ and $b = 1$, $\EX{\norm{D_{2}^{\prime} \widetilde{\nabla^{\mathrm{r}} \psi} D_{3} / \sqrt{N_{1}}}_{2}^{20 + \check{\nu}}} = \mathcal{O}((\sqrt{\log T})^{20 + \check{\nu}} T^{(22 + \check{\nu})/2})$ a.\,s., so $\norm{D_{2}^{\prime} \widetilde{\nabla^{\mathrm{r}} \psi} D_{3}}_{2} = \mathcal{O}_{P}(T^{21/20})$ by Markov's inequality. Analogously, $\norm{D_{1}^{\prime} \widetilde{\nabla^{\mathrm{r}} \psi} D_{3}}_{2} = \mathcal{O}_{P}(T^{21/20})$ and $\norm{D_{1}^{\prime} \widetilde{\nabla^{\mathrm{r}} \psi} D_{2}}_{2} = \mathcal{O}_{P}(T^{21/20})$ (using Lemmas \ref{lemma:moment_bounds_mixing} and \ref{lemma:latalas_theorem}). Hence, $\norm{D^{\prime} \widetilde{\nabla^{\mathrm{r}} \psi} D}_{2} = \mathcal{O}_{P}(T^{21/20})$. By the triangle inequality, Jensen's inequality, Loeve's $c_r$ inequality, and the union bound, $\EX{\norm{D^{\prime} \widetilde{\nabla^{\mathrm{r}} \psi} D}_{\infty}^{20 + \nu}} = \mathcal{O}(T^{1 + 2(20 + \nu)})$ a.\,s., so $\norm{D^{\prime} \widetilde{\nabla^{\mathrm{r}} \psi} D}_{\infty} = o_{P}(T^{41/20})$ by Markov's inequality. By Lemma \ref{lemma:matrix_norm_inequalties}, for $2 < p < \infty$, $\norm{D^{\prime} \widetilde{\nabla^{\mathrm{r}} \psi} D}_{p} \leq \norm{D^{\prime} \widetilde{\nabla^{\mathrm{r}} \psi} D}_{2}^{2 / p} \norm{D^{\prime} \widetilde{\nabla^{\mathrm{r}} \psi} D}_{\infty}^{1 - 2 / p} = o_{P}(T^{41 / 20 - 2 / p})$.
\vspace{0.5em}

\noindent\# vii) For $1 \leq \mathrm{r} \leq 5$, by the triangle inequality, Jensen's inequality, Loeve's $c_r$ inequality, and the union bound, $\EX{\sup_{(\beta, \phi)} \, \norm{D^{\prime} \abs{\nabla^{\mathrm{r}} \psi(\beta, \phi)} D}_{\infty}^{20 + \nu}} = \mathcal{O}(T^{1 + 2(20 + \nu)})$ a.\,s., so \linebreak $\sup_{(\beta, \phi)} \norm{D^{\prime} \abs{\nabla^{\mathrm{r}} \psi(\beta, \phi)} D}_{\infty} = o_{P}(T^{41/20})$ by Markov's inequality. By similar arguments and the Cauchy-Schwarz inequality, $\sup_{(\beta, \phi)} \max_{k} \norm{D^{\prime} \abs{\diag(\dop{\mathrm{r}} \psi(\beta, \phi) \odot X e_{k})} D}_{\infty} = o_{P}(T^{41/20})$.
\vspace{0.5em}

\noindent\# viii) For all $(\beta, \phi) \in \mathfrak{B}(\varepsilon) \times \mathfrak{P}(\eta, q)$, by a Taylor expansion of $H(\beta, \phi) = T^{-2} D^{\prime} \nabla^{2} \psi(\beta, \phi) D$ around $(\beta^0, \phi^0)$, Lemma \ref{lemma:matrix_norm_inequalties}, the triangle inequality, vii), and $\sup_\beta\norm{\beta-\beta^0}_2\leq\varepsilon$, $\sup_\phi\norm{\phi-\phi^0}_q\leq\eta$, $\sup_{(\beta, \phi)} \, \norm{H(\beta, \phi) - H}_{p} \leq \sup_{(\beta, \phi)} \, \norm{H(\beta, \phi) - H}_{\infty} = o_{P}(1)$. By ii) and vi), the conditions of Lemma \ref{lemma:invertibility} with $\bar{p} = 40$ hold wpa1. Thus $H(\beta, \phi)$ is invertible wpa1 and $\sup_{(\beta, \phi)} \norm{(H(\beta, \phi))^{-1}}_{p} = \mathcal{O}_{P}(1)$ for $2 \leq p \leq 40$. By i), $\sup_{(\beta, \phi)} \, \norm{Q(\beta, \phi)}_{p} \leq T^{- 2} \, \sup_{(\beta, \phi)} \, \norm{\big(H(\beta, \phi)\big)^{- 1}}_{p} \norm{D}_{p} \norm{D^{\prime}}_{p} = \mathcal{O}_{P}(1)$.
\vspace{0.5em}

\noindent\# ix) Consider $H = \overline{H} + \widetilde{H}$. By ii) and vi), the conditions of Lemma \ref{lemma:inverse_neumann_series} hold wpa1. By the triangle inequality and vi), $\norm{H^{- 1} - \overline{H}^{- 1}}_{2} = \mathcal{O}_{P}(T^{- 19/20})$ and $\norm{H^{- 1} - \overline{H}^{- 1}}_{p} = o_{P}(T^{1/20 - 2/p})$.
\vspace{0.5em}

\noindent\# x) For $1 \leq \mathrm{r} \leq 5$, by Jensen's inequality and the union bound, $\sup_{(\beta, \phi)} \norm{\dop{\mathrm{r}} \psi(\beta, \phi)}_{20 + \nu} = \mathcal{O}_{P}(T^{3/(20+\nu)})$ by Markov's inequality, and for $1 \leq p \leq 20 + \nu$, $\sup_{(\beta, \phi)} \norm{\dop{\mathrm{r}} \psi(\beta, \phi)}_{p} = \mathcal{O}_{P}\big(T^{\frac{3}{p}}\big)$. By similar arguments and iii), $\sup_{(\beta, \phi)} \max_{k} \norm{\dop{\mathrm{r}} \psi(\beta, \phi) \odot X e_{k}}_{p} = \mathcal{O}_{P}(T^{3/p})$, $\max_{k} \norm{X e_{k}}_{p} = \mathcal{O}_{P}(T^{3/p})$, and $\max_{k} \norm{\mathfrak{D}_{\pi}^{\mathrm{r}} e_{k}}_{p} = \mathcal{O}_{P}(T^{3/p})$. By i),
\begin{equation*}
	\sup_{(\beta, \phi)} \max_{k} \bignorm{D^{\prime} \nabla^{\mathrm{r}} \psi(\beta, \phi) X e_{k}}_{p} \leq \norm{D^{\prime}}_{p} \sup_{(\beta, \phi)} \max_{k} \bignorm{\dop{\mathrm{r}} \psi(\beta, \phi) \odot X e_{k}}_{p} = \mathcal{O}_{P}\big(T^{2 + \frac{1}{p}}\big) \, .
\end{equation*}

\noindent\# xi) By viii) and x), $\sup_{(\beta, \phi)} \max_{k} \norm{P(\beta, \phi) X e_{k}}_{p} = \mathcal{O}_{P}(T^{3/p})$ for $1 \leq p \leq 20+\nu$, and
\begin{equation*}
	\sup_{(\beta, \phi)} \max_{k} \, \bignorm{M(\beta, \phi) X e_{k}}_{p} \leq \max_{k} \, \norm{X e_{k}}_{p} + \sup_{(\beta, \phi)} \max_{k} \, \bignorm{P(\beta, \phi) X e_{k}}_{p} = \mathcal{O}_{P}\big(T^{\frac{3}{p}}\big) \, .
\end{equation*}
By i), v), and viii), $\sup_{(\beta, \phi)} \norm{Q(\beta, \phi) \dop{1} \psi}_{20} = \mathcal{O}_{P}(T^{-17/20})$ and thus $\sup_{(\beta, \phi)} \norm{Q(\beta, \phi) \dop{1} \psi}_{p} = \mathcal{O}_{P}(T^{-1+3/p})$ for $1 \leq p \leq 20$.
\vspace{0.5em}

\noindent\# xii) For $1 \leq \mathrm{r} \leq 2$ and each $k$, decomposing $\nabla^{\mathrm{r}} \psi M X e_{k}$,
\begin{align*}
	&\nabla^{\mathrm{r}} \psi M X e_{k} - \mathfrak{D}_{\pi}^{\mathrm{r}} e_{k} = - T^{- 2} \nabla^{\mathrm{r}} \psi D \overline{H}^{- 1} D^{\prime} \widetilde{\nabla^{2} \psi X} e_{k} - T^{- 2} \nabla^{\mathrm{r}} \psi D (H^{- 1} - \overline{H}^{- 1}) D^{\prime} \nabla^{2} \psi X e_{k} \\
	& \quad \eqqcolon \mathfrak{E}_{1, k} + \mathfrak{E}_{2, k} \, .
\end{align*}
By i), ii), v), ix), and x),
\begin{align*}
	&\max_{k} \bignorm{\mathfrak{E}_{1, k}}_{2} \leq T^{- 2} \bignorm{\overline{H}^{- 1}}_{\infty} \bignorm{\dop{\mathrm{r}} \psi}_{20 + \nu} \norm{D}_{2} \max_{k} \bignorm{D^{\prime} \widetilde{\nabla^{2} \psi X} e_{k}}_{2} = o_{P}\big(T^{\frac{3}{2}}\big) \, , \\
	&\max_{k} \bignorm{\mathfrak{E}_{2, k}}_{2} \leq T^{- 2} \bignorm{H^{- 1} - \overline{H}^{- 1}}_{2} \bignorm{\dop{\mathrm{r}} \psi}_{20 + \nu} \norm{D}_{2} \max_{k} \bignorm{D^{\prime} \nabla^{2} \psi X e_{k}}_{2} = o_{P}\big(T^{\frac{3}{2}}\big) \, .
\end{align*}
Hence, for $1 \leq p \leq 2$, $\max_{k} \norm{\nabla^{\mathrm{r}} \psi M X e_{k} - \mathfrak{D}_{\pi}^{\mathrm{r}} e_{k}}_{p} = o_{P}(T^{3/p})$.

\noindent\# xiii) For any $\beta \in \mathfrak{B}(\varepsilon)$, let $\tilde\phi(\beta)\coloneqq\argmin_{\phi\in\mathfrak{P}(\eta,q)}\norm{u(\beta,\phi)}_q$ where \linebreak $u(\beta,\phi)=T^{-2}D^{\prime}d^1\psi(\beta,\phi)$. By the Legendre transformation and a Taylor expansion around $\upsilon=u(\beta,\phi^0)$,
\begin{equation*}
	\tilde{\phi}(\beta) = \phi^{0} + \big(H(\beta, \phi^{\ast}(\beta, \check{\upsilon}))\big)^{- 1} \big(u(\beta, \tilde{\phi}(\beta)) - u(\beta, \phi^{0})\big) \, ,
\end{equation*}
with $\sup_\beta\norm{u(\beta,\tilde\phi(\beta))-u(\beta,\phi^0)}_q\leq2\sup_\beta\norm{u(\beta,\phi^0)}_q$. By a Taylor expansion of $u(\beta,\phi^0)$ around $\beta^0$, v), x), and $\sup_\beta\norm{\beta-\beta^0}_2\leq\varepsilon=o(T^{-1})$,
\begin{equation*}
	\sup_{\beta} \bignorm{u(\beta, \phi^{0})}_{q} \leq T^{- 2} \norm{D^{\prime} \dop{1} \psi}_{q} + K T^{- 2} \sup_{(\beta, \phi)} \max_{k} \bignorm{D^{\prime} \nabla^{2} \psi(\beta, \phi) X e_{k}}_{q} \varepsilon = \mathcal{O}_{P}\big(T^{- 1 + \frac{1}{q}}\big) \, .
\end{equation*}
Hence, for $q > 20/19$, by viii) with the Riesz-Thorin Theorem,
\begin{equation*}
	\sup_{\beta} \bignorm{\tilde{\phi}(\beta) - \phi^{0}}_{q} \leq 2 \sup_{(\beta, \phi)} \bignorm{(H(\beta, \phi))^{- 1}}_{q} \sup_{\beta} \bignorm{u(\beta, \phi^{0})}_{q} = \mathcal{O}_{P}\big(T^{- 1 + \frac{1}{q}}\big) = o_{P}(\eta) \, ,
\end{equation*}
so $\tilde\phi(\beta)$ is interior wpa1. Strict convexity of $\mathcal{L}_n$ in $\phi$ by viii) then gives $u(\beta,\tilde\phi(\beta))=\mathbf{0}_L$ wpa1, so $\mathbf{0}_L\in u(\beta,\mathfrak{P}(\eta,q))$ for all $\beta\in\mathfrak{B}(\varepsilon)$ wpa1.
\vspace{0.5em}

\noindent\# xiv) By the Courant–Fischer–Weyl min-max principle,
\begin{equation*}
	\lambda_{\min}(\overline{W}) = \min_{\norm{v}_2=1} \frac{1}{N_1N_2T}\sum_{i,j,t}\EX{(d^2\psi)_{ijt}\{(\ddot{X}v)_{ijt}\}^2} \geq c_W > 0 \quad \text{a.\,s.}
\end{equation*}
Hence $\overline{W}>0$ a.\,s. and $\norm{\overline{W}^{-1}}_2=\mathcal{O}_P(1)$. By Lemma \ref{lemma:moment_bounds_mixing} and iii),
\begin{equation*}
	\EX{\max_{k} \bigabs{U e_{k}}^{2}} \leq \sup_{ij} \EX{\bigg(\frac{1}{\sqrt{T}} \sum_{t = 1}^{T} \max_{k} (\mathfrak{D}_{\pi}^{1} e_{k})_{ijt}\bigg)^{2}} = \mathcal{O}(1) \quad \text{a.\,s.} \, ,
\end{equation*}
so $\norm{U}_2=\mathcal{O}_P(1)$ by Markov's inequality.
\vspace{0.5em}

\noindent\# xv) Recall $W = (N_{1} N_{2} T)^{- 1} (M X)^{\prime} \nabla^{2} \psi M X$. Decomposing
\begin{align*}
	&W - \overline{W} = \frac{\widetilde{X^{\prime} \nabla^{2} \psi X}}{N_{1} N_{2} T} - \frac{\mathfrak{X}^{\prime} \widetilde{\nabla^{2} \psi X}}{N_{1} N_{2} T} - \frac{(\widetilde{\nabla^{2} \psi X})^{\prime} \mathfrak{X}}{N_{1} N_{2} T} + \frac{\mathfrak{X}^{\prime} \widetilde{\nabla^{2} \psi} \mathfrak{X}}{N_{1} N_{2} T} + \frac{(M X)^{\prime} (\nabla^{2} \psi M X  - \mathfrak{D}_{\pi}^{2})}{N_{1} N_{2} T} \, + \\
	&\quad \frac{(\nabla^{2} \psi M X - \mathfrak{D}_{\pi}^{2})^{\prime} M X}{N_{1} N_{2} T} \eqqcolon \mathfrak{E}_{1} + \ldots + \mathfrak{E}_{6} \, .
\end{align*}
By the Cauchy-Schwarz inequality, x), and xii), $\norm{\mathfrak{E}_5}_2 = o_P(1)$ and $\norm{\mathfrak{E}_6}_2=o_P(1)$. By Lemma \ref{lemma:moment_bounds_mixing}, $\norm{\mathfrak{E}_1}_2=o_P(1)$. By Jensen's inequality, the Cauchy-Schwarz inequality, the Lyapunov inequality, and v), $\norm{\mathfrak{E}_r}_2=o_P(1)$ for $r\in\{2,3,4\}$. Hence, $\norm{W-\overline{W}}=o_P(1)$.
\vspace{0.5em}

\noindent\# xvi) For all $(\beta, \phi) \in \mathfrak{B}(\varepsilon) \times \mathfrak{P}(\eta, q)$, recall $W(\beta, \phi) = (N_{1} N_{2} T)^{- 1} (\nabla^{2} \psi(\beta, \phi) X)^{\prime} M(\beta, \phi) X$. By a Taylor expansion around $(\beta^0, \phi^0)$ (see \textcite{fwct2014}),
\begin{align*}
	&e_{k}^{\prime} W(\beta, \phi) - W e_{k^{\prime}} = \\
	&\quad\frac{e_{k}^{\prime} X^{\prime} \check{\nabla^{3} \psi} \diag(X (\beta - \beta^{0})) \check{M} X e_{k^{\prime}}}{N_{1} N_{2} T} - \frac{e_{k}^{\prime} X^{\prime} \check{\nabla^{2} \psi} \check{Q} \diag(X (\beta - \beta^{0})) \check{\nabla^{3} \psi} \check{M} X e_{k^{\prime}}}{N_{1} N_{2} T} \, + \\
	&\quad \frac{e_{k}^{\prime} X^{\prime} \check{\nabla^{3} \psi} \diag(D (\phi - \phi^{0})) \check{M} X e_{k^{\prime}}}{N_{1} N_{2} T} - \frac{e_{k}^{\prime} X^{\prime} \check{\nabla^{2} \psi} \check{Q} \diag(D (\phi - \phi^{0})) \check{\nabla^{3} \psi} \check{M} X e_{k^{\prime}}}{N_{1} N_{2} T} \\
	&\quad \eqqcolon \mathfrak{E}_{1, k, k^{\prime}}(\beta) + \ldots + \mathfrak{E}_{4, k, k^{\prime}}(\phi) \, .
\end{align*}
By H\"older's inequality, viii), x), xi), $\sup_\beta\norm{\beta-\beta^0}_2\leq\varepsilon$, and $\sup_\phi\norm{\phi-\phi^0}_q\leq\eta$,
\begin{align*}
	&\sup_{\beta} \max_{k, k^{\prime}} \bigabs{\mathfrak{E}_{1, k, k^{\prime}}(\beta)} \leq \\
	&\quad \bigg(\frac{T}{N_{1}}\bigg) \bigg(\frac{T}{N_{2}}\bigg) T^{- 3} \max_{k} \bignorm{\check{\nabla^{3} \psi} X e_{k}}_{2} \max_{k} \bignorm{\check{M} X e_{k}}_{2} \max_{k} \bignorm{X e_{k}}_{20 + \nu} \varepsilon = o_{P}(1) \, , \\
	&\sup_{\beta} \max_{k, k^{\prime}} \bigabs{\mathfrak{E}_{2, k, k^{\prime}}(\beta)} \leq \\
	&\quad \bigg(\frac{T}{N_{1}}\bigg) \bigg(\frac{T}{N_{2}}\bigg) T^{- 3} \bignorm{\check{Q}}_{2} \max_{k} \bignorm{\check{\nabla^{2} \psi} X e_{k}}_{2} \bignorm{\check{\dop{3} \psi}}_{4} \max_{k} \bignorm{\check{M} X e_{k}}_{4} \max_{k} \bignorm{X e_{k}}_{20 + \nu} \varepsilon = o_{P}(1) \, , \\
	&\sup_{\phi} \max_{k, k^{\prime}} \bigabs{\mathfrak{E}_{3, k, k^{\prime}}(\phi)} \leq 3 \bigg(\frac{T}{N_{1}}\bigg) \bigg(\frac{T}{N_{2}}\bigg) T^{- 3} \max_{k} \bignorm{\check{\nabla^{3} \psi} X e_{k}}_{2} \max_{k} \bignorm{\check{M} X e_{k}}_{2} \eta = o_{P}(1) \, , \\
	&\sup_{\phi} \max_{k, k^{\prime}} \bigabs{\mathfrak{E}_{4, k, k^{\prime}}(\phi)} \leq \\
	&\quad 3 \bigg(\frac{T}{N_{1}}\bigg) \bigg(\frac{T}{N_{2}}\bigg) T^{- 3} \bignorm{\check{Q}}_{2} \max_{k} \bignorm{\check{\nabla^{2} \psi} X e_{k}}_{2} \bignorm{\check{\dop{3} \psi}}_{4} \max_{k} \bignorm{\check{M} X e_{k}}_{4} \eta = o_{P}(1) \, .
\end{align*}
Hence, $\sup_{(\beta, \phi)} \norm{W(\beta, \phi) - W}_{2} = o_{P}(1)$. By xiv) and xv), the conditions of Lemma \ref{lemma:invertibility} with $\bar{p} = 2$ hold wpa1. Thus $W(\beta, \phi)$ is invertible for all $(\beta, \phi) \in \mathfrak{B}(\varepsilon) \times \mathfrak{P}(\eta, q)$ wpa1.\hfill\qedsymbol

\paragraph{Proof of Lemma \ref{lemma:regularity_conditions2_noninteracted}.} By a Taylor expansion of $\widehat{\mathfrak{D}_{\pi}^{\mathrm{r}}} e_{k} = \nabla^{\mathrm{r}} \psi(\hat{\beta}, \hat{\phi}) M(\hat{\beta}, \hat{\phi}) X e_{k}$, $\mathrm{r} \in \{1, 2\}$, around $(\beta^{0}, \phi^{0})$ for each $k$ (see \textcite{fwct2014}),
\begin{align*}
	&\widehat{\mathfrak{D}_{\pi}^{\mathrm{r}}} e_{k} - \mathfrak{D}_{\pi}^{\mathrm{r}} e_{k} = \big(\nabla^{\mathrm{r}} \psi M X e_{k} - \mathfrak{D}_{\pi}^{\mathrm{r}} e_{k}\big) + \check{\nabla}^{\mathrm{r} + 1} \psi \diag(\check{M} X e_{k}) (\hat{\pi} - \pi^{0}) \, - \\
	&\qquad \check{\nabla}^{\mathrm{r}} \psi \check{Q} \check{\nabla^{3} \psi} \diag(\check{M} X e_{k}) (\hat{\pi} - \pi^{0}) \eqqcolon \mathfrak{E}_{1, k} + \mathfrak{E}_{2, k} + \mathfrak{E}_{3, k} \, ,
\end{align*}
where $\check{\beta}$, $\check{\phi}$ can differ for each element and each $k$. By H\"older's inequality and Corollary \ref{corollary:consistency_noninteracted},
\begin{align*}
	&\max_{k} \bignorm{\mathfrak{E}_{1, k}}_{2} = \max_{k}\bignorm{\nabla^{\mathrm{r}} \psi M X e_{k} - \mathfrak{D}_{\pi}^{\mathrm{r}} e_{k}}_{2} = o_{P}\big(T^{\frac{3}{2}}\big) \, , \\
	&\max_{k} \bignorm{\mathfrak{E}_{2, k}}_{2} \leq \bignorm{\check{\dop{\mathrm{r} +1} \psi}}_{6} \, \max_{k} \bignorm{\check{M} X e_{k}}_{6} \, \bignorm{\hat{\pi} - \pi^{0}}_{6} = o_{P}\big(T^{\frac{3}{2}}\big) \, , \\
	&\max_{k} \bignorm{\mathfrak{E}_{3, k}}_{2} \leq \bignorm{\check{Q}}_{2} \, \bignorm{\check{\dop{3} \psi}}_{6} \, \bignorm{\check{\dop{\mathrm{r}} \psi}}_{20 + \nu} \, \max_{k} \bignorm{\check{M} X e_{k}}_{6} \, \bignorm{\hat{\pi} - \pi^{0}}_{6} = o_{P}\big(T^{\frac{3}{2}}\big) \, .
\end{align*}
By the triangle inequality, for $1 \leq p \leq 2$ and $1 \leq \mathrm{r} \leq 2$,
\begin{equation*}
	\max_{k} \bignorm{(\widehat{\mathfrak{D}_{\pi}^{\mathrm{r}}} - \mathfrak{D}_{\pi}^{\mathrm{r}}) e_{k}}_{p} \leq \bigg(\frac{N_{1} N_{2}}{T^{2}}\bigg)^{\frac{1}{p} - \frac{1}{2}} \, T^{\frac{3}{p} - \frac{3}{2}} \max_{k} \bignorm{(\widehat{\mathfrak{D}_{\pi}^{\mathrm{r}}} - \mathfrak{D}_{\pi}^{\mathrm{r}}) e_{k}}_{2} = o_{P}\big(T^{\frac{3}{p}}\big) \, .
\end{equation*}
\hfill\qedsymbol

\section{Preliminary Lemmas}
\label{supplement:preliminary_lemmas}

\begin{lemma}[Matrix Norm Inequalities]
	\label{lemma:matrix_norm_inequalties}
	Let $A \in \Real^{M \times N}$ denote a $(M \times N)$ matrix. Then, (i) $\norm{A}_{p} \leq \norm{A}_{1}^{1 / p} \norm{A}_{\infty}^{1 - 1 / p}$ for $1 \leq p$; (ii) $\norm{A}_{p} \leq \norm{A}_{2}^{2 / p} \norm{A}_{\infty}^{1 - 2 / p}$ for $2 \leq p$.
\end{lemma}

\paragraph{Proof of Lemma \ref{lemma:matrix_norm_inequalties}.} See Proof of Lemma S.1 in \textcite{fw2016}.\hfill\qedsymbol

\begin{lemma}[Covariance Inequality for Strong Mixing Processes]
	\label{lemma:covariance_inequality_mixing}
	Let $\{\vartheta_{t}\}$ be an $\alpha$-mixing process with mixing coefficients $\alpha(q)$. Assume that $\EX{\abs{\vartheta_{t}}^{p}} < \infty$ and $\EX{\abs{\vartheta_{t + q}}^{p^{\prime}}} < \infty$ for some $1 \leq p, p^{\prime}$ and $1 / p + 1 / p^{\prime} < 1$. Then, $\abs{\text{Cov}(\vartheta_{t}, \vartheta_{t + q})} \leq 8 \, \alpha(\abs{q})^{1 / r} (\EX{\abs{\vartheta_{t}}^{p}})^{1 / p} (\EX{\abs{\vartheta_{t + q}}^{p^{\prime}}})^{1 / p^{\prime}}$, where $r = (1 - 1 / p - 1 / p^{\prime})^{- 1}$. 
\end{lemma}

\paragraph{Proof of Lemma \ref{lemma:covariance_inequality_mixing}.} See Proposition 2.5 in \textcite{fy2008}.\hfill\qedsymbol

\begin{lemma}[Moment bounds for strong mixing processes]
	\label{lemma:moment_bounds_mixing}
	Let $\{\vartheta_{t}\}$ be a mean zero $\alpha$-mixing process with mixing coefficients $\alpha(q)$. Let $r$ be a positive integer and let $2 r < \delta$, $r / (1 - 2 r / \delta) < \varphi$, $C_{1} < \infty$, and $C_{2} < \infty$. Assume that $\sup_{t} \EX{\abs{\vartheta_{t}}^{\delta}} \leq C_{1}$ and that $\alpha(q) \leq C_{2} \, q^{- \varphi}$ for any positive integer $q$. Then, there exists a constant $B < \infty$ depending on $r$, $\delta$, $\varphi$, $C_{1}$, and $C_{2}$, but not depending on $T$ or any other distributional characteristics of $\{\vartheta_{t}\}$, such that for any positive integer $T$, $\EX{\big( \tfrac{1}{\sqrt{T}} \sum_{t = 1}^{T} \vartheta_{t} \big)^{2r}} \leq B$.
\end{lemma}

\paragraph{Proof of Lemma \ref{lemma:moment_bounds_mixing}.} See \textcite{ck1995}.\hfill\qedsymbol

\begin{lemma}[Matrix Khintchine Inequality]
	\label{lemma:matrix_khintchine_inequality}
	Let $\epsilon_{1}, \ldots, \epsilon_{N}$ be independent Rademacher random variables and let $A_{1}, \ldots, A_{N}$ be any (fixed) symmetric $T \times T$ matrices. Then, for $p \geq 1$, $\big(\EX{\bignorm{\sum_{i = 1}^{N} \epsilon_{i} \, A_{i}}_{2}^{p}}\big)^{1 / p} \leq C \, \sqrt{p + \log(T)} \, \bignorm{\sum_{i = 1}^{N} A_{i}}_{2}^{1 / 2}$, where $C$ is an absolute constant.
\end{lemma}

\paragraph{Proof of Lemma \ref{lemma:matrix_khintchine_inequality}.} See Theorem 5.4.14 in \textcite{v2025}.\hfill\qedsymbol

\begin{lemma}[Latala's Theorem]
	\label{lemma:latalas_theorem}
	Let $A$ be a random matrix whose entries $a_{ij}$ are independent centered random variables with $\sup_{ij} \EX{\abs{a_{ij}}^{4}} \leq C_{1}$. Then, $\EX{\norm{A}_{2}} \leq C \, \big(\sup_{i} \big(\sum_{j} \mathbb{E}[a_{ij}^{2}]\big)^{1 / 2} + \sup_{j} \big(\sum_{i} \mathbb{E}[a_{ij}^{2}]\big)^{1 / 2} + \big(\sum_{i} \sum_{j} \mathbb{E}[a_{ij}^{4}]\big)^{1 / 4} \big)$, where $C$ is an absolute constant.
\end{lemma}

\paragraph{Proof of Lemma \ref{lemma:latalas_theorem}.} See \textcite{l2005}.\hfill\qedsymbol

\begin{lemma}[Modified \textcite{r1999}'s Inequality]
	\label{lemma:rudelson_inequality}
	Let $x_{1}, \ldots, x_{N}$ be vectors in $\Real^{T}$ and $\epsilon_{1}, \ldots, \epsilon_{N}$ be independent Rademacher random variables. Then, for $p \geq 1$, \linebreak $\big(\EX{\bignorm{\sum_{i = 1}^{N} \epsilon_{i} \, x_{i} x_{i}^{\prime}}_{2}^{p}}\big)^{1 / p} \leq C \, \sqrt{p + \log(T)} \, \max_{i = 1, \ldots, N} \norm{x_{i}}_{2} \, \bignorm{\sum_{i = 1}^{N} x_{i} x_{i}^{\prime}}_{2}^{1 / 2}$, where $C$ is an absolute constant.
\end{lemma}

\paragraph{Proof of Lemma \ref{lemma:rudelson_inequality}.} Apply Lemma \ref{lemma:matrix_khintchine_inequality} with $A_{i} = x_{i} x_{i}^{\prime}$, then use \linebreak $\norm{\sum_{i} x_{i} x_{i}^{\prime}}_{2}^{1/2} = \norm{\sum_{i} \norm{x_{i}}_{2}^{2} x_{i} x_{i}^{\prime}}_{2}^{1/2} \leq \max_{i} \norm{x_{i}}_{2} \norm{\sum_{i} x_{i} x_{i}^{\prime}}_{2}^{1/2}$.\hfill\qedsymbol

\begin{lemma}[Symmetrization]
	\label{lemma:symmetrization}
	Let $x_{1}, \ldots, x_{N}$ be independent mean zero random vectors, and let $\epsilon_{1}, \ldots, \epsilon_{N}$ be independent Rademacher random variables. Then, for $p \geq 1$, $\big(\EX{\bignorm{\sum_{i = 1}^{N} x_{i}}_{2}^{p}}\big)^{1 / p} \leq 2 \, \big(\EX{\bignorm{\sum_{i = 1}^{N} \epsilon_{i} x_{i}}_{2}^{p}}\big)^{1 / p}$.
\end{lemma}

\paragraph{Proof of Lemma \ref{lemma:symmetrization}.} Let $\{y_{i}\}$ be an independent copy of $\{x_{i}\}$. Jensen and $\EX{y_{i} \mid x} = 0$ give $\norm{\sum_{i} x_{i}}_{2}^{p} \leq \EX{\norm{\sum_{i}(x_{i} - y_{i})}_{2}^{p} \mid x}$. Since $\{x_{i} - y_{i}\}$ has the same law as $\{\epsilon_{i}(x_{i} - y_{i})\}$, the triangle and Loeve's $c_{r}$ inequality yield the result.\hfill\qedsymbol

\begin{lemma}[Bound for $p$-th Moment of Spectral Norm]
	\label{lemma:bound_moments_spectral_norm}
	Let $A$ be $N \times T$ with independent rows $a_{i}$, $\Sigma \coloneqq N^{- 1} \sum_{i} \EX{a_{i} a_{i}^{\prime}}$, $m_{p} \coloneqq (\EX{\max_{i} \norm{a_{i}}_{2}^{p}})^{1/p}$. Then for $p \geq 2$, $\big(\EX{\norm{A}_{2}^{p}}\big)^{1 / p} \leq C \sqrt{N} \norm{\Sigma}_{2}^{1/2} + C \sqrt{p + \log(T)} m_{p}$.
\end{lemma}

\paragraph{Proof of Lemma \ref{lemma:bound_moments_spectral_norm}.} $\EX{\norm{A}_{2}^{p}} = N^{p/2} \EX{\norm{N^{- 1} A^{\prime} A}_{2}^{p/2}} \leq 2^{p/2 - 1} N^{p/2}(\norm{\Sigma}_{2}^{p/2} + E)$ with $E \coloneqq \EX{\norm{N^{- 1} A^{\prime} A - \Sigma}_{2}^{p/2}}$. By Lemma \ref{lemma:symmetrization}, $E \leq (2/N)^{p/2} \EX{\norm{\sum_{i} \epsilon_{i} a_{i} a_{i}^{\prime}}_{2}^{p/2}}$. Lemma \ref{lemma:rudelson_inequality} and Cauchy-Schwarz give $E \leq (2 C / \sqrt{N})^{p / 2}(\sqrt{p + \log T})^{p / 2} m_{p}^{p / 2} (E + \norm{\Sigma}_{2}^{p / 2})^{1 / 2}$. Setting $\delta_{p} \coloneqq (4 C \sqrt{p + \log T} m_{p}/\sqrt{N})^{p}$ and solving the quadratic inequality in $\sqrt{E}$ yields $E \leq 2 \sqrt{\delta_{p}} \norm{\Sigma}_{2}^{p/4} + \delta_{p}$. Substituting back gives the claim.\hfill\qedsymbol
\vspace{0.5em}

\begin{lemma}[Asymptotic Bound for $p$-th Moment of Spectral Norm]
	\label{lemma:asymptotic_bound_spectral_norm}
	Let $A$ be a $N \times T$ matrix with entries $a_{it}$, whose rows $a_{i} \coloneqq (a_{i1}, \ldots, a_{iT})$ are independent vectors. Let $p \geq 2$, $\delta > p$, $C_{1} < \infty$, $C_{2} < \infty$, $0 < C_{3} < \infty$, $\varphi > \delta / (\delta - 2)$, and $b > 0$. Assume that $\sup_{it} \big(\EX{\abs{a_{it}}^{\delta}}\big)^{1 / \delta} \leq C_{1}$ and that, for each $i \in \{1, \ldots, N\}$, $\{a_{it}\}$ is a mean zero $\alpha$-mixing process with mixing coefficients satisfying $\sup_{i} \alpha_{i}(\abs{q}) \leq C_{2} \, q^{- \varphi}$ for any integer $q$. In addition, assume that $N / T^{b} \rightarrow C_{3}$ as $N, T \rightarrow \infty$. Then, $\big(\EX{\norm{A}_{2}^{p}}\big)^{1/p} = \mathcal{O}(T^{b/2}) + \mathcal{O}(\sqrt{\log T} \, T^{1/2 + b/p})$ as $N, T \rightarrow \infty$.
\end{lemma}

\paragraph{Proof of Lemma \ref{lemma:asymptotic_bound_spectral_norm}.} Apply Lemma \ref{lemma:bound_moments_spectral_norm}. Lemma \ref{lemma:covariance_inequality_mixing} with $p = p^{\prime} = \delta$ gives $\norm{\Sigma}_{2} = \mathcal{O}(1)$. The union bound, Jensen's inequality, and the Lyapunov inequality give $m_{p} = \mathcal{O}(T^{1/2 + b/p})$.\hfill\qedsymbol
\vspace{0.5em}

\begin{lemma}[Invertibility]
	\label{lemma:invertibility}
	Let $2 \leq q \leq p \leq \bar{p}$ and let $A(\beta, \phi)$ be symmetric on $\mathfrak{B}(\varepsilon) \times \mathfrak{P}(\eta, q)$. If $\overline{A}$ is invertible with $\norm{\overline{A}^{- 1}}_{\bar{p}} = \mathcal{O}(1)$, $\norm{\widetilde{A}}_{\bar{p}} = o_{P}(1)$, and $\sup \norm{A(\beta, \phi) - A}_{\bar{p}} = o_{P}(1)$, then $A(\beta, \phi) > 0$ wpa1 and $\sup \norm{(A(\beta, \phi))^{- 1}}_{p} = \mathcal{O}_{P}(1)$ for $2 \leq p \leq \bar{p}$.
\end{lemma}

\paragraph{Proof of Lemma \ref{lemma:invertibility}.} $A(\beta, \phi) = (I + \Delta A(\beta, \phi) \overline{A}^{- 1}) \overline{A}$ with $\Delta A \coloneqq A(\beta, \phi) - A + \widetilde{A}$, and $\sup \norm{\Delta A}_{\bar{p}} \norm{\overline{A}^{- 1}}_{\bar{p}} = o_{P}(1) < 1$ wpa1. Corollary 5.6.16 in \textcite{hj2012} yields invertibility, and $\sup \norm{(A(\beta, \phi))^{- 1}}_{\bar{p}} \leq \norm{\overline{A}^{- 1}}_{\bar{p}} (1 - o_{P}(1))^{- 1} = \mathcal{O}_{P}(1)$. Symmetry and Lemma \ref{lemma:matrix_norm_inequalties} extend to $2 \leq p \leq \bar{p}$.\hfill\qedsymbol

\begin{lemma}[Neumann Series]
	\label{lemma:inverse_neumann_series}
	For symmetric $A = \overline{A} + \widetilde{A}$ with $\overline{A}$ invertible, $\norm{\overline{A}^{- 1}}_{p} = \mathcal{O}(1)$, and $\norm{\widetilde{A}}_{p} = o_{P}(1)$ for some $p \geq 2$, $A^{- 1} = \overline{A}^{- 1} \sum_{r \geq 0}(- \widetilde{A} \, \overline{A}^{- 1})^{r}$ wpa1.
\end{lemma}

\paragraph{Proof of Lemma \ref{lemma:inverse_neumann_series}.} $\norm{\widetilde{A} \, \overline{A}^{- 1}}_{p} = o_{P}(1) < 1$ wpa1, so the Neumann series for $(I + \widetilde{A} \, \overline{A}^{- 1})^{- 1}$ converges (Corollary 5.6.16 in \textcite{hj2012}).\hfill\qedsymbol

\begin{lemma}[Approximation of Inverse]
	\label{lemma:inverse_approximation}
	For symmetric invertible $A = B + E$ with $B$ invertible: (i) $A^{- 1} = B^{- 1} - A^{- 1} E B^{- 1}$; (ii) $A^{- 1} = B^{- 1} - B^{- 1} E B^{- 1} + B^{- 1} E A^{- 1} E B^{- 1}$.
\end{lemma}

\paragraph{Proof of Lemma \ref{lemma:inverse_approximation}.} (i) follows from $A^{- 1} A = I$ and $A = B + E$. Substituting (i) into itself (using symmetry $A^{- 1} = B^{- 1} - B^{- 1} E A^{- 1}$) gives (ii).\hfill\qedsymbol
\vspace{0.5em}

\begin{lemma}[Asymptotic Bound for Triple Sum of Joint Expectations of a Mixing Process]
	\label{lemma:asymptotic_bound_triple_sum}
	Let $\{\vartheta_{t}\}$ be an $\alpha$-mixing process with mixing coefficients $\alpha(q)$. Let $f, g, h \colon \Real \mapsto \Real$ be measurable functions and define $x_{t} \coloneqq f(\vartheta_{t}) - \EX{f(\vartheta_{t})}$, $y_{t} \coloneqq g(\vartheta_{t})$, and $z_{t} \coloneqq h(\vartheta_{t})$. Let $\delta > 3$, $C_{1} < \infty$, $C_{2} < \infty$, $C_{3} < \infty$, $C_{4} < \infty$, and $\varphi > 2 \, \delta / (\delta - 3)$. Assume that $\sup_{t} \EX{\abs{x_{t}}^{\delta}} \leq C_{1}$, $\sup_{t} \EX{\abs{y_{t}}^{\delta}} \leq C_{2}$, $\sup_{t} \EX{\abs{z_{t}}^{\delta}} \leq C_{3}$, and $\alpha(q) \leq C_{4} \, q^{- \varphi}$ for any positive integer $q$. Then, $\sum_{t = 1}^{T} \sum_{t^{\prime} = 1}^{T} \sum_{t^{\prime\prime} = 1}^{T} \bigabs{\EX{z_{t^{\prime\prime}} y_{t^{\prime}} x_{t}}} = \mathcal{O}(T)$ as $T \rightarrow \infty$.
\end{lemma}

\paragraph{Proof of Lemma \ref{lemma:asymptotic_bound_triple_sum}.} Measurable transformations of $\{\vartheta_{t}\}$ inherit its mixing property, and $\EX{x_{t}} = 0$ by construction. Let $\mathcal{S}$ denote the set of all triplets $(t, t^{\prime}, t^{\prime\prime})$ spanning the $T^{3}$ index space. For any triplet, define the order statistics $t_{(1)} \leq t_{(2)} \leq t_{(3)}$ and the span $m \coloneqq t_{(3)} - t_{(1)} \in \{0, \ldots, T - 1\}$. Define $\mathcal{P}_{m} \coloneqq \{(t, t^{\prime}, t^{\prime\prime}) \colon t_{(3)} - t_{(1)} = m\}$, which partitions $\mathcal{S}$ by span, so that $\mathcal{S} = \bigsqcup_{m = 0}^{T - 1} \mathcal{P}_{m}$. The triple sum can therefore be written as $S_{T} \coloneqq \sum_{m = 0}^{T - 1} \sum_{(t, t^{\prime}, t^{\prime\prime}) \in \mathcal{P}_{m}} \bigabs{\EX{z_{t^{\prime\prime}} y_{t^{\prime}} x_{t}}}$. For $m = 0$, $\lvert \mathcal{P}_{0} \rvert = T$. For $m \geq 1$, let $s \coloneqq \min(t, t^{\prime}, t^{\prime\prime})$. Since $s \in \{1, \ldots, T - m\}$, let $N(m)$ denote the number of triplets with all indices in $\{s, \ldots, s + m\}$, so $\lvert \mathcal{P}_{m} \rvert = (T - m) N(m)$. Applying the inclusion–exclusion principle gives $N(m) = (m + 1)^{3} - 2 m^{3} + (m - 1)^3 = 6 m$ for $1 \leq m \leq T - 1$. Hence, $\mathcal{P}_{0} = T$ and $\mathcal{P}_{m} = (T - m) \, 6 \, m$ for $m \geq 1$. Since $\EX{x_{t}} = 0$, $\EX{z_{t^{\prime\prime}} y_{t^{\prime}} x_{t}} = \text{Cov}(x_{t}, y_{t^{\prime}} z_{t^{\prime\prime}})$. Let $C_5 \coloneqq (C_1 C_2 C_3)^{1/\delta}$. By H\"older's inequality, Jensen's inequality, and the Lyapunov inequality, $\sup_{t} \abs{\EX{x_{t} y_{t} z_{t}}} \leq C_{5}$ and \linebreak $(\sup_{t} \EX{\abs{x_{t}}^{\delta}})^{1 / \delta} \big(\sup_{t^{\prime}t^{\prime\prime}} \EX{\abs{y_{t^{\prime}} z_{t^{\prime\prime}}}^{\delta / 2}}\big)^{2 \ \delta} \leq C_{5}$. Applying Lemma \ref{lemma:covariance_inequality_mixing} with $p = \delta$, $p^{\prime} = \delta / 2$, $\alpha(m) \leq C_{4} m^{-\varphi}$, and $T \geq T - m$, $S_{T} \leq T \, C_{5} \big(1 + 48 \, C_{4} \sum_{m = 1}^{\infty} m^{1 - \varphi ((\delta - 3) / \delta)} \big) = \mathcal{O}(T)$, where the series converges by $\varphi > 2\delta/(\delta - 3)$.\hfill\qedsymbol

\end{document}